\documentclass[aps,pra,twocolumn,notitlepage]{revtex4-1}
\usepackage{amsmath,amssymb}
\usepackage{natbib}
\usepackage{mathtools}
\usepackage{braket}
\usepackage{hyperref}
\usepackage{graphicx}
\usepackage{color}
\usepackage[caption=false]{subfig}
\usepackage{epstopdf}
\usepackage{empheq}

\DeclareMathOperator{\Tr}{Tr}

\newcommand*{\rom}[1]{\expandafter\@slowromancap\romannumeral #1@}
\newcommand{\HO}{\hat{\mathcal{H}}}

\definecolor{amethyst}{rgb}{0.6, 0.4, 0.8}
\definecolor{byzantine}{rgb}{0.74, 0.2, 0.64}
\begin{document}
\title{First-principles analysis of cross-resonance gate operation}
\author{Moein Malekakhlagh}\email{Electronic address: moein.malekakhlagh@ibm.com}
\author{Easwar Magesan}\email{Electronic address: emagesa@us.ibm.com}
\author{David C McKay}\email{Electronic address: dcmckay@us.ibm.com}
\affiliation{IBM Thomas J. Watson Research Center, 1101 Kitchawan Rd, Yorktown Heights, NY 10598}
\date{\today}
\begin{abstract}
We present a comprehensive theoretical study of the cross-resonance gate operation covering estimates for gate parameters and gate error as well as analyzing spectator qubits and multi-qubit frequency collisions. We start by revisiting the derivation of effective Hamiltonian models following Magesan et al. (\href{https://arxiv.org/abs/1804.04073}{arXiv:1804.04073}). Transmon qubits are commonly modeled as a weakly anharmonic Kerr oscillator. Kerr theory only accounts for qubit frequency renormalization, while adopting number states as the eigenstates of the bare qubit Hamiltonian. Starting from the Josephson nonlinearity and by accounting for the eigenstates renormalization, due to counter-rotating terms, we derive a new starting model for the cross-resonance gate with modified qubit-qubit interaction and drive matrix elements. Employing time-dependent Schrieffer-Wolff perturbation theory, we derive an effective Hamiltonian for the cross-resonance gate with estimates for the gate parameters calculated up to the fourth order in drive amplitude. The new model with renormalized eigenstates lead to 10--15 percent relative correction of the effective gate parameters compared to Kerr theory. We find that gate operation is strongly dependent on the ratio of qubit-qubit detuning and anharmonicity. In particular, we characterize five distinct regions of operation, and propose candidate parameter choices for achieving high gate speed and low coherent gate error when the cross-resonance tone is equipped with an echo pulse sequence. Furthermore, we generalize our method to include a third spectator qubit and characterize possible detrimental multi-qubit frequency collisions.
\end{abstract}
\maketitle

\section{Introduction}
\label{Sec:Introduction}

Fault-tolerant quantum computation schemes \cite{Shor_Fault_1996, Gottesman_Theory_1998, Kitaev_Fault_2003, Raussendorf_Fault_2007} require a universal set of high-fidelity quantum gates, where an arbitrary quantum operation is decomposed in terms of a set of single- and two-qubit gates \cite{Nielsen_Quantum_2002}. Architectures based on superconducting qubits \cite{Nakamura_Coherent_1999, Blais_Cavity_2004, Wallraff_Strong_2004, Majer_Coupling_2007} provide a promising platform for this purpose due to an optimal combination of quantum control, coherence and flexibility in design. In particular, the transmon design \cite{Koch_Charge_2007, Schreier_Suppressing_2008} is a common choice because it strongly suppresses charge noise at the expense of a weaker anharmonicity. Transmon qubits are controlled via dipole coupled microwave drives, which allows for arbitrary rotations in the Bloch sphere by varying the envelope and phase of the microwave field. State of the art implementations for single qubit gates can reach very low errors, close to $10^{-4}$ \cite{McKay_Efficient_2017}. 

The difficult task, however, is entangling qubits together in such a way as to realize a low-error two-qubit gate with a high on/off contrast. One method is to add flux tunability to the circuit, where gates can be enabled by dynamically tuning qubits into resonance conditions \cite{Dicarlo_Demonstration_2009, Barends_Coherent_2013} or by parametric modulation of the tunable elements \cite{Mckay_Universal_2016, Caldwell_Parametrically_2018}. However, tunability comes at a definite cost in terms of coherence and scalability. By contrast, an architecture based on fixed-frequency transmons with exchange coupling has a high degree of coherence, stability, and ease of control \cite{Chow_Universal_2012, Corcoles_Demonstration_2015, Takita_Demonstration_2016, Gambetta_Building_2017}. In such a design, only two-qubit gates enabled by microwave drives (all-microwave gates) are possible. Although there are fewer options for all-microwave gates, there is one type -- cross-resonance (CR)-- which has demonstrated tremendous promise in multi-qubit setups and can achieve errors below $10^{-2}$ \cite{Sheldon_Procedure_2016}.

The idea of the CR gate is rather simple; apply a microwave drive to one qubit (the control) at the frequency of another qubit (the target) \cite{Paraoanu_Microwave_2006, Rigetti_Fully_2010, Magesan_Effective_2018, Tripathi_Operation_2019}. Due to the static interaction between the qubits, a Rabi oscillation will occur on the target, where the axis of rotation will depend on the state of the control. The ideal CR effect generates a $ZX$ interaction term. The first experimental attempt of the CR gate \cite{Chow_Simple_2011} achieved a gate fidelity of 81\%. This was later improved to 90\% by the introducing an echo pulse sequence canceling unwanted single qubit terms during the gate \cite{Corcoles_Process_2013}. Currently, gate fidelities exceeding 99\% are possible using a combination of an echo sequence and a secondary active cancellation tone on the target qubit \cite{Sheldon_Procedure_2016}. To improve the CR gate even further, a better theoretical understanding of the gate is required. 		 

Theoretical analysis of CR began with modeling physical qubits as two-level systems \cite{Paraoanu_Microwave_2006, Rigetti_Fully_2010}. Such models provide a general understanding of the dominant $ZX$ interaction for the gate,  while failing to capture the entirety of two-qubit interactions such as the parasitic $ZZ$, and the large single qubit terms, such as $IX$, which are the same order of magnitude as $ZX$ in true multi-level transmons. Despite continuous experimental effort to improve the gate operation, there has been a gap on the theoretical part until very recently \cite{Magesan_Effective_2018, Tripathi_Operation_2019, Kirchhoff_Optimized_2018}. In Ref.~\cite{Magesan_Effective_2018}, it was argued that higher transmon states have non-negligible impact on the \textit{effective} dynamics in the computational basis. In particular, using Schrieffer-Wolff perturbation theory (SWPT), analytical estimates for gate parameters were derived that agrees well with experiment \cite{Sheldon_Procedure_2016}.
 
SWPT is a powerful analytical method for studying effective low-energy physics of an underlying more complex physical interaction \cite{Schrieffer_Relation_1966, Bravyi_Schrieffer_2011}. This method provides great flexibility in determining the form of the desired effective model depending on the nature of the problem. A very common application is to find the eigenenergies and eigenstates of a Hamiltonian by obtaining the transformation to the diagonal frame. A prime example in circuit-QED is the effective dispersive Jaynes-Cummings model used to describe the dispersive readout scheme \cite{Boissonneault_Dispersive_2009}. In the context of the CR gate, the dominant (resonant) interactions occur between the lowest two states of the target qubit when the control qubit is in a well-defined quantum state. Therefore, we would like to devise a Schrieffer-Wolff transformation, from the lab frame to a new frame, in which the effective Hamiltonian becomes \textit{block-diagonal} with respect to the Hilbert space of the target qubit \cite{Magesan_Effective_2018}. Since the CR gate time (100 ns) is much shorter than the typical coherence times (100 $\mu$s), we focus entirely on the Hermetian dynamics of the gate. Nevertheless, if needed, dissipation can also be studied within a SWPT framework. This was shown in a recent study, by one of the authors, where SWPT was generalized to open quantum systems by deriving effective master equations with renormalized eigenenergies as well as dissipators \cite{Malekakhlagh_Lifetime_2020, Petrescu_Lifetime_2020}. 

In this paper, we follow a similar SWPT as in Ref.~\cite{Magesan_Effective_2018} to derive an effective Hamiltonian for the CR gate. We propose a new starting Hamiltonian based on what we call \textit{energy-basis representation} of a transmon qubit. In contrast to Kerr theory, the new model captures the renormalization of the interaction rates between the two-qubit states due to the counter-rotating contributions in the Josephson potential. Using the new model, we obtain estimates for CR gate parameters that deviate up to 15\% from the ones predicted by Kerr theory, but converge to the old estimates as a limiting case. Using the perturbative results, we provide an analytical understanding of the CR gate parameters with an echo pulse \cite{Corcoles_Process_2013, Sheldon_Procedure_2016} devised to mitigate the unwanted two-qubit interactions. Furthermore, we calculate the gate fidelity and provide optimal operating parameters to achieve coherent gate error between $10^{-4}$ and $10^{-3}$. In order to consider a more realistic scenario of CR gate operation, we generalize our model to include a third spectator qubit which could be coupled to either the control or the target qubits. The goal is to quantify the impact of spectator qubits on the intended CR gate operation and summarize various multi-qubit frequency collisions that may occur between the control qubit and the drive, between the control and target qubits, or between the control, target and spectator qubits.

The rest of the main text is organized as follows: In Sec.~\ref{Sec:Model}, we demonstrate a new starting Hamiltonian for the CR gate based on energy-basis representation of transmon qubits. Section~\ref{Sec:CREffHam} discusses the derivation of an effective Hamiltonian for the CR gate via a time-dependent SWPT method. In Sec.~\ref{Sec:CREcho}, using the perturbative results, we provide an analytical understanding of the CR gate parameters with an echo pulse \cite{Sheldon_Procedure_2016} devised to mitigate the unwanted two-qubit interactions. In Sec.~\ref{Sec:SpecQu}, we generalize our model to include a third spectator qubit. In Sec.~\ref{Sec:SumFreqCol}, we summarize various multi-qubit frequency collisions that arise from perturbative analysis of CR. Sections~\ref{Sec:Conclusion} and~\ref{Sec:Acknowledgements} are devoted to conclusion and acknowledgements, respectively.

There are seven appendices, which will be referred to in the main text when necessary. Appendix~\ref{App:TransSpectrum} provides the derivation of energy-basis representation of a transmon qubit. In Appendix~\ref{App:DressedBasis}, we discuss the transformation to the basis that is dressed by the exchange interaction between the qubits. Appendix~\ref{App:SWPT} discusses the derivation of a time-dependent SWPT used to obtain an effective Hamiltonian for the CR gate. Appendix~\ref{App:CrossTalk} revisits the CR gate parameters in the presence of classical cross talk in the circuit. In Appendix~\ref{App:NonLocInv}, we provide an approximate estimate for the non-local properties of the CR echo operation in terms of the Makhlin invariants \cite{Makhlin_Nonlocal_2002}. In Appendix~\ref{App:Saturation}, we apply the semi-analytical method of Ref.~\cite{Tripathi_Operation_2019} to our revised starting model for the CR gate and study saturation behavior of gate parameters in strong drive regime. In Appendix~\ref{App:ExpSign}, we look at prospects for observing the difference between the Kerr and energy basis in experimentally measured quantities. 

\begin{figure}[t!]
\centering
\includegraphics[scale=0.425]{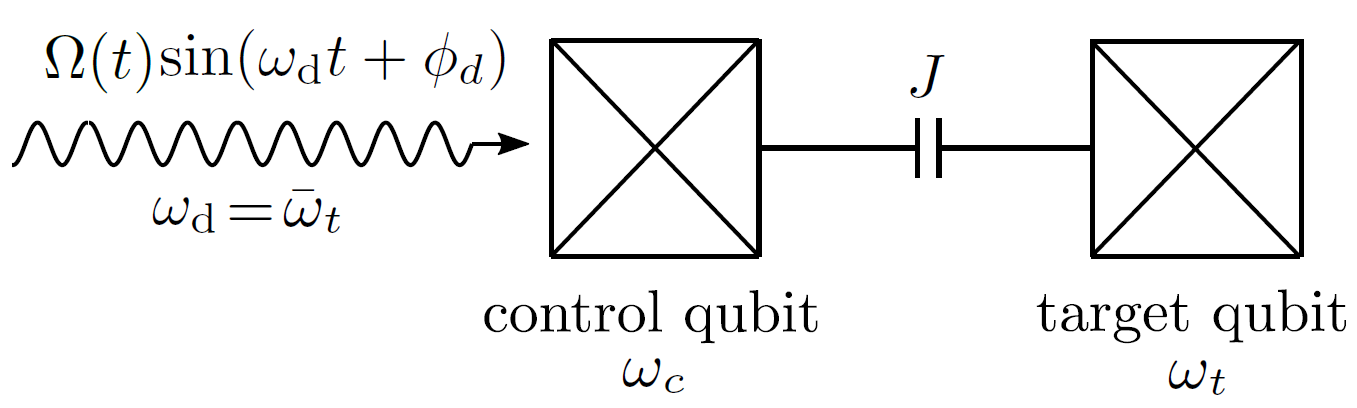}
\caption{Schematic circuit of the CR gate. The control qubit is driven via a charge line by a continuous-wave tone of amplitude $\Omega(t)$ and frequency $\omega_d = \bar{\omega}_t$ that is set to the lowest two-level transition frequency of the \textit{dressed} target qubit. In our analysis, we neglect the pulse shape of the drive amplitude for simplicity and set $\Omega(t) \rightarrow \Omega$. The phase $\phi_d$ determines the type of effective interaction in the equator of the target qubit. For the most part, we assume $\phi_d=\pi$ which results in an effective $ZX$ interaction. Deviation from this assumption can happen due to classical cross-talk, resulting in additional $ZY$ interaction, that is studied in Appendix~\ref{App:CrossTalk}. Moreover, for simplicity, we consider a direct exchange interaction between the qubits, although the interaction is commonly mediated via a common bus resonator \cite{Majer_Coupling_2007}.}
\label{fig:SchematicCR}
\end{figure}

\section{Model}
\label{Sec:Model}

Modeling the transmon circuit as a weakly anharmonic oscillator, as opposed to a two-level system, provides a broader possibility of interactions leading to a renormalization of the parameters of the effective CR Hamiltonian in the computational basis. For instance, besides the desired $ZX$ interaction, there also appears unwanted $ZZ$ interaction that would not exist if the qubits are modeled as two-level systems \cite{Magesan_Effective_2018}. Here, we introduce a new model that is slightly modified with respect to the Kerr theory used in Refs.~\cite{Magesan_Effective_2018, Tripathi_Operation_2019}. In particular, we quantify the error in approximating the qubit as a Kerr oscillator. Note that Kerr theory only accounts for the co-rotating contributions in the lowest order expansion of the Josephson nonlinearity (i.e. $\hat{\varphi}^4$), and hence provides \text{relative} correction of $O(\sqrt{2E_C/E_J})$ to eigenenergies, where $E_C$ and $E_J$ are the charging and the Josephson tunneling energy of the junction. Under this approximation, the number states remain eigenstates of the bare qubit Hamiltonian. In reality, however, transmon eigenstates are also renormalized due to the counter-rotating terms in the Josephson potential and accounting for this correction results in modified qubit-qubit and qubit-drive interaction, which in turn leads to renormalized gate parameters. In Sec.~\ref{SubSec:EnBasis}, we quantify the correction to Kerr theory that arises from eigenstate renormalization. We then employ the result of Sec.~\ref{SubSec:EnBasis} to come up with a new starting Hamiltonian for the CR gate in Sec.~\ref{SubSec:CRHamInEnBasis}.

\subsection{Modified interaction in energy basis}
\label{SubSec:EnBasis}
In this section, instead of making a Kerr approximation, we start from the cosine potential describing the Josephson junction and solve for the renormalized eigenstates. Note that the spectrum of a transmon qubit can be fully described with two alternative parameters: a harmonic frequency $\omega_{\text{h}} \equiv \sqrt{8E_CE_J}$ and a unitless anharmonicity measure $\epsilon\equiv\sqrt{2E_C/E_J}$. This alternative form can be derived from the common transmon Hamiltonian $\HO=4E_C\hat{N}^2-E_J\cos(\hat{\varphi})$
by replacing the phase and number operators in terms of their zero-point fluctuation amplitudes as
\begin{subequations}
\begin{align}
&\hat{\varphi}= \varphi_{\text{zpf}}\hat{x}=\left(\frac{2E_C}{E_J}\right)^{1/4}\left(\hat{b}+\hat{b}^{\dag}\right) \;,
\label{eqn:EnBasis-Phi=Phizpf*X}\\
&\hat{N}= N_{\text{zpf}}\hat{y}=\frac{1}{2}\left(\frac{E_J}{2E_C}\right)^{1/4}\left[-i\left(\hat{b}-\hat{b}^{\dag}\right)\right] \;,
\label{eqn:EnBasis-n=nzpf*Y}
\end{align}
\end{subequations}
resulting in a unitless form for the transmon Hamiltonian as \cite{Malekakhlagh_Cutoff-Free_2017, Didier_Analytical_2018, Malekakhlagh_Lifetime_2020, Petrescu_Lifetime_2020}
\begin{align}
\HO_{\text{q}}=\frac{\omega_{\text{h}}}{4}\left[\hat{y}^2-\frac{2}{\epsilon}\cos(\sqrt{\epsilon}\hat{x})\right] \;.
\label{eqn:EnBasis-Hq ITO W eps}
\end{align}
In Eq.~(\ref{eqn:EnBasis-Hq ITO W eps}), we have denoted $\hat{x}\equiv \hat{b}+\hat{b}^{\dag}$ and $\hat{y}\equiv -i(\hat{b}-\hat{b}^{\dag})$ as the unitless phase (flux) and number (charge) quadratures for the qubit.

We note that the spectrum of Hamiltonian~(\ref{eqn:EnBasis-Hq ITO W eps}) is  in principle exactly solvable in terms of Mathieu functions \cite{Koch_Charge_2007}. In practice, however, the unitless anharmonicity measure is small in the transmon regime ($E_J\gg E_c$) so that we can develop a perturbative correction to the eigenenergies and eigenstates of the transmon (See Appendix~\ref{App:TransSpectrum} and Ref.~\cite{Didier_Analytical_2018}). The first four eigenenergies up to $O(\epsilon^3)$ are found as
\begin{subequations}
\begin{align}
&\frac{E_1-E_0}{\omega_h} =1-\frac{1}{4}\epsilon-\frac{1}{16}\epsilon^2+O(\epsilon^3) \;,
\label{eqn:EnBasis-PTSol of omega_10}\\
&\frac{E_2-E_0}{\omega_h} =2-\frac{3}{4}\epsilon-\frac{17}{64}\epsilon^2+O(\epsilon^3) \;,
\label{eqn:EnBasis-PTSol of omega_20}\\
&\frac{E_3-E_0}{\omega_h} =3-\frac{3}{2}\epsilon-\frac{45}{64}\epsilon^2+O(\epsilon^3) \;, 
\label{eqn:EnBasis-PTSol of omega_30}
\end{align}
\end{subequations}
with $E_n$ denoting the eigenenergy of the $n^{\text{th}}$ eigenstate and the ground state is indexed as $0$. The corresponding lowest four eigenstate are also found up to $O(\epsilon^3)$ in Appendix~\ref{App:TransSpectrum} [See Eqs.~(\ref{Eq:TransSpec-PTSol of psi_0}--\ref{Eq:TransSpec-PTSol of psi_3})]. The qubit frequency in this notation equals the energy difference between the first two eigenstates as $\omega\equiv E_{1}-E_{0}$. Moreover, from Eqs.~(\ref{eqn:EnBasis-PTSol of omega_10}) and~(\ref{eqn:EnBasis-PTSol of omega_20}), we can find the qubit anharmonicity as
\begin{subequations}
\begin{align}
\begin{split}
\frac{\alpha}{\omega_h} \equiv \frac{(E_{2}-E_{1})-(E_{1}-E_{0})}{\omega_h}=-\frac{1}{4}\epsilon-\frac{9}{64}\epsilon^2+O(\epsilon^3) \;.
\end{split}
\label{eqn:EnBasis-PTSol of alpha}
\end{align}
Note that although we here express qubit quantities in powers of $\epsilon$, it is not feasible to measure $\epsilon$ directly. However, we can infer $\epsilon$ from qubit frequency $\omega$ and anharmonicity $\alpha$. Dividing Eq.~(\ref{eqn:EnBasis-PTSol of alpha}) by Eq.~(\ref{eqn:EnBasis-PTSol of omega_10}) we find the following approximate equation for $\epsilon$ as
\begin{align}
\left[9-4\left(\frac{\alpha}{\omega}\right)\right]\epsilon^2+16\left[1-\left(\frac{\alpha}{\omega}\right)\right]\epsilon+64\left(\frac{\alpha}{\omega}\right)=0, \ 	\epsilon>0 \;.
\label{eqn:EnBasis-eps ITO alpha}
\end{align}
\end{subequations}
For typical IBM transmons with $\omega\approx 5 \ \text{GHz}$ and $\alpha\approx -330 \ \text{MHz}$ \cite{Sheldon_Procedure_2016}, one finds $\epsilon\approx 0.2$ .

Moreover, based on Eqs.~(\ref{eqn:EnBasis-PTSol of omega_10}--\ref{eqn:EnBasis-PTSol of alpha}), we can express the transmon qubit Hamiltonian under a four-level approximation in the energy basis as
\begin{align}
\begin{split}
\HO_{\text{q}} &= \omega\ket{\psi_1}\bra{\psi_1}+ (2\omega+\alpha) \ket{\psi_2}\bra{\psi_2}\\
&+ (3\omega+3\alpha+\beta) \ket{\psi_3}\bra{\psi_3},
\label{eqn:EnBasis-EnBasis rep of H}
\end{split}
\end{align}
where $\ket{\psi_n}$ denotes the $n^{\text{th}}$ energy eigenstate. Furthermore, $\beta\equiv -(6/64)\epsilon^2 \omega_h$ provides the deviation of the energy of the third excited state from Kerr theory. For IBM transmons, $\beta$ can be as large as 20 MHz. We neglect such a correction and use the Kerr eigenenergies in our perturbative result for clarity. This would become more relevant when the drive frequency is directly resonant with the $\ket{\psi_2}\leftrightarrow \ket{\psi_3}$ transition.

A more important consequence of eigenstate renormalization is that the resulting interactions between the qubits are also modified. To quantify this correction, we need to project the interactions into the transmon energy eigenstates. Assuming that the interaction Hamiltonian is linear in quadratures of each qubit, which is the case for a capacitive or inductive interaction, it is sufficient to find the matrix representation of the unitless flux and charge operators in the new basis as $\mu_{mn}\equiv \bra{\psi_m}\hat{x}\ket{\psi_n}$ and $\nu_{mn}\equiv \bra{\psi_m}\hat{y}\ket{\psi_n}$. For simplicity, we separate the lowering (-) and raising (+) parts of the quadratures as $\hat{x}=\hat{x}^{-}+\hat{x}^{+}$ and $\hat{y}=-i\left(\hat{y}^{-}-\hat{y}^{+}\right)$, where $\hat{x}^{+}=(\hat{x}^{-})^{\dag}$ and $\hat{y}^{+}=(\hat{y}^{-})^{\dag}$. Using the perturbative solutions for the first four eigenstates, we find the following matrix representations for $\hat{x}^-$ up to the fourth level of transmon as (See Appendix~\ref{App:TransSpectrum}) 
\begin{subequations}
\begin{align}
\hat{x}^- \approx
\begin{bmatrix}
0 & \mu_{01} & 0 & \mu_{03}\\
0 & 0 & \mu_{12} & 0\\
0 & 0 & 0 & \mu_{23}\\
0 & 0 & 0 & 0\\
\end{bmatrix} \;,
\label{eqn:EnBasis-MatRep of X^-}
\end{align}
where $\mu_{mn}$ are found up to $O(\epsilon^3)$ as
\begin{align}
&\mu_{01}=1+\frac{1}{8}\epsilon+\frac{13}{256}\epsilon^2+O(\epsilon^3) \;,
\label{eqn:EnBasis-PTSol of mu_01}\\
&\mu_{12}=\left(1+\frac{1}{4}\epsilon+\frac{95}{512}\epsilon^2\right)\sqrt{2}+O(\epsilon^3) \;,
\label{eqn:EnBasis-PTSol of mu_12}\\
&\mu_{23}=\left(1+\frac{3}{8}\epsilon+\frac{105}{256}\epsilon^2\right)\sqrt{3}+O(\epsilon^3) \;,
\label{eqn:EnBasis-PTSol of mu_23}\\
&\mu_{03}=-\frac{\sqrt{6}}{48}\epsilon-\frac{3\sqrt{6}}{128}\epsilon^2+O(\epsilon^3) \;.
\label{eqn:EnBasis-PTSol of mu_03}
\end{align}
\end{subequations}
Similarly, for the lowering part of charge operator $\hat{y}^{-}$ we find the following matrix representation in the energy basis:
\begin{subequations}
\begin{align}
\hat{y}^- \approx
\begin{bmatrix}
0 & \nu_{01} & 0 & \nu_{03}\\
0 & 0 & \nu_{12} & 0\\
0 & 0 & 0 & \nu_{23}\\
0 & 0 & 0 & 0\\
\end{bmatrix} \;,
\label{eqn:EnBasis-MatRep of Y^-}
\end{align}
where $\nu_{mn}$ read
\begin{align}
&\nu_{01}=1-\frac{1}{8}\epsilon-\frac{11}{256}\epsilon^2+O(\epsilon^3) \;,
\label{eqn:EnBasis-PTSol of nu_01}\\
&\nu_{12}=\left(1-\frac{1}{4}\epsilon-\frac{73}{512}\epsilon^2\right)\sqrt{2}+O(\epsilon^3) \;,
\label{eqn:EnBasis-PTSol of nu_12}\\
&\nu_{23}=\left(1-\frac{3}{8}\epsilon-\frac{79}{256}\epsilon^2\right)\sqrt{3}+O(\epsilon^3) \;,
\label{eqn:EnBasis-PTSol of nu_23}\\
&\nu_{03}=-\frac{\sqrt{6}}{16}\epsilon-\frac{5\sqrt{6}}{128}\epsilon^2+O(\epsilon^3) \;.
\label{eqn:EnBasis-PTSol of nu_03}
\end{align}
\end{subequations}
Setting $\epsilon=0$ in Eqs.~(\ref{eqn:EnBasis-PTSol of mu_01}--\ref{eqn:EnBasis-PTSol of mu_03}) and~(\ref{eqn:EnBasis-PTSol of nu_01}--\ref{eqn:EnBasis-PTSol of nu_03}), we recover the harmonic/Kerr limit for the lowering operators, i.e. $\lim\limits_{\epsilon\rightarrow 0} \hat{x}^{-}=\lim\limits_{\epsilon\rightarrow 0} \hat{y}^{-}=\hat{b}$.	

Based on Eqs.~(\ref{eqn:EnBasis-PTSol of mu_01}--\ref{eqn:EnBasis-PTSol of mu_03}) and~(\ref{eqn:EnBasis-PTSol of nu_01}--\ref{eqn:EnBasis-PTSol of nu_03}), we find that the matrix elements of the flux (charge) operator are enhanced (suppressed) with respect to Kerr theory. Moreover, there is also a direct interaction between the ground and the third excited state proportional to $\nu_{03}$ $(\mu_{03})$ depending on the nature of interaction. Processes involving such a transition, if kept in our perturbation, contribute very little (1-10 Hz) to the gate parameters and hence are dropped to achieve simpler expressions. Since the qubit-qubit interaction and the drive are commonly implemented capacitively, we expect the resulting interactions to be suppressed, i.e. Kerr theory overestimates the interaction rates. In Sec.~\ref{SubSec:CRHamInEnBasis}, we express the CR gate Hamiltonian in the energy eigenstate of the qubits.

\subsection{Cross-resonance Hamiltonian in energy basis} 
\label{SubSec:CRHamInEnBasis}

Following representation~(\ref{eqn:EnBasis-Hq ITO W eps}), our starting Hamiltonian for the CR gate can be written as
\begin{subequations}
\begin{align}
&\HO_{0}=\sum\limits_{j=c,t}\frac{\omega_{\text{jh}}}{4}\left[\hat{y}_j^2-\frac{2}{\epsilon_j}\cos(\sqrt{\epsilon_j}\hat{x}_j)\right] \;,
\label{eqn:CRHamInEnBasis-H0}\\
&\HO_{\text{int}}(t)=J\hat{y}_c\hat{y}_t+\Omega\hat{y}_c\sin(\omega_{\text{d}} t+\phi_d) \;,
\label{eqn:CRHamInEnBasis-H_int}
\end{align}
\end{subequations}
where $\hat{x}_{c,t}\equiv(\hat{b}_{c,t}+\hat{b}_{c,t}^{\dag})$ and $\hat{y}_{c,t}\equiv -i(\hat{b}_{c,t}-\hat{b}_{c,t}^{\dag})$ denote the unitless flux and charge quadratures for the control and the target qubits, respectively. Moreover, we have considered a capacitive interaction Hamiltonian of the form $J\hat{y}_c\hat{y}_t$ as well as a drive tone that couples capacitively to the control qubit with strength $\Omega$. Cross resonance is achieved when the drive frequency is set to be equal to the lowest transition of the dressed target qubit denoted as $\omega_d=\bar{\omega}_t$, where a bar notation is employed to distinguish between bare and dressed quantities (See also Appendix~\ref{App:DressedBasis}). In our analysis, we keep the drive phase as $\phi_d=\pi$ in order to implement an effective $ZX$ interaction in the computational basis. At $\phi_d=\pi/2$, the situation is reverse and an effective $ZY$ is found, and anywhere in between both types of interactions exist.  

\begin{figure}[t!]
\centering
\includegraphics[scale=0.49]{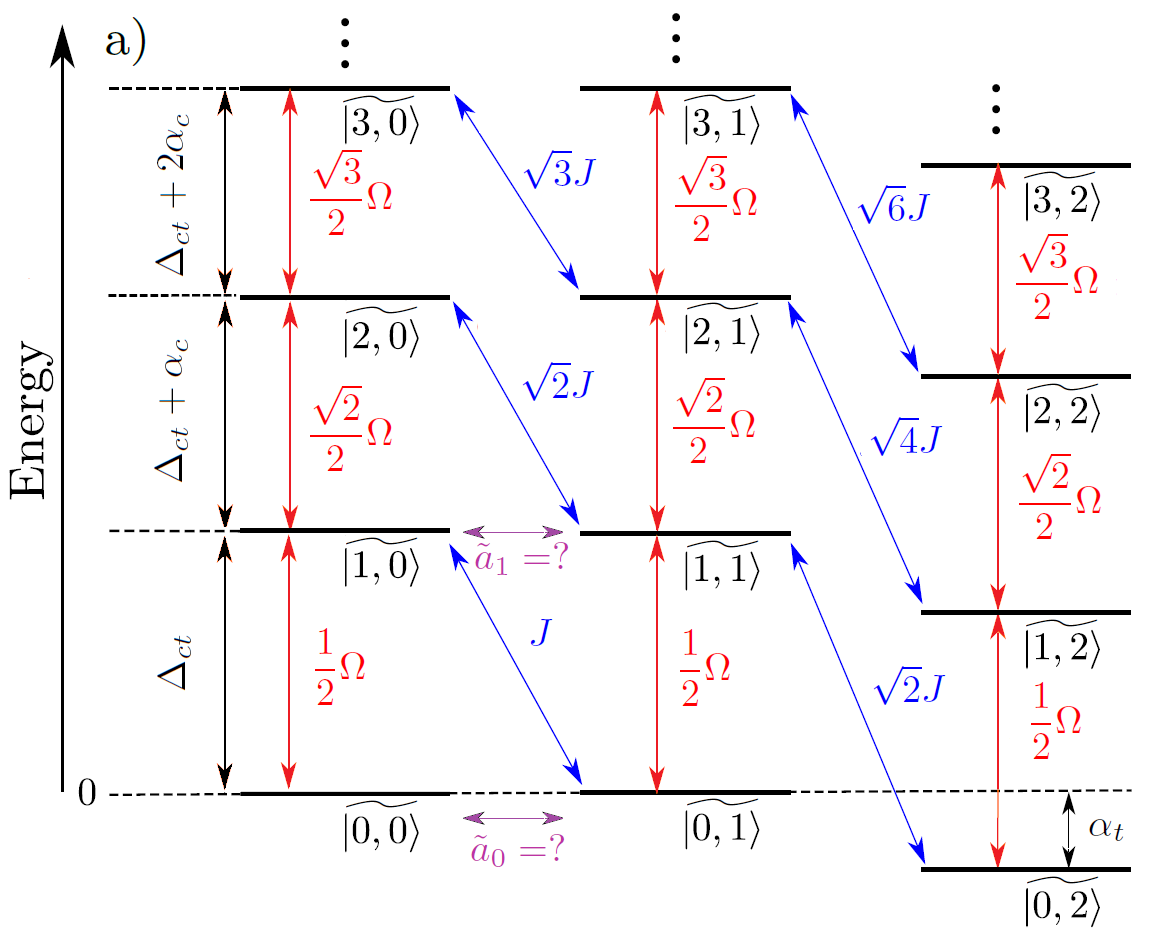}\\
\includegraphics[scale=0.49]{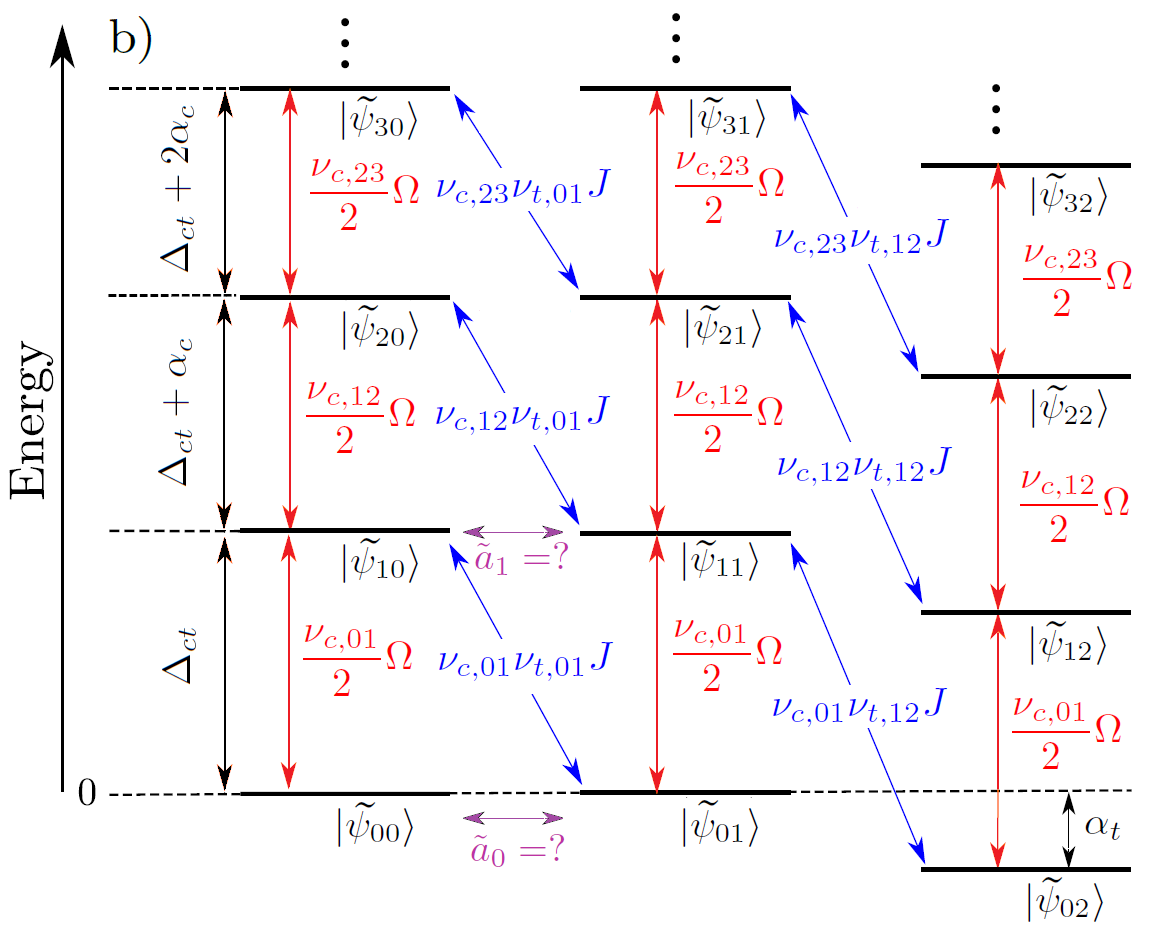}
\caption{Two-qubit energy ladder for the CR gate: a) Kerr representation, and b) energy-basis representation that accounts for eigenstate renormalization according to Eqs.~(\ref{eqn:CRHamInEnBasis-H0 2}--\ref{eqn:CRHamInEnBasis-Hd in new basis}). For clarity, the ladder is shown in the rotating frame of the drive (denoted by tilde), and we have used the shorthand notation $\ket{\tilde{\psi}_{mn}}\equiv \ket{\tilde{\psi}_{c,m}}\ket{\tilde{\psi}_{t,n}}$ to denote the two-qubit states. Compared to the Kerr ladder, the interaction rates are modified with the matrix elements of the charge operator in the energy basis, i.e. $\nu_{i,mn}$. Importantly, replacing $\ket{\tilde{\psi}_{mn}}\rightarrow \widetilde{\ket{m,n}}$, $\nu_{i,01}\rightarrow 1$, $\nu_{i,12}\rightarrow \sqrt{2}$ and $\nu_{i,23}\rightarrow \sqrt{3}$ for $i=c,t$ yields back the Kerr model. Note that we have neglected the direct interaction between the ground and third excited state of strength proportional to $\nu_{i,03}$, since the resulting contributions to gate parameters are small and keeping them would unnecessarily complicate the perturbative expressions at higher order. Furthermore, to compare our new model to Kerr theory, we denote the energy of the third excited level as $3\Delta_{ct}+3\alpha_c$, despite having found a small deviation from Kerr level structure in Eq.~(\ref{eqn:EnBasis-EnBasis rep of H}). Such a deviation become more relevant only when the drive frequency is directly resonant with the $\ket{\psi_{c,2}}\leftrightarrow \ket{\psi_{c,3}}$ transition.} 	
\label{fig:CRHamInEnBasis-JCLadderForCR}
\end{figure}

In writing Hamiltonian~(\ref{eqn:CRHamInEnBasis-H0}--\ref{eqn:CRHamInEnBasis-H_int}), we have considered the bare qubit Hamiltonian as the zeroth-order and kept the qubit-qubit exchange interaction and the drive as the interaction part (See Sec.~\ref{Sec:CREffHam} for details). Following Sec.~\ref{SubSec:EnBasis}, we can solve for the spectrum of each qubit independently. Therefore, the bare qubit Hamiltonian $\HO_0$ can be represented as
\begin{align}
\HO_{0}=\sum\limits_{m,n=0}^{\infty}(E_{c,m}+E_{t,n})\ket{\psi_{c,m}}\ket{\psi_{t,n}} \;,
\label{eqn:CRHamInEnBasis-H0 2}
\end{align}
where the first index denotes the qubit and the second denotes the corresponding state. Depending on the context and for clarity, we may also use a shorthand notation for the two-qubit state as $\ket{\psi_{mn}}\equiv\ket{\psi_{c,m}}\ket{\psi_{t,n}}$ (See e.g. Fig.~\ref{fig:CRHamInEnBasis-JCLadderForCR} and Tables~\ref{tab:NeOrAnal-Resonances} and~\ref{tab:SpecQu-Resonances}). The main distinction with respect to Kerr theory appears in the renormalization of the interaction Hamiltonian, where the transition rates between the states are modified in terms of the matrix elements $\nu_{mn}$ of Eqs.~(\ref{eqn:EnBasis-PTSol of nu_01}--\ref{eqn:EnBasis-PTSol of nu_03}). For instance, the exchange interaction between the qubits can be expressed as
\begin{align}
\begin{split}
\hat{\mathcal{H}}_J &=-J(\hat{y}_c^{-}-\hat{y}_c^{+})(\hat{y}_t^{-}-\hat{y}_t^{+})\approx J(\hat{y}_c^{-}\hat{y}_t^{+}+\hat{y}_c^{+}\hat{y}_t^{-})\\
&\approx \sum\limits_{m,n=0 \atop m<n}^{3}\sum\limits_{l,r=0 \atop l<r}^{3}\nu_{c,mn}\nu_{t,lr}J (\hat{P}_{c,mn}\hat{P}_{t,rl}+\hat{P}_{c,nm}\hat{P}_{t,lr}) \;,
\end{split}
\label{eqn:CRHamInEnBasis-HJ in new basis}
\end{align}
where we have defined the projection operators $\hat{P}_{i,mn}\equiv \ket{\psi_{i,m}}\bra{\psi_{i,n}}$ into the subspace $mn$ of qubit $i=c, t$. Similarly, the drive Hamiltonian in the energy basis reads
\begin{align}
\begin{split}
\hat{\mathcal{H}}_d &=\frac{\Omega}{2}\left(\hat{y}_c^{-}-\hat{y}_c^{+}\right)\left(e^{i\omega_d t}-e^{-i\omega_d t}\right) \\
& \approx \frac{\Omega}{2}\left(\hat{y}_c^{-}e^{i\omega_d t}+\hat{y}_c^{+}e^{-i\omega_d t}\right) \\
&\approx\sum\limits_{m,n=0 \atop m<n}^{3}\frac{1}{2}\nu_{mn}\Omega \left(\hat{P}_{c,mn}e^{i\omega_d t}+\hat{P}_{c,nm}e^{-i\omega_d t}\right) \;.
\end{split}
\label{eqn:CRHamInEnBasis-Hd in new basis}
\end{align}
Note that, for simplicity, we have only kept the co-rotating contributions in both the exchange and the drive interactions. We have checked, using our time-dependent SWPT, that the impact of these counter-rotating terms on gate parameters is negligible. Figure~\ref{fig:CRHamInEnBasis-JCLadderForCR} provides a comparison between the new and the initial Kerr theory under approximating transmon as a four-level system (See also Fig.~\ref{fig:NeOrAnal-HilSpCutOff} for a comparison of different Hilbert space cut-off numbers).	 

\section{Effective CR Gate Hamiltonian}
\label{Sec:CREffHam}	

In this section, we discuss the derivation of an effective Hamiltonian for the CR gate based on our modified model given by Eqs.~(\ref{eqn:CRHamInEnBasis-H0 2}--\ref{eqn:CRHamInEnBasis-Hd in new basis}). First, we discuss the method of time-dependent SWPT that is used to obtain the effective Hamiltonian (See also Appendix~\ref{App:SWPT}). Next, in Secs.~\ref{SubSec:LoOrAnal} and~\ref{SubSec:NeOrAnal}, we provide the lowest and the next order \textit{non-zero} corrections to the CR gate parameters, respectively. 

Note that, in principle, we first need to obtain the renormalization of qubit states due to the exchange interaction $J$, and study the gate parameters when the drive frequency is set to the \textit{dressed} frequency of the target qubit. In other words, the perturbation needs to be applied in two steps: 1) in exchange coupling $J$, and 2) in the drive amplitude $\Omega$ as in Refs.~\cite{Magesan_Effective_2018, Tripathi_Operation_2019} (See also Appendix~\ref{App:DressedBasis} for the dressed eignenstates and eigenenergies). On the other hand, here, we implement a SWPT that provides corrections jointly in $J$ and $\Omega$ for two practical reasons. Firstly, the exchange coupling for the CR gate is typically of the order of a few MHz and is at least one order of magnitude smaller than the drive amplitude $\Omega$. Secondly, performing perturbation in two stages makes the bookkeeping of corrections more difficult. This will become more challenging especially for larger network of qubits, and in particular for our analysis of the spectator qubit physics. We find that as long as the block-diagonalization is done consistently, for each of the perturbative methods, the main difference between the two calculations is the \text{static} frequency shifts of the qubits ($IZ$ or $ZI$	) proportional to $J^2$ up to the lowest order. In experiment, however, only the \text{dynamic} (caused by $\Omega$) part of the frequency shifts are observable. Hence, in our predictions for the gate parameters and especially the gate error we have to be careful to exclude the static parts. 

To develop a perturbation theory simultaneously in $J$ and $\Omega$, we introduce a fictitious expansion parameter $\lambda$ such that the total Hamiltonain is expressed as  
\begin{align}
\HO_{s}(t)=\HO_{0}+\lambda\HO_{\text{int}}(t) \;,
\label{eqn:CREffHam-Def of lambda}
\end{align}
with $\HO_0$ and $\HO_{\text{int}}= \HO_J +\HO_d$ given in Eqs.~(\ref{eqn:CRHamInEnBasis-H0 2}--\ref{eqn:CRHamInEnBasis-Hd in new basis}). Having the additional parameter $\lambda$ helps to collect consistent powers of the interaction during perturbation, while it is eventually set to $\lambda=1$. Since the interaction Hamiltonian consists of both the exchange interaction and the drive simultaneously, $O(\lambda^p)$ contributions contain any combination of the form $J^{m}\Omega^n$ such that $m+n=p$. To simplify the implementation of perturbation, we move to the interaction frame with respect to $\HO_{0}$ as
\begin{align}
\lambda\HO_{\text{I}}(t)\equiv e^{i\HO_0 t} \left[\lambda \HO_{\text{int}}(t)\right]e^{-i\HO_0 t} \;.
\label{eqn:CREffHam-Def of H_I(t)}
\end{align}

Equation~(\ref{eqn:CREffHam-Def of H_I(t)}) is the starting Hamiltonian for our perturbation theory. In order to find an effective Hamiltonian, we employ a time-dependent SWPT as
\begin{align}
\hat{\mathcal{H}}_{\text{I,eff}}(t)\equiv e^{i\hat{G}(t)}\left[\lambda \hat{\mathcal{H}}_{\text{I}}(t)-i \partial_t \right]e^{-i\hat{G}(t)} \;,
\label{eqn:CREffHam-Def of H_I,eff}
\end{align}
where $\hat{G}(t)$ is the generator of SW transformation that needs to be solved for order by order such that the effective Hamiltonian becomes block-diagonal with respect to the Hilbert space of the target qubit. We follow a series solution for the generator $\hat{G}(t)$ and the effective Hamiltonian $\HO_{\text{I,eff}}(t)$ by expanding in terms of $\lambda$ as
\begin{subequations}
\begin{align}
&\hat{G}(t)=\sum\limits_{\lambda=1}^{\infty}\lambda^n \hat{G}_n(t) \;,
\label{eqn:CREffHam-G exp}\\
&\HO_{\text{I,eff}}(t)=\sum\limits_{\lambda=1}^{\infty}\lambda^n \HO_{\text{I,eff}}^{(n)}(t) \;.
\label{eqn:CREffHam-H_I,eff exp}	
\end{align}
\end{subequations}
Inserting Eqs.~(\ref{eqn:CREffHam-G exp}--\ref{eqn:CREffHam-H_I,eff exp}) into Eq.~(\ref{eqn:CREffHam-Def of H_I,eff}), employing the BCH lemma, one finds operator-valued ordinary differential equations (ODE) for the successive orders of SWPT up to $O(\lambda^4)$ (See Appendix~\ref{App:SWPT}). 

Since the interaction terms are not in the desired block-diagonal form from the outset, the effective Hamiltonian is zero up to the first order, while the generator $\hat{G}_1(t)$ is solved for in order to remove all the non-block-diagonal parts as
\begin{subequations}
\begin{align}
O(\lambda):
\begin{cases}
&\hat{\mathcal{H}}_{\text{I,eff}}^{(1)}=0 \;,\\
&\dot{\hat{G}}_1=\hat{\mathcal{H}}_I \;.
\end{cases}
\label{eqn:CREffHam-O(lambda) HIeff}
\end{align}
Following the same procedure, up to the second order, we obtain
\begin{align}
O(\lambda^2):
\begin{cases}
&\hat{\mathcal{H}}_{\text{I,eff}}^{(2)}=\mathcal{B}(\frac{i}{2}[\hat{G}_1,\hat{\mathcal{H}}_I]) \;,\\
&\dot{\hat{G}}_2=\mathcal{N}(\frac{i}{2}[\hat{G}_1,\hat{\mathcal{H}}_I]) \;,
\end{cases}
\label{eqn:CREffHam-O(lambda^2) HIeff}
\end{align}
where now the commutator $[\hat{G}_1,\hat{\mathcal{H}}_I]$ can produce both block-diagonal and non-block-diagonal contributions, hence $\mathcal{B}(\bullet)$ and~$\mathcal{N}(\bullet)$ denote these two parts with respect to the energy basis of the target qubit, respectively. The result for the third order can be summarized as
\begin{align}
O(\lambda^3):
\begin{cases}
\begin{split}
\hat{\mathcal{H}}_{\text{I,eff}}^{(3)}&=\mathcal{B}\left(-\frac{i}{2}[\hat{G}_1,\dot{\hat{G}}_2]+\frac{i}{2}[\hat{G}_2,\hat{\mathcal{H}}_I]\right.\\
&\left.-\frac{1}{3}[\hat{G}_1,[\hat{G}_1,\hat{\mathcal{H}}_I]]\right) \;,
\end{split}\\
\begin{split}
\dot{\hat{G}}_3&=\mathcal{N}\left(-\frac{i}{2}[\hat{G}_1,\dot{\hat{G}}_2]+\frac{i}{2}[\hat{G}_2,\hat{\mathcal{H}}_I]\right.\\
&\left.-\frac{1}{3}[\hat{G}_1,[\hat{G}_1,\hat{\mathcal{H}}_I]]\right) \;.
\end{split}
\end{cases}
\label{eqn:CREffHam-O(lambda^3) HIeff}
\end{align}
Finally, the fourth order reads
\begin{align}
O(\lambda^4):
\begin{cases}
\begin{split}
\hat{\mathcal{H}}_{\text{I,eff}}^{(4)}&=\mathcal{B}\left(-\frac{i}{2}[\hat{G}_1,\dot{\hat{G}}_3]-\frac{i}{2}[\hat{G}_2,\dot{\hat{G}}_2]\right. \\
&+\frac{1}{6}[\hat{G}_1,[\hat{G}_1,\dot{\hat{G}}_2]]+\frac{i}{2}[\hat{G}_3,\hat{\mathcal{H}}_I]\\
&-\frac{1}{3}[\hat{G}_1,[\hat{G}_2,\hat{\mathcal{H}}_I]]-\frac{1}{3}[\hat{G}_2,[\hat{G}_1,\hat{\mathcal{H}}_I]]\\
&-\left. \frac{i}{8}[\hat{G}_1,[\hat{G}_1,[\hat{G}_1,\hat{\mathcal{H}}_I]]]\right) \;,
\end{split}\\
\begin{split}
\dot{\hat{G}}_4 &=\mathcal{N}\left(-\frac{i}{2}[\hat{G}_1,\dot{\hat{G}}_3]-\frac{i}{2}[\hat{G}_2,\dot{\hat{G}}_2]\right. \\
&+\frac{1}{6}[\hat{G}_1,[\hat{G}_1,\dot{\hat{G}}_2]]+\frac{i}{2}[\hat{G}_3,\hat{\mathcal{H}}_I]\\
&-\frac{1}{3}[\hat{G}_1,[\hat{G}_2,\hat{\mathcal{H}}_I]]-\frac{1}{3}[\hat{G}_2,[\hat{G}_1,\hat{\mathcal{H}}_I]]\\
&-\left. \frac{i}{8}[\hat{G}_1,[\hat{G}_1,[\hat{G}_1,\hat{\mathcal{H}}_I]]]\right) \;.
\end{split}
\end{cases}
\label{eqn:CREffHam-O(lambda^4) HIeff}
\end{align}
\end{subequations}

Equations~(\ref{eqn:CREffHam-O(lambda) HIeff}--\ref{eqn:CREffHam-O(lambda^4) HIeff}) provide the main results for SWPT that can be solved iteratively by finding the generator that removes the non-block-diagonal part at each order. In practice, by keeping $N_{c,t}$ levels for the control and target qubits, we can calculate the generator and the corresponding effective Hamiltonian by solving a system of ODEs of dimension $(N_cN_t)^2$. We find that keeping energy states beyond the computational space of transmon induce non-negligible renormalization of the gate parameters compared to a two-level model. Therefore, in our analytical calculation of the gate parameters, we make a four-level approximation ($N_c=N_t=4$) and justify this choice by quantifying the resulting error compared to three- and two-level models (See Fig.~\ref{fig:NeOrAnal-HilSpCutOff}). 

Once the effective Hamiltonian in the \textit{extended} Hilbert space is obtained, we can infer the effective CR gate Hamiltonian in the computational space as 
\begin{subequations}
\begin{align}
&\HO_{\text{CR,eff}}\equiv \sum\limits_{m,n=i,x,y,z}\frac{1}{2}\omega_{\sigma_m \sigma_n } \hat{\sigma}_m \otimes \hat{\sigma}_n \;,
\label{eqn:CREffHam-Def of H_CR,eff}\\
&\omega_{\sigma_m \sigma_n}\equiv \frac{1}{2}\Tr \left(\left(\hat{\sigma}_m \otimes \hat{\sigma}_n \right)\HO_{\text{I,eff}}\right) \;,
\label{eqn:CREffHam-Def of w_(sigma)}
\end{align}
\end{subequations}
where $\sigma_{i, x, y, z}$ are the Pauli operators, $\omega_{\sigma_m\sigma_n}$ is the corresponding two-qubit gate parameter and the order of Hilbert space components is $\text{control}\otimes \text{target}$. In what follows, for simplicity, we use the shorthand notation $\hat{\sigma}_i = \hat{I}$, $\hat{\sigma}_x = \hat{X}$, $\hat{\sigma}_y = \hat{Y}$ and $\hat{\sigma}_z = \hat{Z}$ and relax the explicit tensor product notation as well. Following this procedure, we find 5 non-zero two-qubit gate parameters as	
\begin{align}
\begin{split}
\HO_{\text{CR,eff}}&=\omega_{ix}\frac{\hat{I}\hat{X}}{2}+\omega_{iz}\frac{\hat{I}\hat{Z}}{2}+\omega_{zi}\frac{\hat{Z}\hat{I}}{2}\\
&+\omega_{zx}\frac{\hat{Z}\hat{X}}{2}+\omega_{zz}\frac{\hat{Z}\hat{Z}}{2}\;,
\end{split}
\label{eqn:CREffHam-H_CR,eff}
\end{align}
indicating that additional unwanted two-qubit interactions are also induced on top of the desired $ZX$ term. In Sec.~\ref{Sec:CREcho}, we discuss how the effect of unwanted terms can be mitigated via the echo sequence commonly used in experiments, e.g. Ref.~\cite{Sheldon_Procedure_2016}. Furthermore, note that there are indeed two conventions for the definition of the $ZZ$ rate. The one according to Eq.~(\ref{eqn:CREffHam-H_CR,eff}) can be related to the difference in the target qubit frequency conditioned on the state of the control qubit as
\begin{align}
\omega_{zz} \equiv \frac{1}{2}\Big(\left.\omega_{t}\right|_{c=1}- \left.\omega_{t}\right|_{c=0}\Big)	\;.
\label{eqn:CREffHam-Def of wzz}
\end{align}
It is more common in experiment, however, to call the total frequency shift of the target as the $ZZ$ rate, which would be twice the value we quote in this paper. 

In the following, we provide the lowest and the next order estimates for the gate parameters in Secs.~\ref{SubSec:LoOrAnal} and~\ref{SubSec:NeOrAnal}, respectively.
\begin{table*}
  \begin{tabular}{|c|c|c|c|c|}
  \hline	
   Operator & Coefficient (Kerr) & Estimate (MHz) & Coefficient (energy basis) & Estimate (MHz) \\
  \hline\hline
    $\frac{1}{2}\hat{I}\hat{X}$ & $ -\frac{1}{\Delta_{ct}+\alpha_c}J \Omega$ & 1.462 & $ -\frac{\nu_{t,01}\nu_{c,12}^2}{2 \left(\Delta_{ct}+\alpha_c\right)}J\Omega$ & 1.250\\
   \hline\hline
     $\frac{1}{2}\hat{Z}\hat{I}$ & $\Bigg[\frac{1}{2\left(\Delta_{ct}+\alpha_c\right)}-\frac{1}{2 \Delta_{ct}}\Bigg]\Omega^2	$ & $-15.865$ & $\Bigg[\frac{\nu _{c,12}^2}{4 \left(\Delta_{ct}+\alpha_c\right)}-\frac{\nu _{c,01}^2}{2 \Delta_{ct}}\Bigg]\Omega^2$ & $-14.371$  \\
  \hline\hline
  $\frac{1}{2}\hat{Z}\hat{X}$ & $\left(\frac{1}{\Delta_{ct}+\alpha_c}-\frac{1}{\Delta_{ct}}\right)J \Omega $ & -2.411 & $\frac{1}{2} \left(\frac{\nu_{t,01}\nu_{c,12}^2}{\Delta_{ct}+\alpha_c}-\frac{2 \nu_{t,01}\nu_{c,01}^2}{\Delta_{ct}}\right)J \Omega$ & -2.118	\\
  \hline\hline
  $\frac{1}{2}\hat{Z}\hat{Z}$ & $\left(\frac{1}{\Delta_{ct}-\alpha_t}-\frac{1}{\Delta_{ct}+\alpha_c}\right)J^2$ & 0.138 & $\frac{1}{2}\left(\frac{\nu _{c,01}^2 \nu_{t,12}^2}{\Delta_{ct}-\alpha_t}-\frac{\nu_{t,01}^2 \nu_{c,12}^2}{\Delta_{ct}+\alpha_c}\right) J^2$ & 0.114 \\
  \hline
  \end{tabular}
  \caption{Lowest order estimates for CR gate parameters. The left two columns summarize the result from Kerr theory \cite{Magesan_Effective_2018}, while the right two columns present expressions using energy-basis representation. System parameters are chosen from Refs.~\cite{Sheldon_Procedure_2016, Magesan_Effective_2018} as $\omega_c=5114$ MHz, $\omega_t=4914$, $\Delta_{ct}=200$ MHz, $\alpha_c=\alpha_t=-330$ MHz, $J=3.8$ MHz and $\Omega=50$ MHz. From Eq.~(\ref{eqn:EnBasis-eps ITO alpha}), one obtains an estimate for unitless anharmonicity measures as $\epsilon_c=0.217$ and $\epsilon_t=0.224$. Importantly, setting $\nu_{i,01}\rightarrow 1$ and $\nu_{i,12}\rightarrow \sqrt{2}$ for $i=c,t$ yields the lowest order estimates from Kerr theory.}
\label{tab:LoOrAnal-CRGateParams}
\end{table*}

\subsection{Lowest order analytics}
\label{SubSec:LoOrAnal}

The lowest non-zero estimate for the gate parameters arise from the $O(\lambda^2)$ contribution given in Eq.~(\ref{eqn:CREffHam-O(lambda^2) HIeff}). Solving for $\hat{G}_1(t)$ from the $O(\lambda)$ Eq.~(\ref{eqn:CREffHam-O(lambda) HIeff}), we find the lowest order effective Hamiltonian as
\begin{align}
\HO_{\text{I,eff}}^{(2)}(t)=\mathcal{B}\left(\frac{i}{2}\left[\int_{0}^{t}dt'\HO_{\text{I}}(t'),\HO_{\text{I}}(t)\right]\right) \;,
\label{eqn:LoOrAnal:Explicit H_I,eff 2nd}
\end{align} 
where $\mathcal{B}(\bullet)$ denotes the block-diagonal part with respect to the target qubit Hilbert space. Since we have accounted for the exchange and the drive Hamiltonian on equal footing through $\lambda$, at this order, we anticipate quadratic corrections in $J$ and $\Omega$ of either of the following forms: $\Omega^2$, $J\Omega$ and $J^2$. 

Next, we discuss the effective gate parameters introduced in Eq.~(\ref{eqn:CREffHam-H_CR,eff}). The $ZX$ and $IX$ rates are obtained as
\begin{subequations}
\begin{align}
&\omega_{zx}^{(2)}=\frac{1}{2}\left(\frac{\nu_{t,01}\nu_{c,12}^2}{\Delta_{ct}+\alpha_c}-\frac{2 \nu_{t,01}\nu_{c,01}^2}{\Delta_{ct}}\right)J \Omega \;,
\label{eqn:LoOrAnal:ZX2nd}\\
&\omega_{ix}^{(2)}=-\frac{\nu_{t,01}\nu_{c,12}^2}{2 \left(\Delta_{ct}+\alpha_c\right)}J\Omega \;,
\label{eqn:LoOrAnal:IX2nd}
\end{align}
which are proportional to $J\Omega$. An alternative and more heuristic derivation of these rates can be understood from the interaction rates $\tilde{a}_{0}$ and $\tilde{a}_1$ in Fig.~\ref{fig:CRHamInEnBasis-JCLadderForCR}, in terms of which $\omega_{zx}=\tilde{a}_0-\tilde{a}_1$ and $\omega_{ix}=\tilde{a}_0+\tilde{a}_1$ \cite{Tripathi_Operation_2019}. There are multiple processes that contribute to these rates up to the lowest non-zero order in perturbation. The only contribution to $\tilde{a}_0$ is the transition from $\ket{\psi_{00}}$ to $\ket{\psi_{01}}$ via $\ket{\psi_{10}}$ with the net rate $(\nu_{c,01}\Omega/2)(-1/\Delta_{ct})(\nu_{c,01}\nu_{t,01}J)$. The first contribution to $\tilde{a}_1$ comes from the transition $\ket{\psi_{10}}$ to $\ket{\psi_{11}}$ via $\ket{\psi_{01}}$ leading to the rate $(\nu_{c,01}\nu_{t,01}J)(1/\Delta_{ct})(\nu_{c,01}\Omega/2)$. The second contribution to $\tilde{a}_1$ comes from the transition $\ket{\psi_{10}}$ to $\ket{\psi_{11}}$ via $\ket{\psi_{20}}$ resulting $(\nu_{c,12}\Omega/2)[-1/(\Delta_{ct}+\alpha_c)](\nu_{c,12}\nu_{t,01}J)$. Adding these contributions accordingly recovers expressions~(\ref{eqn:LoOrAnal:ZX2nd}) and~(\ref{eqn:LoOrAnal:IX2nd}) for the $ZX$ and the $IX$ rates. Importantly, note that neglecting the eigenstate renormalization of qubits, i.e. setting $\nu_{i,01}\rightarrow 1$ and $\nu_{i,12}\rightarrow \sqrt{2}$, yields the old results from Kerr theory in Ref~\cite{Magesan_Effective_2018}.  

The $ZZ$ interaction is understood as half of the difference in the target qubit frequency, when the control is in the excited or the ground state and is obtained as
\begin{align}
&\omega_{zz}^{(2)}=\frac{1}{2}\left(\frac{\nu _{c,01}^2 \nu_{t,12}^2}{\Delta_{ct}-\alpha_t}-\frac{\nu_{t,01}^2 \nu_{c,12}^2}{\Delta_{ct}+\alpha_c}\right) J^2 \;.
\label{eqn:LoOrAnal:ZZ2nd}   
\end{align}
In terms of the two-qubit ladder of Fig.~\ref{fig:CRHamInEnBasis-JCLadderForCR}, the first term in Eq.~(\ref{eqn:LoOrAnal:ZZ2nd}) is just a frequency renormalization of state $\ket{\psi_{11}}$ due to repulsion from the state $\ket{\psi_{02}}$ with the corresponding interaction matrix element $\nu_{c,01}\nu_{t,12}J$ and frequency difference $\Delta_{ct}-\alpha_t$. The second term is due to state $\ket{\psi_{20}}$ with the corresponding interaction matrix element $\nu_{c,12}\nu_{t,01}J$ and frequency difference $-(\Delta_{ct}+\alpha_c)$. Thus, the lowest order contribution to the $ZZ$ rate comes from the third level of each qubit. In other words, a two-level model is unable to predict a $ZZ$ rate for the CR gate.

As brought up earlier, the qubit frequency shifts have two distinct sources. The first static contribution comes from the exchange interaction between the qubits resulting in dressed qubit frequencies (See also Appendix~\ref{App:DressedBasis})
\begin{align}
&\bar{\omega}_c \equiv \omega_c+\frac{1}{2}\left(\frac{\nu_{c,01}^2\nu_{t,12}^2}{\Delta_{ct}-\alpha_t}-\frac{\nu_{c,12}^2 \nu_{t,01}^2}{\Delta_{ct}+\alpha_c}+\frac{2\nu_{c,01}^2 \nu_{t,01}^2}{\Delta_{ct}}\right)J^2 ,
\label{eqn:LoOrAnal:Def of bar(w)_c}\\
&\bar{\omega}_t \equiv \omega_t+\frac{1}{2}\left(\frac{\nu_{c,01}^2\nu_{t,12}^2}{\Delta_{ct}-\alpha_t}-\frac{\nu_{c,12}^2 \nu_{t,01}^2}{\Delta_{ct}+\alpha_c}-\frac{2\nu_{c,01}^2 \nu_{t,01}^2}{\Delta_{ct}}\right)J^2  ,
\label{eqn:LoOrAnal:Def of bar(w)_t}
\end{align}
where we have denoted the dressed frequencies with a bar. We will not include these contributions in our estimate for $ZI$ and $IZ$ rate as such static terms are not measurable. On top of this, there is a dynamic Stark shift that is induced by the drive. At this order in perturbation, the Stark shift only appears in the frequency of the control qubit as
\begin{align}
\omega_{zi}^{(2)}=\Bigg[\frac{\nu _{c,12}^2}{4 \left(\Delta_{ct}+\alpha_c\right)}-\frac{\nu _{c,01}^2}{2 \Delta_{ct}}\Bigg]\Omega^2 \;.
\label{eqn:LoOrAnal:ZI2nd}
\end{align}
\end{subequations}
For typical CR parameters the dynamic Stark shift on the control is much larger (at least two orders of magnitude) compared to the static contributions. Table~\ref{tab:LoOrAnal-CRGateParams} summarizes the lowest order perturbative result for the gate parameters as well as an experimental estimate based on circuit parameters of Refs.~\cite{Sheldon_Procedure_2016, Magesan_Effective_2018}. 

\subsection{Higher order analytics}
\label{SubSec:NeOrAnal}

In this section, we summarize the next order contributions to the gate parameters. We find that \textit{non-zero} terms in the effective Hamiltonian of the CR gate come in alternating orders in perturbation, i.e. the $O(\lambda^3)$ Eq.~(\ref{eqn:CREffHam-O(lambda^3) HIeff}) leads to an indirect contribution through a non-zero $\hat{G}_3$ while $\HO_{\text{I,eff}}^{(3)}=0$. Therefore, the next non-zero correction comes from the $O(\lambda^4)$ Eq.~(\ref{eqn:CREffHam-O(lambda^4) HIeff}), leading to dominant terms of the form $J^2\Omega^2$, $J\Omega^3$ and $\Omega^4$. Generally speaking, the diagonal gate parameters ($ZI$, $IZ$ and $ZZ$) will only adopt even powers of $\Omega$, while the off-diagonal gate parameters ($ZX$ and $IX$) contain odd powers. 

Such higher order contributions, in particular to the diagonal rates, contain a large number of independent physical processes. Here, for simplicity, we only quote the result for the the $ZX$ rate as 
\begin{align}
\begin{split}
\omega_{zx}^{(4)}&=\Bigg[\frac{\nu_{c,01}^4 \nu_{t,01}}{2 \Delta_{ct}^3}+\frac{-\nu_{c,01}^2 \nu_{c,12}^2 \nu_{t,01}-3 \nu_{c,12}^2 \nu_{c,23}^2\nu_{t,01}}{4\Delta_{ct}^2 \left(\Delta_{ct}+\alpha_c\right)}\\
&+\frac{\nu_{c,01}^2\nu_{c,12}^2 \nu_{t,01}-\nu_{c,12}^2\nu_{c,23}^2 \nu_{t,01}}{4\Delta_{ct}\left(\Delta_{ct}+\alpha_c\right)^2}-\frac{\nu_{c,12}^4 \nu_{t,01}}{4 \left(\Delta_{ct}+\alpha_c\right)^3}\\
&-\frac{\nu_{c,01}^2 \nu_{c,12}^2 \nu_{t,01}}{4 \Delta_{ct}^2\left(2 \Delta_{ct}+\alpha_c\right)}+\frac{9\nu_{c,12}^2 \nu_{c,23}^2 \nu_{t,01}}{4\Delta_{ct}^2 \left(2\Delta_{ct}+3\alpha_c\right)}\Bigg]J\Omega^3 \;,
\end{split}
\label{eqn:NeOrAnal-ZX4th}
\end{align}
resulting in a correction proportional to $J\Omega^3$. Importantly, by mapping the interaction matrix elements to the ones from Kerr theory, we recover the higher order estimate found in Refs.~\cite{Magesan_Effective_2018, Tripathi_Operation_2019} as
\begin{align}
\omega_{zx, \text{Kerr}}^{(4)}=\frac{\left(3 \alpha_c^5+11 \alpha_c^4 \Delta_{ct}+15\alpha_c^3 \Delta_{ct}^2+9 \alpha _c^2 \Delta_{ct}^3 \right) J\Omega^3}{2 \Delta_{ct}^3 \left(\Delta_{ct}+\alpha_c\right){}^3 \left(2 \Delta_{ct}+\alpha_c\right) \left(2 \Delta_{ct}+3 \alpha_c\right)} \;.
\label{eqn:NeOrAnal-ZX4th Kerr}   
\end{align}

\begin{figure}[t!]
\centering
\subfloat[\label{subfig:CREffHam-IXCompMineVsKerrSarahsParams}]{%
\includegraphics[scale=0.210]{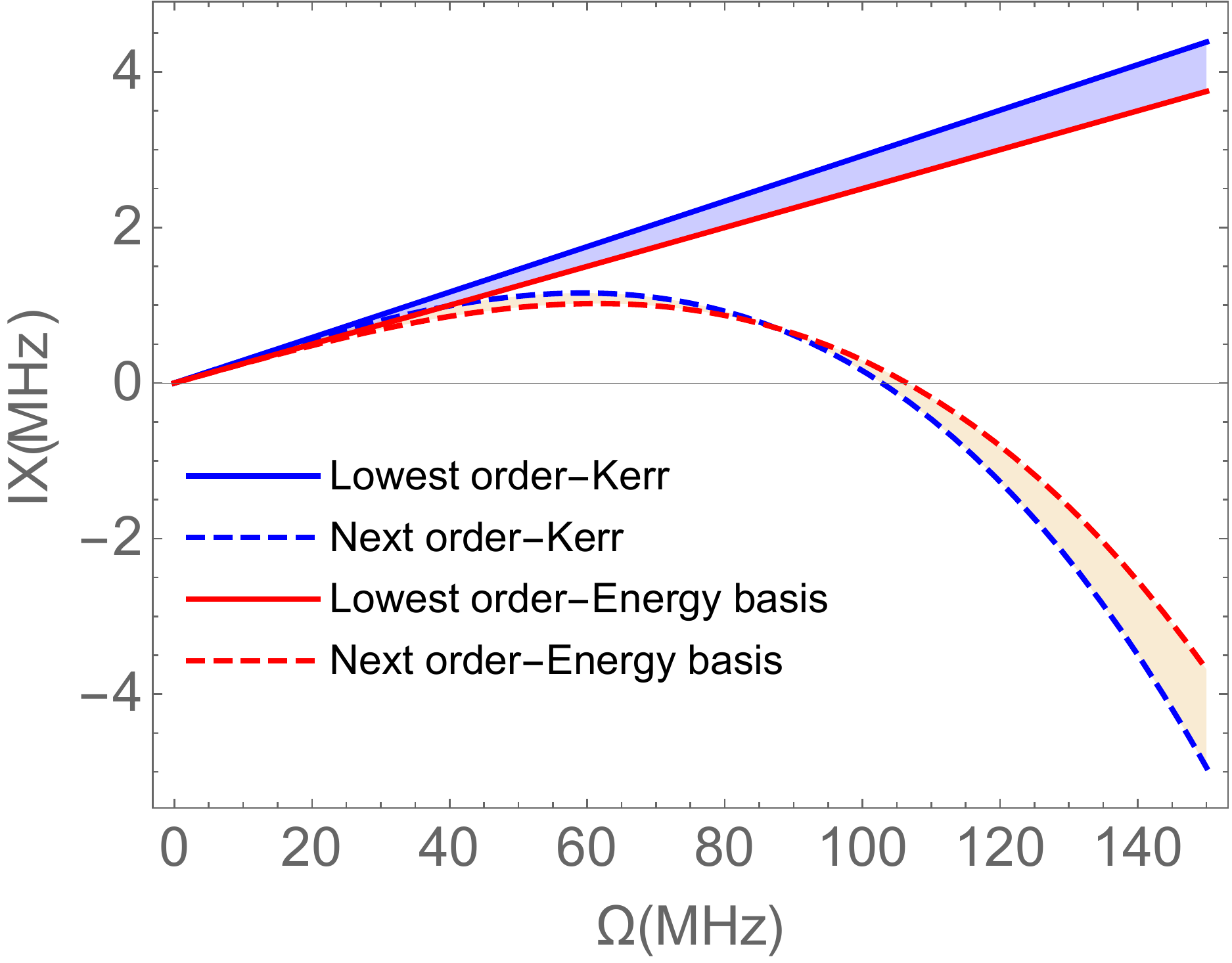}%
} 
\subfloat[\label{subfig:CREffHam-IZCompMineVsKerrSarahsParams}]{%
\includegraphics[scale=0.220]{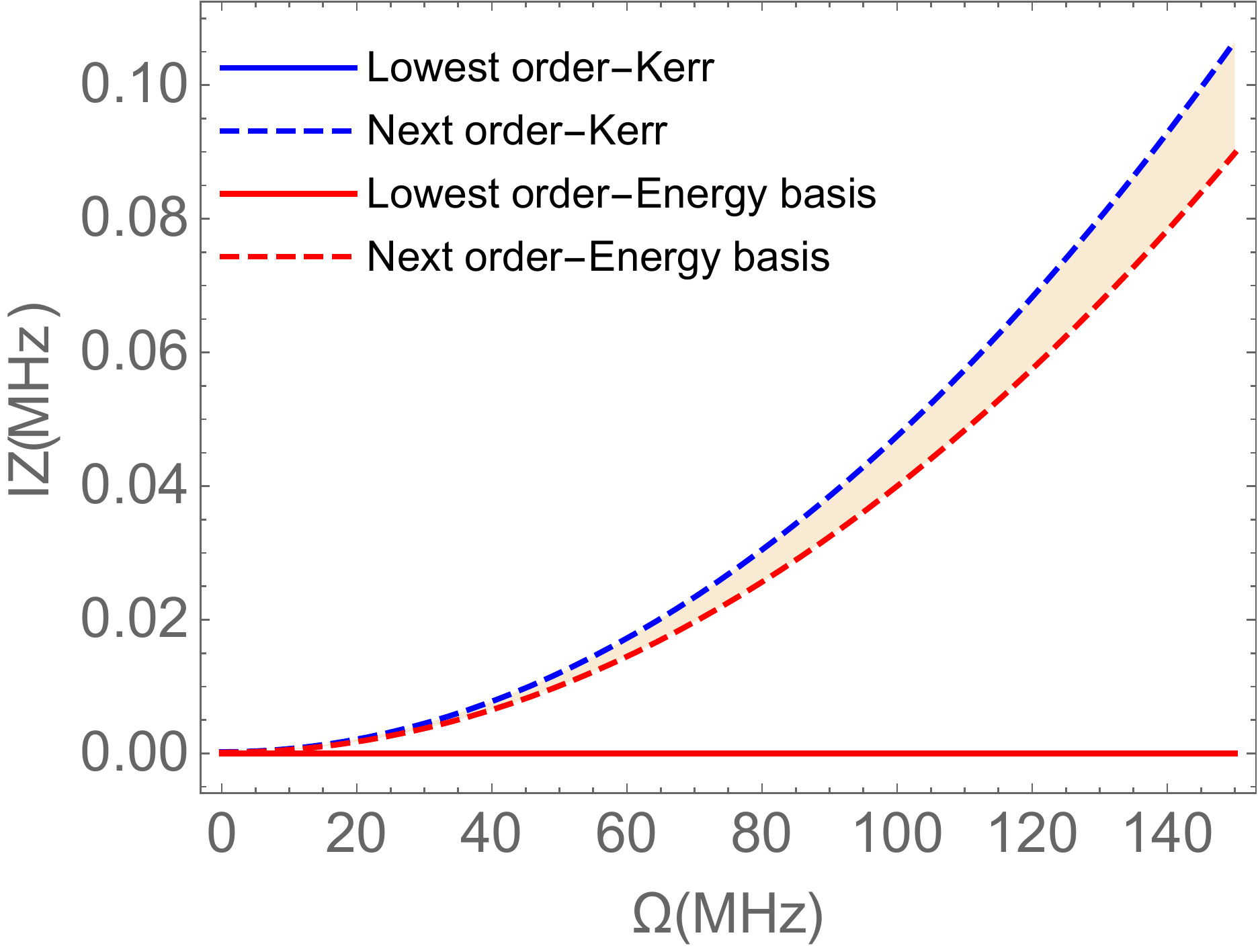}%
} \\ 
\subfloat[\label{subfig:CREffHam-ZICompMineVsKerrSarahsParams}]{%
\includegraphics[scale=0.220]{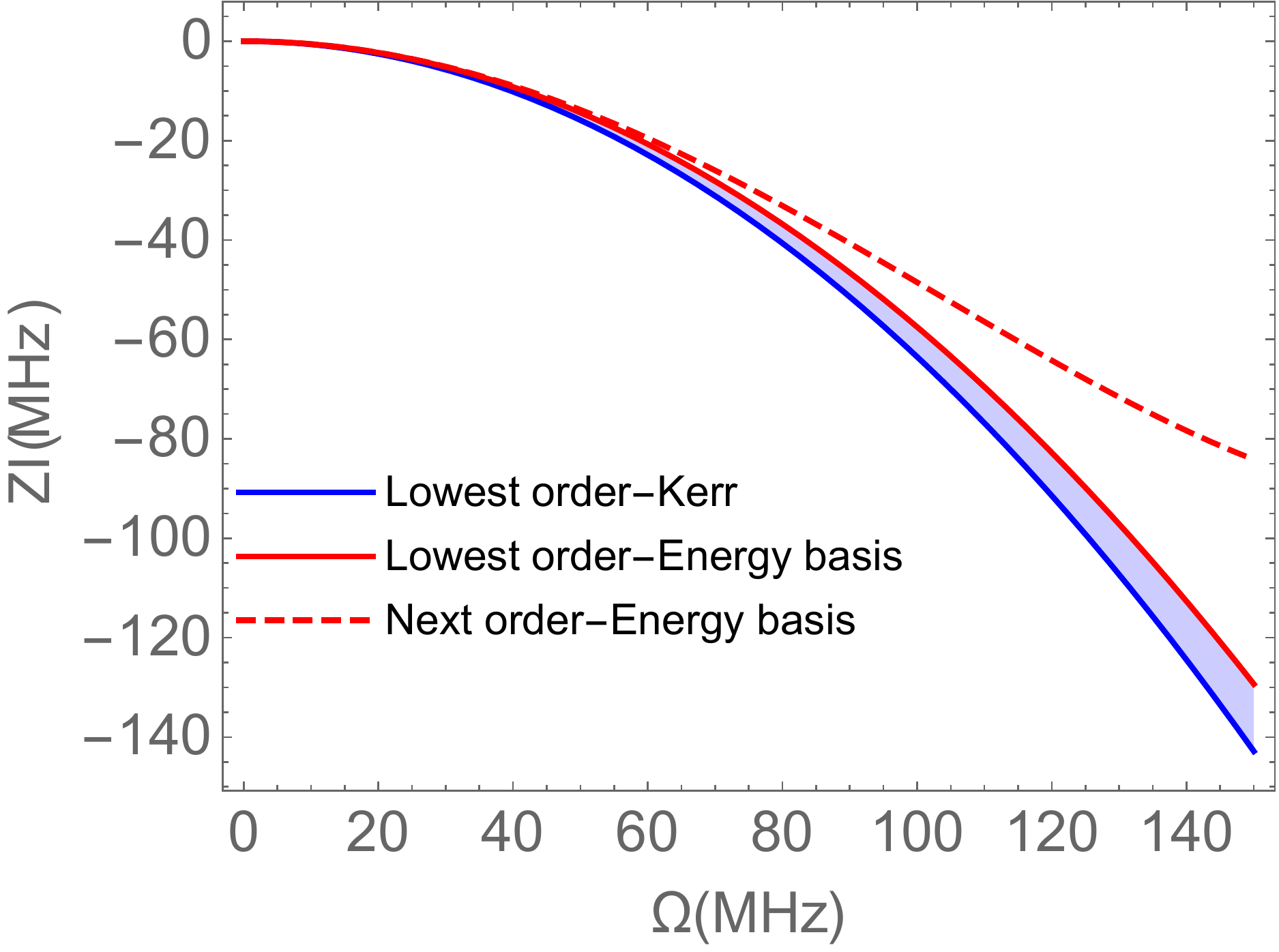}%
} 
\subfloat[\label{subfig:CREffHam-ZXCompMineVsKerrSarahsParams}]{%
\includegraphics[scale=0.210]{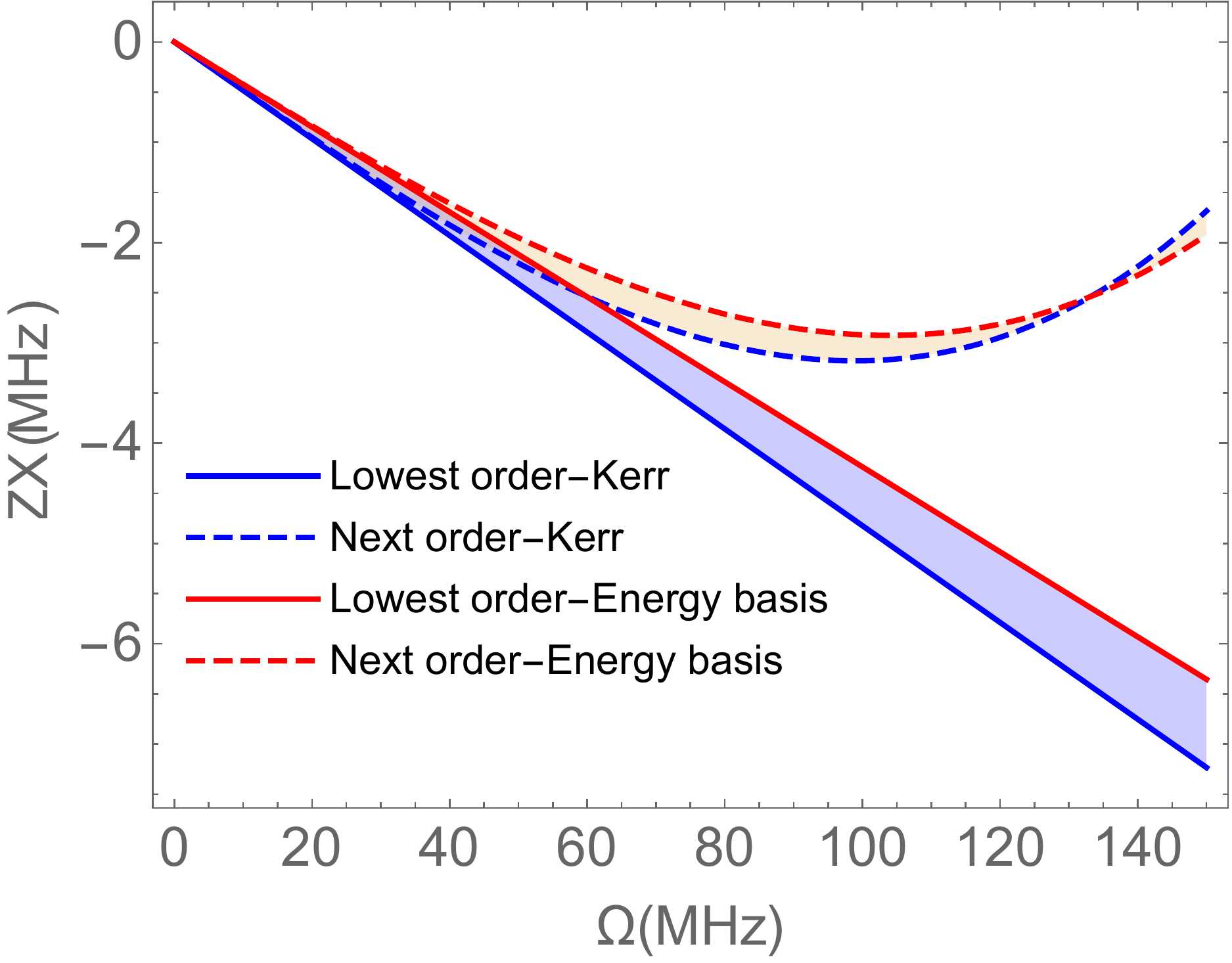}%
} \\
\subfloat[\label{subfig:CREffHam-ZZCompMineVsKerrSarahsParams}]{%
\includegraphics[scale=0.250]{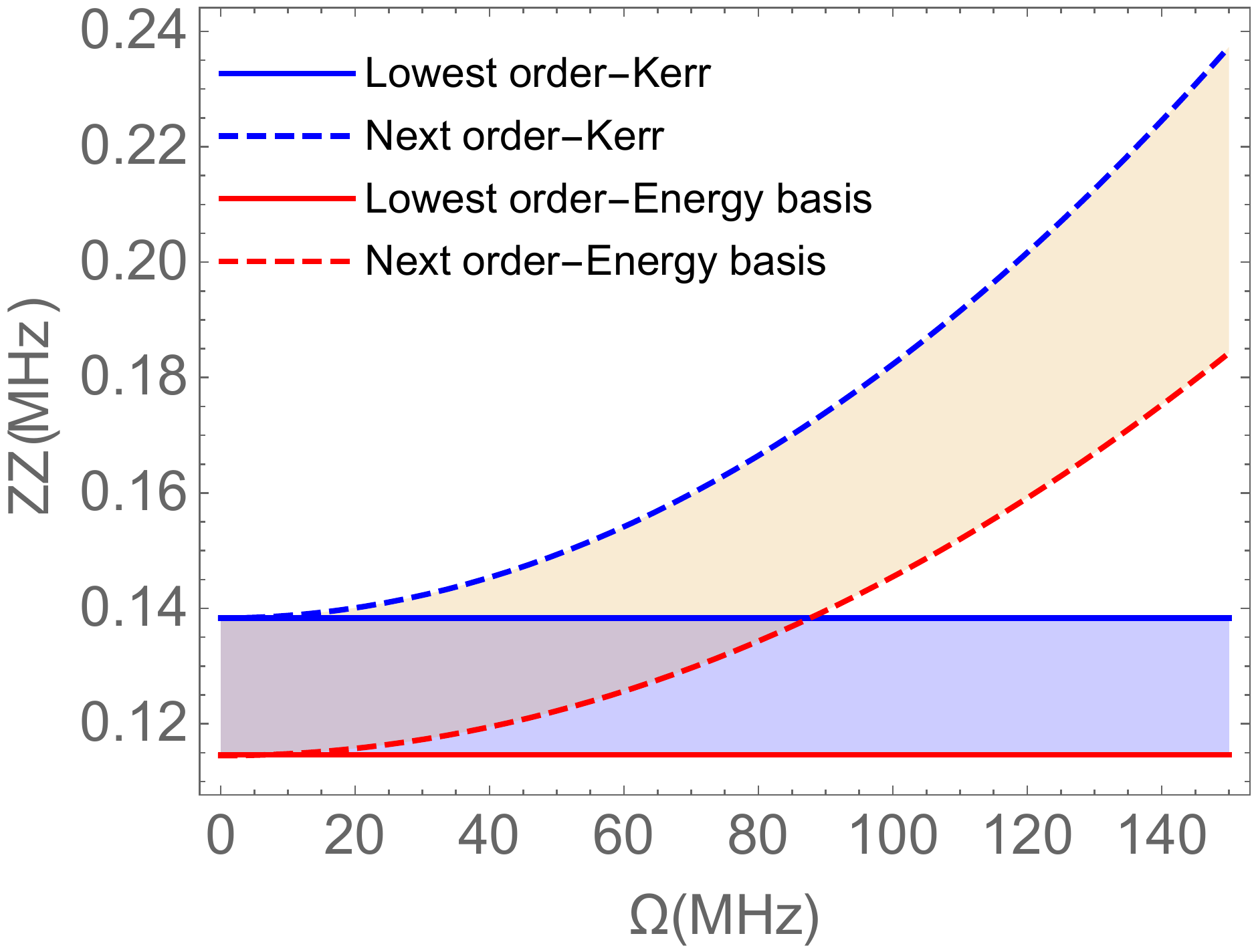}%
} 	
\caption{Gate parameters as a function of drive amplitude $\Omega$: a) $IX$ rate, b) $IZ$ rate, c) $ZI$ rate, d) $ZX$ rate and e) $ZZ$ rate. System parameters are the same as Table~\ref{tab:LoOrAnal-CRGateParams}. The solid blue curve shows the result for the lowest order Kerr theory, the dashed blue shows the higher order Kerr theory, the solid red shows the lowest order in the energy basis and the dashed red shows the higher order in the energy basis. Note that we have implemented the perturbation up to $O(\lambda^4)$, therefore it will capture terms proportional to $\Omega^4$, if any, in the gate parameters. We find such behavior only in the $ZI$ rate (dashed red) of Fig.~\ref{subfig:CREffHam-ZICompMineVsKerrSarahsParams}. On the other hand, the available results from Kerr theory were presented up to $O(\Omega^3)$ in Refs.~\cite{Magesan_Effective_2018, Tripathi_Operation_2019}.}
\label{fig:CREffHam-EnBasisVsKerr}
\end{figure}

Figure~\ref{fig:CREffHam-EnBasisVsKerr} compares the lowest and the next order perturbative estimates between Kerr and energy-basis representations for the parameters of Refs.~\cite{Sheldon_Procedure_2016, Magesan_Effective_2018} (same as Table~\ref{tab:LoOrAnal-CRGateParams}). We find that the next order corrections to $ZX$ and $IX$ rates come in opposite sign compared to the previous order, hence suppressing the rates at stronger drive. Furthermore, we observe that the $ZZ$ rate is slightly increased with a correction proportional to $J^2\Omega^2$ (Fig.~\ref{subfig:CREffHam-ZZCompMineVsKerrSarahsParams}). The Stark shift on the control qubit ($ZI$ rate) is suppressed by a correction proportional to $\Omega^4$ at higher drive (dashed red curve in Fig.~\ref{subfig:CREffHam-ZICompMineVsKerrSarahsParams}). In comparison, the Stark shift on the target qubit is much smaller and of the order of 10 KHz for medium drive power (Fig.~\ref{subfig:CREffHam-IZCompMineVsKerrSarahsParams}). All in all, roughly speaking, one observes a relative difference of up to 15\% between the two theories where in all instances the energy basis predicts smaller rates in absolute value.

In Fig.~\ref{fig:CREffHam-EnBasisVsKerr} and Table~\ref{tab:LoOrAnal-CRGateParams}, we have considered a specific choice of parameters with control-target detuning $\Delta_{ct}=200$ MHz that translates in terms of anharmonicity as $\Delta_{ct}\approx -0.61\alpha_c$. We find, however, that the gate parameters are extremely sensitive to the qubit-qubit detuning. The underlying reason is numerous possibilities for a resonance between the two-qubit states in a rather narrow frequency interval. These two-qubit resonances can be better understood from the energy ladder of Fig.~\ref{fig:CRHamInEnBasis-JCLadderForCR} and are summarized in Table~\ref{tab:NeOrAnal-Resonances}.
\begin{table}
  \begin{tabular}{|c|c|c|c|}
  \hline
  States ($\ket{CT}\ket{D}$) & Condition &  $\Delta_{ct}$ & Type\\
  \hline\hline
  $\ket{\psi_{11}}\ket{n_d}\sim \ket{\psi_{02}}\ket{n_d}$ & $\Delta_{ct}=\alpha_t$ & $-330$ & $\text{II}_B$ \\
  \hline
  $\ket{\psi_{01}}\ket{n_d}\sim \ket{\psi_{10}}\ket{n_d}$ & $\Delta_{ct}=0$ & 0 & $\text{II}_A$ \\
  \hline
  $\ket{\psi_{20}}\ket{n_d}\sim \ket{\psi_{02}}\ket{n_d}$ & $2\Delta_{ct}=\alpha_t-\alpha_c$ & 0 & $\text{II}_C$\\
  \hline
  $\ket{\psi_{21}}\ket{n_d}\sim \ket{\psi_{12}}\ket{n_d}$ & $\Delta_{ct}=\alpha_t-\alpha_c$ & 0 & $\text{II}_D$\\
  \hline
  $\ket{\psi_{00}}\ket{n_d+2}\sim \ket{\psi_{20}}\ket{n_d}$& $2\Delta_{ct}=-\alpha_c $ & 165 & $\text{I}_A$ \\
  \hline
  $\ket{\psi_{11}}\ket{n_d}\sim \ket{\psi_{20}}\ket{n_d}$ & $\Delta_{ct}=-\alpha_c$ & 330 & $\text{II}_B$\\
  \hline
  $\ket{\psi_{10}}\ket{n_d+1}\sim \ket{\psi_{20}}\ket{n_d}$ & $\Delta_{ct}=-\alpha_c$ & 330 & $\text{I}_C$ \\
  \hline
  $\ket{\psi_{00}}\ket{n_d+3}\sim \ket{\psi_{30}}\ket{n_d}$ & $3\Delta_{ct}\approx -3\alpha_c$ & 330 & $\text{I}_B$ \\
  \hline
  $\ket{\psi_{10}}\ket{n_d+2}\sim \ket{\psi_{30}}\ket{n_d}$ & $2\Delta_{ct}=-3\alpha_c$ & 495 & $\text{I}_D$\\
  \hline
  $\ket{\psi_{21}}\ket{n_d}\sim \ket{\psi_{30}}\ket{n_d}$ & $\Delta_{ct}=-2\alpha_c$ & 660 & $\text{II}_E$\\
  \hline
  $\ket{\psi_{20}}\ket{n_d+1}\sim \ket{\psi_{30}}\ket{n_d}$ & $\Delta_{ct}=-2\alpha_c$ & 660 & $\text{I}_E$\\
  \hline
  \end{tabular}
  \caption{Summary of resonances that emerge under the assumption of four energy states for each qubit and up to the fourth-order perturbation in $\Omega$ and $J$. From left to right, the first column denotes the underlying physical process in terms of qubit and drive photon states, the second show the corresponding resonance condition in terms of qubit-qubit detuning, the third gives the experimental estimate in MHz assuming $\alpha_c=\alpha_t=-330 \ \text{MHz}$, and the fourth labels such resonances in terms of broader categories for multi-qubit resonances (See Sec.~\ref{Sec:SumFreqCol}). It is important to note that the resonances are classified in terms of the underlying physical process and the degeneracy between some of them is merely due to $\omega_d \approx \omega_t$. The above resonances translate as poles in our perturbative solution and divide the parameter space for the qubit-qubit detuning into distinct regions of operation.}
\label{tab:NeOrAnal-Resonances}
\end{table}
To make a connection with our perturbative result, these resonances translate as poles in our expressions for the gate parameters and hence break the landscape for the qubit-qubit detuning $\Delta_{ct}$ into multiple regions. Under a four-level model for each qubit, we recognize the following five distinct regions in our calculation: I) $-\alpha_t<\Delta_{ct}<0$, II) $0<\Delta_{ct}<-\alpha_c/2$, III) $-\alpha_c/2<\Delta_{ct}<-\alpha_c$, IV) $-\alpha_c<\Delta_{ct}<-3\alpha_c/2$ and V) $-3\alpha_c/2<\Delta_{ct}<-2\alpha_c$. Assuming $\alpha_c=\alpha_t =-330$ MHz, the detuning range translates as $-330 \ \text{MHz}<\Delta_{ct}<660 \ \text{MHz}$. 

\begin{figure}[t!]
\centering
\subfloat[\label{subfig:CREffHam-IX2DPlot4thConstDriveContoursSarahsParams}]{%
\includegraphics[scale=0.145]{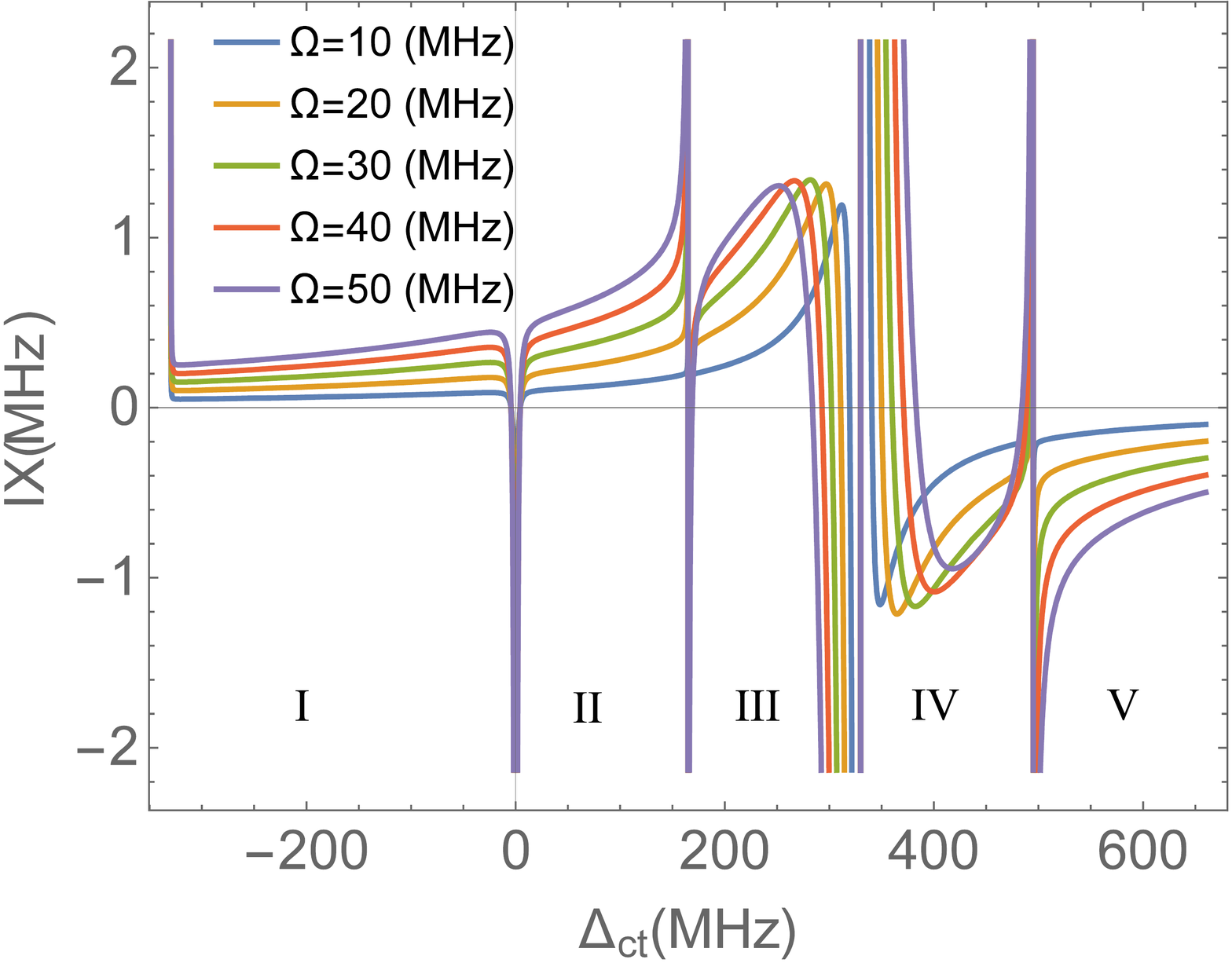}%
}
\subfloat[\label{subfig:CREffHam-IZ2DPlot4thConstDriveContoursSarahsParams}]{%
\includegraphics[scale=0.15]{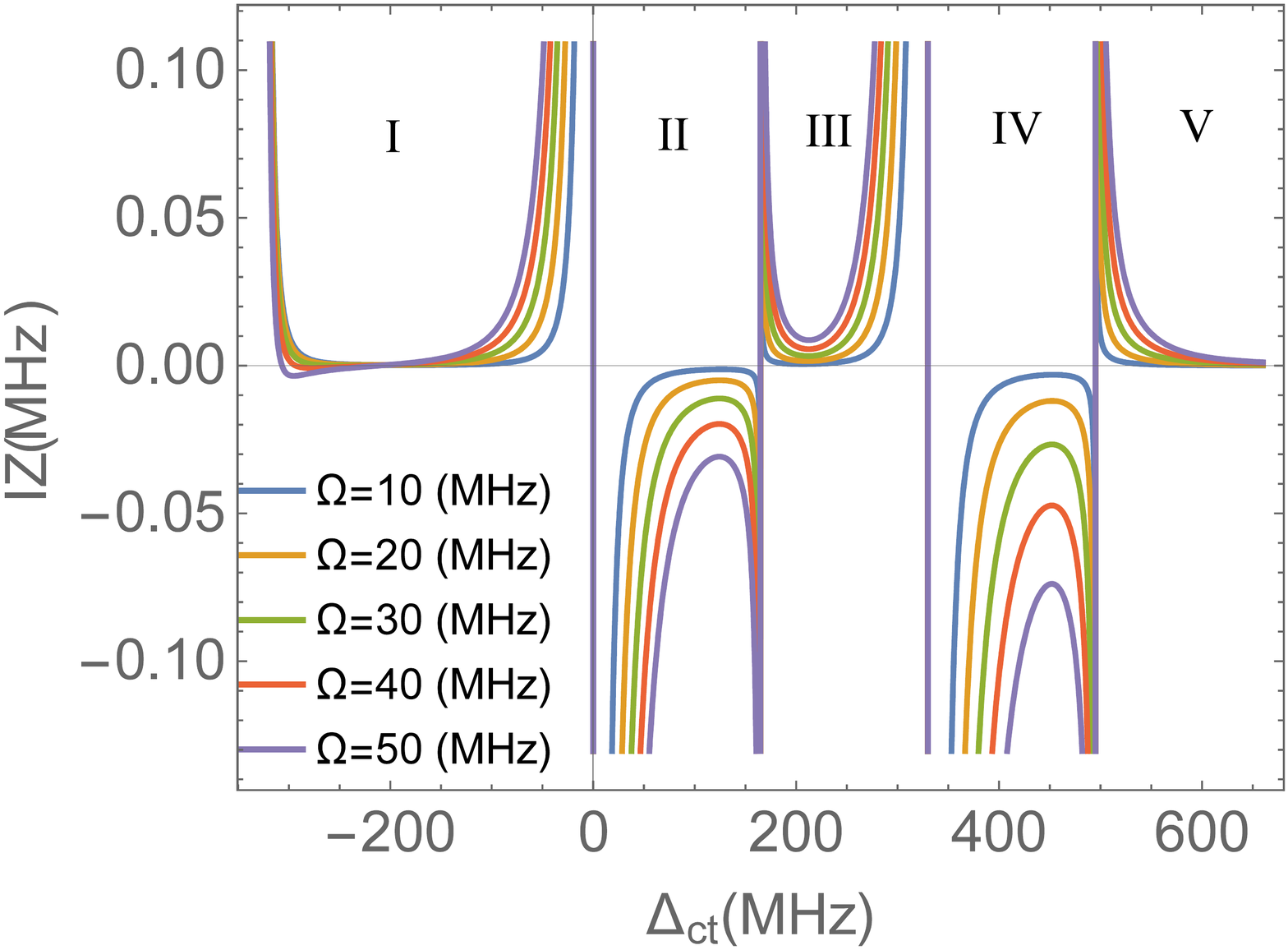}%
}\\
\subfloat[\label{subfig:CREffHam-ZI2DPlot4thConstDriveContoursSarahsParams}]{%
\includegraphics[scale=0.15]{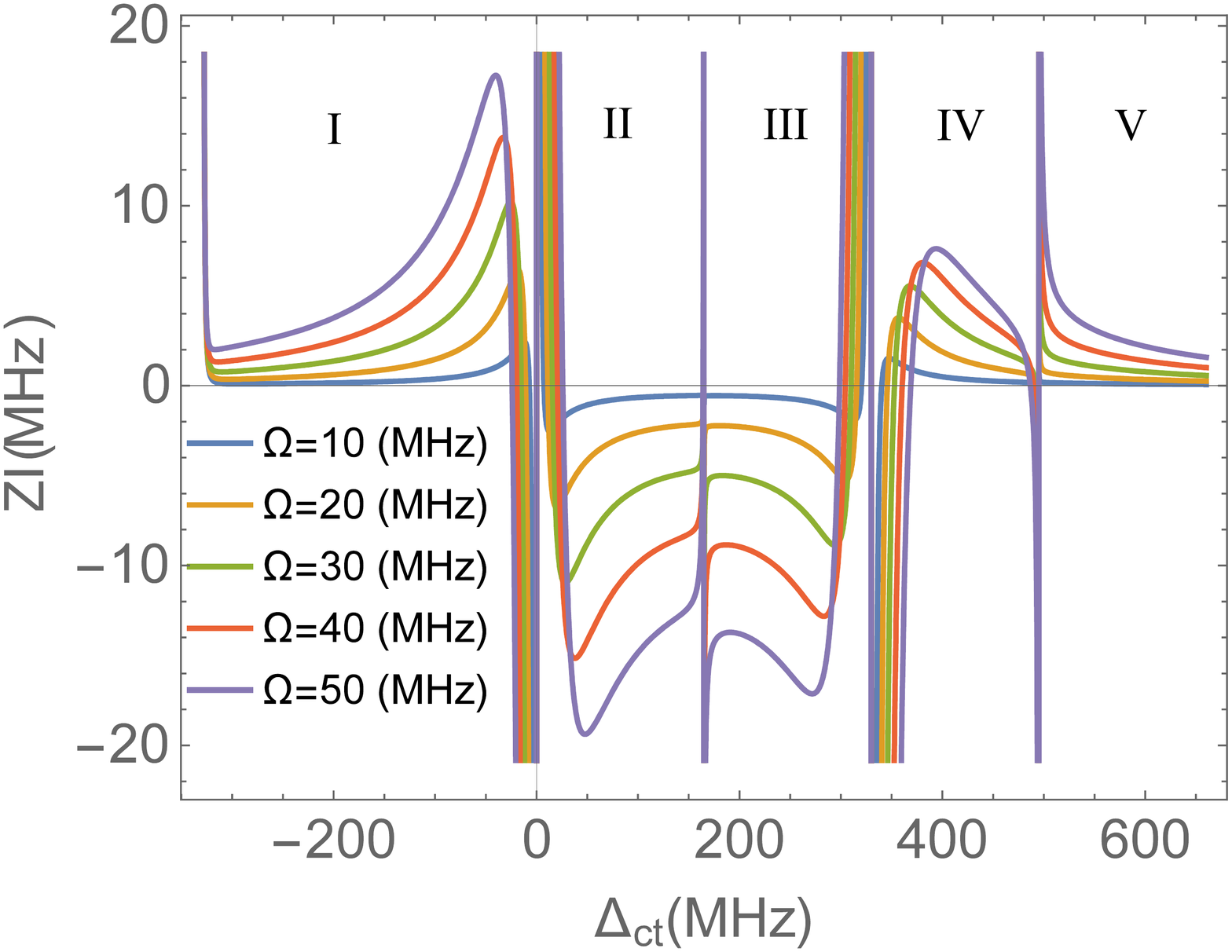}%
}
\subfloat[\label{subfig:CREffHam-ZX2DPlot4thConstDriveContoursSarahsParams}]{%
\includegraphics[scale=0.1475]{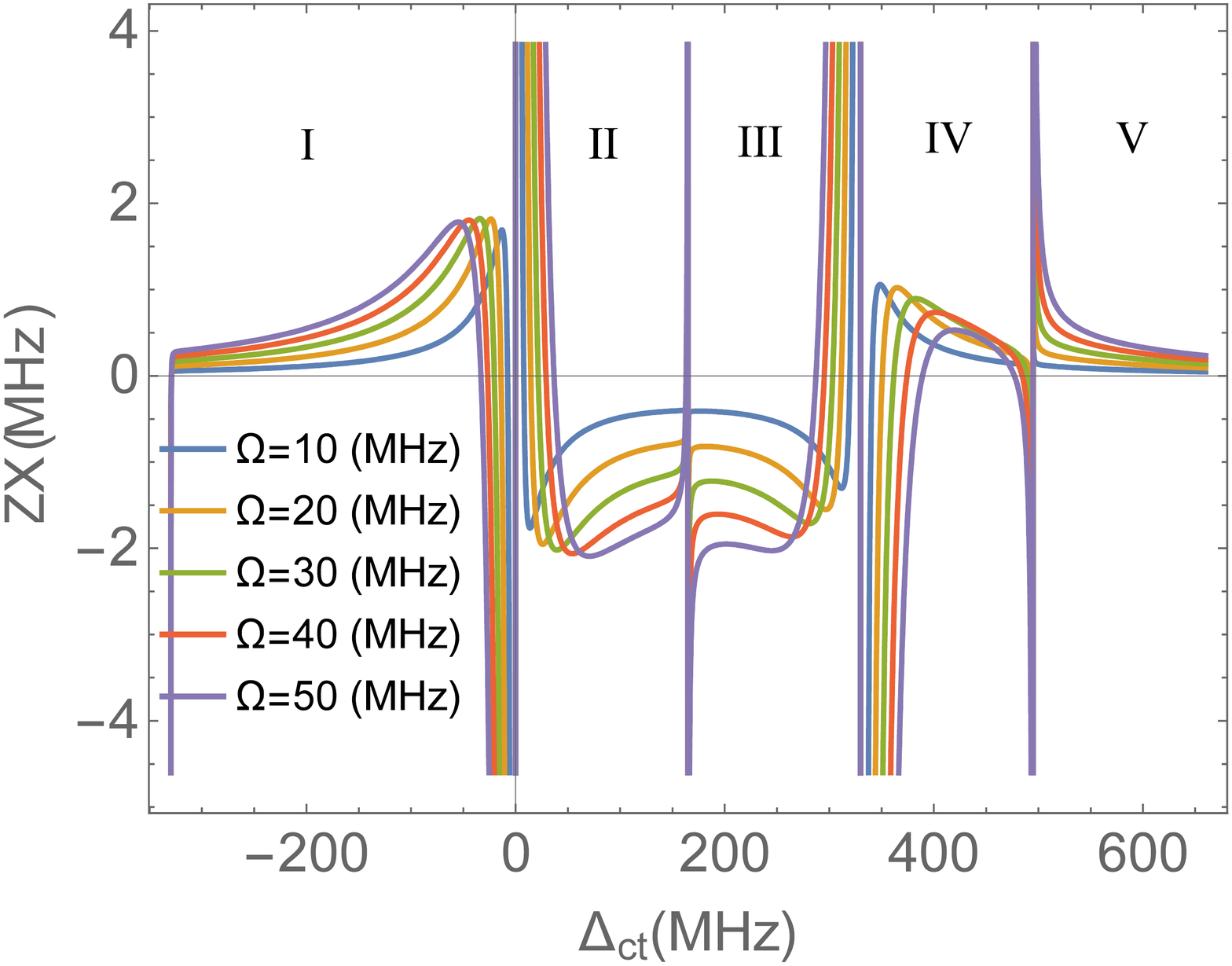}%
}\\
\subfloat[\label{subfig:CREffHam-ZZ2DPlot4thConstDriveContoursSarahsParams}]{%
\includegraphics[scale=0.15]{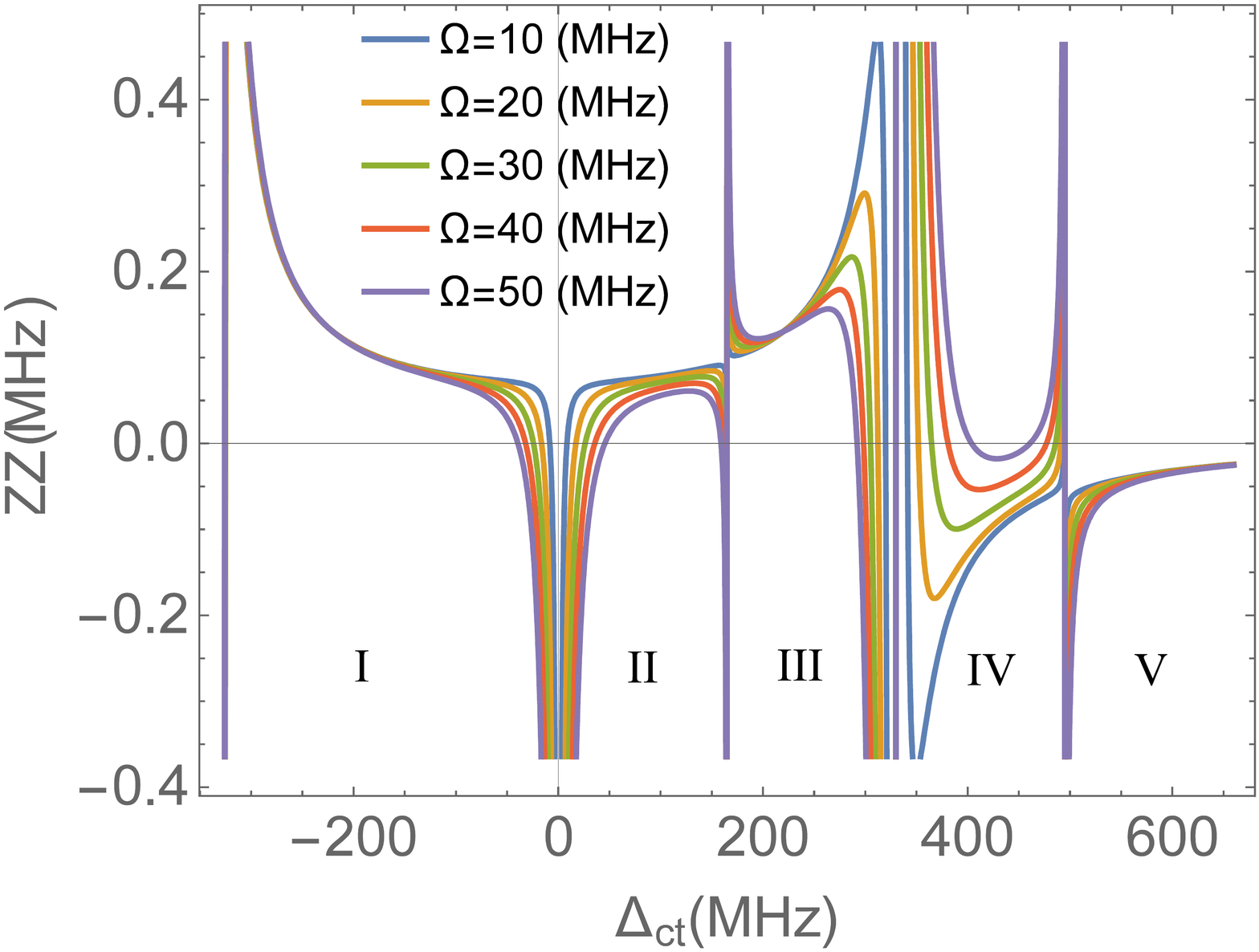}%
}
\subfloat[\label{subfig:CREffHam-ZXZZRatioFuncOfDriveSarahsParams}]{%
\includegraphics[scale=0.15]{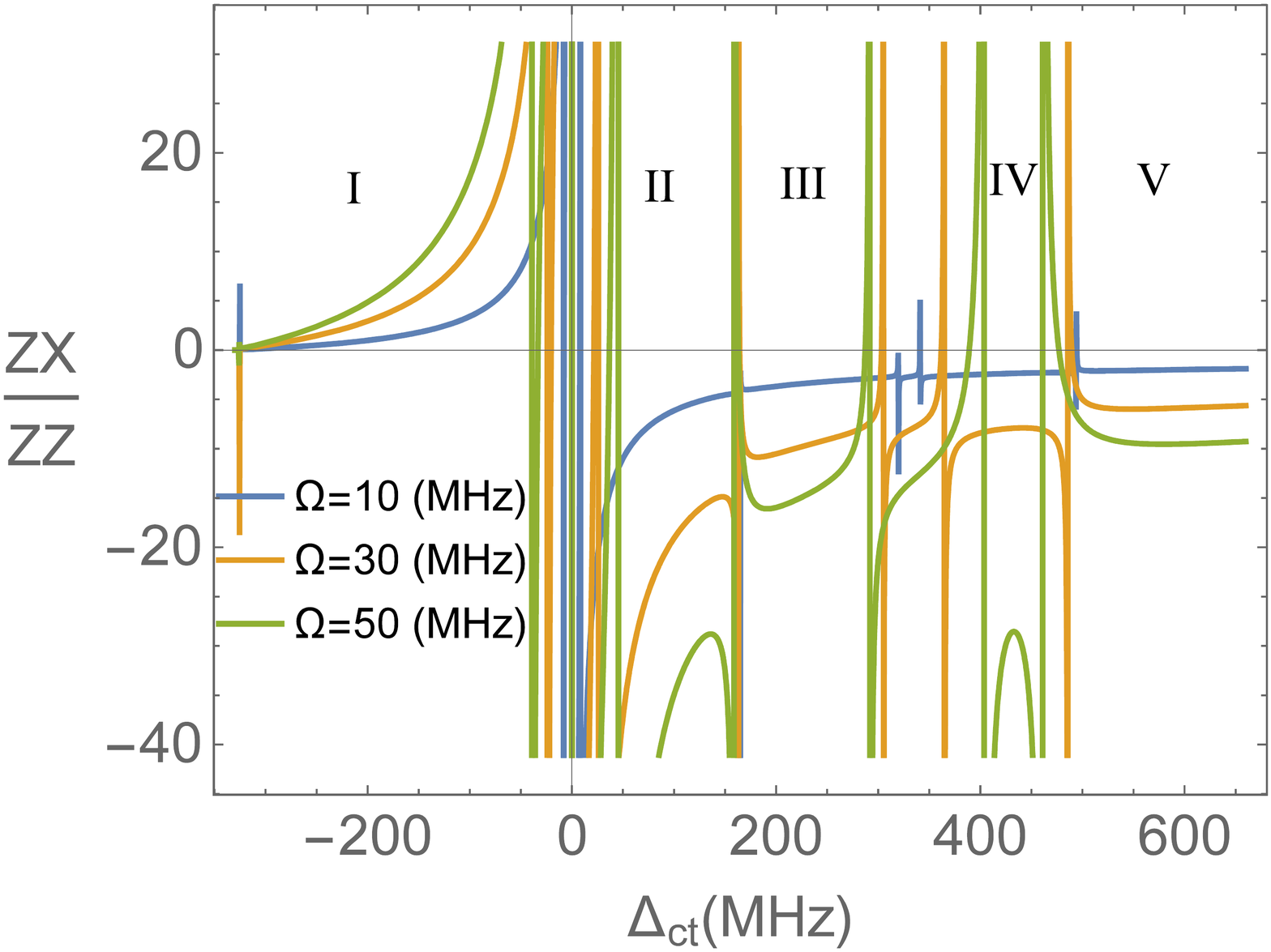}%
}
\caption{Two-qubit rates as a function of qubit-qubit detuning $\Delta_{ct}$ and drive amplitude $\Omega$ with other parameters the same as Table~\ref{tab:LoOrAnal-CRGateParams}. According to the perturbative result, there exist poles at detuning values $\Delta_{ct}=\alpha_t,\ 0,\ -\alpha_c/2, \ -\alpha_c, \ -3\alpha_c/2$, which naturally divides the values for detuning into five regions: I) $-\alpha_t<\Delta_{ct}<0$, II) $0<\Delta_{ct}<-\alpha_c/2$, III) $-\alpha_c/2<\Delta_{ct}<-\alpha_c$, IV) $-\alpha_c<\Delta_{ct}<-3\alpha_c/2$ and V) $-3\alpha_c/2<\Delta_{ct}<-2\alpha_c$.  a) $IX$, b) $IZ$, c) $ZI$, d) $ZX$, e) $ZZ$ and f) $ZX/ZZ$ ratio.}
\label{fig:NeOrAnal-TwoQubitRatesFuncOfDrDetSarahsParams}
\end{figure}

Next, in Fig.~\ref{fig:NeOrAnal-TwoQubitRatesFuncOfDrDetSarahsParams}, we take a closer look into the gate parameters as a function of both drive amplitude $\Omega$ and qubit-qubit detuning $\Delta_{ct}$. The aformentioned detuning regions can be clearly distinguished with their distinct behavior. Note that we expect the perturbation to be valid close to the middle of each region and away from the poles. In terms of achieving the largest ZX rate (fastest gate), we find from Fig.~\ref{subfig:CREffHam-ZX2DPlot4thConstDriveContoursSarahsParams} that the best operating point is region III for $\Delta_{ct}\approx -0.61\alpha_c \ (200 \ \text{MHz})$. This is further confirmed in Appendix~\ref{App:Saturation}, where we numerically calculate $ZX$ rate following the semi-analytical method of Ref.~\cite{Tripathi_Operation_2019}. Up to medium drive power, region II results in a $ZX$ rate that is comparable to region III (blue, yellow and green curves in Fig.~\ref{subfig:CREffHam-ZX2DPlot4thConstDriveContoursSarahsParams}), while saturates to a smaller maximum rate of approximately $0.6J$ (See also Appendix~\ref{App:Saturation}). In terms of achieving the lowest $ZZ$ rate, however, region II has an important advantage, where the static $ZZ$ is comparably small and increasing the drive seems to further decrease the rate as shown in Fig.~\ref{subfig:CREffHam-ZZ2DPlot4thConstDriveContoursSarahsParams}. On the other hand, region III has a larger static $ZZ$ to begin with and depending on the detuning can exhibit distinct dependences on the drive: i) close to the pole at $-\alpha_c/2$, $ZZ$ rate is slightly increased, ii) in the middle close to $\Delta_{ct}\approx -0.61\alpha_c$, the rate becomes insensitive to drive and iii) close to the pole at $-\alpha_c$, the rate becomes substantially large. The ratio of the \textit{desired} $ZX$ rate to the \textit{unwanted} $ZZ$ rate is a heuristic measure for finding candidate detuning spots for achieving low coherent error for the CR gate. This is shown in Fig.~\ref{subfig:CREffHam-ZXZZRatioFuncOfDriveSarahsParams}, where we find $ZX/ZZ$ is maximized in the middle of regions II, I and III, respectively, and hence would expect to get reasonable two-qubit coherent error. This intuitive understanding will be confirmed in Sec.~\ref{Sec:CREcho}, where we quantify the gate error with a CR echo pulse sequence.    

\begin{figure}[t!]
\centering
\subfloat[\label{subfig:CREffHam-ZXCompHilbertSpCutOffSarahsParams}]{%
\includegraphics[scale=0.225]{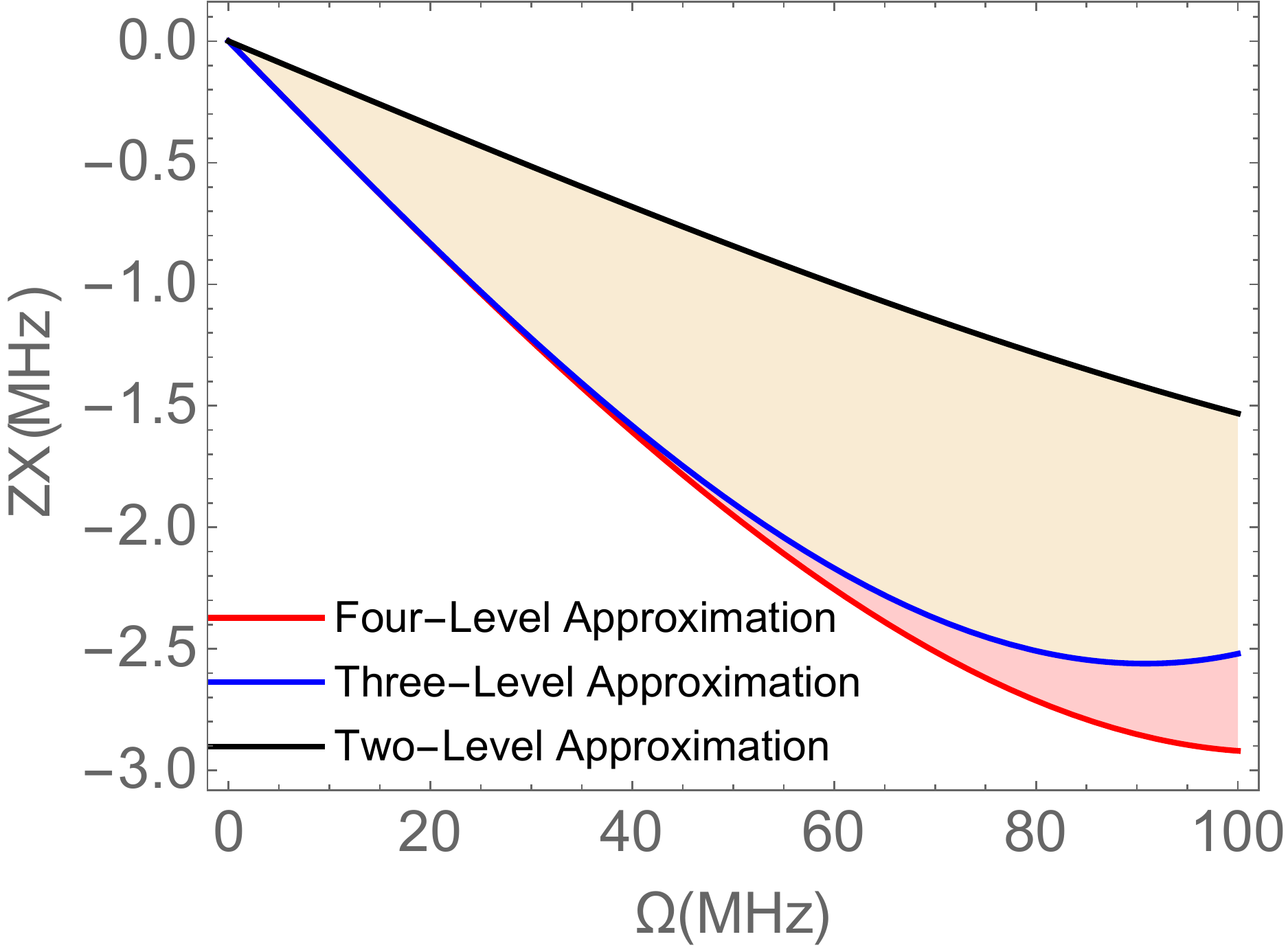}%
}
\subfloat[\label{subfig:CREffHam-ZZCompHilbertSpCutOffSarahsParams}]{%
\includegraphics[scale=0.225]{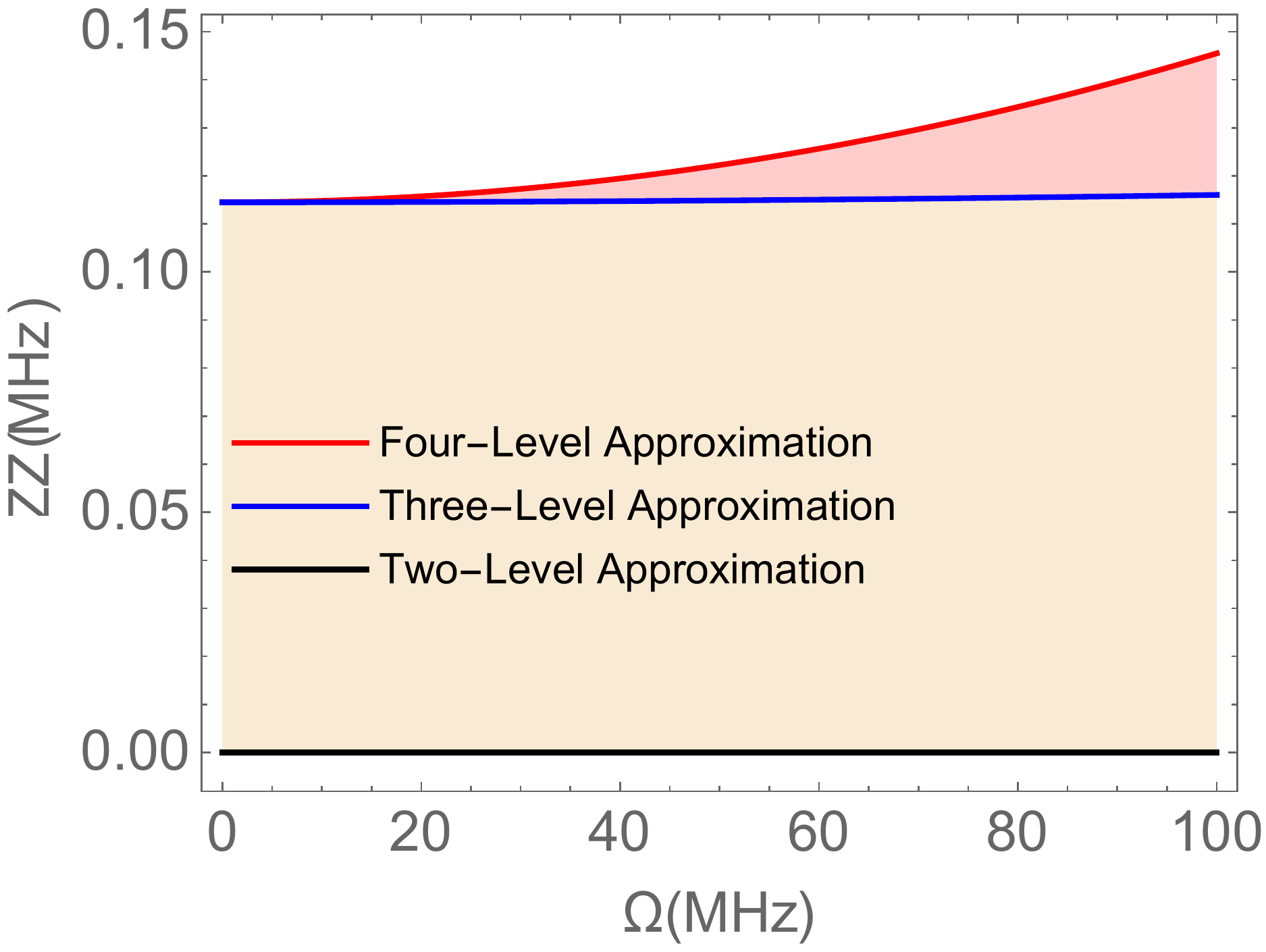}%
}
\caption{Dependence of gate parameters on Hilbert space cut-off: a) $ZX$ rate and b) $ZZ$ rate. The beige and the pink area represents the correction in going from two to three and three to four energy states, respectively.}
\label{fig:NeOrAnal-HilSpCutOff}
\end{figure}

Lastly, we revisit the impact of Hilbert space cut-off number for the qubits on the gate parameters in Fig.~\ref{fig:NeOrAnal-HilSpCutOff}. The results presented so far is with the assumption of keeping the first four energy eigenstates for each qubit. From a two- (solid black) to a three-level model (solid blue), one observes a large correction for both $ZX$ and $ZZ$ rates. In the case of $ZZ$, the two-level model is unable to predict a non-zero rate. Adding the fourth level (solid red) brings corrections that become significant at higher drive amplitudes.

\section{Effective Hamiltonian and gate fidelity with Echo pulse}
\label{Sec:CREcho}

In this section, we study the CR gate operation combined with an echo pulse sequence introduced in Refs.~\cite{Corcoles_Process_2013, Sheldon_Procedure_2016}. The echo pulse is designed to suppress the unwanted gate parameters, while leaving the intended $ZX$ term intact. Figure~\ref{fig:CREchoPulseSequence} demonstrates the pulse sequence that is applied on the control qubit. In Sec.~\ref{SubSec:CREchoEffHam}, we derive the resulting time-evolution operator and the corresponding effective echo Hamiltonian. In Sec.~\ref{SubSec:GateFid}, we provide an estimate for the coherent error of CR echo and characterize optimal parameters (qubit-qubit detuning in particular). Furthermore, in Sec.~\ref{SubSec:NonLocInv}, we discuss local equivalence of two-qubit operations \cite{Zhang_Geometric_2003} and provide an estimate for the non-local CR echo fidelity \cite{Watts_Optimizing_2015} and entangling power \cite{Zanardi_Entangling_2000} in terms of Makhlin invariants \cite{Makhlin_Nonlocal_2002}. 

\subsection{Effective echo Hamiltonian}
\label{SubSec:CREchoEffHam}

The CR echo sequence consists of two CR tones with flipped amplitudes accompanied with intermediate $\pi$ rotations of the control qubit around its $X$ axis as shown schematically in Fig.~\ref{fig:CREchoPulseSequence}. In the following, we derive analytical expressions for the time evolution operator of the echo sequence as well as an approximate CR echo Hamiltonian. 

The time-evolution operator with the CR echo pulse can be expressed as
\begin{align}
\begin{split}
\hat{U}_{\text{ech}}(\Omega,\tau_p) \equiv \hat{R}_{X}(-\pi)\hat{U}_{\text{CR,eff}}(-\Omega,\tau_p)\hat{R}_{X}(\pi)\hat{U}_{\text{CR,eff}}(\Omega,\tau_p)\\
=e^{+i\frac{\pi}{2}\hat{X}\hat{I}} e^{-i\HO_{\text{CR,eff}}(-\Omega)\tau_p} e^{-i\frac{\pi}{2}\hat{X}\hat{I}} e^{-i\HO_{\text{CR,eff}}(+\Omega)\tau_p},
\end{split}
\label{eqn:CREcho-Def of U_echo}
\end{align}
with $\tau_p$ being the half-CR pulse duration and $\HO_{\text{CR,eff}}$ has the same form as in Eq.~(\ref{eqn:CREffHam-H_CR,eff}). In order to implement our desired $ZX$ operation of the form $\hat{U}_{\text{ide}}\equiv \exp[-i\pi \hat{Z}\hat{X}/4]$, $\tau_p$ needs to be set as 
\begin{align}
\omega_{zx}\times(2\tau_p) = \frac{\pi}{2} \quad (\text{mod} \ 2\pi)\;.
\label{eqn:CREcho-wzx in terms of tau_p}
\end{align}
The perturbative result for the two-qubit rates revealed that $ZX$ and $IX$ are odd functions of the drive amplitude $\Omega$, while $ZZ$, $ZI$ and $IZ$ rates are even functions. Hence, flipping the drive amplitude in the echo sequence yields
\begin{subequations}
\begin{align}
\begin{split}
\HO_{\text{CR,eff}}(+\Omega)&=\omega_{ix}\frac{\hat{I}\hat{X}}{2}+\omega_{iz}\frac{\hat{I}\hat{Z}}{2}+\omega_{zi}\frac{\hat{Z}\hat{I}}{2}\\
&+\omega_{zx}\frac{\hat{Z}\hat{X}}{2}+\omega_{zz}\frac{\hat{Z}\hat{Z}}{2} \;,
\end{split}
\label{eqn:CREcho-H_I,eff(W)}
\end{align}
\begin{align}
\begin{split}
\HO_{\text{CR,eff}}(-\Omega)&=-\omega_{ix}\frac{\hat{I}\hat{X}}{2}+\omega_{iz}\frac{\hat{I}\hat{Z}}{2}+\omega_{zi}\frac{\hat{Z}\hat{I}}{2}\\
&-\omega_{zx}\frac{\hat{Z}\hat{X}}{2}+\omega_{zz}\frac{\hat{Z}\hat{Z}}{2} \;.
\end{split}
\label{eqn:CREcho-H_I,eff(-W)}
\end{align}

\begin{figure}
\centering
\includegraphics[scale=0.33]{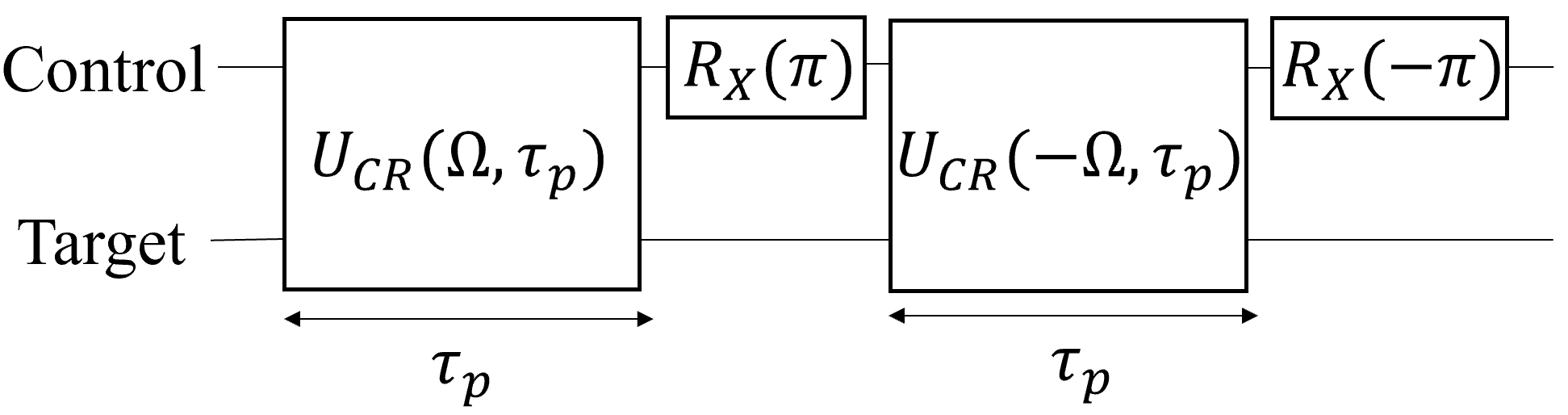}
\caption{Schematic of the echo pulse sequence, where the original CR pulse is broken into two parts of duration $\tau_p$ with positive and negative amplitudes, i.e. a $\pi$ phase difference in the CR tone, accompanied by two $\pi$ rotations around the $X$ axis in between. Equation~\ref{eqn:CREcho-Def of U_echo} shows the time-evolution operator under this pulse sequence.}
\label{fig:CREchoPulseSequence}
\end{figure}

Next, we obtain an explicit solution for the time-evolution operator by replacing Eqs.~(\ref{eqn:CREcho-H_I,eff(W)}--\ref{eqn:CREcho-H_I,eff(-W)}) into Eq.~(\ref{eqn:CREcho-Def of U_echo}). We find that the echo pulse removes $ZI$, $IX$ and $ZZ$ interaction, but in turn induces effective echoed $IY$, $IZ$ and $ZX$ interactions as
\end{subequations}
\begin{align}
\hat{U}_{\text{ech}}(\Omega,\tau_p)=u_{ii}\hat{I}\hat{I}+u_{iy}\hat{I}\hat{Y}+u_{iz}\hat{I}\hat{Z}+u_{zx}\hat{Z}\hat{X} \;.
\label{eqn:CREcho-Uecho(W,tau)}
\end{align}
The corresponding coefficients $u_{ii}$, $u_{iy}$, $u_{iz}$ and $u_{zx}$ can be found as
\begin{subequations}
\begin{align}
\begin{split}
&u_{ii} \equiv \cos \Big(\frac{1}{2}\omega_{+} \tau_p \Big)\cos \Big(\frac{1}{2}\omega_{-} \tau_p \Big)\\
&+\frac{[\omega_{ix}^2-\omega_{iz}^2-\omega_{zx}^2+\omega_{zz}^2]}{\omega_{+}\omega_{-}} \\
& \times \sin\Big(\frac{1}{2}\omega_{+}\tau_p\Big)\sin\Big(\frac{1}{2}\omega_{-}\tau_p\Big)\; ,
\end{split}
\label{eqn:CREcho-Def of uii}
\end{align}
\begin{align}
\begin{split}
u_{iy} &\equiv \frac{2i(\omega_{zx}\omega_{zz}-\omega_{ix}\omega_{iz})}{\omega_{+}\omega_{-}} \\
& \times \sin \Big(\frac{1}{2}\omega_{+}\tau_p \Big)\sin \Big(\frac{1}{2}\omega_{-}\tau_p \Big)\; ,
\end{split}
\label{eqn:CREcho-Def of uiy}
\end{align}
\begin{align}
\begin{split}
u_{iz} &\equiv i\frac{\omega_{zz}-\omega_{iz}}{\omega_{-}}\cos \Big(\frac{1}{2}\omega_{+} \tau_p \Big)\sin \Big(\frac{1}{2}\omega_{-} \tau_p \Big)\\
&-i\frac{\omega_{zz}+\omega_{iz}}{\omega_{+}}\sin \Big(\frac{1}{2}\omega_{+} \tau_p \Big)\cos \Big(\frac{1}{2}\omega_{-} \tau_p \Big)\; ,
\end{split}
\label{eqn:CREcho-Def of uiz}
\end{align}
\begin{align}
\begin{split}
u_{zx}&\equiv i\frac{\omega_{ix}-\omega_{zx}}{\omega_{-}}\cos \Big(\frac{1}{2}\omega_{+} \tau_p \Big)\sin \Big(\frac{1}{2}\omega_{-} \tau_p \Big) \\
&-i\frac{\omega_{ix}+\omega_{zx}}{\omega_{+}}\sin \Big(\frac{1}{2}\omega_{+} \tau_p \Big)\cos \Big(\frac{1}{2}\omega_{-} \tau_p \Big)\;.
\end{split}
\label{eqn:CREcho-Def of uzx}
\end{align}
According to Eqs.~(\ref{eqn:CREcho-Def of uii}--\ref{eqn:CREcho-Def of uzx}), the echo dynamics can be understood as a beating between two collective two-qubit frequencies $\omega_{\pm}$ that are found in term of the bare rates as
\begin{align}
\omega_{+} \equiv \sqrt{(\omega_{zx}+\omega_{ix})^2+(\omega_{iz}+\omega_{zz})^2} \;,
\label{eqn:CREcho-Def of w+}\\
\omega_{-}\equiv \sqrt{(\omega_{zx}-\omega_{ix})^2+(\omega_{iz}-\omega_{zz})^2} \;.
\label{eqn:CREcho-Def of w-}
\end{align}
\end{subequations}
These collective frequencies slightly deviate from the intended $ZX$ frequency depending on the strength of unwanted terms in the effective Hamiltonian ($\omega_{ix}$, $\omega_{zz}$ and $\omega_{iz}$). In Sec.~\ref{SubSec:GateFid}, we compare the explicit solutions~(\ref{eqn:CREcho-Uecho(W,tau)}--\ref{eqn:CREcho-Def of w-}) to the ideal CR unitary and provide estimate for the error.

\begin{figure}
\centering
\subfloat[\label{subfig:CREffHam-IYEchoFuncOfDrDetSarahsParams}]{%
\includegraphics[scale=0.154]{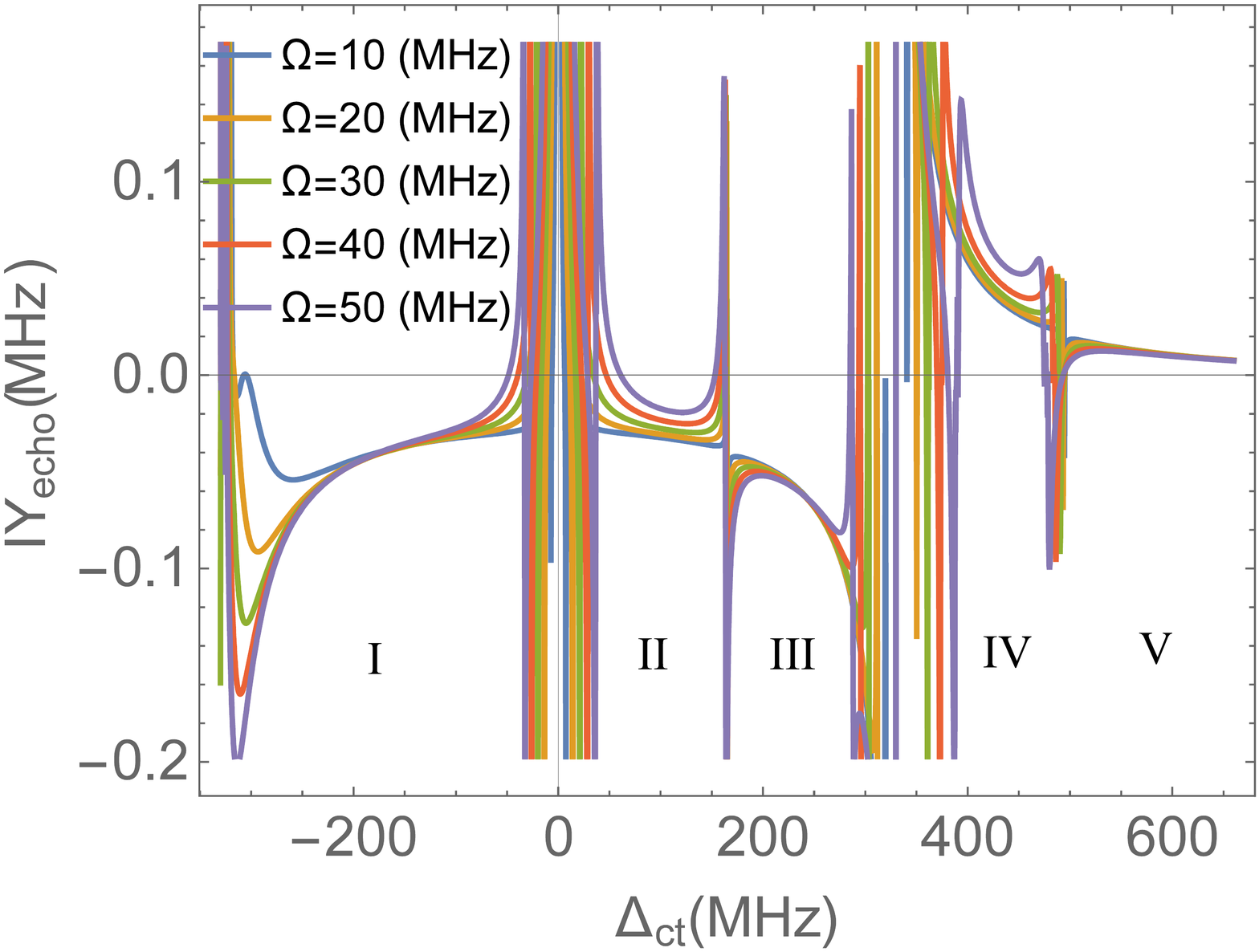}%
}
\subfloat[\label{subfig:CREffHam-IZEchoFuncOfDrDetSarahsParams}]{%
\includegraphics[scale=0.157]{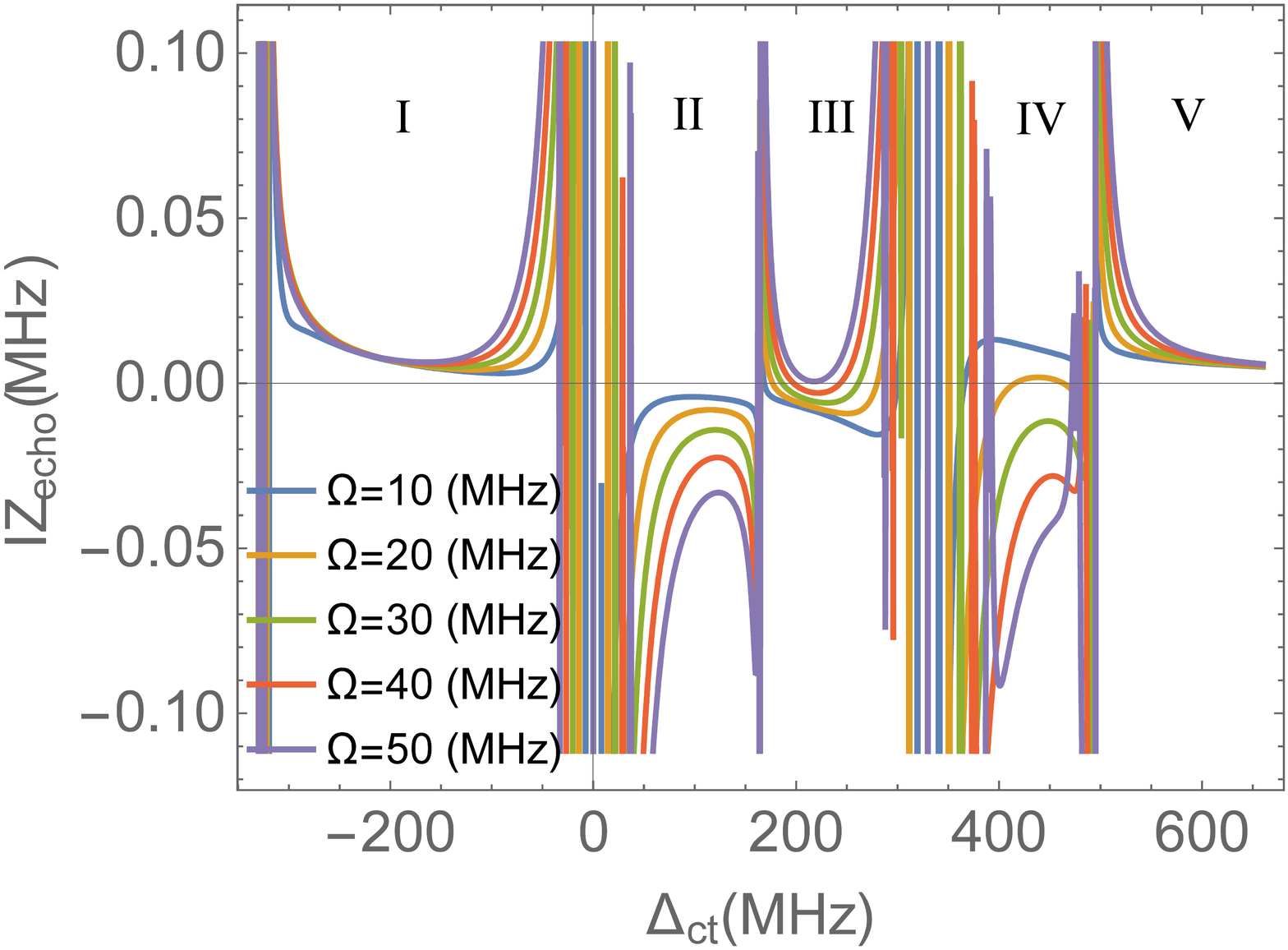}%
}\\
\subfloat[\label{subfig:CREffHam-ZXEchoFuncOfDrDetSarahsParams}]{%
\includegraphics[scale=0.155]{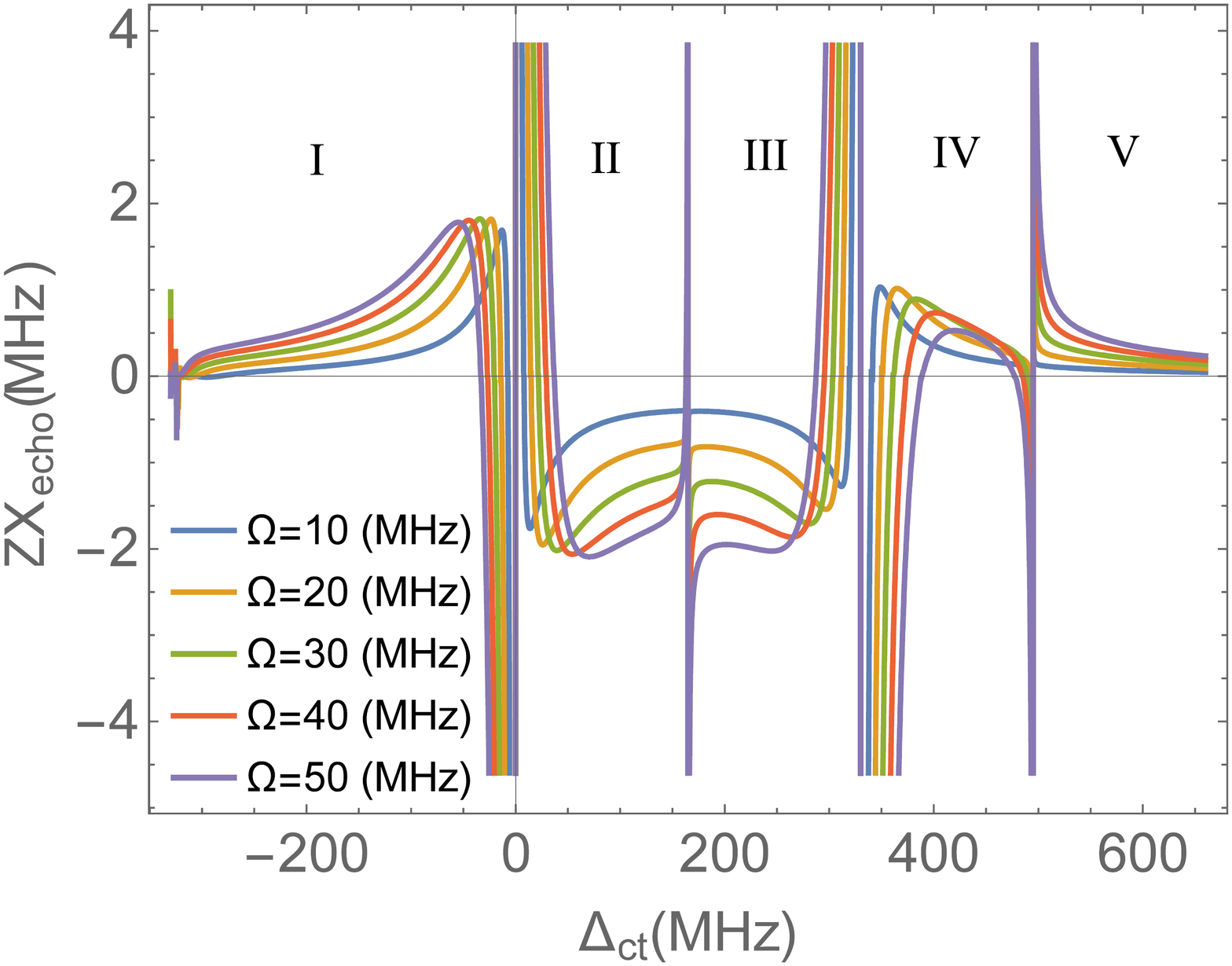}%
}
\caption{Echoed gate parameters based on Eqs.~(\ref{eqn:CREcho-Def of uii}--\ref{eqn:CREcho-Def of uzx}) and~(\ref{eqn:CREcho-Def of w_ii,echo}--\ref{eqn:CREcho-Def of w_zx,echo}). a) Echoed $IY$, b) echoed $IZ$ and c) echoed $ZX$. System parameters are the same as those of Fig.~\ref{fig:NeOrAnal-TwoQubitRatesFuncOfDrDetSarahsParams} and Table~\ref{tab:LoOrAnal-CRGateParams}. Note that the splitting of the poles around $\Delta_{ct}=0$ and $\Delta_{ct}=-\alpha_c$ is an artifact of using the perturbative result for $ZX$ rate (Fig.~\ref{subfig:CREffHam-ZX2DPlot4thConstDriveContoursSarahsParams}) to set the pulse time $\tau_p$ according to Eq.~(\ref{eqn:CREcho-wzx in terms of tau_p}). Hence, the result is valid only in the middle of each region.}
\label{fig:CREcho-EchoedRates}
\end{figure}

To visualize the impact of echo pulse on the two-qubit rates, we can obtain an approximate effective echo Hamiltonian by writing 
\begin{align}
\hat{U}_{\text{ech}}(\Omega,\tau_p)\approx \exp[-i\HO_{\text{ech}}(\Omega,\tau_p)(2\tau_p)] \;.
\label{eqn:CREcho-Def of H_ech}
\end{align}
Equation~(\ref{eqn:CREcho-Def of H_ech}) is true under the assumptions that firstly the single qubit rotations happen on a time scale much smaller than $\tau_p$ of the CR tone, and secondly that all pulses have constant amplitude and transient effects are negligible. Substituting Eq.~(\ref{eqn:CREcho-Uecho(W,tau)}) for $\hat{U}_{\text{ech}}$ into Eq.~(\ref{eqn:CREcho-Def of H_ech}) we find
\begin{align}
\begin{split}
\HO_{\text{ech}}(\Omega,\tau_p) &\approx \omega_{ii,\text{ech}}\frac{\hat{I}\hat{I}}{2}+\omega_{iy,\text{ech}}\frac{\hat{I}\hat{Y}}{2}\\
&+\omega_{iz,\text{ech}}\frac{\hat{I}\hat{Z}}{2}+\omega_{zx,\text{ech}}\frac{\hat{Z}\hat{X}}{2}\;,
\end{split}
\label{eqn:CREcho-H_echo}
\end{align}
where the echoed gate parameters are obtained as
\begin{subequations}
\begin{align}
&\omega_{ii,\text{ech}}\equiv \frac{i}{2\tau_p}\left[\ln(u_{ii}+u)+\ln(u_{ii}-u)\right] \;,
\label{eqn:CREcho-Def of w_ii,echo}\\
&\omega_{iy,\text{ech}}\equiv \frac{i}{2\tau_p}\frac{u_{iy}}{u}\left[\ln(u_{ii}+u)-\ln(u_{ii}-u)\right] \;,
\label{eqn:CREcho-Def of w_iy,echo}\\
&\omega_{iz,\text{ech}}\equiv \frac{i}{2\tau_p}\frac{u_{iz}}{u}\left[\ln(u_{ii}+u)-\ln(u_{ii}-u)\right] \;,
\label{eqn:CREcho-Def of w_iz,echo}\\
&\omega_{zx,\text{ech}}\equiv \frac{i}{2\tau_p}\frac{u_{zx}}{u}\left[\ln(u_{ii}+u)-\ln(u_{ii}-u)\right] \;,
\label{eqn:CREcho-Def of w_zx,echo}
\end{align}
\end{subequations}
with $u\equiv\sqrt{u_{iy}^2+u_{iz}^2+u_{zx}^2}$. 

The resulting echoed gate parameters are studied further in Fig.~\ref{fig:CREcho-EchoedRates}. Comparing Figs.~\ref{subfig:CREffHam-ZXEchoFuncOfDrDetSarahsParams} and~\ref{subfig:CREffHam-ZX2DPlot4thConstDriveContoursSarahsParams}, we observe that the echoed and the bare $ZX$ rates are more or less equal, indicating that the pulse sequence barely touches the intended $ZX$ rate. On top of this, one finds residual echoed $IY$ and $IZ$ rates. Looking at Fig.~\ref{subfig:CREffHam-IYEchoFuncOfDrDetSarahsParams} we find that, in terms of lowest $IY_{\text{ech}}$, regions II, I and III provide the most optimal detuning values, respectively. On the other hand, Fig.~\ref{subfig:CREffHam-IZEchoFuncOfDrDetSarahsParams} suggests that smallest $IZ_{\text{ech}}$ is achieved in regions I, III and II, respectively. All in all, we expect to observe reasonable coherent error in the middle of each of those aforementioned regions. This is studied in more detail in the following subsection.  	

\subsection{Gate fidelity}
\label{SubSec:GateFid}

We define the CR echo Gate fidelity as the overlap between $\hat{U}_{\text{ech}}$ that is implemented via CR echo in Eq.~(\ref{eqn:CREcho-Uecho(W,tau)}) and the ideal $ZX$ operation $\hat{U}_{\text{ide}}$ as \cite{Pedersen_Fidelity_2007}
\begin{figure}[h!]
\centering
\includegraphics[scale=0.26]{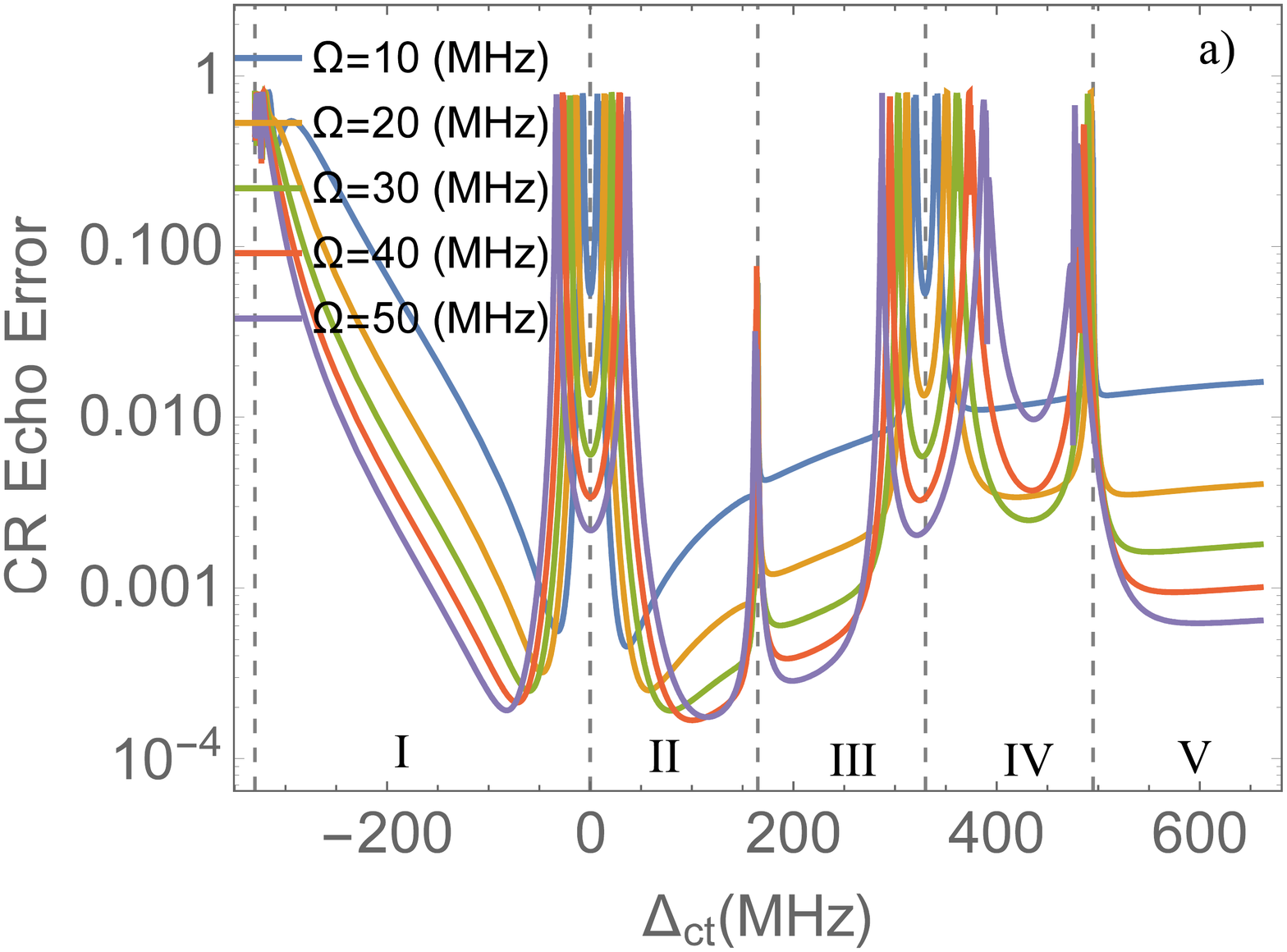}\\
\includegraphics[scale=0.26]{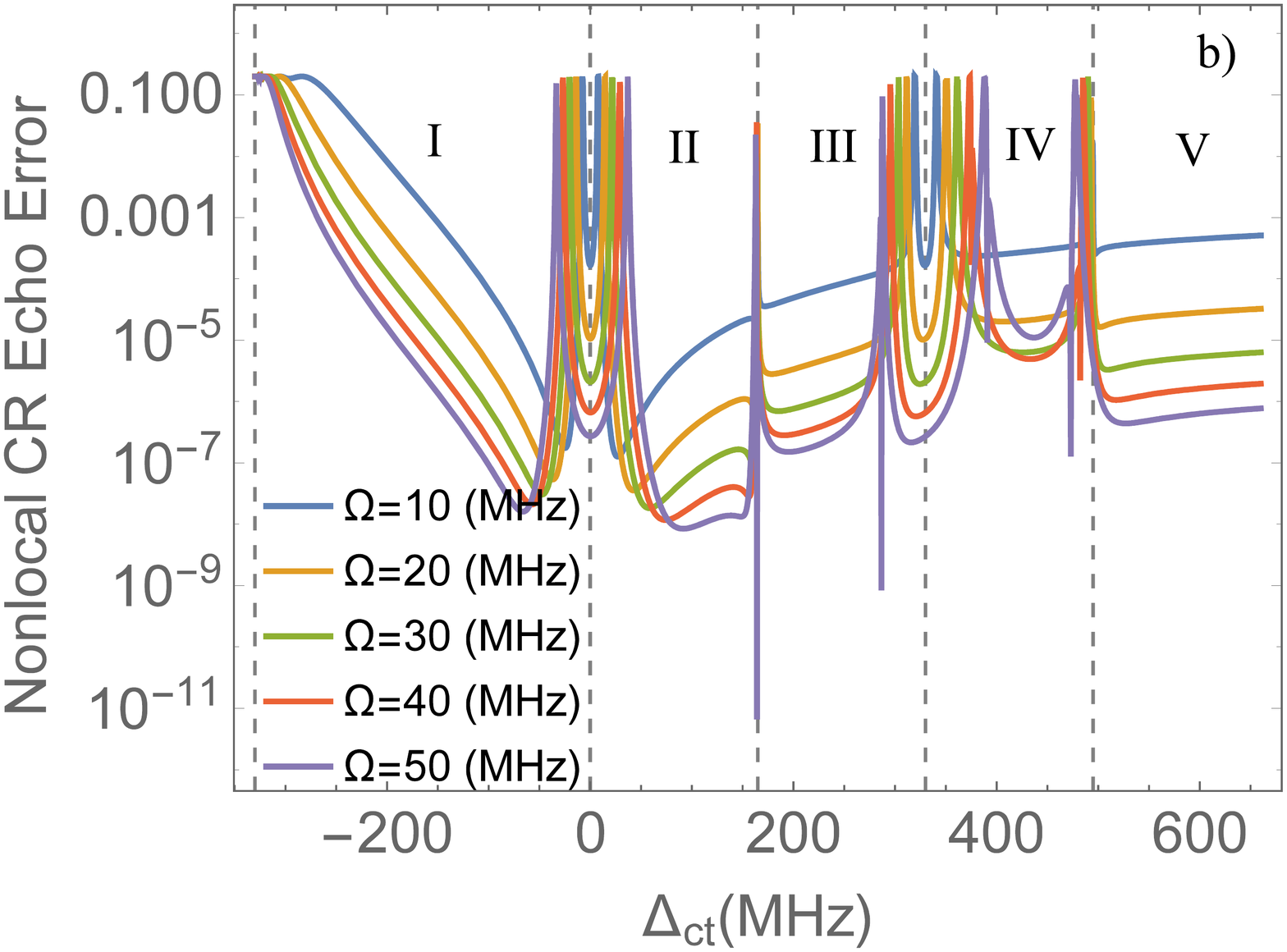}\\
\includegraphics[scale=0.26]{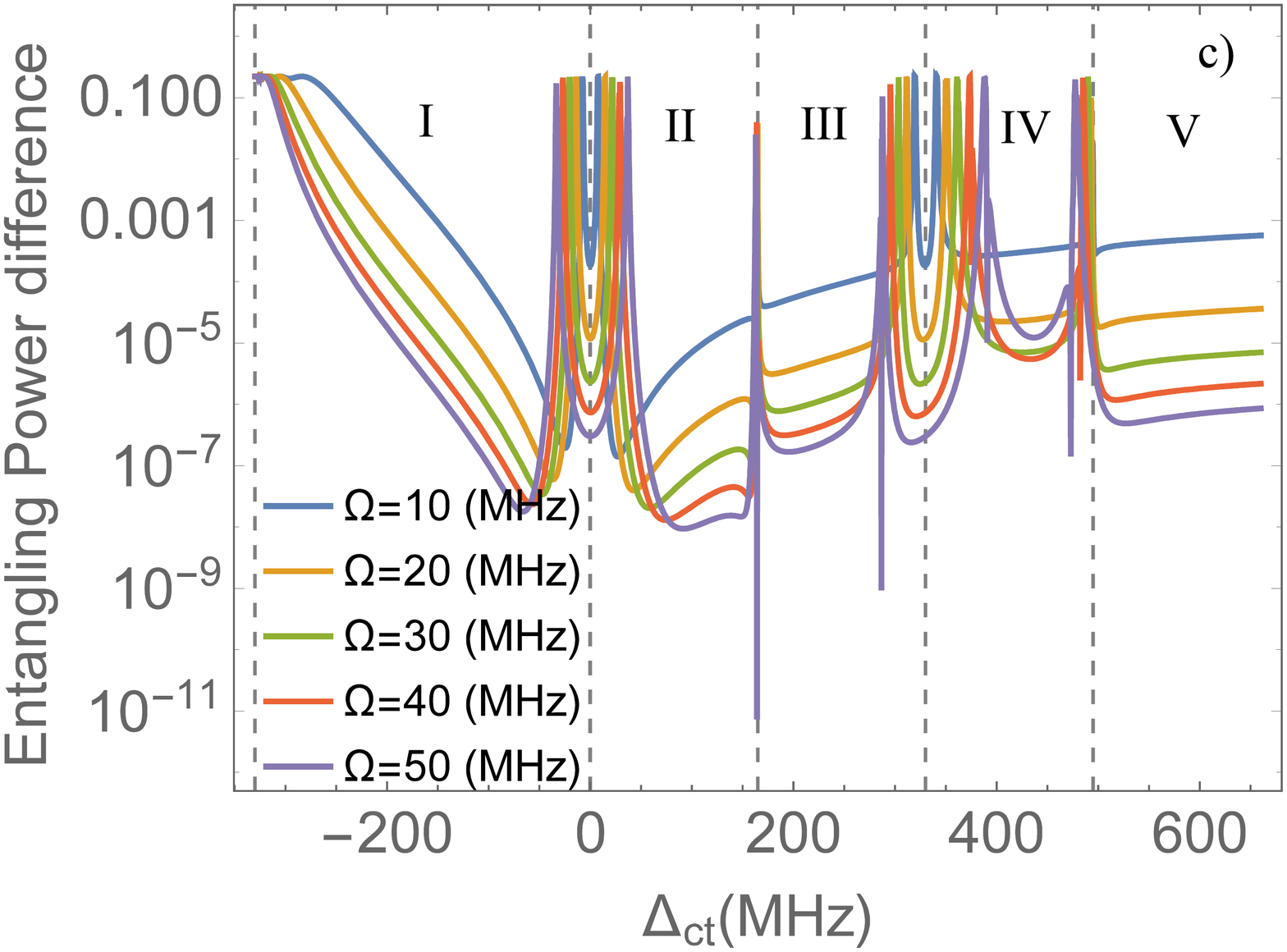}
\caption{a) CR echo error obtained from Eqs.~(\ref{eqn:CREcho-Def of Fid}--\ref{eqn:CREcho-Def of GateError}), b) Nonlocal CR echo error obtained from Eq.~(\ref{eqn:NonLocInv-Estimate for E_ech^(nl)}), c) entangling power of CR echo found from Eq.~(\ref{eqn:NonLocInv-[e_p(U_ide)-e_p(U_ech)]}) as a function of qubit-qubit detuning $\Delta_{ct}$ and drive amplitude $\Omega$. Other parameters  are the same as those in Fig.~\ref{fig:NeOrAnal-TwoQubitRatesFuncOfDrDetSarahsParams} and Table~\ref{tab:LoOrAnal-CRGateParams}. The dashed vertical lines separates the five regions of qubit-qubit detuning. Note that the splitting of the poles around $\Delta_{ct}=0$ and $\Delta_{ct}=-\alpha_c$ is an artifact of using the perturbative result for $ZX$ rate (Fig.~\ref{subfig:CREffHam-ZX2DPlot4thConstDriveContoursSarahsParams}) to set the pulse time $\tau_p$ based on Eq.~(\ref{eqn:CREcho-wzx in terms of tau_p}). Hence, the result is valid only in the middle of each region.}
\label{fig:CREcho-GateError}
\end{figure}
\begin{align}
F(\hat{U}_{\text{ech}},\hat{U}_{\text{ide}})\equiv \frac{\Tr\left(\hat{U}_{\text{ech}}^{\dag}\hat{U}_{\text{ech}}\right)}{d(d+1)}+\frac{\left|\Tr\left(\hat{U}_{\text{ech}}^{\dag}\hat{U}_{\text{ide}}\right)\right|^2}{d(d+1)},
\label{eqn:CREcho-Def of Fid}
\end{align}
where $d=4$ is the dimension of the two-qubit Hilbert space. Two-qubit gate error is hence defined as the distance between the implemented and the ideal time-evolution operators and is expressed in terms of the fidelity measure~(\ref{eqn:CREcho-Def of Fid}) as 	 
\begin{align}
E_{\text{ech}} \equiv 1-F(\hat{U}_{\text{ech}},\hat{U}_{\text{ide}}) \;.
\label{eqn:CREcho-Def of GateError}
\end{align}

Figure~\ref{fig:CREcho-GateError}a shows the resulting gate error as a function of both qubit-qubit detuning and drive. As expected, we observe a suppression of gate error in the middle of each region. Generally speaking, increasing the drive amplitude improves the coherent error and widens the optimal interval in certain regions as the ratio of $ZX$ over other unwanted terms is enhanced. However, this continues until a certain drive power in which the $ZX$ rate start to saturate. Furthermore, we find that the most optimal qubit-qubit detuning lies in the middle of region II ($\Delta_{ct}\approx 100 \ \text{MHz}\approx -0.30 \alpha_c$), for which an error of the order of $10^{-4}$ is predicted. A similar error is also observed in region I, centered around $\Delta_{ct}\approx -100 \ \text{MHz} \approx 0.30 \alpha_c$, with the caveat that the optimal region is much narrower and the error increases significantly as the detuning gets closer to the pole at $\Delta_{ct}=\alpha_t$. This increase in error for larger negative detuning can be traced back to two simultaneous detrimental effects. Firstly, the $ZZ$ rate is noticeably enhanced as a result of resonance between states $\ket{\psi_{11}}$ and $\ket{\psi_{02}}$ that occurs at $\Delta_{ct}=\alpha_t$ (Fig.~\ref{subfig:CREffHam-ZZ2DPlot4thConstDriveContoursSarahsParams}), and secondly the $ZX$ rate is noticeably suppressed outside the detuning interval $0<\Delta_{ct}<-\alpha_c$ (Fig.~\ref{subfig:CREffHam-ZX2DPlot4thConstDriveContoursSarahsParams}). Region III ($\Delta_{ct}\approx 200 \ \text{MHz}\approx -0.61 \alpha_c$) provides a slightly larger error compared to regions II and I, but with the important advantage that the optimal detuning interval is wide for any drive amplitude. This flexibility is crucial for fixed frequency transmons, for which there is not a precise control over the fabricated qubit frequency.

\subsection{Non-local gate fidelity and entangling power}
\label{SubSec:NonLocInv}

Two-qubit unitary operators can be categorized in terms of local equivalence classes, where two operators belong to the same class if they can be transformed into one another merely by single-qubit operations. In this section, we provide the error budget of CR echo sequence in terms local and non-local contributions. The local contribution can in principle be corrected by designing a series of single-qubit operations. The non-local part, however, provides a lower bound for the optimal CR echo error and shows how close the implemented unitary is to a perfect CNOT entangler \cite{Zanardi_Entangling_2000, Zhang_Geometric_2003, Zhang_Minimum_2004, Rezakhani_Characterization_2004, Watts_Optimizing_2015}. 

An arbitrary two-qubit unitary operator is uniquely determined in terms of 15 independent parameters. Based on the above equivalence relation, there are only 3 independent parameters that determine the local equivalence class, known as non-local invariants \cite{Makhlin_Nonlocal_2002, Zhang_Geometric_2003}. In the canonical form, a two-qubit unitary operator can be uniquely represented in terms of its Cartan coordinates $c_x$, $c_y$ and $c_z$ as
\begin{subequations}
\begin{align}
&\hat{U}=\hat{K}_\text{L}\hat{A} \hat{K}_R  \;,
\label{eqn:NonLocInv-Canonical Form of U}\\
&\hat{A}\equiv e^{-\frac{i}{2}\left(c_x\hat{X}_c\hat{X}_t+c_y\hat{Y}_c\hat{Y}_t+c_z\hat{Z}_c\hat{Z}_t\right)} \;,
\label{eqn:NonLocInv-Def of A}
\end{align}
\end{subequations}
with $\hat{K}_L \equiv \hat{L}_c \otimes \hat{L}_t$ and $\hat{K}_R \equiv \hat{R}_c \otimes \hat{R}_t$ acting locally on the Hilbert space of each qubit. 

Calculating the Cartan representation directly can be challenging. Makhlin invariants \cite{Makhlin_Nonlocal_2002}, on the other hand, can be computed more conveniently by rewriting the unitary in the Bell (magic) frame, defined by the unitary change of basis 
\begin{align}
\hat{Q} \equiv \frac{1}{\sqrt{2}}\begin{bmatrix}
1 & 0 & 0 & i\\
0 & i & 1 & 0\\
0 & i & -1 & 0\\
1 & 0 & 0 & -i
\end{bmatrix} \;,
\label{eqn:NonLocInv-Def of Q}
\end{align}
as $\hat{U}_{M,\text{ech}}\equiv\hat{Q}^{\dag}\hat{U}_{\text{ech}}\hat{Q}$. In particular, Makhlin proved that the spectrum of the operator $\hat{M}_{\text{ech}}\equiv\hat{U}_{M,\text{ech}}^{T}\hat{U}_{M,\text{ech}}$ remains invariant under single-qubit operations \cite{Makhlin_Nonlocal_2002}, resulting in an alternative set of invariants as 	
\begin{subequations}
\begin{align}
&g_x \equiv \text{Re}\Big\{\frac{\Tr^2(\hat{M}_{\text{ech}})}{16\text{Det}(\hat{U}_{M,\text{ech}})}\Big\} \;,
\label{eqn:NonLocInv-Def of Makhlin gx}\\
&g_y \equiv \text{Im}\Big\{\frac{\Tr^2(\hat{M}_{\text{ech}})}{16\text{Det}(\hat{U}_{M,\text{ech}})}\Big\} \;,
\label{eqn:NonLocInv-Def of Makhlin gy}\\
&g_z \equiv \frac{\Tr^2(\hat{M}_{\text{ech}})-\Tr(\hat{M}_{\text{ech}}^2)}{4\text{Det}(\hat{U}_{M,\text{ech}})} \;.
\label{eqn:NonLocInv-Def of Makhlin gz}
\end{align}
\end{subequations}
There is a one-to-one correspondence between Makhlin invariants and Cartan coordinates and they uniquely determine the equivalence class of an arbitrary two-qubit operation. For instance, the ideal $ZX$ unitary $\hat{U}_{\text{ide}}$ belong to the CNOT class identified with $g_x=g_y=0$, $g_z=1$ and corresponding Cartan coordinates $c_x=\pi/2$ and $c_y=c_z=0$.  

We define the non-local CR echo fidelity in terms of the overlap between the non-local parts of $\hat{U}_{\text{ech}}$ and $\hat{U}_{\text{ide}}$ [Eqs.~(\ref{eqn:NonLocInv-Canonical Form of U}--\ref{eqn:NonLocInv-Def of A})] as \cite{Watts_Optimizing_2015}
\begin{align}
\begin{split}
F_{\text{ech}}^{\text{(nl)}} \equiv F(\hat{A}_{\text{ech}},\hat{A}_{\text{ide}})&=\frac{\Tr\left(\hat{A}_{\text{ech}}^{\dag}\hat{A}_{\text{ech}}\right)}{d(d+1)}\\
&+\frac{\left|\Tr\left(\hat{A}_{\text{ech}}^{\dag}\hat{A}_{\text{ide}}\right)\right|^2}{d(d+1)}\;.
\end{split}
\label{eqn:NonLocInv-Def of F_ech^{nl}}
\end{align}
Given that the echo pulse produces a unitary evolution that is sufficiently close to the CNOT class in the non-local coordinates, it is possible to derive a rather simple estimate for the non-local CR echo error only in terms of the Makhlin invariants as (See Appendix~\ref{SubApp:NonLocGateFid})  
\begin{align}
E_{\text{ech}}^{(\text{nl})}\equiv 1-F_{\text{ech}}^{(\text{nl})}=\frac{1}{10} (4g_x - g_z +1)+O(\Delta \textbf{c}^4) \;.
\label{eqn:NonLocInv-Estimate for E_ech^(nl)}
\end{align}
This measure is shown in Fig.~\ref{fig:CREcho-GateError}b alongside the regular CR echo error. We observe that the generic behavior is similar in terms of qubit-qubit detuning and drive, and the result indicates that there is room for improvement in all possible detuning values. In particular, in regions II and I, the lower bound on the coherent error can be as small as $10^{-8}$, while in region III it is $10^{-7}$.

Entangling power is another important measure that quantifies the average entanglement that a unitary operator can produce when acting on separable states \cite{Zanardi_Entangling_2000}. The entangling power of a two-qubit unitary operator depends only on its non-local properties and can be written directly in terms of the Cartan coordinates as \cite{Rezakhani_Characterization_2004}
\begin{align}
\begin{split}
e_p(\hat{U})&=\frac{1}{18}\big[3-\cos(2c_x)\cos(2c_y)\\
&-\cos(2c_y)\cos(2c_z)-\cos(2c_z)\cos(2c_x)\big] \;.
\end{split}
\label{eqn:NonLocInv-Def of e_p(U)}
\end{align}
The maximum value of entangling power is $2/9$ and is achieved if and only if the two-qubit unitary operator belongs to the set of \textit{special perfect entanglers} such as CNOT, DCNOT and B classes \cite{Rezakhani_Characterization_2004, Zhang_Minimum_2004}. Given that $\hat{U}_{\text{echo}}$ and $\hat{U}_{\text{ide}}$ are sufficiently close in the Cartan space, a modified measure can be defined as the difference between the entangling powers of the ideal CR and the implemented CR echo unitary operators as
\begin{align}
e_p(\hat{U}_\text{ide})-e_p(\hat{U}_{\text{ech}})=\frac{2}{9}g_x+O(\Delta \textbf{c}^4) \;,
\label{eqn:NonLocInv-[e_p(U_ide)-e_p(U_ech)]}
\end{align}
where we have expressed the result in terms of Makhlin invariant $g_x$ up to $O(\Delta \textbf{c}^4)$ in the Cartan coordinate difference of $\hat{U}_{\text{echo}}$ and $\hat{U}_{\text{ide}}$ (See Appendix~\ref{SubApp:EntPow}). Figure~\ref{fig:CREcho-GateError}c shows the entangling power based on Eq.~(\ref{eqn:NonLocInv-[e_p(U_ide)-e_p(U_ech)]}), where we observe a similar behavior as the non-local error in Fig.~\ref{fig:CREcho-GateError}b. 

All in all, we conclude that there is a correlation between the \textit{regular} gate fidelity, non-local gate fidelity and entangling power of CR echo unitary such that the optimal qubit-qubit detuning spots are more or less the same for all considered measures. We note that finding the exact single qubit rotations that map the local to non-local coordinates, i.e. $\hat{K}_{L,R}$, is in general challenging and beyond the scope of this paper.

\section{Spectator qubits}
\label{Sec:SpecQu}

\begin{figure}
\centering
\subfloat[\label{subfig:SpecQu-ControlSpectatorSchematic}]{%
\includegraphics[scale=0.38]{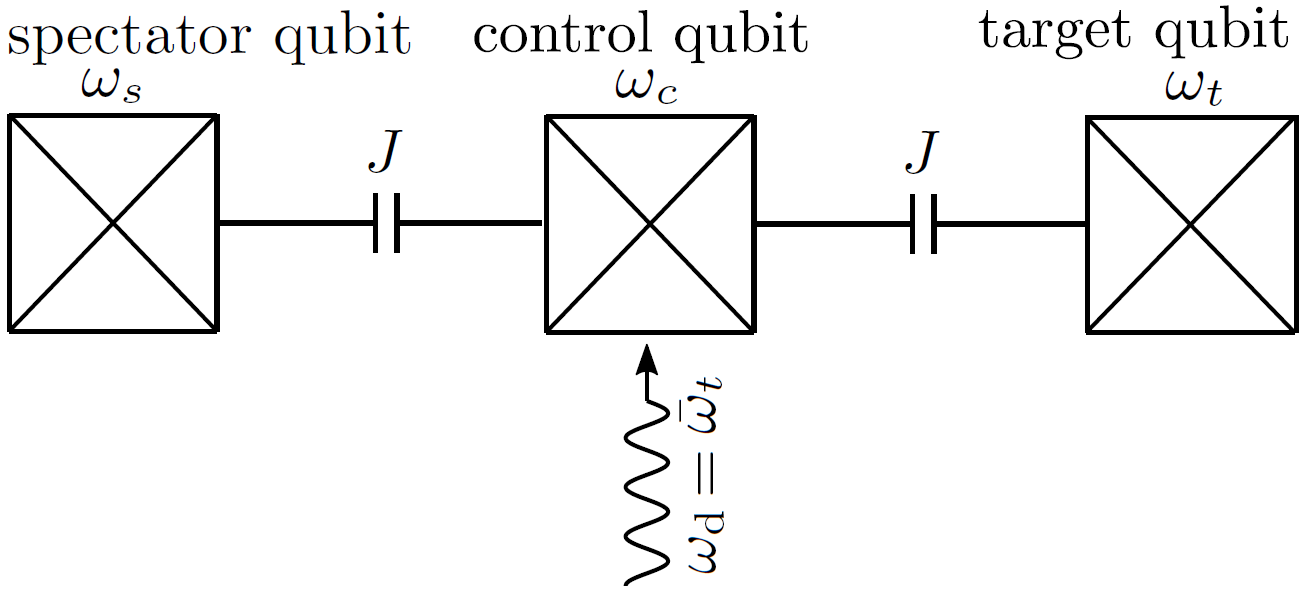}%
} \quad
\subfloat[\label{subfig:SpecQu-TargetSpectatorSchematic}]{%
\includegraphics[scale=0.38]{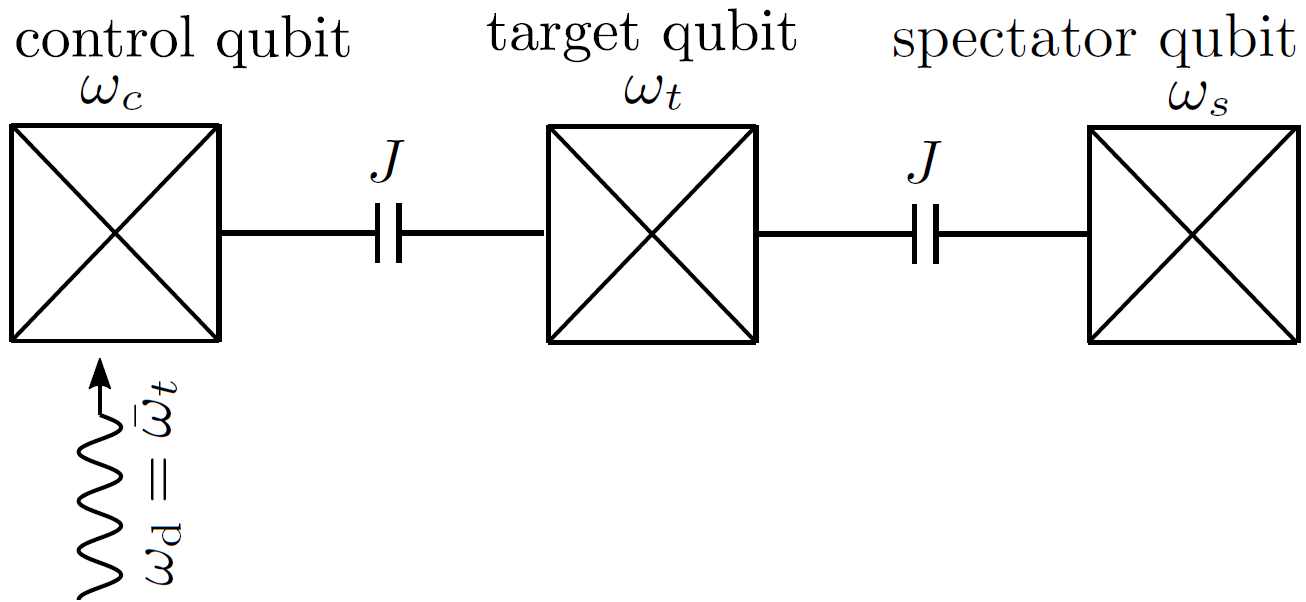}%
}
\caption{Schematic circuit for CR with a) a control spectator, and b) a target spectator qubit. For simplicity, we assume the exchange interaction between all qubit pairs is the same.}
\label{fig:SpectatorQubits}
\end{figure}

So far, we have analyzed the operation of an \textit{isolated} CR gate. In reality, however, both the control and the target qubits may be connected to neighboring qubits in a quantum processor. The goal of this section is to understand how the original two-qubit gate parameters are influenced by the presence of a third spectator qubit. We consider a minimal extension of our original model for CR with either a control or a target spectator qubit as shown in Fig.~\ref{fig:SpectatorQubits}. We study each case independently in Secs.~\ref{SubSec:ConSpecQu} and~\ref{SubSec:TarSpecQu}, respectively. 

\begin{figure*}[t!]
\centering
\includegraphics[scale=0.22]{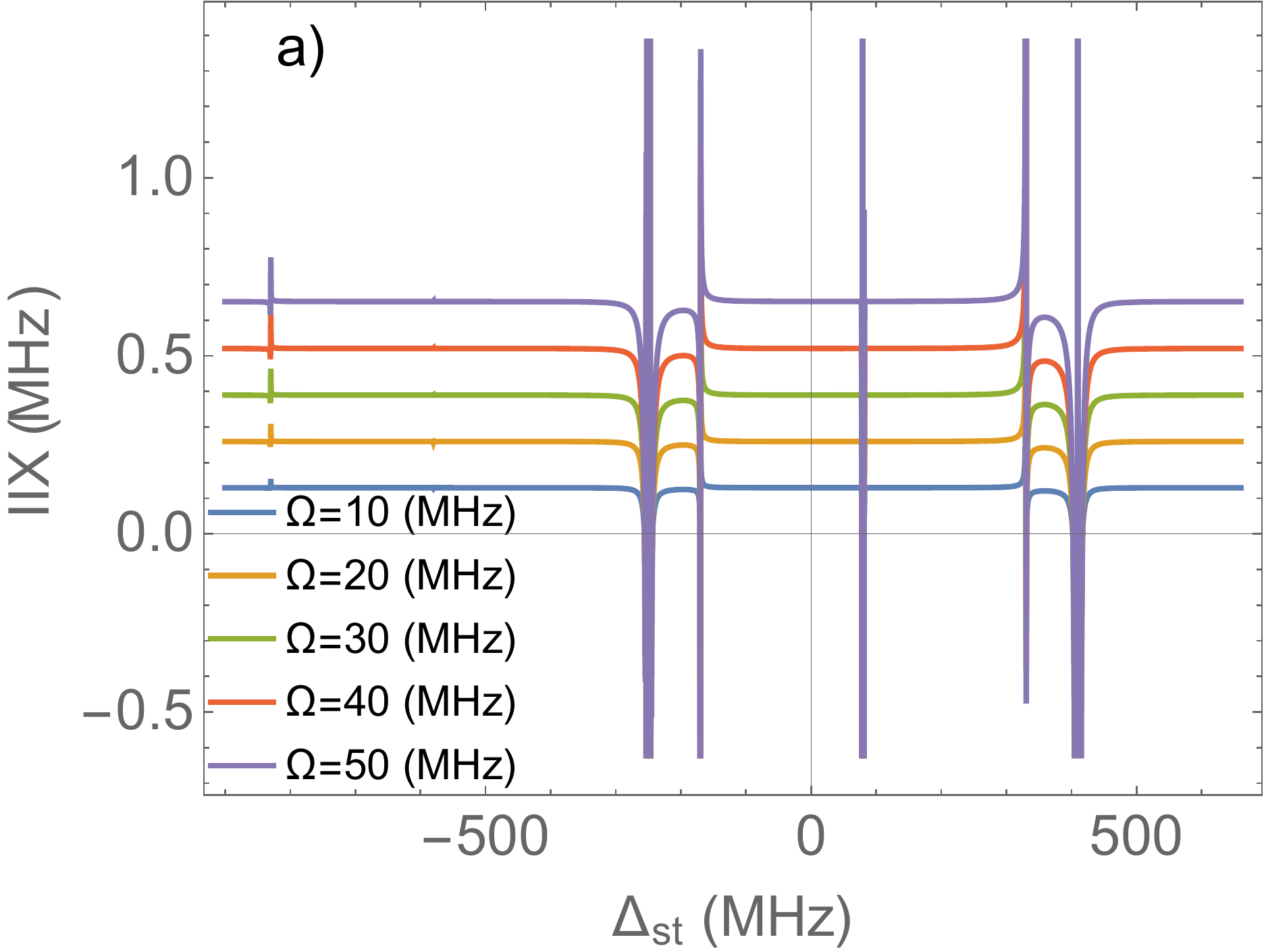}
\includegraphics[scale=0.23]{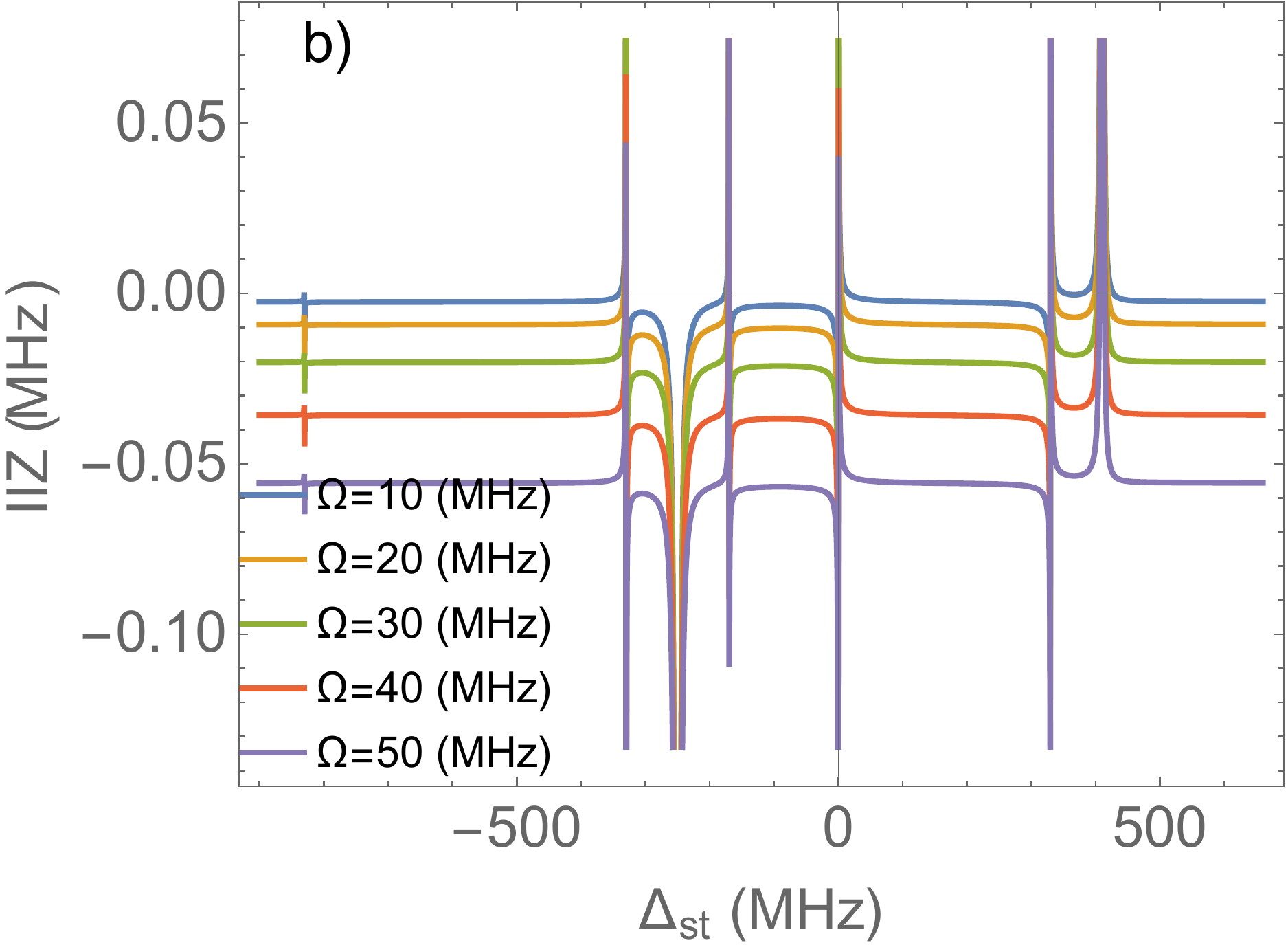}
\includegraphics[scale=0.22]{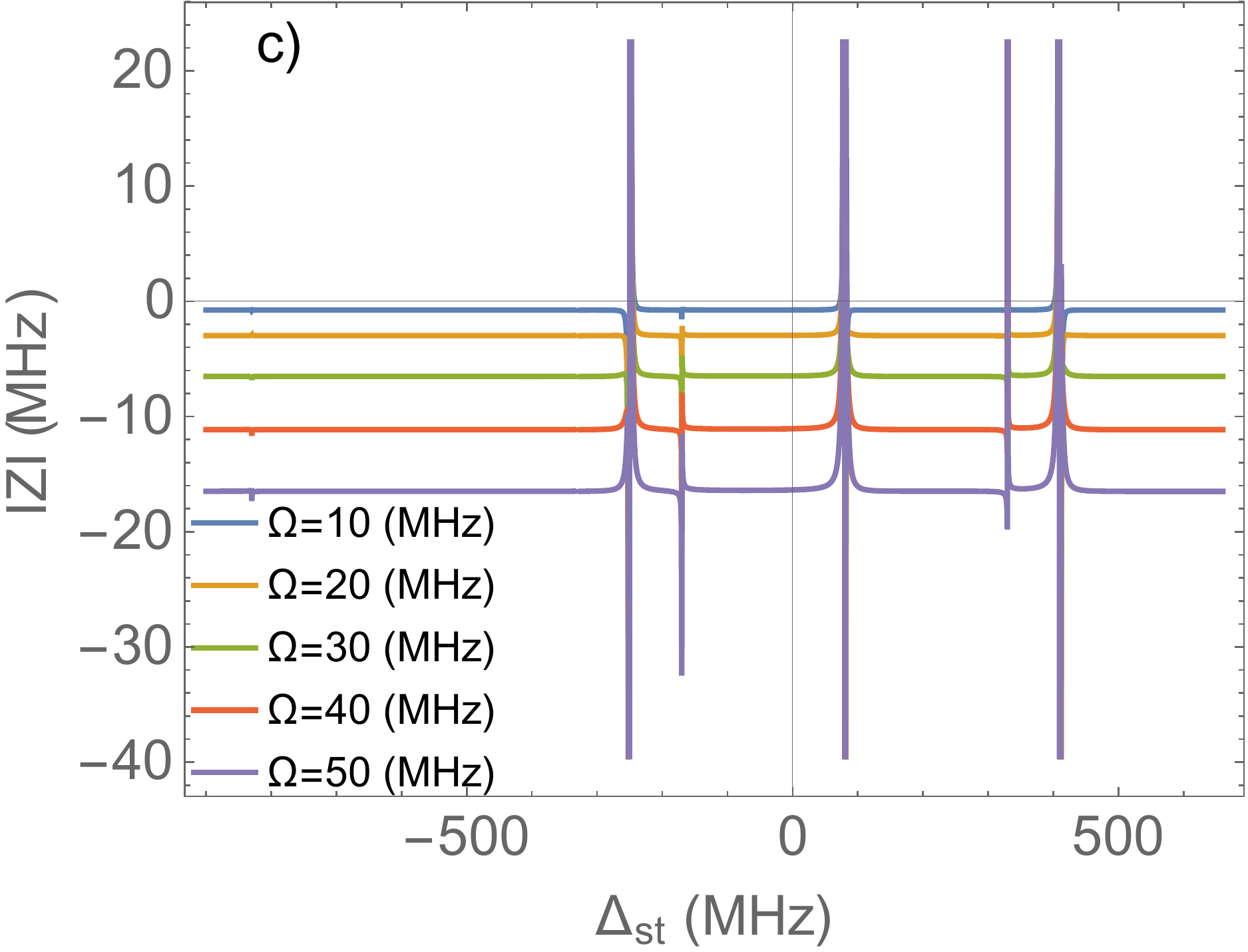}
\includegraphics[scale=0.215]{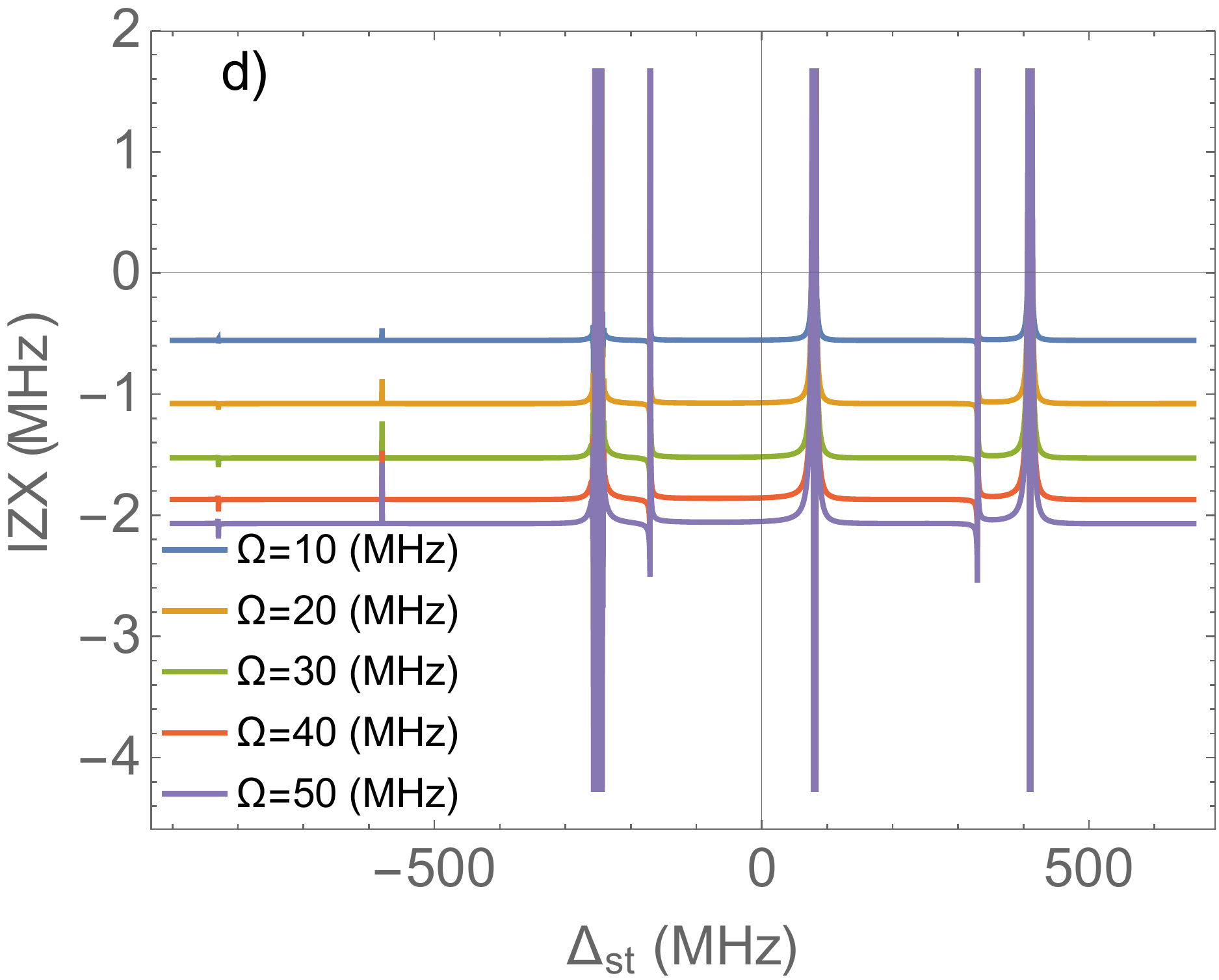}\\
\includegraphics[scale=0.22]{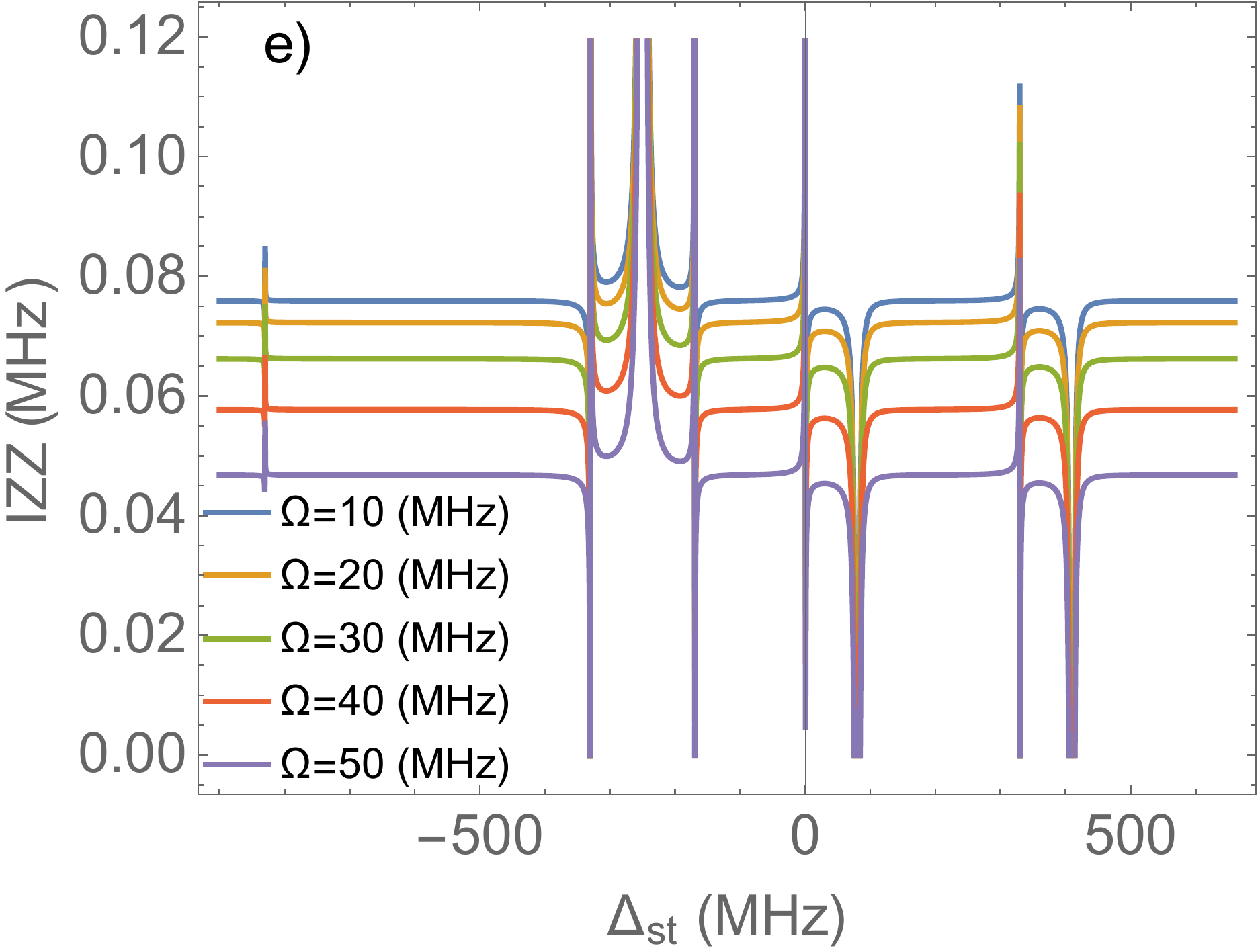}
\includegraphics[scale=0.22]{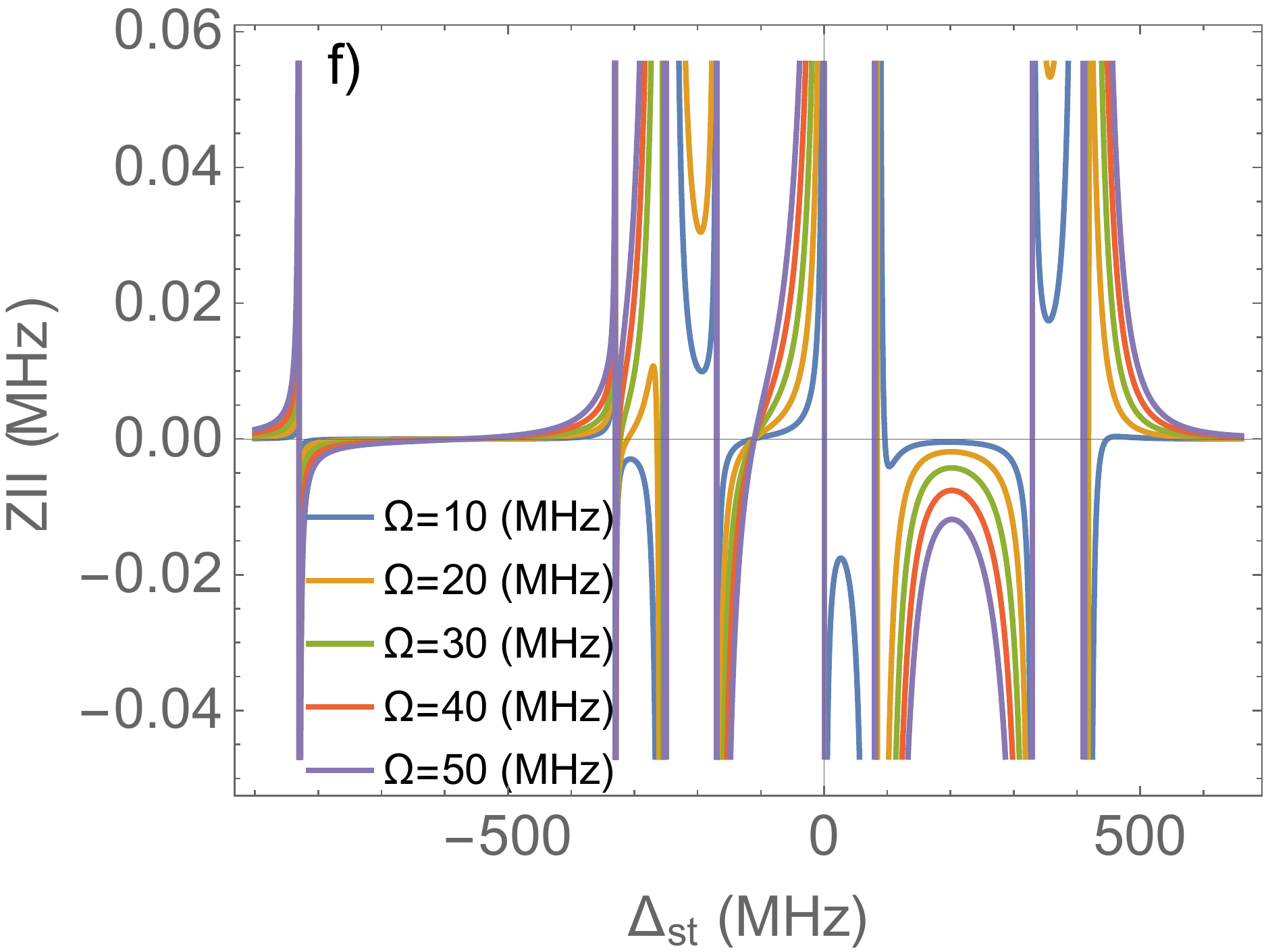} 
\includegraphics[scale=0.23]{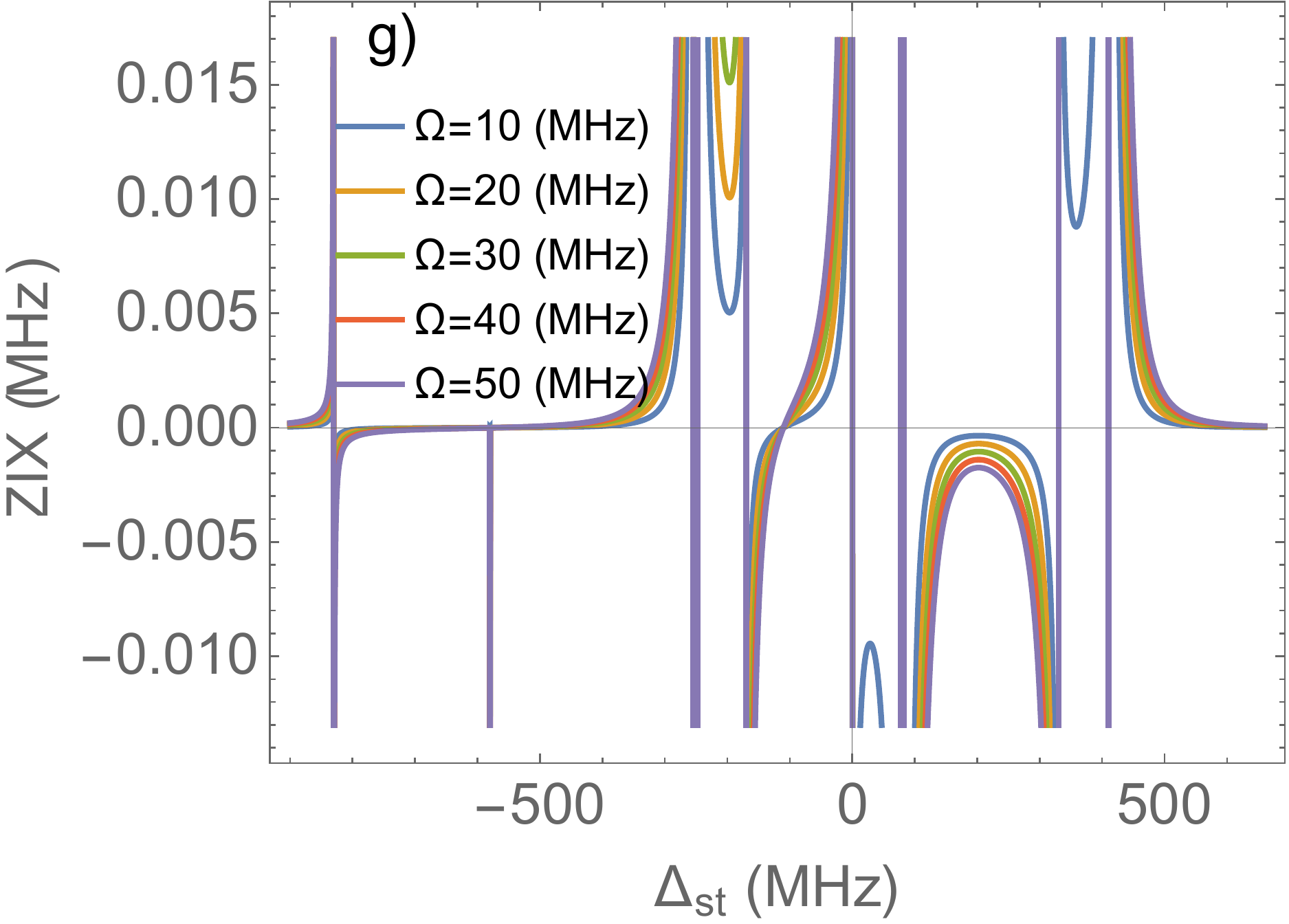} 
\includegraphics[scale=0.23]{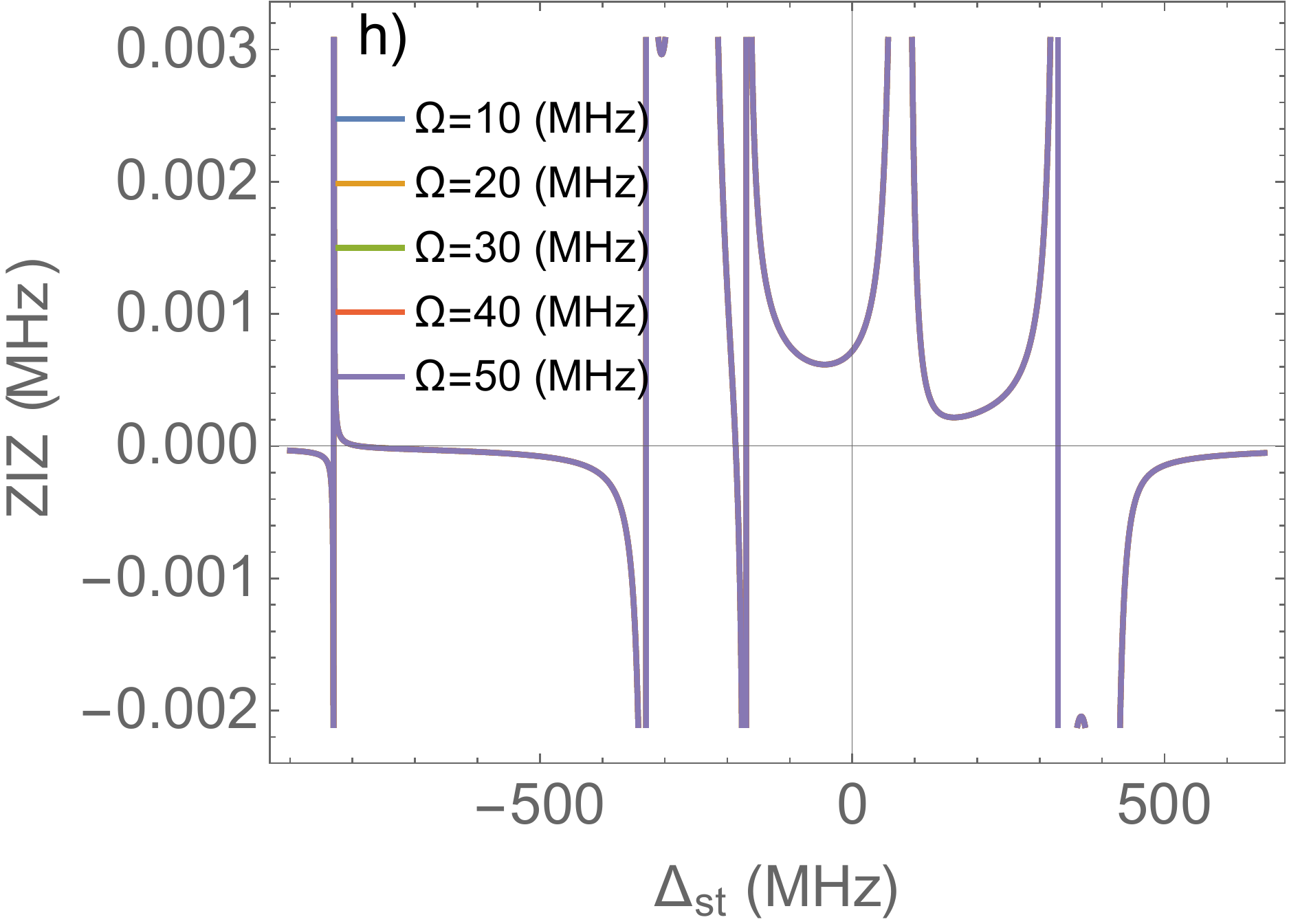}\\
\includegraphics[scale=0.22]{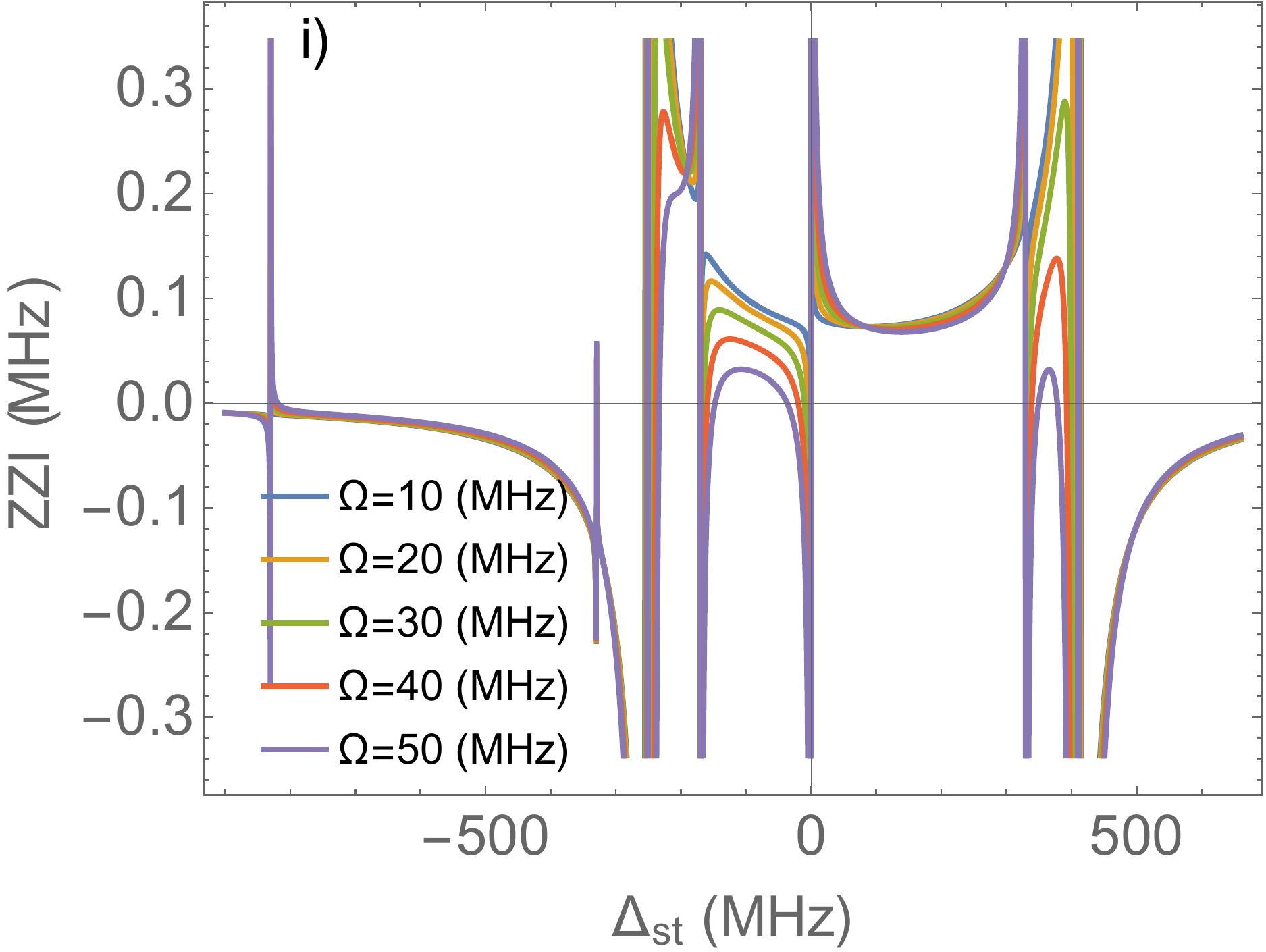}
\includegraphics[scale=0.23]{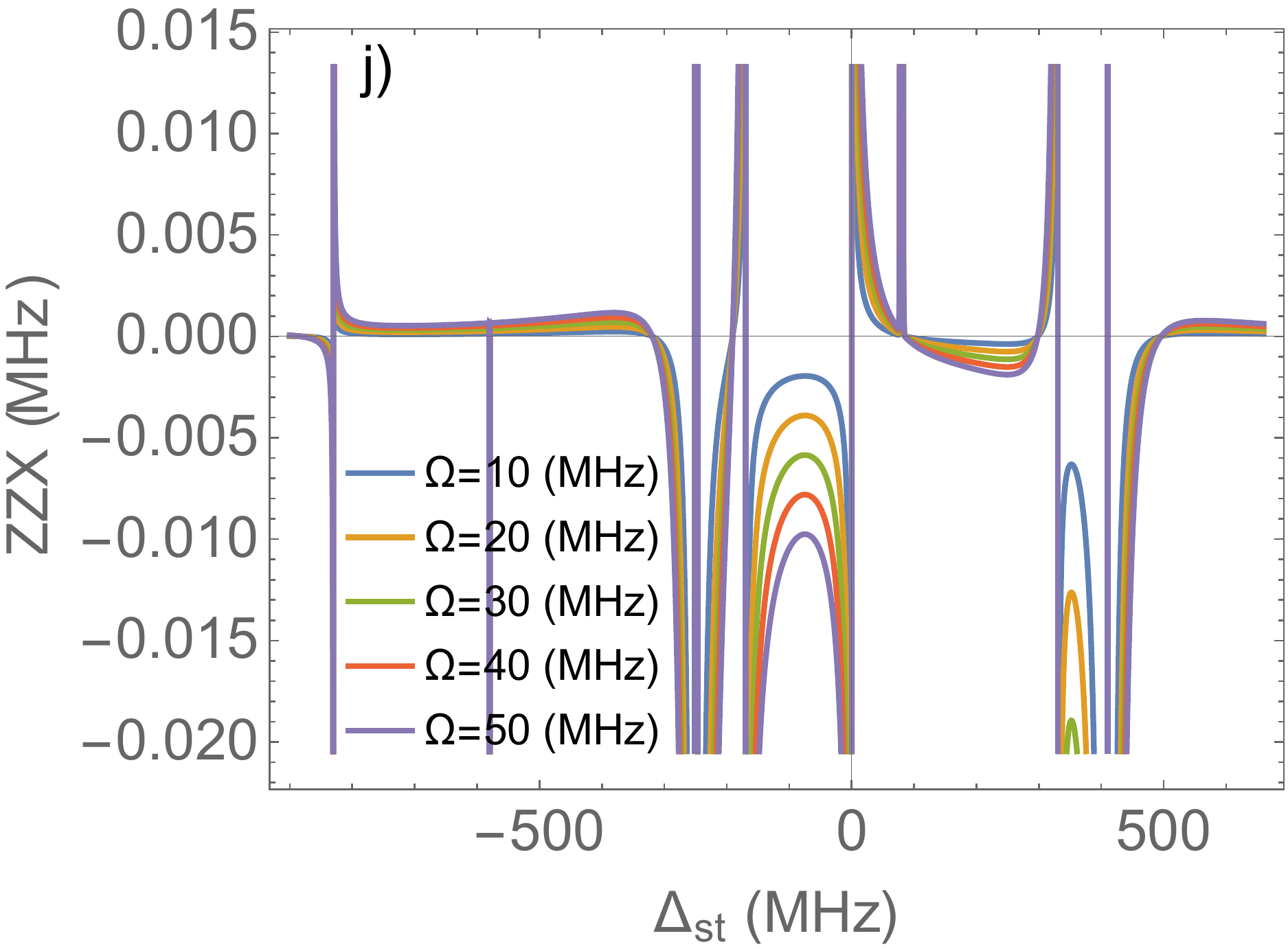} 
\includegraphics[scale=0.235]{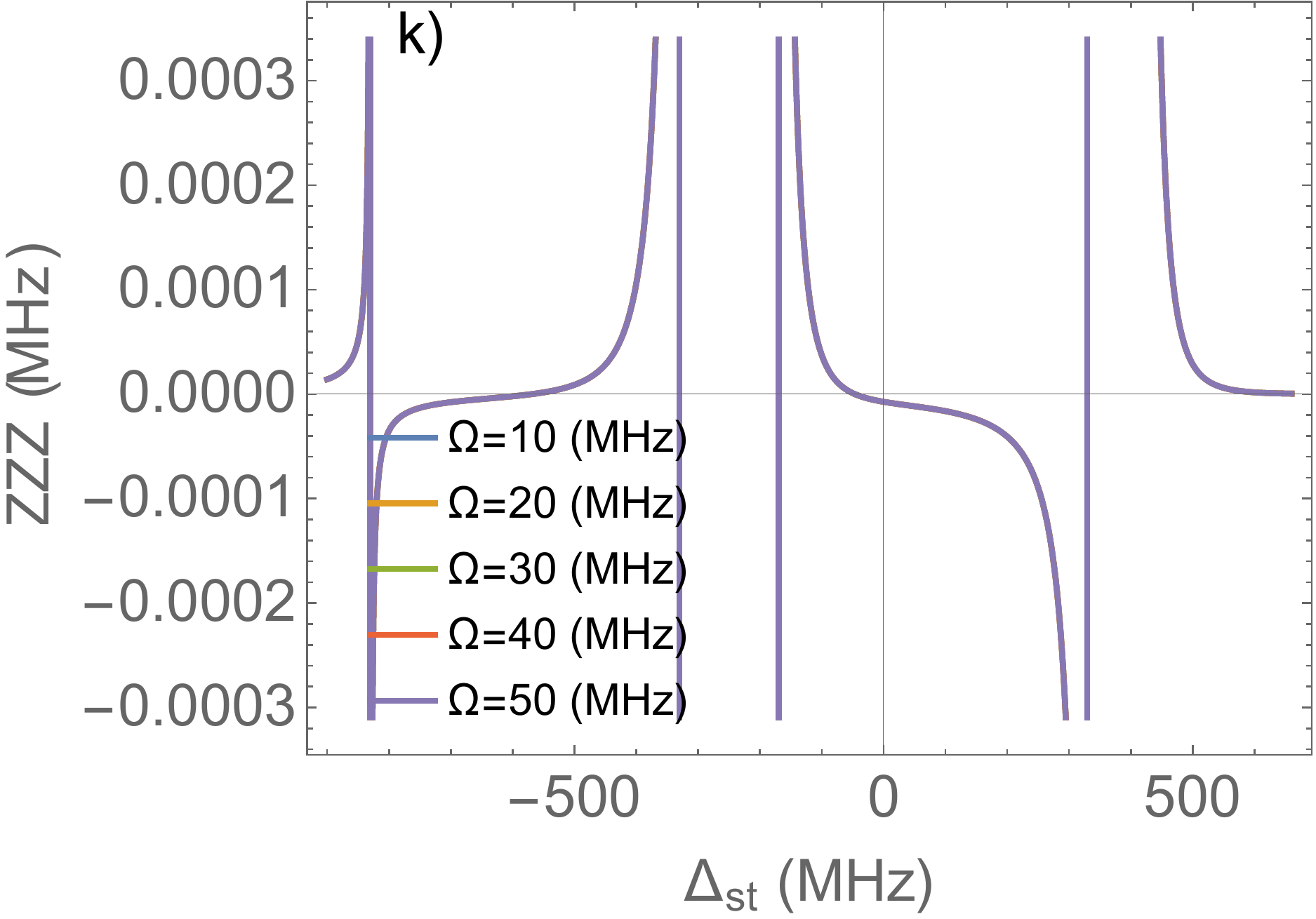}
\caption{Three-qubit gate parameters as a function of spectator-target detuning $\Delta_{st}$ with fixed control-target detuning $\Delta_{ct}=80$ MHz for the circuit of Fig.~\ref{subfig:SpecQu-ControlSpectatorSchematic}. Other parameters are set as $J=3.8$ MHz and $\alpha_c=\alpha_t=\alpha_s=-330$ MHz. The order of gate operators is taken as $\text{spectator} \otimes \text{control} \otimes \text{target}$. The observed resonances are understood in terms of two- and three-qubit processes and are summarized in the top panel of Table~\ref{tab:SpecQu-Resonances}.}
\label{fig:ConSpecQu-GateParamsFuncOfDrDet}
\end{figure*}

\begin{figure*}[t!]
\centering
\includegraphics[scale=0.22]{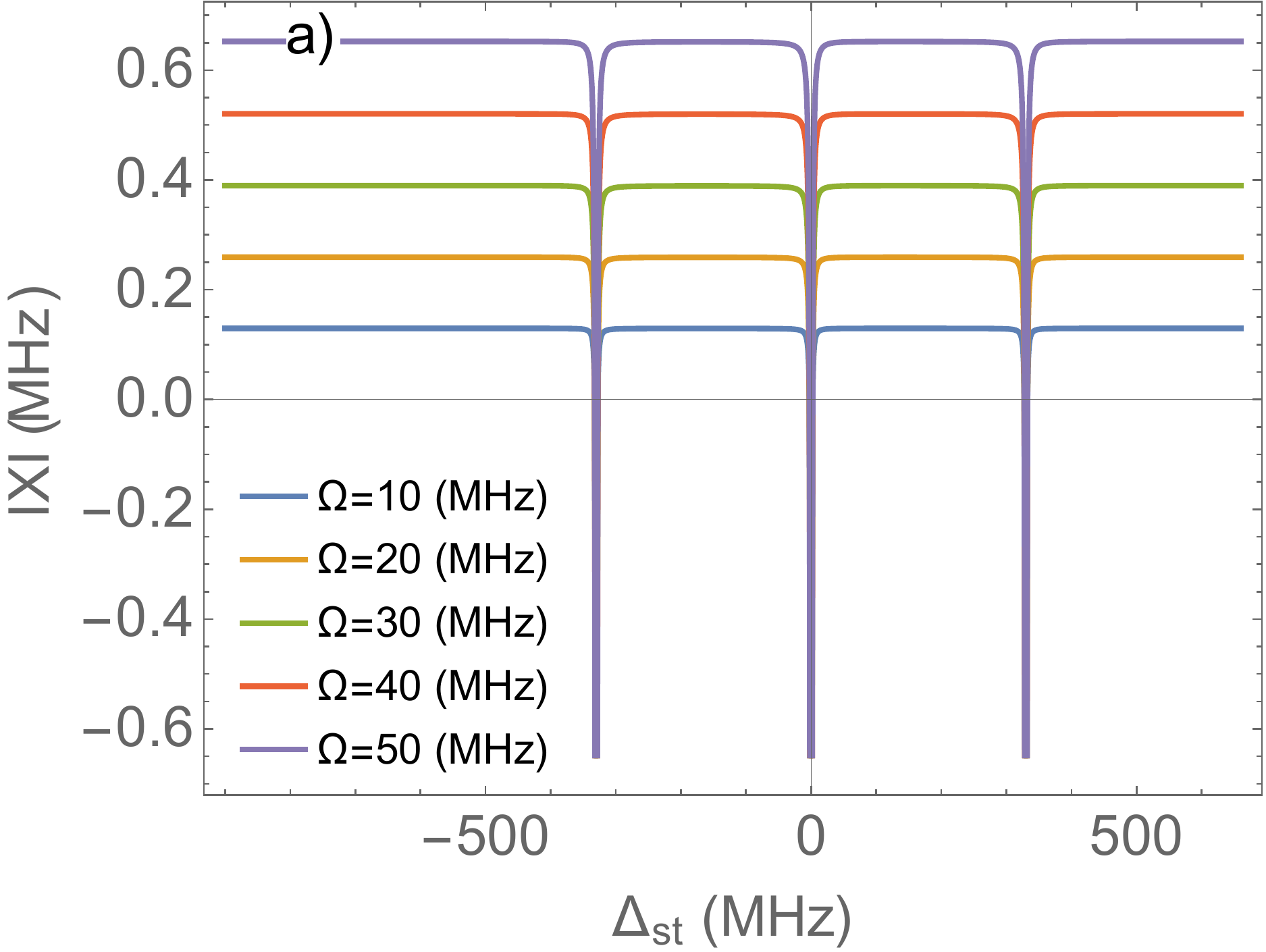}
\includegraphics[scale=0.225]{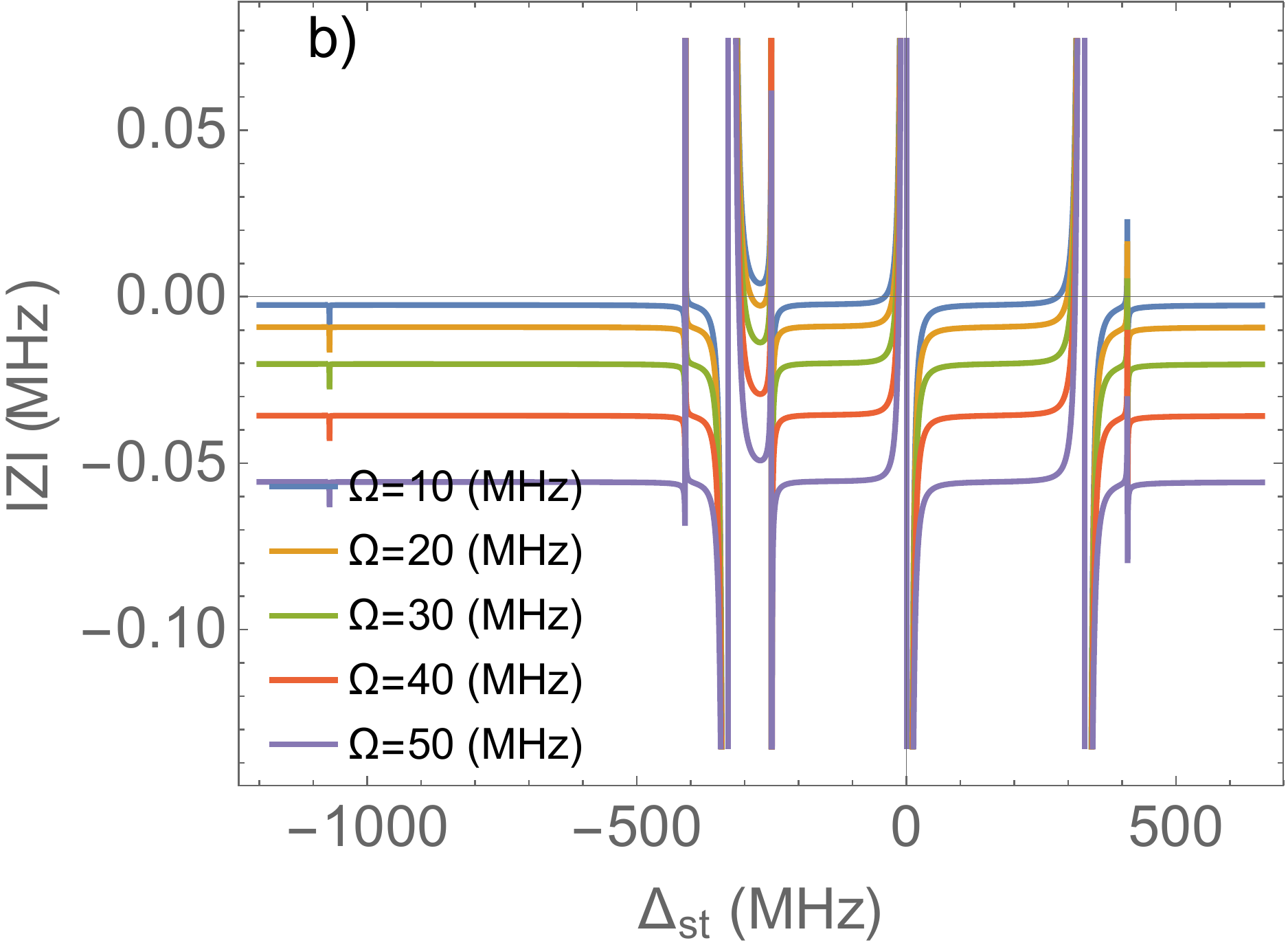}
\includegraphics[scale=0.22]{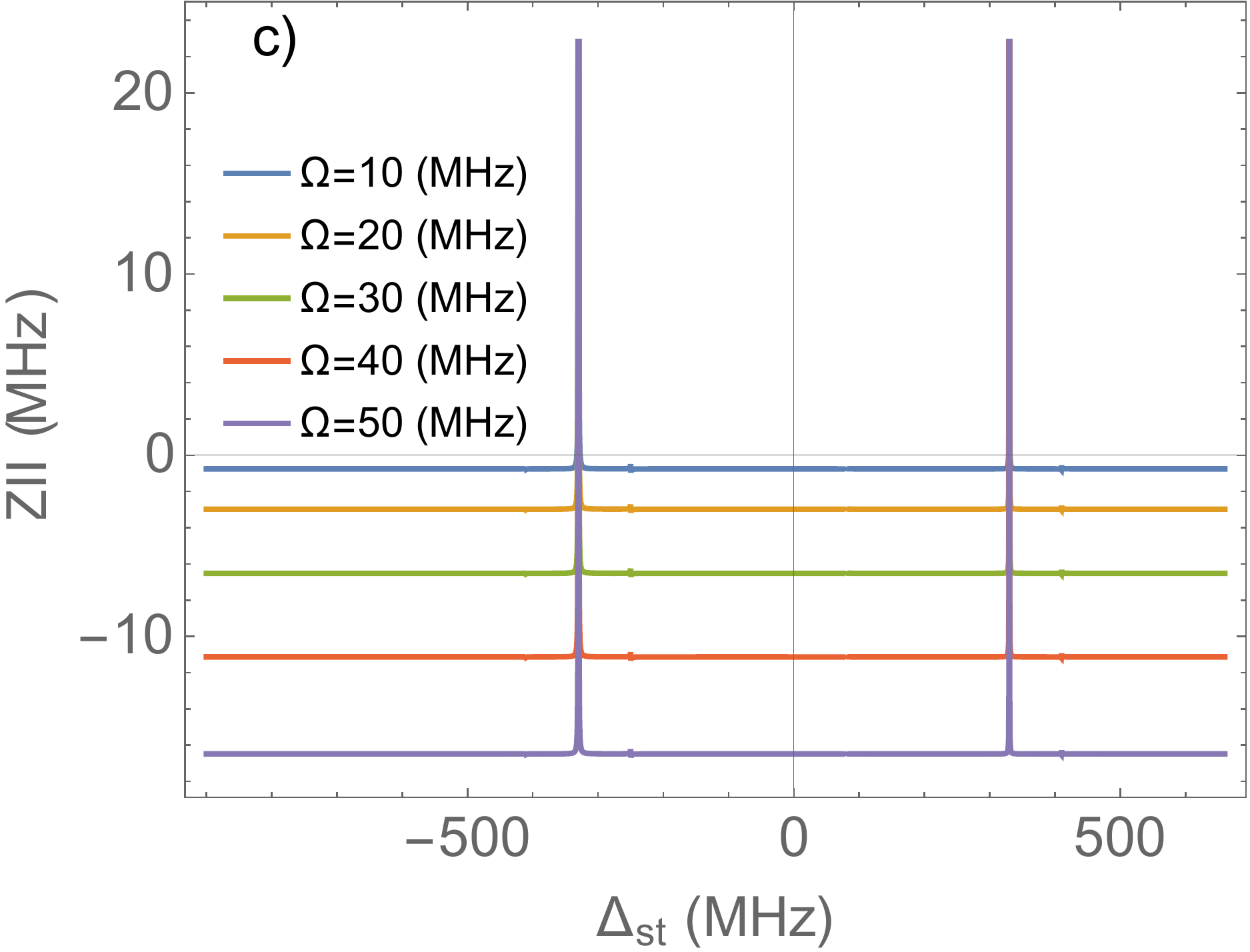} 
\includegraphics[scale=0.22]{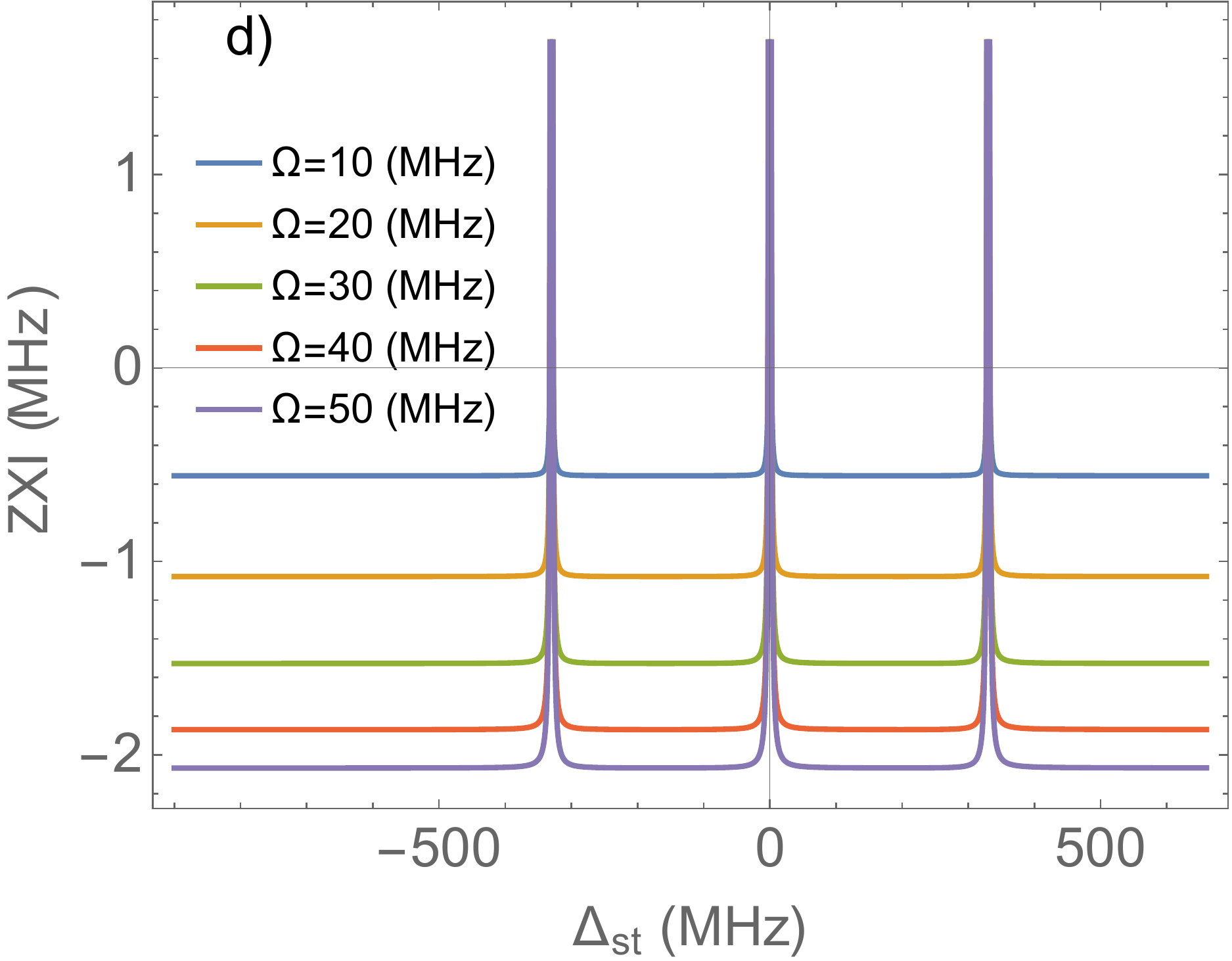}\\
\includegraphics[scale=0.22]{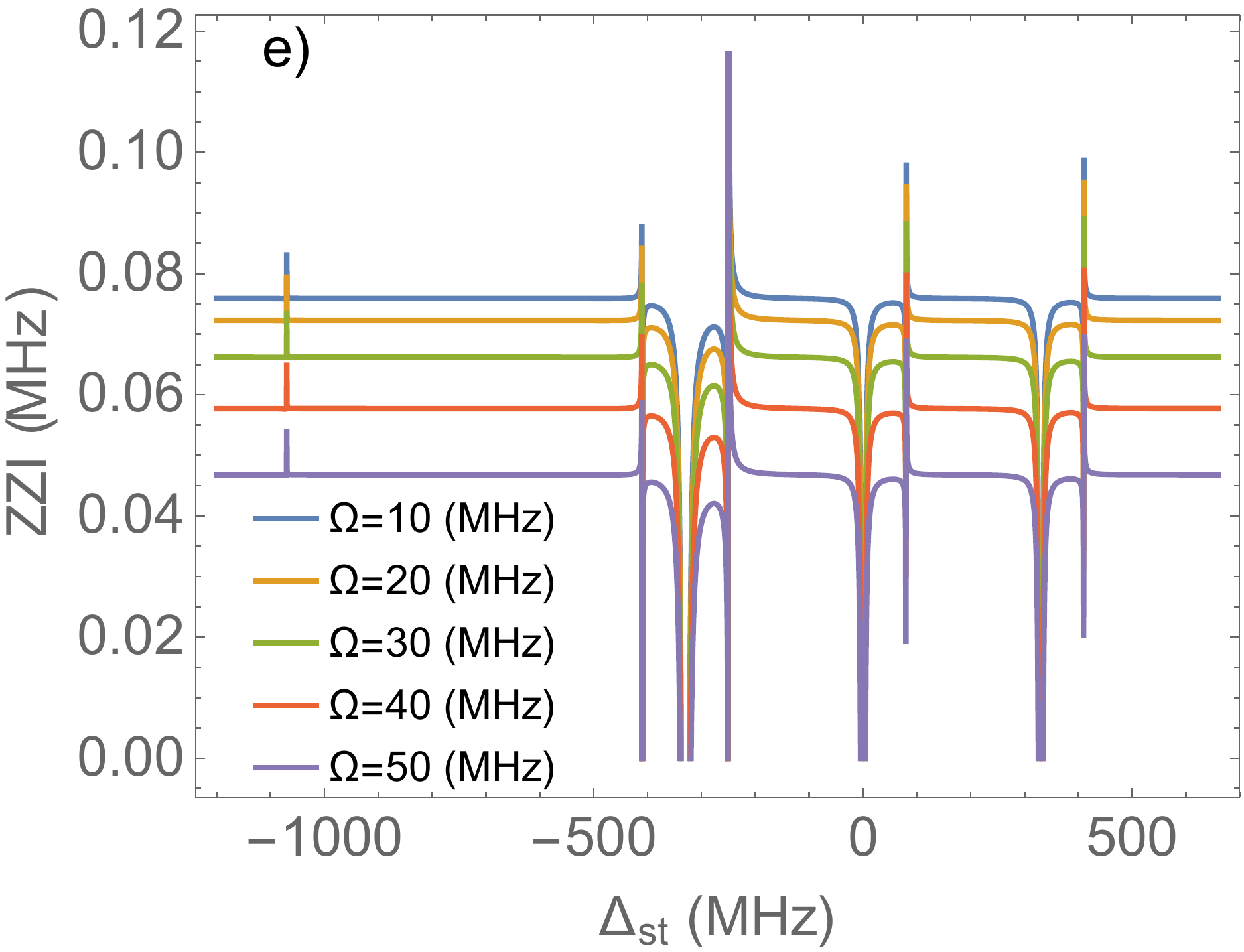}
\includegraphics[scale=0.22]{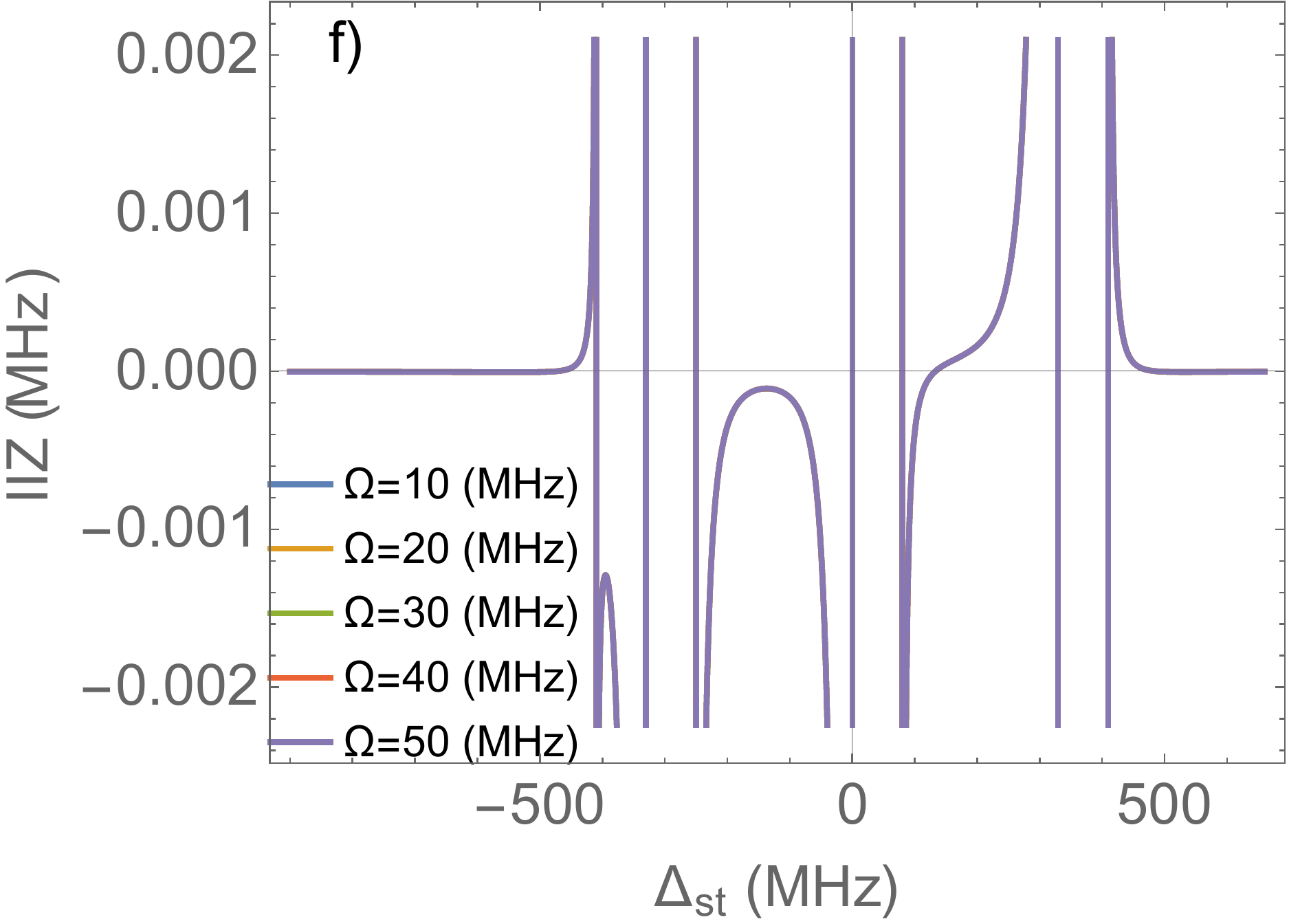}
\includegraphics[scale=0.23]{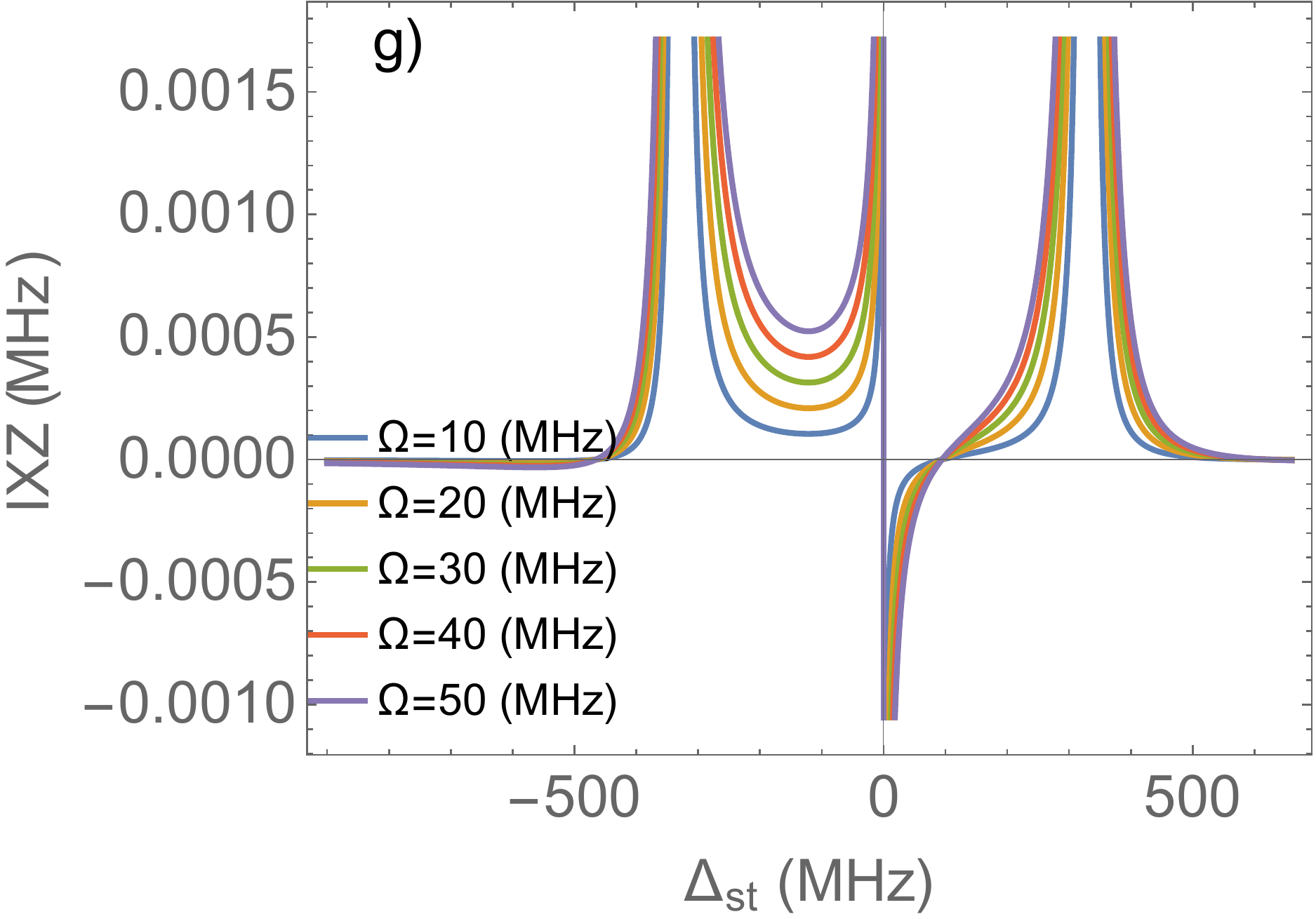}
\includegraphics[scale=0.22]{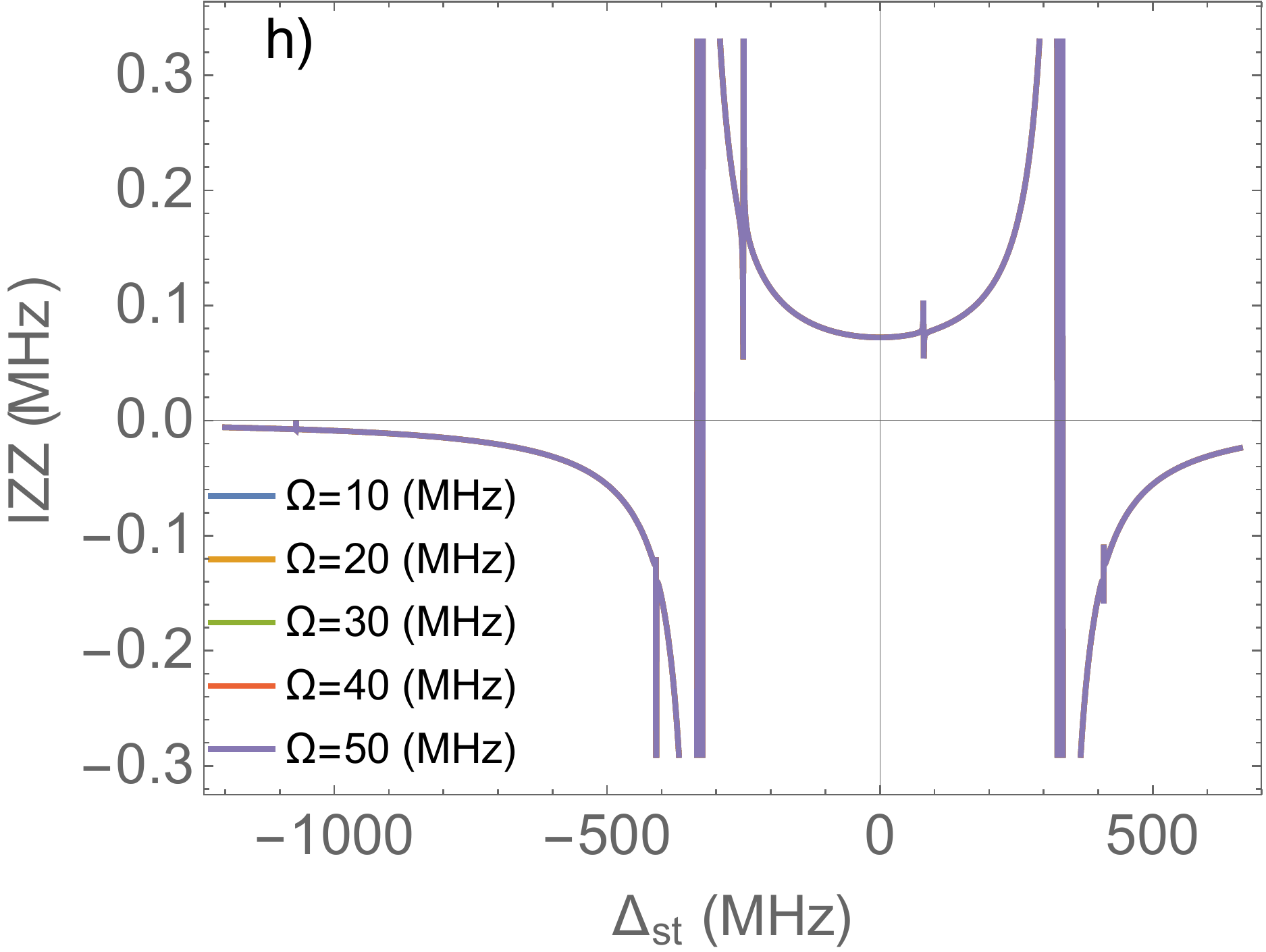}\\
\includegraphics[scale=0.23]{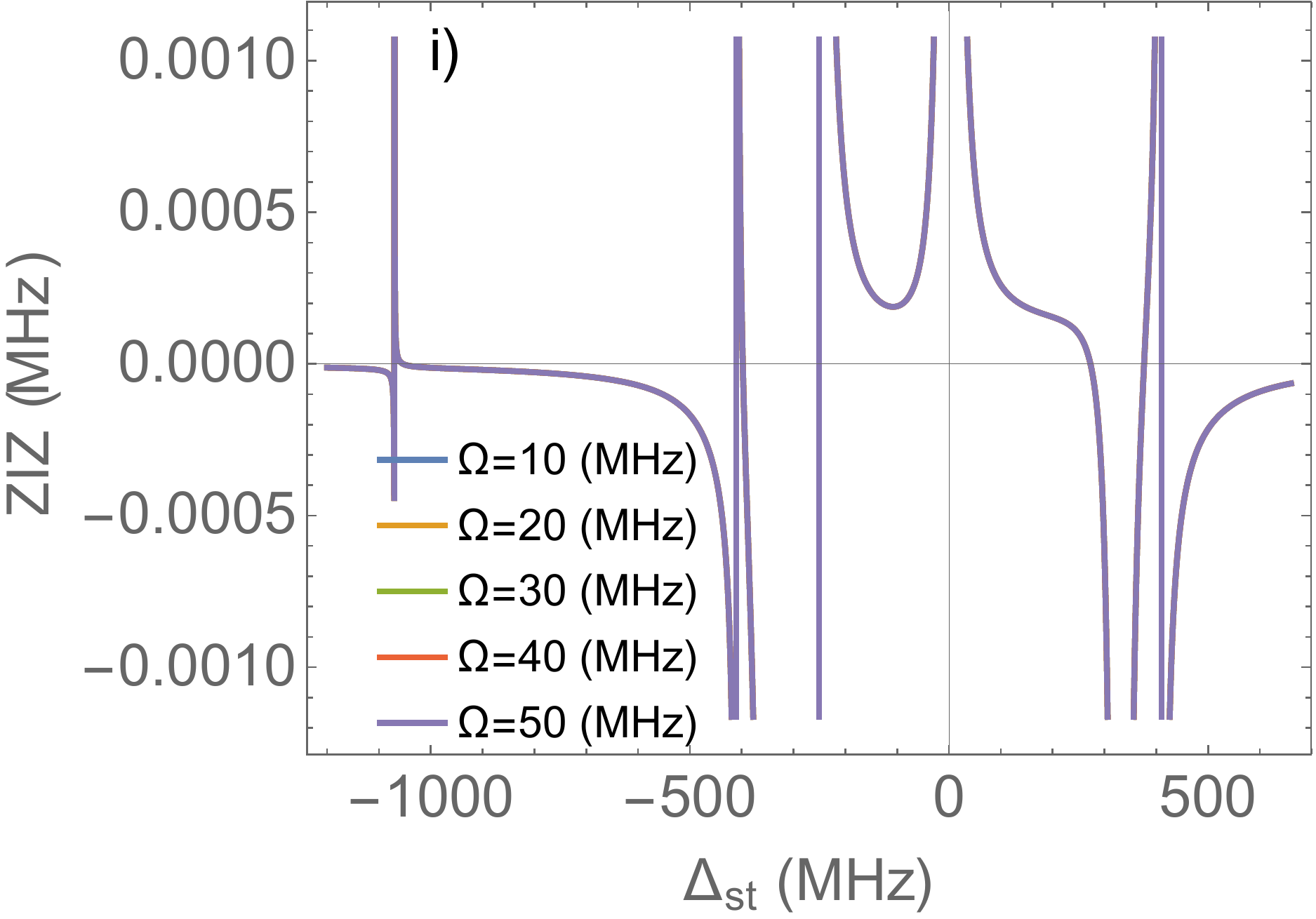}
\includegraphics[scale=0.225]{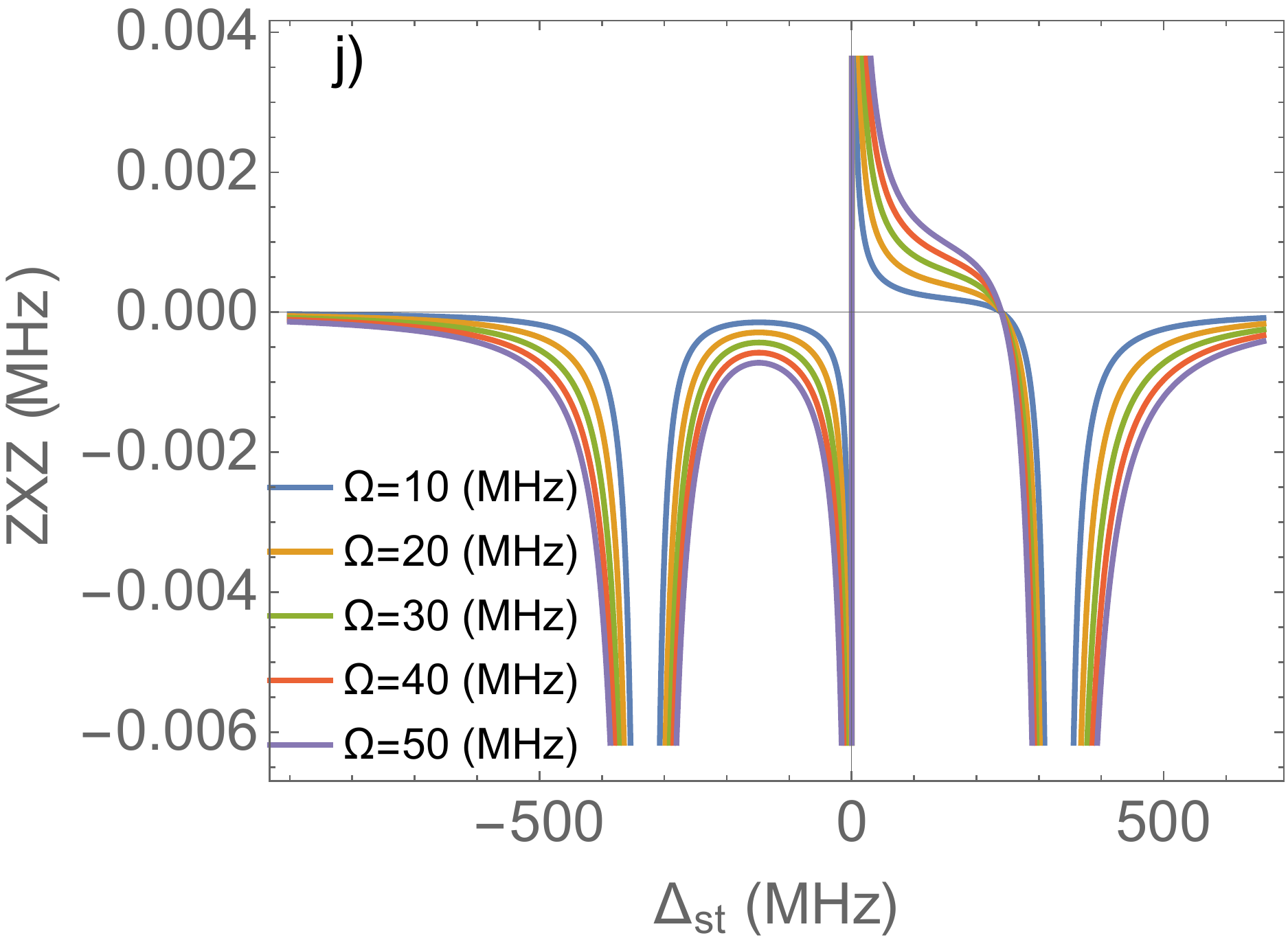}
\includegraphics[scale=0.23]{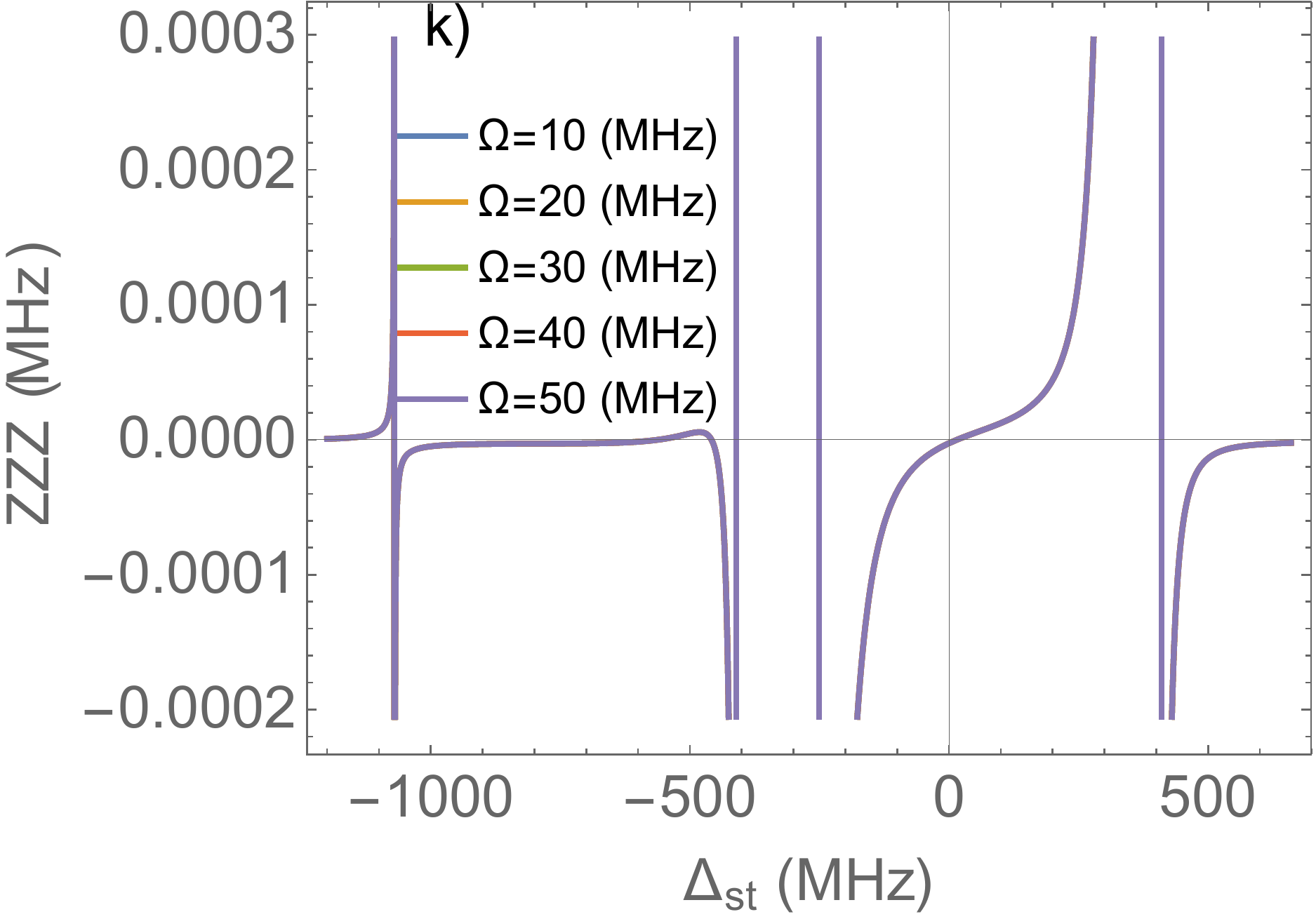}
\caption{Three-qubit gate parameters as a function of spectator-target detuning $\Delta_{st}$ with fixed control-target detuning $\Delta_{ct}=80$ MHz for the circuit of Fig.~\ref{subfig:SpecQu-TargetSpectatorSchematic}. Other parameters are set as $J=3.8$ MHz and $\alpha_c=\alpha_t=\alpha_s=-330$ MHz. We observe that these rates are mainly renormalized only for the values of $\Delta_{st}$ close to a frequency collision as summarized in Table~\ref{tab:SpecQu-Resonances}. Notice the different order of gate operators, $\text{control} \otimes \text{target} \otimes \text{spectator}$, compared to the case of a control spectator in Fig.~\ref{fig:ConSpecQu-GateParamsFuncOfDrDet}}
\label{fig:TarSpecQu-GateParamsFuncOfDrDet}
\end{figure*}

The main results of this section is presented as follows. The three-qubit gate parameters are shown in Figs.~\ref{fig:ConSpecQu-GateParamsFuncOfDrDet} and~\ref{fig:TarSpecQu-GateParamsFuncOfDrDet} for the aforementioned two scenarios up to the fourth-order in perturbation theory. In our analysis, we have fixed the control-target detuning to lie in the optimal detuning interval of region II based on our two-qubit calculation in Section~\ref{SubSec:NeOrAnal} ($\Delta_{ct}= 80$ MHz) and sweep the spectator qubit frequency. The resulting three-qubit resonances are summarized in Table~\ref{tab:SpecQu-Resonances}. 

\begin{table}
  \begin{tabular}{|c|c|c|c|}
  \hline
  States ($\ket{SCT}\ket{D}$) & Condition &  $\Delta_{st}$ & Type \\
  \hline\hline
  $\ket{\psi_{111}}\ket{n_d}\sim \ket{\psi_{030}}\ket{n_d}$ & $\Delta_{st}=2\Delta_{ct}+3\alpha_c$ & -830 & $\text{III}_{E}$\\
  \hline
  $\ket{\psi_{120}}\ket{n_d}\sim \ket{\psi_{030}}\ket{n_d}$ & $\Delta_{st}=\Delta_{ct}+2\alpha_c$ & -580 & $\text{II}_E$\\
  \hline
  $\ket{\psi_{101}}\ket{n_d}\sim \ket{\psi_{002}}\ket{n_d}$ & $\Delta_{st}=\alpha_t$ & -330 & $\text{III}_B$\\
  \hline
  $\ket{\psi_{110}}\ket{n_d}\sim \ket{\psi_{020}}\ket{n_d}$ & $\Delta_{st}=\Delta_{ct}+\alpha_c$ & -250 & $\text{II}_B$\\
  \hline
  $\ket{\psi_{101}}\ket{n_d}\sim \ket{\psi_{020}}\ket{n_d}$ & $\Delta_{st}=2\Delta_{ct}+\alpha_c$ & -170 & $\text{III}_{D}$\\
  \hline
  $\ket{\psi_{100}}\ket{n_d}\sim \ket{\psi_{001}}\ket{n_d}$ & $\Delta_{st}=0$ & 0 & $\text{III}_A$\\
  \hline
  $\ket{\psi_{100}}\ket{n_d}\sim \ket{\psi_{010}}\ket{n_d}$ & $\Delta_{st}=\Delta_{ct}$ & 80 & $\text{II}_A$\\
  \hline
  $\ket{\psi_{210}}\ket{n_d}\sim \ket{\psi_{120}}\ket{n_d}$ & $\Delta_{st}=\Delta_{ct}+\alpha_c-\alpha_s$ & 80 & $\text{II}_D$\\
  \hline
  $\ket{\psi_{200}}\ket{n_d}\sim \ket{\psi_{101}}\ket{n_d}$ & $\Delta_{st}=-\alpha_s$ & 330 & $\text{III}_B$\\
  \hline
  $\ket{\psi_{200}}\ket{n_d}\sim \ket{\psi_{110}}\ket{n_d}$ & $\Delta_{st}=\Delta_{ct}-\alpha_s$ & 410 & $\text{II}_B$\\
  \hline\hline
  States ($\ket{CTS}\ket{D}$) & Condition &  $\Delta_{st}$  & Type \\
  \hline\hline
  $\ket{\psi_{111}}\ket{n_d}\sim \ket{\psi_{030}}\ket{n_d}$ & $\Delta_{st}=-\Delta_{ct}+3\alpha_t$ & -1070 & $\text{III}_{E}$\\
  \hline
  $\ket{\psi_{101}}\ket{n_d}\sim \ket{\psi_{020}}\ket{n_d}$ & $\Delta_{st}=-\Delta_{ct}+\alpha_t$ & -410 & $\text{III}_{D}$\\
  \hline
  $\ket{\psi_{011}}\ket{n_d}\sim \ket{\psi_{020}}\ket{n_d}$ & $\Delta_{st}=\alpha_t$ & -330 & $\text{II}_B$\\
  \hline
  $\ket{\psi_{101}}\ket{n_d}\sim \ket{\psi_{200}}\ket{n_d}$ & $\Delta_{st}=\Delta_{ct}+\alpha_c$ & -250 & $\text{III}_B$\\
  \hline
   $\ket{\psi_{001}}\ket{n_d}\sim \ket{\psi_{010}}\ket{n_d}$ & $\Delta_{st}=0$ & 0 & $\text{II}_A$\\
  \hline
  $\ket{\psi_{002}}\ket{n_d}\sim \ket{\psi_{020}}\ket{n_d}$ & $2\Delta_{st}=\alpha_t-\alpha_s$ & 0 & $\text{II}_C$\\
  \hline
  $\ket{\psi_{001}}\ket{n_d}\sim \ket{\psi_{100}}\ket{n_d}$ & $\Delta_{st}=\Delta_{ct}$ & 80 & $\text{III}_A$\\
  \hline
  $\ket{\psi_{102}}\ket{n_d}\sim \ket{\psi_{201}}\ket{n_d}$ & $\Delta_{st}=\Delta_{ct}+\alpha_c-\alpha_s$ & 80 & $\text{III}_C$\\
  \hline
  $\ket{\psi_{002}}\ket{n_d}\sim \ket{\psi_{011}}\ket{n_d}$ & $\Delta_{st}=-\alpha_s$ & 330 & $\text{II}_B$\\
  \hline
  $\ket{\psi_{002}}\ket{n_d}\sim \ket{\psi_{101}}\ket{n_d}$ & $\Delta_{st}=\Delta_{ct}-\alpha_s$ & 410 & $\text{III}_B$\\
  \hline
  \end{tabular}
  \caption{Summary of three-qubit resonances with a control spectator qubit (top table) and a target spectator (bottom table) that appear up to the fourth order in perturbation theory and with four energy eigenstates for each qubit. From left to right, the first column denotes the underlying physical process in terms of qubit and drive photon states, the second show the corresponding resonance condition in terms of qubit-qubit detuning, the third gives an experimental estimate in MHz for spectator-target detuning $\Delta_{st}$ in terms of a fixed control-traget detuning as $\Delta_{ct}=80$ MHz, and the fourth labels such resonances in terms of broader categories for multi-qubit resonances (See Sec.~\ref{Sec:SumFreqCol}). The resonances are ordered increasingly for this particular choice of control-target detuning. However, note that if the parameters change there is possibility for the resonances to move around.}	
\label{tab:SpecQu-Resonances}
\end{table}

\subsection{Control spectator}
\label{SubSec:ConSpecQu}

Here, we consider the impact of a control spectator qubit on the CR gate parameters. The starting CR Hamiltonian~(\ref{eqn:CRHamInEnBasis-H0}--\ref{eqn:CRHamInEnBasis-H_int}) is now modified according to Fig.~\ref{subfig:SpecQu-ControlSpectatorSchematic} as
\begin{subequations}
\begin{align}
&\HO_{0}=\sum\limits_{j=s,c,t}\frac{\omega_{\text{jh}}}{4}\left[\hat{y}_j^2-\frac{2}{\epsilon_j}\cos(\sqrt{\epsilon_j}\hat{x}_j)\right]\;,
\label{eqn:ConSpecQu-H0}\\
&\HO_{\text{int}}(t)=J\hat{y}_s\hat{y}_c+J\hat{y}_c\hat{y}_t-\Omega\hat{y}_c\sin(\omega_{\text{d}} t) \;,
\label{eqn:ConSpecQu-H_int}
\end{align}
\end{subequations}
where the spectator qubit operators and parameters are labeled with $s$ and we have considered a direct interaction between the control and the spectator qubit of the same strength $J$ for simplicity. Furthermore, the order of subsystems in the composite Hilbert space is taken as $\text{spectator}\otimes \text{control} \otimes \text{target}$ consistent with Fig.~\ref{subfig:SpecQu-ControlSpectatorSchematic}.

Starting from Hamiltonian~(\ref{eqn:ConSpecQu-H0}--\ref{eqn:ConSpecQu-H0}) and following SWPT Eqs.~(\ref{eqn:CREffHam-O(lambda) HIeff}--\ref{eqn:CREffHam-O(lambda^4) HIeff}), we obtain an effective Hamiltonian for this extended model via block-diagonalization with respect to the Hilbert space of the target qubit. The corresponding three-qubit gate parameters are then read off of the effective Hamiltonian in the computational basis. Note that, in principle, there are 64 distinct gate parameters for the three-qubit problem under consideration. However, only a few contain dominant resonant processes. Since the drive is only resonant with the target qubit frequency, the dominant interactions may only involve $\hat{I}$, $\hat{Z}$ for the control and the spectator sectors, and all four Pauli matrices for the target sector. Furthermore, we assume that the phase of the CR drive is set such that it only induces a resonant $\hat{X}$ interaction on the target. With all these considerations, there are $2\times 2\times 3-1=11$ independent \textit{non-zero} gate parameters, where we neglect the irrelevant energy shift due to the $III$ term. 

The lowest order estimates for the gate parameters recover the original dominant gate parameters in the $\text{control} \otimes \text{target}$ sector as those in Table~\ref{tab:LoOrAnal-CRGateParams}, as well as a $ZZ$ interaction between the control and the spectator qubits. Fourth order expressions include a plethora of independent multi-qubit multi-photon processes and hence are not given explicitly. In Fig.~\ref{fig:ConSpecQu-GateParamsFuncOfDrDet}, we study the behavior of all non-zero three-qubit gate parameters as a function spectator-target detuning $\Delta_{st}$ and up to the fourth order in perturbation. To better understand the result, we can break the gate parameters into two sub-categories. 

\begin{figure}[t!]
\centering
\includegraphics[scale=0.47]{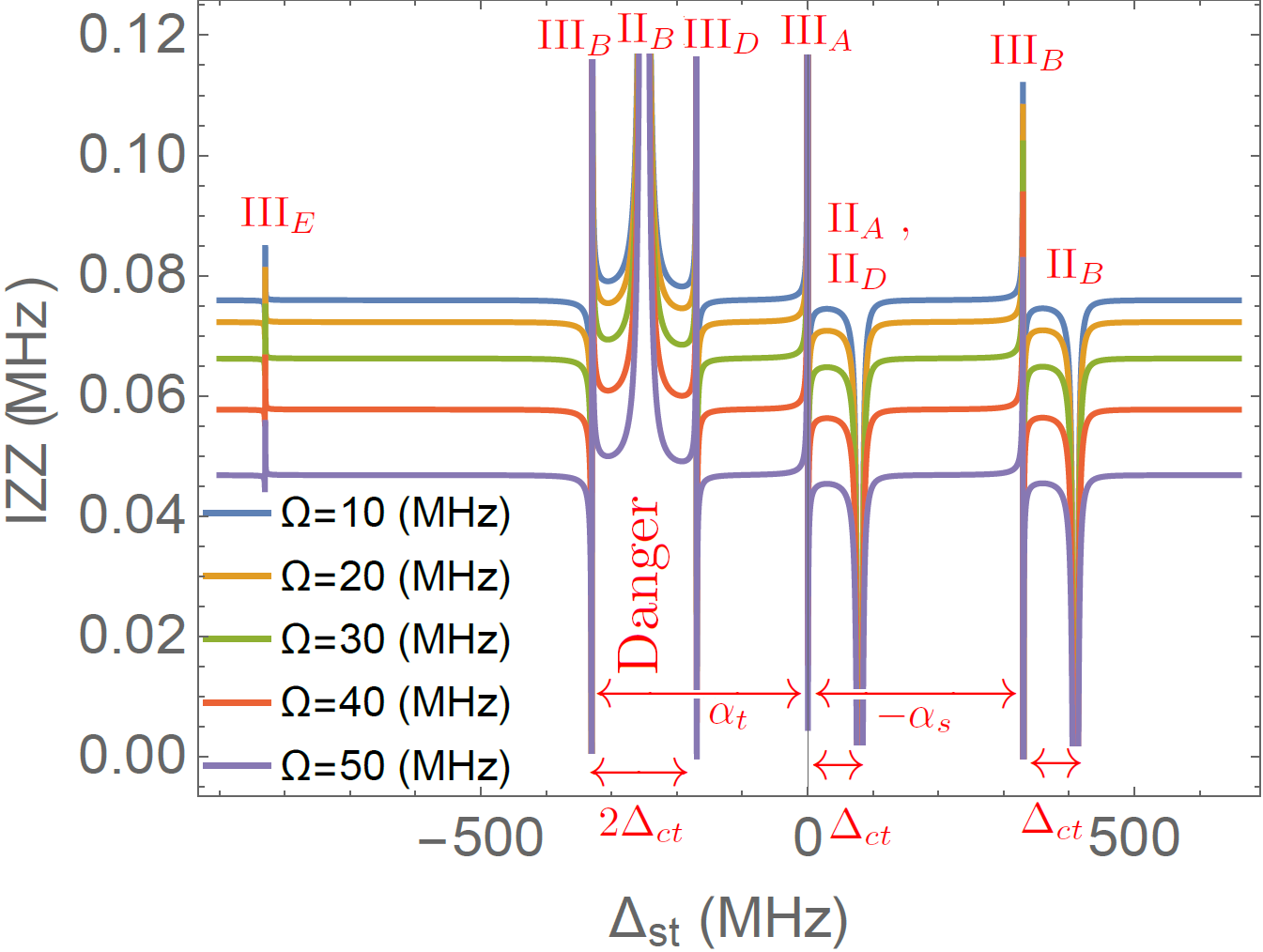}
\caption{A closer look into the $IZZ$ term for the case of a control spectator (same as Fig.~\ref{fig:ConSpecQu-GateParamsFuncOfDrDet}e). The induced resonances by the spectator qubit are labeled with the underlying physical process according to Table~\ref{tab:SpecQu-Resonances} and Sec.~\ref{Sec:SumFreqCol}. Ideally, the spectator frequency should be tuned such that it is sufficiently far from these resonances. The boundaries of the resulting detuning regions can move around dependent on the control-target detuning.}
\label{fig:TarSpecQu-ZZConSpecQuWithLabeledCollisions}
\end{figure}	

The first category includes those gate parameters with the spectator sector being idle such as $IIX$, $IIZ$, $IZI$, $IZX$ and $IZZ$ as shown in Figs.~\ref{fig:ConSpecQu-GateParamsFuncOfDrDet}a--\ref{fig:ConSpecQu-GateParamsFuncOfDrDet}e, respectively. We observe that these rates remain almost intact compared to the ones found without a spectator qubit, unless the spectator frequency is sufficiently close to specific values that are understood in terms of two- or three-qubit resonances (See top panel of Table~\ref{tab:SpecQu-Resonances}). Moreover, the strength of each resonance, i.e. the numerator in the perturbative expansion, determines its effective width in frequency. Generally speaking, we find that resonances involving nearest neighbors result in stronger and hence wider peaks and are in turn more detrimental from design perspective. This has been illustrated in Fig.~\ref{fig:TarSpecQu-ZZConSpecQuWithLabeledCollisions} in terms of observed resonances in the $IZZ$ gate parameter. In particular, we find that the strongest resonance occurs when the control spectator qubit frequency, transition $\ket{\psi_{s,0}}\leftrightarrow \ket{\psi_{s,1}}$, is equal with transition $\ket{\psi_{c,1}}\leftrightarrow \ket{\psi_{c,2}}$ of the control qubit, which translates as $\Delta_{st}=\Delta_{ct}+\alpha_c$.

The second category contains gate parameters with spectator sector set to the $Z$ Pauli matrix as given in Figs.~\ref{fig:ConSpecQu-GateParamsFuncOfDrDet}f--\ref{fig:ConSpecQu-GateParamsFuncOfDrDet}k. In particular, $ZII$ and $ZZI$, Figs.~\ref{fig:ConSpecQu-GateParamsFuncOfDrDet}f and~\ref{fig:ConSpecQu-GateParamsFuncOfDrDet}i, belong to the reduced $\text{spectator}\otimes\text{control}$ sector and hence exhibit a more pronounced dependence on $\Delta_{st}$. On the other hand, gate parameters $ZIX$, $ZIZ$, $ZZX$ and $ZZZ$ describe an effective mediated interaction between the target and the spectator qubits via the control. The underlying processes are fourth-order in nature, and the estimates can range from 0.1--10 KHz (Figs.~\ref{fig:ConSpecQu-GateParamsFuncOfDrDet}g,~\ref{fig:ConSpecQu-GateParamsFuncOfDrDet}h, \ref{fig:ConSpecQu-GateParamsFuncOfDrDet}j, ~\ref{fig:ConSpecQu-GateParamsFuncOfDrDet}k). 

Lastly, we note that a specific gate parameter contains specific combination of multi-qubit multi-photon processes. For instance, the $IIZ$ rate exhibits a resonance at $\Delta_{st}=0$, while the $IIX$ rate does not. The entirety of such multi-qubit resonances that emerge in Fig.~\ref{fig:ConSpecQu-GateParamsFuncOfDrDet}, as a result of a control spectator, have been explained in terms of their underlying physical process and summarized in the top panel of Table~\ref{tab:SpecQu-Resonances}.


\subsection{Target spectator}
\label{SubSec:TarSpecQu}

The case of a target spectator is shown schematically in Fig.~\ref{subfig:SpecQu-TargetSpectatorSchematic}, with the corresponding bare and interaction Hamiltonian defined as
\begin{subequations}
\begin{align}
&\HO_{0}=\sum\limits_{j=c,t,s}\frac{\omega_{\text{jh}}}{4}\left[\hat{y}_j^2-\frac{2}{\epsilon_j}\cos(\sqrt{\epsilon_j}\hat{x}_j)\right]\;,
\label{eqn:TarSpecQu-H0}\\
&\HO_{\text{int}}(t)=J\hat{y}_c\hat{y}_t+J\hat{y}_t\hat{y}_s-\Omega\hat{y}_c\sin(\omega_{\text{d}} t) \;.
\label{eqn:TarSpecQu-H_int}
\end{align}
\end{subequations}
In contrast to the case of a control spectator, the order of the composite Hilbert space is taken as $\text{control} \otimes \text{target} \otimes \text{spectator}$ in agreement with Fig.~\ref{subfig:SpecQu-TargetSpectatorSchematic}.

Most of our discussions from the previous section applies to the target spectator as well. Hence, to avoid repetition, we briefly summarize our main findings. Up to the second order, we recover the results of two-qubit calculation in addition to a $ZZ$ interaction between the target and the spectator qubits. There are again 11 non-zero gate parameters up to the fourth order in perturbation, which are shown in Fig.~\ref{fig:TarSpecQu-GateParamsFuncOfDrDet} as a function of spectator-target detuning $\Delta_{st}$ and drive amplitude $\Omega$. The observed resonances have been explained in terms of two- and three-qubit processes in the bottom panel of Table~\ref{tab:SpecQu-Resonances}.

\section{Summary of multi-qubit resonances}
\label{Sec:SumFreqCol}

The two-qubit and three-qubit perturbative calculations in Secs.~\ref{Sec:CREffHam} and~\ref{Sec:SpecQu} revealed various possibilities for multi-qubit resonances. Here, we summarize such resonances in terms of three broader categories which have been used to label and understand the similarity between those particular resonances that appeared in Tables~\ref{tab:NeOrAnal-Resonances} and~\ref{tab:SpecQu-Resonances}. Two key factors in our categorization are the number of qubits involved in a particular resonance and the underlying qubit states or the physical process. In general, there are infinite possibilities and the purpose of this section is to summarize only those resonances that emerge within a four-level approximation for each qubit and up to the fourth-order perturbation in drive amplitude. 

To make a connection with our perturbative results, the analysis in this section only categorizes the distinct energy denominators. The numerator of a process, on the other hand, is determined by the matrix elements of the underlying interaction Hamiltonian. Whether or not a particular process appears in a physical quantity depends on whether those states that are involved in that process are occupied in the time-evolution of that physical quantity. A general rule of thumb is that the resonances involving higher energy states and next-nearest neighboring qubits appear to be weaker. In the following, we enumerate three resonance types, where the first index shows how many qubits are involved and the second labels the sub-type.

\textbf{Type I}. This category includes the single-qubit resonances that occur as a result of drive frequency $\omega_d$ (approximately the same as $\omega_t$) being resonant with different transition frequencies of the control qubit.
\begin{itemize}

\item[$\text{I}_{A}$)]
A two-photon process, where twice the drive frequency is resonant with the transition $\ket{\psi_{c,0}}\to\ket{\psi_{c,2}}$ of the control qubit, resulting in the effective resonance condition $2\Delta_{ct}=-\alpha_c$.

\item[$\text{I}_{B}$)]
A three-photon process, where three times the drive frequency is resonant with the transition $\ket{\psi_{c,0}}\to\ket{\psi_{c,3}}$ of the control qubit, resulting in the effective resonance condition $3\Delta_{ct} \approx -3\alpha_c$ [approximate sign is due to neglecting $\beta_c$ in Eq.~(\ref{eqn:EnBasis-EnBasis rep of H})].

\item[$\text{I}_{C}$)]
A single-photon process, where the drive frequency is resonant with the transition $\ket{\psi_{c,1}}\to\ket{\psi_{c,2}}$ of the control qubit, resulting in the effective resonance condition $\Delta_{ct}= -\alpha_c$.

\item[$\text{I}_{D}$)]
A two-photon process, where twice the drive frequency is resonant with the transition $\ket{\psi_{c,1}}\to\ket{\psi_{c,3}}$ of the control qubit, resulting in the effective resonance condition $2\Delta_{ct}=-3\alpha_c$.

\item[$\text{I}_{E}$)]
A single-photon process, where the drive frequency is resonant with the transition $\ket{\psi_{c,2}}\to\ket{\psi_{c,3}}$ of the control qubit, resulting in the effective resonance condition $\Delta_{ct}=-2\alpha_c$.
\end{itemize}

\textbf{Type II}. This category contains the resonances between any two neighboring qubits in the network denoted by index m and n.
\begin{itemize}

\item[$\text{II}_{A}$)]
Transition $\ket{\psi_{m,0}}\to\ket{\psi_{m,1}}$ of qubit $m$ is resonant with transition $\ket{\psi_{n,0}}\to\ket{\psi_{n,1}}$ of qubit $n$ resulting in the resonance condition $\Delta_{mn}=0$. 

\item[$\text{II}_{B}$)]
Transition $\ket{\psi_{m,0}}\to\ket{\psi_{m,1}}$ of qubit $m$ is resonant with transition $\ket{\psi_{n,1}}\to\ket{\psi_{n,2}}$ of qubit $n$, or vice versa, resulting in the resonance condition $\Delta_{mn}=\alpha_n$ or $\Delta_{mn}=-\alpha_m$, respectively. 
 
\item[$\text{II}_{C}$)]
Transition $\ket{\psi_{m,0}}\to\ket{\psi_{m,2}}$ of qubit $m$ is resonant with transition $\ket{\psi_{n,0}}\to\ket{\psi_{n,2}}$ of qubit $n$ resulting in the resonance condition $2\Delta_{mn}=\alpha_n-\alpha_m$.

\item[$\text{II}_{D}$)]
Transition $\ket{\psi_{m,1}}\to\ket{\psi_{m,2}}$ of qubit $m$ is resonant with transition $\ket{\psi_{n,1}}\to\ket{\psi_{n,2}}$ of qubit $n$ resulting in the resonance condition $\Delta_{mn}=\alpha_n-\alpha_m$.

\item[$\text{II}_{E}$)]
Transition $\ket{\psi_{m,2}}\to\ket{\psi_{m,3}}$ of qubit $m$ is resonant with transition $\ket{\psi_{n,0}}\to\ket{\psi_{n,1}}$ of qubit $n$, or vice versa, resulting in the resonance condition $\Delta_{mn}=-2\alpha_m$ or $\Delta_{mn}=2\alpha_n$, respectively.
\end{itemize}

\textbf{Type III}. This category contains possible three-qubit resonances. We label the qubits as $m$, $n$ and $l$ and without loss of generality we assume that $m$ is connected to $n$, and $n$ is connected to $l$. Some of the types enumerated below resemble the ones in Type II as it involves only two of the qubits. The important distinction, however, is that here those resonance happen between \textit{next-nearest} neighboring qubits.   
\begin{itemize}

\item[$\text{III}_{A}$)]
Transition $\ket{\psi_{m,0}}\to\ket{\psi_{m,1}}$ of qubit $m$ is resonant with transition $\ket{\psi_{l,0}}\to\ket{\psi_{l,1}}$ of qubit $l$ resulting in the resonance condition $\Delta_{ml}=0$.   

\item[$\text{III}_{B}$)]
Transition $\ket{\psi_{m,0}}\to\ket{\psi_{m,1}}$ of qubit $m$ is resonant with transition $\ket{\psi_{l,1}}\to\ket{\psi_{l,2}}$ of qubit $l$, or vice versa, resulting in the resonance condition $\Delta_{ml}=\alpha_l$ or $\Delta_{ml}=-\alpha_m$.

\item[$\text{III}_{C}$)]
Transition $\ket{\psi_{m,1}}\to\ket{\psi_{m,2}}$ of qubit $m$ is resonant with transition $\ket{\psi_{l,1}}\to\ket{\psi_{l,2}}$ of qubit $l$ resulting in the resonance condition $\Delta_{ml}=\alpha_l-\alpha_m$.

\item[$\text{III}_{D}$)]
Sum of transition frequencies $\ket{\psi_{m,0}}\to\ket{\psi_{m,1}}$ and $\ket{\psi_{l,0}}\to\ket{\psi_{l,1}}$ of qubits $m$ and $l$ is resonant with transition $\ket{\psi_{n,0}}\to\ket{\psi_{0,2}}$ of qubit $n$ resulting in the resonance condition $\Delta_{mn}+\Delta_{ln}=\alpha_n$.

\item[$\text{III}_{E}$)]
Sum of transition frequencies $\ket{\psi_{m,0}}\to\ket{\psi_{m,1}}$, $\ket{\psi_{n,0}}\to\ket{\psi_{n,1}}$ and $\ket{\psi_{l,0}}\to\ket{\psi_{l,1}}$ of qubits $m$, $n$ and $l$ is resonant with transition $\ket{\psi_{n,0}}\to\ket{\psi_{0,3}}$ of qubit $n$ resulting in the resonance condition $\Delta_{mn}+\Delta_{ln}=3\alpha_n$.
\end{itemize}

Despite the fact that the broad resonance types enumerated above are found for only two- or three-qubit systems, they lay out a realistic guideline for avoiding frequency collisions and frequency crowding in a larger network of qubits. This is true since the resonances involving qubits beyond next-nearest neighbors only appear in higher order perturbation and tend to be orders of magnitude weaker, unless for very strong drive regime.
\section{Conclusion}
\label{Sec:Conclusion}

In this work, we followed a bottom-up approach in our analysis of CR gate operation. Starting from a slightly modified model that accounts for qubit eigenstate renormalization, compared to previous theoretical studies \cite{Magesan_Effective_2018, Tripathi_Operation_2019}, we first analyzed an isolated CR gate and characterized candidate parameters to achieve reasonable gate speed and gate error. Our calculations confirm that for drive amplitude close to 50 MHz, a gate speed of the order of 2 MHz can be achieved when control-target detuning is in the straddling regime $0<\Delta_{ct}<-\alpha_c$. In particular, detuning region III ($-\alpha_c/2<\Delta_{ct}<-\alpha_c$) results in the largest $ZX$ rate and hence the fastest gate. Analysis of the gate error with an echo pulse sequence revealed that there are optimal spots in the middle of detuning regions II ($0 <\Delta_{ct}<-\alpha_c/2$), I ($\alpha_t <\Delta_{ct}<0 $) and III, where the coherent error can range in $10^{-4}<E_{\text{ech}}<10^{-3}$. Splitting the error into local (single-qubit) and non-local (two-qubit) parts shows a wide room for error improvement provided that the echo pulse is amended with additional single-qubit rotations. Having understood the two-qubit physics, we considered the simplest generalization consisting of three-qubits, in which either the control or the target is coupled to a spectator qubit. Spectator qubit analysis reveals a series of multi-qubit processes causing detrimental frequency collisions, which are crucial to avoid in designing a network of qubits. In summary, our analysis lays out the groundwork and provides a roadmap for designing optimal CR gate operation in a quantum processor.    
\section{Acknowledgements}
\label{Sec:Acknowledgements} 
We appreciate helpful conversations with Emily Pritchett, John Timmerwilke, Zlatko K. Minev, Abhinav Kandala, Hanhee Paik, Jared Hertzberg, Antonio Corcoles, Petar Jurcevic, Seth Merkel, Oliver Dial, Jerry M. Chow and Jay M. Gambetta. This work was in part supported by the Army Research Office (ARO) under contract W911NF-14-1-0124 and by the Intelligence Advanced Research Projects Activity (IARPA) under contract W911NF-16-0114.

\clearpage
\appendix
\section{Transmon spectrum and modified two-qubit interactions}
\label{App:TransSpectrum}

In this appendix, we revisit the spectrum of a transmon qubit and provide perturbative results for its eigenenergies and eigenstates in terms of the unitless anharmonicity scale $\epsilon\equiv\sqrt{2E_C/E_J}$. The main difference with respect to Kerr theory appears in the renormalization of the eigenstates, which leads to modified matrix elements for qubit-qubit interaction and drive. 

We start from the transmon Hamiltonian in terms of the charging ($E_C$) and Jopsephson ($E_J$) energies as
\begin{align}
\HO_q=4E_C\hat{N}^2-E_J\cos(\hat{\varphi}) \;,
\label{Eq:TransSpec-Def of Hq 1}
\end{align}
where $\hat{\varphi}$ and $\hat{N}$ are the phase and number operators, respectively. Next, we replace the quadratures in terms of their zero-point fluctuation amplitudes as
\begin{subequations}
\begin{align}
&\hat{\varphi}=\varphi_{\text{zpf}}\hat{x}=\left(\frac{2E_C}{E_J}\right)^{1/4}\left(\hat{b}+\hat{b}^{\dag}\right),\\
&\hat{N}=N_{\text{zpf}}\hat{y}=\frac{1}{2}\left(\frac{E_J}{2E_C}\right)^{1/4}\left[-i\left(\hat{b}-\hat{b}^{\dag}\right)\right] \;,
\end{align}
\end{subequations}
in terms of which we obtain a new representation of the transmon Hamiltonian as
\begin{align}
\HO_{\text{q}}=\frac{\omega_h}{4}\left[\hat{y}^2-\frac{2}{\epsilon}\cos(\sqrt{\epsilon}\hat{x})\right] \;.
\label{Eq:TransSpec-Def of Hq 2}
\end{align}
In Eq.~(\ref{Eq:TransSpec-Def of Hq 2}), $\omega_h\equiv \sqrt{8E_CE_J}$ is the harmonic frequency of the qubit and $\epsilon\equiv\sqrt{2E_C/E_J}$ is a unitless anharmonicty measure in terms of which we can solve for the spectrum perturbatively. Moreover, $\hat{x}\equiv \hat{b}+\hat{b}^{\dag}$ and $\hat{y}\equiv -i(\hat{b}-\hat{b}^{\dag})$ are the unitless phase and number operators.

Next, we expand Hamiltonian~(\ref{Eq:TransSpec-Def of Hq 2}) in powers of $\epsilon$ as
\begin{align}
\hat{\mathcal{H}}_{\text{q}}=\sum\limits_{p=0}^{\infty}\epsilon^p\hat{\mathcal{H}}_{\text{q}}^{(p)} \;,
\label{Eq:TransSpec-SymbExpansion of Hq}
\end{align}
where the harmonic part is given as $\hat{\mathcal{H}}_{\text{q}}^{(0)}=\omega_h\hat{b}^{\dag}\hat{b}$. The nonlinear contributions for $p \geq 1$ read	
\begin{align}
\begin{split}
\hat{\mathcal{H}}_{\text{q}}^{(p)} &\equiv \omega_h \frac{(-1)^{p}}{2(2p+2)!}\left(\hat{b}+\hat{b}^{\dag}\right)^{2p+2}\\
&=\omega_h\sum\limits_{m=0}^{p}\sum\limits_{l=-(m+1)}^{l=m+1}\Big[\frac{(-1)^p}{2^{p-m+1}(p-m)!}\\
&\times\frac{\left(\hat{b}^{\dag}\right)^{m+1+l}}{(m+1+l)!}\frac{\hat{b}^{m+1-l}}{(m+1-l)!}\Big] \;,	
\end{split}
\label{Eq:TransSpec-Def of Hq^(n)}
\end{align}
where the first expression shows the Taylor expansion of the cosine potential and the last step shows the normal-ordered form. We then develop a perturbative expansion of the eigenenergies and eigenstates of the transmon in powers of $\epsilon$ as
\begin{subequations}
\begin{align}
&E_n=\sum\limits_{p=0}^{\infty}\epsilon^p E_n^{(p)} \;,
\label{Eq:TransSpec-Def of omega_n}\\
&\ket{\psi_n}=\sum\limits_{p=0}^{\infty}\epsilon^p \ket{\psi_n^{(p)}} \;.
\label{Eq:TransSpec-Def of Psi_n}
\end{align}
\end{subequations}
Replacing Eqs.~(\ref{Eq:TransSpec-SymbExpansion of Hq}), (\ref{Eq:TransSpec-Def of omega_n}) and (\ref{Eq:TransSpec-Def of Psi_n}) into the eigenvalue problem $\hat{\mathcal{H}}_{\text{q}}\ket{\psi_n}=E_n \ket{\psi_n}$, one can solve for the spectrum recursively as (See also Ref.~\cite{Didier_Analytical_2018})
\begin{subequations}
\begin{align}
E_n^{(p)}=\sum\limits_{r=0}^{p-1}\bra{n}\hat{\mathcal{H}}_{\text{q}}^{(p-r)}\ket{\psi_n^{(r)}} \;,
\label{Eq:TransSpec-RecSol of omega_n^(p)}
\end{align}
\begin{align}
\begin{split}
\ket{\psi_n^{(p)}}&=\sum\limits_{m\neq n}\Big\{\frac{1}{(n-m)\omega_h}\Big[\bra{m}\hat{\mathcal{H}}_{\text{q}}^{(p)}\ket{n}\\
&+\sum\limits_{r=0}^{p-1}\bra{m}\left(\hat{\mathcal{H}}_{\text{q}}^{(p-r)}-E_n^{(p-r)}\right)\ket{\psi_n^{(r)}}\Big]\Big\}\ket{m} \;.
\end{split}
\label{Eq:TransSpec-RecSol of psi_n^(p)}
\end{align}
\end{subequations}

For our analytical calculation of the CR gate parameters, we keep four transmon levels which is essential to correctly capture the higher order behavior of gate parameters in terms of drive amplitude $\Omega$. Therefore, the eigenenergies up to $O(\epsilon^3)$ are found as
\begin{subequations}
\begin{align}
&\frac{E_1-E_0}{\omega_h} =1-\frac{1}{4}\epsilon-\frac{1}{16}\epsilon^2+O(\epsilon^3) \;,
\label{Eq:TransSpec-PTSol of omega_10}\\
&\frac{E_2-E_0}{\omega_h} =2-\frac{3}{4}\epsilon-\frac{17}{64}\epsilon^2+O(\epsilon^3) \;,
\label{Eq:TransSpec-PTSol of omega_20}\\
&\frac{E_3-E_0}{\omega_h} =3-\frac{3}{2}\epsilon-\frac{45}{64}\epsilon^2+O(\epsilon^3) \;,
\label{Eq:TransSpec-PTSol of omega_30}
\end{align}
\end{subequations}
where Eq.~(\ref{Eq:TransSpec-PTSol of omega_10}) provides the expression for qubit frequency $\omega\equiv E_1-E_0$. Furthermore, from Eqs.~(\ref{Eq:TransSpec-PTSol of omega_10}) and~(\ref{Eq:TransSpec-PTSol of omega_20}) we obtain qubit anharmonicity $\alpha$ as
\begin{align}
\frac{\alpha}{\omega_h} \equiv \frac{(E_{2}-E_{1})-(E_{1}-E_{0})}{\omega_h}=-\frac{1}{4}\epsilon-\frac{9}{64}\epsilon^2+O(\epsilon^3) \;.
\label{Eq:TransSpec-PTSol of alpha}
\end{align}
The first four transmon eigenstates read
\begin{subequations}
\begin{align}
\begin{split}
\ket{\psi_0}&=\left(1-\frac{13}{3072}\epsilon^2\right)\ket{0}+\left(\frac{1}{8\sqrt{2}}\epsilon+\frac{13}{384 \sqrt{2}}\epsilon ^2\right)\ket{2}\\
&+\left(\frac{\sqrt{6}}{96}\epsilon+\frac{\sqrt{6}}{96}\epsilon^2\right)\ket{4}+\frac{23}{768\sqrt{5}}\epsilon^2\ket{6}\\
&+\frac{\sqrt{\frac{35}{2}}}{1536}\epsilon^2\ket{8}+O(\epsilon^3) \;,
\end{split}
\label{Eq:TransSpec-PTSol of psi_0}
\end{align}
\begin{align}
\begin{split}
\ket{\psi_1}&=\left(1-\frac{35}{1024}\epsilon^2\right)\ket{1}+\left(\frac{5}{8\sqrt{6}}\epsilon +\frac{37}{128 \sqrt{6}}\epsilon ^2\right)\ket{3}\\
&+\left(\frac{1}{16}\sqrt{\frac{5}{6}} \epsilon+\frac{41 }{64\sqrt{30}}\epsilon^2\right)\ket{5}+\frac{11}{256}\sqrt{\frac{7}{5}}\epsilon^2\ket{7}\\
&+\frac{1}{512} \sqrt{\frac{35}{2}} \epsilon ^2\ket{9}+O(\epsilon^3) \;,
\end{split}
\label{Eq:TransSpec-PTSol of psi_1}
\end{align}
\begin{align}
\begin{split}
\ket{\psi_2}&=\left(-\frac{1}{8\sqrt{2}}\epsilon-\frac{5 }{96 \sqrt{2}}\epsilon ^2\right)\ket{0}+\left(1-\frac{419}{3072}\epsilon ^2\right)\ket{2}\\
&+\left(\frac{7}{8\sqrt{3}}\epsilon+\frac{145}{256 \sqrt{3}}\epsilon^2\right)\ket{4}\\
&+\left(\frac{1}{16}\sqrt{\frac{5}{2}}\epsilon+\frac{103}{96\sqrt{10}}\epsilon^2\right)\ket{6}\\
&+\frac{43}{384} \sqrt{\frac{7}{5}} \epsilon ^2\ket{8}+\frac{5}{512} \sqrt{\frac{7}{2}} \epsilon ^2 \ket{10}+O(\epsilon^3) \;,
\end{split}
\label{Eq:TransSpec-PTSol of psi_2}
\end{align}
\begin{align}
\begin{split}
\ket{\psi_3}&=\left(-\frac{5}{8\sqrt{6}}\epsilon-\frac{13}{32\sqrt{6}}\epsilon^2\right)\ket{1}+\left(1-\frac{405}{1024}\epsilon ^2\right)\ket{3}\\
&+\left(\frac{3\sqrt{5}}{8}\epsilon+\frac{79\sqrt{5}}{256}\epsilon ^2\right)\ket{5}\\
&+\left(\frac{1}{16}\sqrt{\frac{35}{6}}\epsilon+\frac{103}{64}\sqrt{\frac{7}{30}}\epsilon^2\right)\ket{7}\\
&+\frac{53}{128} \sqrt{\frac{7}{15}} \epsilon ^2 \ket{9}+ \frac{5}{512} \sqrt{\frac{77}{6}} \epsilon ^2\ket{11}+O(\epsilon^3) \;.
\end{split}
\label{Eq:TransSpec-PTSol of psi_3}
\end{align}
\end{subequations}
Based on Eqs.~(\ref{Eq:TransSpec-PTSol of omega_10}--\ref{Eq:TransSpec-PTSol of psi_3}), we can write the qubit Hamiltonian in the energy basis as
\begin{align}
\begin{split}
\HO_{\text{q}} &=\omega\ket{\psi_1}\bra{\psi_1}+ (2\omega+\alpha) \ket{\psi_2}\bra{\psi_2}\\
&+ (3\omega+3\alpha+\beta) \ket{\psi_3}\bra{\psi_3},
\label{Eq:TransSpec-EnBasis rep of Ham}
\end{split}
\end{align}
where $\beta\equiv -(6/64)\epsilon^2 \omega_h$ provides the deviation from the Kerr level structure for the third excited state of transmon, which is negligible unless the drive frequency is directly resonant with the $\ket{\psi_2}\leftrightarrow \ket{\psi_3}$ transition.

Eigenstate renormalization leads to modified interactions between the qubits. To see this explicitly, we need to project the interaction Hamiltonian into the energy basis. If the interactions between the qubits is linear (capacitive or inductive), it is sufficient to first obtain the matrix elements of $\hat{x}$ and $\hat{y}$ in the new basis as
\begin{subequations}
\begin{align}
&\mu_{mn}\equiv \bra{\psi_m}\hat{x}\ket{\psi_n} \;,
\label{Eq:TransSpec-Def of mu_mn}\\
&\nu_{mn}\equiv \bra{\psi_m}\hat{y}\ket{\psi_n} \;.
\label{Eq:TransSpec-Def of nu_mn}
\end{align}
\end{subequations}
For simplicity, we separate the lowering (-) and raising (+) parts of the quadratures as
\begin{subequations}
\begin{align}
&\hat{x}=\hat{x}^{-}+\hat{x}^{+} \;,
\label{Eq:TransSpec-Def of X^-}\\
&\hat{y}=-i\left(\hat{y}^{-}-\hat{y}^{+}\right) \;,
\label{Eq:TransSpec-Def of Y^-}
\end{align}
\end{subequations}
where $\hat{x}^{+}=(\hat{x}^{-})^{\dag}$ and $\hat{y}^{+}=(\hat{y}^{-})^{\dag}$. Note that in the harmonic limit of the problem one finds $\lim\limits_{\epsilon\rightarrow 0} \hat{x}^{-}=\lim\limits_{\epsilon\rightarrow 0} \hat{y}^{-}=\hat{b}$. Using Eqs.~(\ref{Eq:TransSpec-PTSol of psi_0}--\ref{Eq:TransSpec-PTSol of psi_3}) we find the following matrix representations for $\hat{x}^-$ up to the fourth level of transmon as 
\begin{subequations}
\begin{align}
\hat{x}^- \approx
\begin{bmatrix}
0 & \mu_{01} & 0 & \mu_{03}\\
0 & 0 & \mu_{12} & 0\\
0 & 0 & 0 & \mu_{23}\\
0 & 0 & 0 & 0\\
\end{bmatrix} \;,
\label{Eq:TransSpec-MatRep of X^-}
\end{align}
where $\mu_{mn}$ are found up to $O(\epsilon^3)$ as
\begin{align}
&\mu_{01}=1+\frac{1}{8}\epsilon+\frac{13}{256}\epsilon^2+O(\epsilon^3) \;,
\label{Eq:TransSpec-PTSol of mu_01}\\
&\mu_{12}=\left(1+\frac{1}{4}\epsilon+\frac{95}{512}\epsilon^2\right)\sqrt{2}+O(\epsilon^3) \;,
\label{Eq:TransSpec-PTSol of mu_12}\\
&\mu_{23}=\left(1+\frac{3}{8}\epsilon+\frac{105}{256}\epsilon^2\right)\sqrt{3}+O(\epsilon^3) \;,
\label{Eq:TransSpec-PTSol of mu_23}\\
&\mu_{03}=-\frac{\sqrt{6}}{48}\epsilon-\frac{3\sqrt{6}}{128}\epsilon^2+O(\epsilon^3) \;.
\label{Eq:TransSpec-PTSol of mu_03}
\end{align}
\end{subequations}
We find a similar matrix representation for $\hat{y}^{-}$ as
\begin{subequations}
\begin{align}
\hat{y}^- \approx
\begin{bmatrix}
0 & \nu_{01} & 0 & \nu_{03}\\
0 & 0 & \nu_{12} & 0\\
0 & 0 & 0 & \nu_{23}\\
0 & 0 & 0 & 0\\
\end{bmatrix} \;,
\label{Eq:TransSpec-MatRep of Y^-}
\end{align}
where $\nu_{mn}$ read
\begin{align}
&\nu_{01}=1-\frac{1}{8}\epsilon-\frac{11}{256}\epsilon^2+O(\epsilon^3) \;,
\label{Eq:TransSpec-PTSol of nu_01}\\
&\nu_{12}=\left(1-\frac{1}{4}\epsilon-\frac{73}{512}\epsilon^2\right)\sqrt{2}+O(\epsilon^3) \;,
\label{Eq:TransSpec-PTSol of nu_12}\\
&\nu_{23}=\left(1-\frac{3}{8}\epsilon-\frac{79}{256}\epsilon^2\right)\sqrt{3}+O(\epsilon^3) \;,
\label{Eq:TransSpec-PTSol of nu_23}\\
&\nu_{03}=-\frac{\sqrt{6}}{16}\epsilon-\frac{5\sqrt{6}}{128}\epsilon^2+O(\epsilon^3) \;.
\label{Eq:TransSpec-PTSol of nu_03}
\end{align}
\end{subequations}

Equations~(\ref{Eq:TransSpec-EnBasis rep of Ham}) and~(\ref{Eq:TransSpec-MatRep of X^-}--\ref{Eq:TransSpec-PTSol of nu_03}) are the main results of this appendix and are used to construct a new starting Hamiltonian for the CR gate [See Eqs.~(\ref{eqn:CRHamInEnBasis-H0 2}--\ref{eqn:CRHamInEnBasis-Hd in new basis}) and Fig.~\ref{fig:CRHamInEnBasis-JCLadderForCR} of the main text].

\section{Two-qubit dressed basis}
\label{App:DressedBasis}

In this appendix, we obtain the dressing of transmon energy eigenstates due to the exchange interaction $J$. Note that $J$ is at least one order of magnitude smaller than the drive amplitude $\Omega$. Hence, in practice, it is sufficient to only obtain the lowest order correction to eigenenergies and eigenstates due to the exchange interaction. We note that the outcome of this appendix is not directly utilized in the main body of the paper, since we performed a simultaneous perturbation in $J$ and $\Omega$. However, we present the dressed two-qubit states for completeness and for a sanity check on our simultaneous perturbative results. Moreover, for more precise numerical analysis, the knowledge of the dressed frame becomes essential.

The undriven system Hamiltonian for the CR gate can be expressed as	
\begin{align}
\HO_s=\HO_{q_c}+\HO_{q_t}+\HO_J \;,
\label{Eq:DressedBasis-Def of H_0}
\end{align}
where $\HO_{q_c}$ and $\HO_{q_t}$ denote the control and the target qubit Hamiltonians, and $\HO_J$ is the exchange interaction, respectively. We employ a four-level representation of each qubit following our discussion in Appendix~\ref{App:TransSpectrum}. The qubit Hamiltonian is then expressed as
\begin{align}
\begin{split}
\HO_{\text{q}_c} &=\omega_c \ket{\psi_{c,1}}\bra{\psi_{c,1}}+ (2\omega_c+\alpha_c) \ket{\psi_{c,2}}\bra{\psi_{c,2}}\\
&+ (3\omega_c+3\alpha_c) \ket{\psi_{c,3}}\bra{\psi_{c,3}} \;,
\end{split}
\label{Eq:DressedBasis-H_qc}\\
\begin{split}
\HO_{\text{q}_t} &=\omega_t \ket{\psi_{t,1}}\bra{\psi_{t,1}}+ (2\omega_t+\alpha_t) \ket{\psi_{t,2}}\bra{\psi_{t,2}}\\
&+ (3\omega_t+3\alpha_t) \ket{\psi_{t,3}}\bra{\psi_{t,3}} \;,
\end{split}
\label{Eq:DressedBasis-H_qc}
\end{align}
with $\omega_{q/c}$ and $\alpha_{c/t}$ denoting the frequency and the anharmonicity for each qubit. The exchange Hamiltonian is engineered through a charge-charge interaction of the form
\begin{align}
\HO_J=J\hat{y}_c\hat{y}_t\approx J\left(\hat{y}_c^{+}\hat{y}_t^{-}+\hat{y}_t^{+}\hat{y}_c^{-}\right) \;,
\end{align}
with the raising and lowering operators $\hat{y}_{c/t}^{\pm}$ given in terms of Eqs.~(\ref{Eq:TransSpec-MatRep of Y^-}--\ref{Eq:TransSpec-PTSol of nu_03}).

In the following, we apply a time-independent perturbation theory in $J/\Delta_{ct}$ to obtain corrections to the two-qubit eigenenergies and eigenstates (16 in total for a four-level model of each transmon). We note that there are multiple perturbation techniques available, namely either the Rayleigh-Schrodinger perturbation used in Appendix~\ref{App:TransSpectrum} or the SWPT of Appendix~\ref{App:SWPT} [Eqs.~(\ref{Eq:SWPT-O(1) Heff}--\ref{Eq:SWPT-O(lambda^4) Heff})], and we confirm that regardless of the technique the results agree. To visualize the underlying interactions and understand the corrections better, we refer the reader to Fig.~\ref{fig:CRHamInEnBasis-JCLadderForCR}. 	

We set the order of composite Hilbert space as $\text{control}\otimes \text{target}$ and group the results in terms of sectors labeled by the state of the control qubit. The eigenenergies in the $c=0$ sector of the two-qubit Hilbert space read
\begin{subequations}
\begin{align}
&\bar{E}_{00}=E_{00} \;,
\label{Eq:DressedBasis-bar(E)_00}\\
&\bar{E}_{01}=E_{01}-\frac{\nu_{c,01}^2\nu_{t,01}^2J^2}{\Delta_{ct}} \;,
\label{Eq:DressedBasis-bar(E)_01}\\
&\bar{E}_{02}=E_{02}-\frac{\nu_{\text{c,01}}^2\nu_{\text{t,12}}^2 J^2}{\Delta_{\text{ct}}-\alpha_t} \;,
\label{Eq:DressedBasis-bar(E)_02}\\
&\bar{E}_{03}=E_{03}-\frac{\nu_{\text{c,01}}^2\nu_{\text{t,23}}^2 J^2}{\Delta_{\text{ct}}-2\alpha_t} \;, \label{Eq:DressedBasis-bar(E)_03}
\end{align}
where we have used a bar-notation to distinguish between bare and dressed states.
Similarly, when the control is in the first excited state $c=1$, the dressed eigenenergies are found as
\begin{align}
&\bar{E}_{10}=E_{10}+\frac{\nu_{\text{c,01}}^2 \nu_{\text{t,01}}^2 J^2}{\Delta_{\text{ct}}} \;,
\label{Eq:DressedBasis-bar(E)_10}\\
&\bar{E}_{11}=E_{11}+\frac{\nu_{\text{c,01}}^2 \nu_{\text{t,12}}^2 J^2}{\Delta_{\text{ct}}-\alpha_t}-\frac{\nu_{\text{c,12}}^2 \nu_{\text{t,01}}^2 J^2}{\Delta_{\text{ct}}+\alpha_c} \;,
\label{Eq:DressedBasis-bar(E)_11}\\
&\bar{E}_{12}=E_{12}+\frac{\nu_{\text{c,01}}^2 \nu_{\text{t,23}}^2 J^2}{\Delta _{\text{ct}}-2\alpha _t}-\frac{\nu_{\text{c,12}}^2 \nu_{\text{t,12}}^2 J^2}{\Delta_{\text{ct}}+\alpha_c-\alpha_t} \;,
\label{Eq:DressedBasis-bar(E)_12}\\
&\bar{E}_{13}=E_{13}-\frac{\nu_{\text{c,12}}^2 \nu_{\text{t,23}}^2 J^2}{\Delta_{\text{ct}}+\alpha_c-2\alpha_t} \;.
\label{Eq:DressedBasis-bar(E)_13}
\end{align}
The dressed eigenenergies in the $c=2$ sector are obtained as
\begin{align}
&\bar{E}_{20}=E_{20}+\frac{\nu_{\text{c,12}}^2 \nu_{\text{t,01}}^2 J^2}{\Delta_{\text{ct}}+\alpha_c} \;,
\label{Eq:DressedBasis-bar(E)_20}\\
&\bar{E}_{21}=E_{21}+\frac{\nu_{\text{c,12}}^2 \nu_{\text{t,12}}^2 J^2}{\Delta_{\text{ct}}+\alpha_c-\alpha_t}-\frac{\nu_{\text{c,23}}^2 \nu_{\text{t,01}}^2 J^2}{\Delta_{\text{ct}}+2\alpha_c} \;,
\label{Eq:DressedBasis-bar(E)_21}\\
&\bar{E}_{22}=E_{22}+\frac{\nu_{\text{c,12}}^2 \nu_{\text{t23}}^2 J^2}{\Delta_{\text{ct}}+\alpha_c-2\alpha_t}-\frac{J^2\nu_{\text{c,23}}^2 \nu_{\text{t,12}}^2 J^2}{\Delta_{\text{ct}}+2\alpha_c-\alpha_t} \;,
\label{Eq:DressedBasis-bar(E)_22}\\
&\bar{E}_{23}=E_{23}-\frac{J^2 \nu_{\text{c,23}}^2 \nu_{\text{t,23}}^2}{\Delta_{\text{ct}}+2\alpha_c-2 \alpha_t} \;.
\label{Eq:DressedBasis-bar(E)_23}
\end{align}
Lastly, when the control qubit is in the third excited state, i.e. $c=3$ sector, we find
\begin{align}
&\bar{E}_{30}=E_{30}+\frac{\nu_{\text{c,23}}^2 \nu_{\text{t,01}}^2 J^2}{\Delta_{\text{ct}}+2\alpha_c} \;,
\label{Eq:DressedBasis-bar(E)_30}\\
&\bar{E}_{31}=E_{31}+\frac{\nu _{\text{c,23}}^2 \nu_{\text{t,12}}^2 J^2}{\Delta_{\text{ct}}+2\alpha_c-\alpha_t} \;,
\label{Eq:DressedBasis-bar(E)_31}\\
&\bar{E}_{32}=E_{32}+\frac{\nu_{\text{c,23}}^2\nu_{\text{t,23}}^2 J^2}{\Delta_{\text{ct}}+2\alpha_c-2\alpha_t} \;,
\label{Eq:DressedBasis-bar(E)_32}\\
&\bar{E}_{33}=E_{33}	 \;.
\label{Eq:DressedBasis-bar(E)_33}
\end{align}
\end{subequations}

The lowest order corrections to eigenstates are proportional to $J$. The states with $c=0$ are renormalized as
\begin{subequations}
\begin{align}
&\ket{\bar{\psi}_{00}}=\ket{\psi_{00}} \;,
\label{Eq:DressedBasis-bar(psi)_00}\\
&\ket{\bar{\psi}_{01}}=\ket{\psi_{01}}-\frac{\nu_{c,01}\nu_{t,01} J}{\Delta_{ct}}\ket{\psi_{10}} \;,
\label{Eq:DressedBasis-bar(psi)_01}\\
&\ket{\bar{\psi}_{02}}=\ket{\psi_{02}}-\frac{\nu_{c,01}\nu_{t,12} J}{\Delta_{ct}-\alpha_t}\ket{\psi_{11}} \;,
\label{Eq:DressedBasis-bar(psi)_02}\\
&\ket{\bar{\psi}_{03}}=\ket{\psi_{03}}-\frac{\nu_{c,01}\nu_{t,23} J}{\Delta_{ct}-2\alpha_t}\ket{\psi_{12}} \;.
\label{Eq:DressedBasis-bar(psi)_03}
\end{align}
\end{subequations}
In the $c=1$ sector we find
\begin{subequations}
\begin{align}
\ket{\bar{\psi}_{10}}=\ket{\psi_{10}}+\frac{\nu_{c,01}\nu_{t,01} J}{\Delta_{ct}}\ket{\psi_{01}} \;,
\label{Eq:DressedBasis-bar(psi)_10}
\end{align}
\begin{align}
\begin{split}
\ket{\bar{\psi}_{11}}=\ket{\psi_{11}}+\frac{\nu_{c,01}\nu_{t,12} J}{\Delta_{ct}-\alpha_t}\ket{\psi_{02}}\\
-\frac{\nu_{c,12}\nu_{t,01} J}{\Delta_{ct}+\alpha_c}\ket{\psi_{20}} \;,
\end{split}
\label{Eq:DressedBasis-bar(psi)_11}
\end{align}
\begin{align}
\begin{split}
\ket{\bar{\psi}_{12}}=\ket{\psi_{12}}+\frac{\nu_{c,01}\nu_{t,23} J}{\Delta_{ct}-2\alpha_t}\ket{\psi_{03}}\\
-\frac{\nu_{c,12}\nu_{t,12}J}{\Delta_{ct}+\alpha_c-\alpha_t}\ket{\psi_{21}} \;,
\end{split}
\label{Eq:DressedBasis-bar(psi)_12}
\end{align}
\begin{align}
\ket{\bar{\psi}_{13}}=\ket{\psi_{13}}-\frac{\nu_{c,12} \nu_{\text{t,23}}J}{\Delta_{ct}+\alpha_c-2 \alpha_t}\ket{\psi_{22}} \;.
\label{Eq:DressedBasis-bar(psi)_13}
\end{align}	
\end{subequations}
The eigenstates in the $c=2$ sector read
\begin{subequations}
\begin{align}
\ket{\bar{\psi}_{20}}=\ket{\psi_{20}}+\frac{\nu_{c,12}\nu_{t,01}J}{\Delta_{ct}+\alpha_c}\ket{\psi_{11}} \;,
\label{Eq:DressedBasis-bar(psi)_20}
\end{align}
\begin{align}
\begin{split}
\ket{\bar{\psi}_{21}}=\ket{\psi_{21}}+\frac{\nu_{c,12}\nu_{t,12}J}{\Delta_{\text{ct}}+\alpha_c-\alpha_t}\ket{\psi_{12}}\\
-\frac{\nu_{c,23}\nu_{t,01}J}{\Delta_{ct}+2\alpha_c}\ket{\psi_{30}} \;,
\end{split}
\label{Eq:DressedBasis-bar(psi)_21}
\end{align}
\begin{align}
\begin{split}
\ket{\bar{\psi}_{22}}=\ket{\psi_{22}}+\frac{\nu_{c,12}\nu_{t,23}J}{\Delta_{\text{ct}}+\alpha_c-2\alpha_t}\ket{\psi_{13}}\\
-\frac{\nu_{c,23}\nu_{t,12}J}{\Delta_{ct}+2\alpha _c-\alpha_t}\ket{\psi_{31}} \;,
\end{split}
\label{Eq:DressedBasis-bar(psi)_22}
\end{align}
\begin{align}
\ket{\bar{\psi}_{23}}=\ket{\psi_{23}}-\frac{\nu_{c,23}\nu_{t,23}J}{\Delta_{ct}+2\alpha_c-2\alpha_t}\ket{\psi_{32}} \;.
\label{Eq:DressedBasis-bar(psi)_23}
\end{align}	
\end{subequations}
Lastly, the eigenstates in the $c=3$ sector are obtained as
\begin{subequations}
\begin{align}
&\ket{\bar{\psi}_{30}}=\ket{\psi_{30}}+\frac{\nu_{c,23}\nu_{t,01}J}{\Delta_{ct}+2\alpha_c}\ket{\psi_{21}} \;,
\label{Eq:DressedBasis-bar(psi)_30}\\
&\ket{\bar{\psi}_{31}}=\ket{\psi_{31}}+\frac{\nu_{c,23} \nu_{t,12}J}{\Delta_{ct}+2\alpha_c-\alpha_t}\ket{\psi_{22}} \;,
\label{Eq:DressedBasis-bar(psi)_31}\\
&\ket{\bar{\psi}_{32}}=\ket{\psi_{32}}+\frac{\nu_{c,23}\nu_{t,23}J}{\Delta_{ct}+2\alpha_c-2\alpha_t}\ket{\psi_{23}} \;,
\label{Eq:DressedBasis-bar(psi)_32}\\
&\ket{\bar{\psi}_{33}}=\ket{\psi_{33}} \;.
\label{Eq:DressedBasis-bar(psi)_33}
\end{align}	
\end{subequations}

Based on Eqs.~(\ref{Eq:DressedBasis-bar(E)_00}), (\ref{Eq:DressedBasis-bar(E)_01}), (\ref{Eq:DressedBasis-bar(E)_10}) and~(\ref{Eq:DressedBasis-bar(E)_11}) we find \textit{static} renormalizations of the qubit frequencies as well as an effective \textit{static} $ZZ$ interaction. The resulting effective Hamiltonian in the computational basis reads
\begin{subequations}
\begin{align}
\HO_{s,{\text{TLA}}}=\bar{\omega}_{iz}\frac{\hat{I}\hat{Z}}{2}+\bar{\omega}_{zi}\frac{\hat{Z}\hat{I}}{2}+\bar{\omega}_{zz}\frac{\hat{Z}\hat{Z}}{2} \;,
\label{Eq:DressedBasis-Def of H_0,TLA}
\end{align}
with static $\bar{\omega}_{iz}$, $\bar{\omega}_{zi}$ and $\bar{\omega}_{zz}$ given as
\begin{align}
&\bar{\omega}_{zz} \equiv \frac{1}{2}\left(\frac{\nu_{c,01}^2\nu_{t,12}^2}{\Delta_{ct}-\alpha_t}-\frac{\nu_{c,12}^2 \nu _{t,01}^2}{\Delta_{ct}+\alpha_c}\right)J^2  \;,
\label{Eq:DressedBasis-Def of w_zz^(0)}\\
&\bar{\omega}_{iz} \equiv -\omega_t-\bar{\omega}_{zz}+\frac{\nu_{c,01}^2 \nu_{t,01}^2 J^2}{\Delta_{ct}} \;,
\label{Eq:DressedBasis-Def of w_iz^(0)}\\
&\bar{\omega}_{zi} \equiv -\omega_c -\bar{\omega}_{zz}-\frac{\nu_{c,01}^2 \nu_{t,01}^2 J^2}{\Delta_{ct}} \;.
\label{Eq:DressedBasis-Def of w_zi^(0)}
\end{align}
\end{subequations}

Our knowledge of the dressed frame become important when the CR drive is added to the picture. Since only the dressed frequencies are accessible experimentally, the drive frequency is hence tuned to the dressed frequency of the target qubit as 
\begin{align}
\omega_d \approx \omega_t+\frac{1}{2}\left(\frac{\nu_{c,01}^2\nu_{t,12}^2}{\Delta_{ct}-\alpha_t}-\frac{\nu_{c,12}^2 \nu_{t,01}^2}{\Delta_{ct}+\alpha_c}-\frac{2\nu_{c,01}^2 \nu_{t,01}^2}{\Delta_{ct}}\right)J^2  .
\label{Eq:DressedBasis-Cond for wd}
\end{align}
Therefore, the CR Hamiltonian can be expressed as 
\begin{align}
\HO_s(t)=\HO_s+\HO_{d}(t)\;,
\label{Eq:DressedBasis-Def of Hs(t)}
\end{align}
with $\HO_s$ given in Eq.~(\ref{Eq:DressedBasis-Def of H_0}) and $\HO_{d}(t)$ as
\begin{align}
\begin{split}
\hat{\mathcal{H}}_d(t) = -\Omega \hat{y}_c \sin(\omega_d t)\approx \frac{\Omega}{2}\left(\hat{y}_c^{-}e^{i\omega_d t}+\hat{y}_c^{+}e^{-i\omega_d t}\right) \;.
\end{split}
\label{eqn:DressedBasis-Hd in lab frame}
\end{align}
So far, we found the diagonal form for $\HO_s$ in terms of the dressed two-qubit basis. On the other hand, we also need to rotate $\HO_d(t)$ into this new frame. Based on Eq.~(\ref{eqn:DressedBasis-Hd in lab frame}), it is sufficient to find the representation of the lowering charge operator $\hat{y}_c^{-}$ in the dressed basis as
\begin{subequations}
\begin{align}
\hat{y}_c^{-}=\sum\limits_{m,n=0 \atop l,p=0}^{3}\bra{\bar{\psi}_{mn}}\hat{y}_c^{-}\ket{\bar{\psi}_{lp}} \ket{\bar{\psi}_{mn}}\bra{\bar{\psi}_{lp}} \;.
\label{eqn:DressedBasis-Mat rep of y_c^- in dressed frame}
\end{align}
In principle, there are quite a few non-zero matrix elements which can be found from Eqs.~(\ref{Eq:DressedBasis-bar(psi)_00}--\ref{Eq:DressedBasis-bar(psi)_33}) for the dressed states and Eq.~(\ref{Eq:TransSpec-MatRep of Y^-}) for $\hat{y}_c^{-}$. Here, for simplicity, we only quote the non-zero matrix elements in the computational basis as
\begin{align}
&\bra{\bar{\psi}_{00}}\hat{y}_c^{-}\ket{\bar{\psi}_{01}}=-\frac{\nu_{c,01}^2\nu_{t,01} J}{\Delta_{ct}} \;,
\label{eqn:DressedBasis-<00|y_c^-|01>}\\
&\bra{\bar{\psi}_{00}}\hat{y}_c^{-}\ket{\bar{\psi}_{10}}=\nu_{c,01} \;,
\label{eqn:DressedBasis-<00|y_c^-|10>}\\
&\bra{\bar{\psi}_{01}}\hat{y}_c^{-}\ket{\bar{\psi}_{11}}=\nu_{c,01} \;,
\label{eqn:DressedBasis-<01|y_c^-|11>}\\
&\bra{\bar{\psi}_{10}}\hat{y}_c^{-}\ket{\bar{\psi}_{11}}=\frac{\nu_{c,01}^2 \nu_{t,01} J }{\Delta_{ct}}-\frac{\nu_{c,12}^2 \nu_{t,01} J}{\Delta_{ct}+\alpha_c} \;,
\label{eqn:DressedBasis-<10|y_c^-|11>}
\end{align}
\end{subequations}
from which we find how the drive indirectly acts on the target qubit while being mediated by the control. 

To summarize the main results of this appendix, we found the transformation between the energy-basis representation of the CR gate to the basis that is dressed by the exchange coupling $J$ for the main reason that the dressed basis is the one that is probed experimentally. Consequently, we re-expressed the drive Hamiltonian in this new frame, which serves as a perturbation to the system. Note that the result in this section can be trivially generalized to the spectator calculation, since up to the lowest order the renormalizations are pairwise and will not include next-nearest neighbor coupling.

\section{Time-dependent Schrieffer-Wolff perturbation theory}
\label{App:SWPT}
In this appendix, we provide the derivation of a time-dependent SWPT up to the fourth order. The perturbative equations are used to find an effective Hamiltonian for the CR gate. In Sec.~\ref{SubApp:SWPTLabFrame}, we start our analysis in the lab frame and present our results in terms of a set of linear operator-valued ODEs for the generator of the SW transformation. Next, in Sec.~\ref{SubApp:SWPTIntFrame}, we argue that the form of perturbative equations become simpler when we re-express them in the interaction frame. Lastly, in Sec.~\ref{SubApp:SWPTBlockDiag}, we fine-tune the generic perturbative equations for the CR gate, where we are interested in transforming to a frame in which the effective CR Hamiltonian is block-diagonal with respect to the Hilbert space of the target qubit.     
\subsection{Lab frame}
\label{SubApp:SWPTLabFrame}
The Hamiltonian in the lab frame can be written as
\begin{align}
\hat{\mathcal{H}}_s(t)=\hat{\mathcal{H}}_0+\lambda\hat{\mathcal{H}}_{\text{int}}(t) \;,
\label{Eq:SWPT-Hs}
\end{align}
where $\lambda$ is a small expansion parameter and $\hat{\mathcal{H}}_{\text{int}}(t)$ is the time-dependent perturbation that is applied to the system. We assume that the interaction is off-diagonal. Otherwise, one can always add the diagonal parts of $\hat{\mathcal{H}}_{\text{int}}$ to $\hat{\mathcal{H}}_0$ and redefine the zeroth order Hamiltonian.

In order to obtain an effective Hamilotnian, we apply a SW transformation to the Floquet Hamiltonian as
\begin{align}
\hat{\mathcal{H}}_{\text{eff}}(t)=e^{i\hat{G}(t)}\left[\hat{\mathcal{H}}_s(t)-i\partial_t \right]e^{-i\hat{G}(t)} \;,
\label{Eq:SWPT-Def of Heff}
\end{align}
where $\hat{G}(t)$ is the generator of the SW transformation that can be solved for order by order in the small parameter $\lambda$. 

To obtain a perturbative expansion, we use the BCH lemma as
\begin{subequations}
\begin{align}
\begin{split}
e^{\hat{A}}\hat{B}e^{-\hat{A}}&=\sum\limits_{n=0}^{\infty}\frac{1}{n!}\mathcal{C}_n[\hat{A}]\hat{B}\\
&=\hat{B}+[\hat{A},\hat{B}]+\frac{1}{2}[\hat{A},[\hat{A},\hat{B}]]+\ldots \;,
\end{split}
\label{Eq:SWPT-BCH Lemma}
\end{align}
where $\hat{A}$ and $\hat{B}$ are two arbitrary operators and $\mathcal{C}_n[\hat{A}](\bullet)$ is a nested commutator defined as 	
\begin{align}
\mathcal{C}_n[\hat{A}](\bullet)=[\underbrace{\hat{A},[\hat{A},[\hat{A},[\ldots}_{\text{n}},\bullet]]] \;.
\label{Eq:SWPT-BCH Lemma-C_n[A]}
\end{align}
\end{subequations}
According to Eq.~(\ref{Eq:SWPT-Def of Heff}), there are three separate contributions to the effective Hamiltonian found as the transformations of $\hat{\mathcal{H}}_0$, $\lambda\hat{\mathcal{H}}_{\text{int}}(t)$ and the energy operator $-i\partial_t$. In the following, we focus on each term separately. To this aim, we write the generator as
\begin{align}
\hat{G}(t)=\sum\limits_{n=1}^{\infty}\lambda^n\hat{G}_n(t) \;,
\label{Eq:SWPT-Expansion of G}
\end{align}
where $\hat{G}_n(t)$ denotes the solution to the generator at order $\lambda^n$.

We first consider the transformation of $\hat{\mathcal{H}}_0$. Setting $\hat{A}=i\hat{G}(t)$ and $\hat{B}=\hat{\mathcal{H}}_0$ in the BCH lemma~(\ref{Eq:SWPT-BCH Lemma}) and inserting the expansion for $\hat{G}(t)$ from Eq.~(\ref{Eq:SWPT-Expansion of G}) we find
\begin{align}
\begin{split}
&e^{i\hat{G}(t)}\hat{\mathcal{H}}_0(t)e^{-i\hat{G}(t)}=\hat{\mathcal{H}}_0+i[\hat{G},\hat{\mathcal{H}}_0]-\frac{1}{2}[\hat{G},[\hat{G},\hat{\mathcal{H}}_0]]\\
&-\frac{i}{6}[\hat{G},[\hat{G},[\hat{G},\hat{\mathcal{H}}_0]]]+\frac{1}{24}[\hat{G},[\hat{G},[\hat{G},[\hat{G},\hat{\mathcal{H}}_0]]]]+\ldots\\
&=\hat{\mathcal{H}}_0+\lambda\Big(i[\hat{G}_1,\hat{\mathcal{H}}_0]\Big)+\lambda^2\Big(i[\hat{G}_2,\hat{\mathcal{H}}_0]-\frac{1}{2}[\hat{G}_1,[\hat{G}_1,\hat{\mathcal{H}}_0]]\Big)\\
&+\lambda^3\Big(i[\hat{G}_3,\hat{\mathcal{H}}_0]-\frac{1}{2}[\hat{G}_1,[\hat{G}_2,\hat{\mathcal{H}}_0]]-\frac{1}{2}[\hat{G}_2,[\hat{G}_1,\hat{\mathcal{H}}_0]]\\
&-\frac{i}{6}[\hat{G}_1,[\hat{G}_1,[\hat{G}_1,\hat{\mathcal{H}}_0]]]\Big)+\lambda^4\Big(i[\hat{G}_4,\hat{\mathcal{H}}_0]-\frac{1}{2}[\hat{G}_1,[\hat{G}_3,\hat{\mathcal{H}}_0]]\\
&-\frac{1}{2}[\hat{G}_2,[\hat{G}_2,\hat{\mathcal{H}}_0]]-\frac{1}{2}[\hat{G}_3,[\hat{G}_1,\hat{\mathcal{H}}_0]]-\frac{i}{6}[\hat{G}_1,[\hat{G}_1,[\hat{G}_2,\hat{\mathcal{H}}_0]]]\\
&-\frac{i}{6}[\hat{G}_1,[\hat{G}_2,[\hat{G}_1,\hat{\mathcal{H}}_0]]]-\frac{i}{6}[\hat{G}_2,[\hat{G}_1,[\hat{G}_1,\hat{\mathcal{H}}_0]]]\\
&+\frac{1}{24}[\hat{G}_1,[\hat{G}_1,[\hat{G}_1,[\hat{G}_1,\hat{\mathcal{H}}_0]]]\Big)+O(\lambda^5) \;,
\end{split}
\label{Eq:SWPT-Trans of H0}
\end{align}
where in the last step we have collected distinct powers of $\lambda$ and dropped the time-dependence of operators for clarity.

In a similar manner, we can find the transformation of the interaction term $\lambda\hat{\mathcal{H}}_{\text{int}}(t)$ as
\begin{align}
\begin{split}
&e^{i\hat{G}(t)}\left[\lambda\hat{\mathcal{H}}_{\text{int}}(t)\right]e^{-i\hat{G}(t)}=\lambda\hat{\mathcal{H}}_{\text{int}}+i[\hat{G},\lambda\hat{\mathcal{H}}_{\text{int}}]\\
&-\frac{1}{2}[\hat{G},[\hat{G},\lambda\hat{\mathcal{H}}_{\text{int}}]]-\frac{i}{6}[\hat{G},[\hat{G},[\hat{G},\lambda\hat{\mathcal{H}}_{\text{int}}]]]+\ldots\\
&=\lambda\hat{\mathcal{H}}_{\text{int}}(t)+\lambda^2\Big(i[\hat{G}_1,\hat{\mathcal{H}}_{\text{int}}]\Big)+\lambda^3\Big(i[\hat{G}_2,\hat{\mathcal{H}}_{\text{int}}]\\
&-\frac{1}{2}[\hat{G}_1,[\hat{G}_1,\hat{\mathcal{H}}_{\text{int}}]]\Big)+\lambda^4\Big(i[\hat{G}_3,\hat{\mathcal{H}}_{\text{int}}]\\
&-\frac{1}{2}[\hat{G}_1,[\hat{G}_2,\hat{\mathcal{H}}_{\text{int}}]]-\frac{1}{2}[\hat{G}_2,[\hat{G}_1,\hat{\mathcal{H}}_{\text{int}}]]\\
&-\frac{i}{6}[\hat{G}_1,[\hat{G}_1,[\hat{G}_1,\hat{\mathcal{H}}_{\text{int}}]]]\Big)+O(\lambda^5) \;,
\end{split}
\label{Eq:SWPT-Trans of Hint}
\end{align}
where we find fewer terms due to the fact that the interaction Hamiltonian is of order $\lambda$ to begin with.

Lastly, we need to transform the energy operator $-i\partial_t$. We start from the time-derivative of an operator exponential as
\begin{align}
\frac{d}{dt}e^{\hat{\mathcal{O}}(t)}=\int_{0}^{1}dz e^{z\hat{\mathcal{O}}(t)}\dot{\hat{\mathcal{O}}}(t)e^{(1-z)\hat{\mathcal{O}}(t)} \;.
\label{Eq:SWPT-d/dt exp(O)}
\end{align}
Using identity~(\ref{Eq:SWPT-d/dt exp(O)}) we can write
\begin{align}
\begin{split}
&e^{i\hat{G}(t)}(-i\partial_t)e^{-i\hat{G}(t)}\\
&=\int_{0}^{1}dz e^{i(1-z)\hat{G}(t)}\left[-\dot{\hat{G}}(t)\right]e^{-i(1-z)\hat{G}(t)}\\
&=\int_{0}^{1}dz e^{iz\hat{G}(t)}\left[-\dot{\hat{G}}(t)\right]e^{-iz\hat{G}(t)} \;,
\end{split}
\label{Eq:SWPT-IntegForm exp(G)(-id_t)exp(-G)}
\end{align}
where in the last line, we have employed the change of variable $z\rightarrow 1-z$ to simplify the integral. Setting $\hat{A}=iz\hat{G}(t)$ and $\hat{B}=-\dot{\hat{G}}(t)$ in the BCH lemma~(\ref{Eq:SWPT-BCH Lemma}) and taking the resulting $z$-integral in Eq.~(\ref{Eq:SWPT-IntegForm exp(G)(-id_t)exp(-G)}) we find
\begin{align}
\begin{split}
&e^{i\hat{G}(t)}(-i\partial_t)e^{-i\hat{G}(t)}=-\sum\limits_{n=0}^{n}\frac{i^n}{(n+1)!}\mathcal{C}_n[\hat{G}]\dot{\hat{G}}\\
&=-\dot{\hat{G}}-\frac{i}{2!}[\hat{G},\dot{\hat{G}}]+\frac{1}{3!}[\hat{G},[\hat{G},\dot{\hat{G}}]]+\frac{i}{4!}[\hat{G},[\hat{G},[\hat{G},\dot{\hat{G}}]]+\ldots \;.
\end{split}
\label{Eq:SWPT-CommutFor exp(G)(-id_t)exp(-G)}
\end{align}
Inserting the expansion~(\ref{Eq:SWPT-Expansion of G}) in the generic Eq.~(\ref{Eq:SWPT-CommutFor exp(G)(-id_t)exp(-G)}) and collecting powers of $\lambda$ we find
\begin{align}
\begin{split}
&e^{i\hat{G}(t)}(-i\partial_t)e^{-i\hat{G}(t)}=\lambda\Big(-\dot{\hat{G}}_1\Big)+\lambda^2\Big(-\dot{\hat{G}}_2-\frac{i}{2}[\hat{G}_1,\dot{\hat{G}}_1] \Big)\\
&+\lambda^3\Big(-\dot{\hat{G}}_3-\frac{i}{2}[\hat{G}_1,\dot{\hat{G}}_2]-\frac{i}{2}[\hat{G}_2,\dot{\hat{G}}_1]+\frac{1}{6}[\hat{G}_1,[\hat{G}_1,\dot{\hat{G}}_1]]\Big)\\
&+\lambda^4\Big(-\dot{\hat{G}}_4-\frac{i}{2}[\hat{G}_1,\dot{\hat{G}}_3]-\frac{i}{2}[\hat{G}_2,\dot{\hat{G}}_2]-\frac{i}{2}[\hat{G}_3,\dot{\hat{G}}_1] \\
&+\frac{1}{6}[\hat{G}_1,[\hat{G}_1,\dot{\hat{G}}_2]]+\frac{1}{6}[\hat{G}_1,[\hat{G}_2,\dot{\hat{G}}_1]]+\frac{1}{6}[\hat{G}_2,[\hat{G}_1,\dot{\hat{G}}_1]]\\
&+\frac{i}{24}[\hat{G}_1,[\hat{G}_1,[\hat{G}_1,\dot{\hat{G}}_1]]\Big)+O(\lambda^5) \;.
\end{split}
\label{Eq:SWPT-LamPowers exp(G)(-id_t)exp(-G)}
\end{align}

Adding equal powers of $\lambda$ in Eqs.~(\ref{Eq:SWPT-Trans of H0}), (\ref{Eq:SWPT-Trans of Hint}) and (\ref{Eq:SWPT-LamPowers exp(G)(-id_t)exp(-G)}) we find the effective Hamiltonian as
\begin{subequations}
\begin{align}
\hat{\mathcal{H}}_{\text{eff}}(t)=\sum\limits_{n=0}^{\infty}\lambda^n\hat{\mathcal{H}}_{\text{eff}}^{(n)}(t) \;,
\end{align}
where
\begin{align}
\hat{\mathcal{H}}_{\text{eff}}^{(0)}=\hat{\mathcal{H}}_0 \;,
\label{Eq:SWPT-O(1) Heff}
\end{align}
\begin{align}
\hat{\mathcal{H}}_{\text{eff}}^{(1)}=-\dot{\hat{G}}_1+i[\hat{G}_1,\hat{\mathcal{H}}_0]+\hat{\mathcal{H}}_{\text{int}} \;,
\label{Eq:SWPT-O(lambda) Heff}
\end{align}
\begin{align}
\begin{split}
\hat{\mathcal{H}}_{\text{eff}}^{(2)}&=-\dot{\hat{G}}_2-\frac{i}{2}[\hat{G}_1,\dot{\hat{G}}_1]+i[\hat{G}_2,\hat{\mathcal{H}}_0]\\
&-\frac{1}{2}[\hat{G}_1,[\hat{G}_1,\hat{\mathcal{H}}_0]]+i[\hat{G}_1,\hat{\mathcal{H}}_{\text{int}}] \;,
\end{split}
\label{Eq:SWPT-O(lambda^2) Heff}
\end{align}
\begin{align}
\begin{split}
\hat{\mathcal{H}}_{\text{eff}}^{(3)}&=-\dot{\hat{G}}_3-\frac{i}{2}[\hat{G}_1,\dot{\hat{G}}_2]-\frac{i}{2}[\hat{G}_2,\dot{\hat{G}}_1]\\
&+\frac{1}{6}[\hat{G}_1,[\hat{G}_1,\dot{\hat{G}}_1]]+i[\hat{G}_3,\hat{\mathcal{H}}_0]\\
&-\frac{1}{2}[\hat{G}_1,[\hat{G}_2,\hat{\mathcal{H}}_0]]-\frac{1}{2}[\hat{G}_2,[\hat{G}_1,\hat{\mathcal{H}}_0]]\\
&-\frac{i}{6}[\hat{G}_1,[\hat{G}_1,[\hat{G}_1,\hat{\mathcal{H}}_0]]]+i[\hat{G}_2,\hat{\mathcal{H}}_{\text{int}}]\\
&-\frac{1}{2}[\hat{G}_1,[\hat{G}_1,\hat{\mathcal{H}}_{\text{int}}]] \;,
\end{split}
\label{Eq:SWPT-O(lambda^3) Heff}
\end{align}
\begin{align}
\begin{split}
&\hat{\mathcal{H}}_{\text{eff}}^{(4)} = -\dot{\hat{G}}_4-\frac{i}{2}[\hat{G}_1,\dot{\hat{G}}_3]-\frac{i}{2}[\hat{G}_2,\dot{\hat{G}}_2]-\frac{i}{2}[\hat{G}_3,\dot{\hat{G}}_1]\\
&+\frac{1}{6}[\hat{G}_1,[\hat{G}_1,\dot{\hat{G}}_2]]+\frac{1}{6}[\hat{G}_1,[\hat{G}_2,\dot{\hat{G}}_1]]+\frac{1}{6}[\hat{G}_2,[\hat{G}_1,\dot{\hat{G}}_1]]\\
&+\frac{i}{24}[\hat{G}_1,[\hat{G}_1,[\hat{G}_1,\dot{\hat{G}}_1]]+i[\hat{G}_4,\hat{\mathcal{H}}_0]-\frac{1}{2}[\hat{G}_1,[\hat{G}_3,\hat{\mathcal{H}}_0]]\\
&-\frac{1}{2}[\hat{G}_2,[\hat{G}_2,\hat{\mathcal{H}}_0]]-\frac{1}{2}[\hat{G}_3,[\hat{G}_1,\hat{\mathcal{H}}_0]]\\
&-\frac{i}{6}[\hat{G}_1,[\hat{G}_1,[\hat{G}_2,\hat{\mathcal{H}}_0]]]-\frac{i}{6}[\hat{G}_1,[\hat{G}_2,[\hat{G}_1,\hat{\mathcal{H}}_0]]]\\
&-\frac{i}{6}[\hat{G}_2,[\hat{G}_1,[\hat{G}_1,\hat{\mathcal{H}}_0]]]+\frac{1}{24}[\hat{G}_1,[\hat{G}_1,[\hat{G}_1,[\hat{G}_1,\hat{\mathcal{H}}_0]]]\\
&+i[\hat{G}_3,\hat{\mathcal{H}}_{\text{int}}]-\frac{1}{2}[\hat{G}_1,[\hat{G}_2,\hat{\mathcal{H}}_{\text{int}}]]-\frac{1}{2}[\hat{G}_2,[\hat{G}_1,\hat{\mathcal{H}}_{\text{int}}]]\\
&-\frac{i}{6}[\hat{G}_1,[\hat{G}_1,[\hat{G}_1,\hat{\mathcal{H}}_{\text{int}}]]] \;.
\end{split}
\label{Eq:SWPT-O(lambda^4) Heff}
\end{align}
\end{subequations}
Equations~(\ref{Eq:SWPT-O(1) Heff}--\ref{Eq:SWPT-O(lambda^4) Heff}) provide the generic result for the effective Hamiltonian up to the fourth order in perturbation. Depending on the nature of the problem, we determine the successive orders $\hat{G}_n(t)$ to reach a desired form. 

\subsection{Interaction frame}
\label{SubApp:SWPTIntFrame}
It is important to note that the form of corrections become significantly simpler if we apply the perturbation to the interaction frame from the outset. The interaction frame Hamiltonian is defined as
\begin{align}
\begin{split}
\lambda\hat{\mathcal{H}}_I(t) &\equiv e^{i\hat{\mathcal{H}}_0 t}\left[\hat{\mathcal{H}}_0+\lambda\hat{\mathcal{H}}_{\text{int}}(t)-i\partial_t\right]e^{-i\hat{\mathcal{H}}_0 t}\\
&= e^{i\hat{\mathcal{H}}_0 t}\lambda\hat{\mathcal{H}}_{\text{int}}(t) e^{-i\hat{\mathcal{H}}_0 t} \;.
\end{split}
\label{Eq:SWPT-O(lambda^3) HI}
\end{align}
Applying the perturbation theory on Hamiltonian~(\ref{Eq:SWPT-O(lambda^3) HI}) instead leads to
\begin{subequations}
\begin{align}
\hat{\mathcal{H}}_{\text{I,eff}}(t)=\sum\limits_{n=0}^{\infty}\lambda^n\hat{\mathcal{H}}_{\text{I,eff}}^{(n)}(t) \;,
\end{align}
where
\begin{align}
\hat{\mathcal{H}}_{\text{I,eff}}^{(0)}=0 \;,
\label{Eq:SWPT-O(1) HIeff}
\end{align}
\begin{align}
\hat{\mathcal{H}}_{\text{I,eff}}^{(1)}=-\dot{\hat{G}}_1+\hat{\mathcal{H}}_I \;,
\label{Eq:SWPT-O(lambda) HIeff}
\end{align}
\begin{align}
\hat{\mathcal{H}}_{\text{I,eff}}^{(2)}=-\dot{\hat{G}}_2-\frac{i}{2}[\hat{G}_1,\dot{\hat{G}}_1]+i[\hat{G}_1,\hat{\mathcal{H}}_I] \;,
\label{Eq:SWPT-O(lambda^2) HIeff}
\end{align}
\begin{align}
\begin{split}
\hat{\mathcal{H}}_{\text{I,eff}}^{(3)}&=-\dot{\hat{G}}_3-\frac{i}{2}[\hat{G}_1,\dot{\hat{G}}_2]-\frac{i}{2}[\hat{G}_2,\dot{\hat{G}}_1]\\
&+\frac{1}{6}[\hat{G}_1,[\hat{G}_1,\dot{\hat{G}}_1]]+i[\hat{G}_2,\hat{\mathcal{H}}_I]\\
&-\frac{1}{2}[\hat{G}_1,[\hat{G}_1,\hat{\mathcal{H}}_I]] \;,
\end{split}
\label{Eq:SWPT-O(lambda^3) HIeff}
\end{align}
\begin{align}
\begin{split}
& \hat{\mathcal{H}}_{\text{I,eff}}^{(4)}=-\dot{\hat{G}}_4-\frac{i}{2}[\hat{G}_1,\dot{\hat{G}}_3]-\frac{i}{2}[\hat{G}_2,\dot{\hat{G}}_2]-\frac{i}{2}[\hat{G}_3,\dot{\hat{G}}_1] \\
&+\frac{1}{6}[\hat{G}_1,[\hat{G}_1,\dot{\hat{G}}_2]]+\frac{1}{6}[\hat{G}_1,[\hat{G}_2,\dot{\hat{G}}_1]]+\frac{1}{6}[\hat{G}_2,[\hat{G}_1,\dot{\hat{G}}_1]]\\
&+\frac{i}{24}[\hat{G}_1,[\hat{G}_1,[\hat{G}_1,\dot{\hat{G}}_1]]+i[\hat{G}_3,\hat{\mathcal{H}}_I]-\frac{1}{2}[\hat{G}_1,[\hat{G}_2,\hat{\mathcal{H}}_I]]\\
&-\frac{1}{2}[\hat{G}_2,[\hat{G}_1,\hat{\mathcal{H}}_I]]-\frac{i}{6}[\hat{G}_1,[\hat{G}_1,[\hat{G}_1,\hat{\mathcal{H}}_I]]] \;.
\end{split}
\label{Eq:SWPT-O(lambda^4) HIeff}
\end{align}
\end{subequations}

SWPT provides flexibility in determining the desired form for the effective Hamiltonian and the corresponding solution for the generator $\hat{G}(t)$ depending on the nature of the problem. Important examples are obtaining \textit{diagonal} or \textit{block-diagonal} effective Hamiltonian. For CR, since the drive frequency is resonant with target and off-resonant from the control, we are interested in a block-diagonal form with respect to the Hilbert space of the target qubit.

\subsection{Block-diagonalization}
\label{SubApp:SWPTBlockDiag}

The CR interaction Hamiltonian $\HO_{\text{I}}(t)$ is not block-diagonal with respect to the target qubit from the outset. Hence, at the lowest order, we need to solve for $\hat{G}_1$ such that it removes $\hat{\mathcal{H}}_I(t)$
\begin{subequations}
\begin{align}
O(\lambda):
\begin{cases}
&\hat{\mathcal{H}}_{\text{I,eff}}^{(1)}=0 \;,\\
&\dot{\hat{G}}_1=\hat{\mathcal{H}}_I \;,
\end{cases}
\label{Eq:SWPT-Bdiag-O(lambda) HIeff}
\end{align}
Replacing the solution for $\hat{G}_1$ from Eq.~(\ref{Eq:SWPT-Bdiag-O(lambda) HIeff}) into the generic $O(\lambda^2)$ correction~(\ref{Eq:SWPT-O(lambda^2) HIeff}) we obtain
\begin{align}
O(\lambda^2):
\begin{cases}
&\hat{\mathcal{H}}_{\text{I,eff}}^{(2)}=\mathcal{B}(\frac{i}{2}[\hat{G}_1,\hat{\mathcal{H}}_I]) \;,\\
&\dot{\hat{G}}_2=\mathcal{N}(\frac{i}{2}[\hat{G}_1,\hat{\mathcal{H}}_I]) \;,
\end{cases}
\label{Eq:SWPT-Bdiag-O(lambda^2) HIeff}
\end{align}
where $\mathcal{B}(\bullet)$ and $\mathcal{N}(\bullet)$ denote the block-diagonal and non-block-diagonal parts of an operator with respect to the target qubit. Using Eq.~(\ref{Eq:SWPT-Bdiag-O(lambda) HIeff}), we can simplify expressions~(\ref{Eq:SWPT-O(lambda^3) HIeff}--\ref{Eq:SWPT-O(lambda^4) HIeff}) for higher orders as
\begin{align}
O(\lambda^3):
\begin{cases}
\begin{split}
\hat{\mathcal{H}}_{\text{I,eff}}^{(3)}&=\mathcal{B}\left(-\frac{i}{2}[\hat{G}_1,\dot{\hat{G}}_2]+\frac{i}{2}[\hat{G}_2,\hat{\mathcal{H}}_I]\right.\\
&\left.-\frac{1}{3}[\hat{G}_1,[\hat{G}_1,\hat{\mathcal{H}}_I]]\right) \;,
\end{split}\\
\begin{split}
\dot{\hat{G}}_3&=\mathcal{N}\left(-\frac{i}{2}[\hat{G}_1,\dot{\hat{G}}_2]+\frac{i}{2}[\hat{G}_2,\hat{\mathcal{H}}_I]\right.\\
&\left.-\frac{1}{3}[\hat{G}_1,[\hat{G}_1,\hat{\mathcal{H}}_I]]\right) \;,
\end{split}
\end{cases}
\label{Eq:SWPT-Bdiag-O(lambda^3) HIeff}
\end{align}

\begin{align}
O(\lambda^4):
\begin{cases}
\begin{split}
\hat{\mathcal{H}}_{\text{I,eff}}^{(4)}&=\mathcal{B}\left(-\frac{i}{2}[\hat{G}_1,\dot{\hat{G}}_3]-\frac{i}{2}[\hat{G}_2,\dot{\hat{G}}_2]\right. \\
&+\frac{1}{6}[\hat{G}_1,[\hat{G}_1,\dot{\hat{G}}_2]]+\frac{i}{2}[\hat{G}_3,\hat{\mathcal{H}}_I]\\
&-\frac{1}{3}[\hat{G}_1,[\hat{G}_2,\hat{\mathcal{H}}_I]]-\frac{1}{3}[\hat{G}_2,[\hat{G}_1,\hat{\mathcal{H}}_I]]\\
&-\left. \frac{i}{8}[\hat{G}_1,[\hat{G}_1,[\hat{G}_1,\hat{\mathcal{H}}_I]]]\right) \;,
\end{split}\\
\begin{split}
\dot{\hat{G}}_4 &=\mathcal{N}\left(-\frac{i}{2}[\hat{G}_1,\dot{\hat{G}}_3]-\frac{i}{2}[\hat{G}_2,\dot{\hat{G}}_2]\right. \\
&+\frac{1}{6}[\hat{G}_1,[\hat{G}_1,\dot{\hat{G}}_2]]+\frac{i}{2}[\hat{G}_3,\hat{\mathcal{H}}_I]\\
&-\frac{1}{3}[\hat{G}_1,[\hat{G}_2,\hat{\mathcal{H}}_I]]-\frac{1}{3}[\hat{G}_2,[\hat{G}_1,\hat{\mathcal{H}}_I]]\\
&-\left. \frac{i}{8}[\hat{G}_1,[\hat{G}_1,[\hat{G}_1,\hat{\mathcal{H}}_I]]]\right) \;,
\end{split}
\end{cases}
\label{Eq:SWPT-Bdiag-O(lambda^4) HIeff}
\end{align}
\end{subequations}

Equations~(\ref{Eq:SWPT-Bdiag-O(lambda) HIeff}--\ref{Eq:SWPT-Bdiag-O(lambda^4) HIeff}) are the main results of this appendix and have been used in the main text to both study the isolated CR gate as well three-qubit models with a spectator qubit, in Secs~\ref{Sec:CREffHam} and~\ref{Sec:SpecQu}, respectively.   

\section{Classical cross-talk}
\label{App:CrossTalk}

\begin{figure}[t!]
\centering
\subfloat[\label{subfig:CrossTalk-IXCrossTalkFuncOfAWSarahsParams}]{%
\includegraphics[scale=0.210]{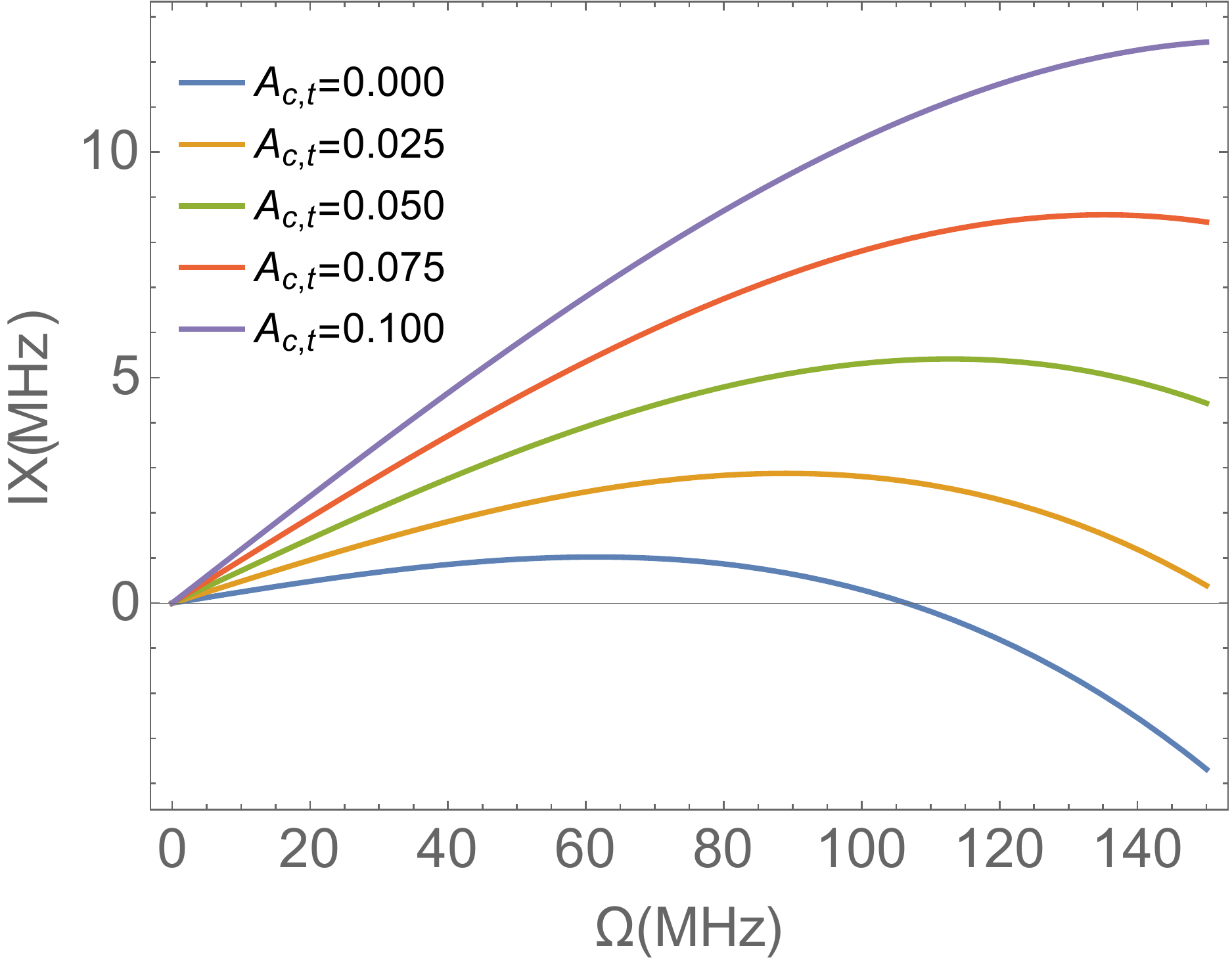}%
} 
\subfloat[\label{subfig:CrossTalk-IYCrossTalkFuncOfAWSarahsParams}]{%
\includegraphics[scale=0.210]{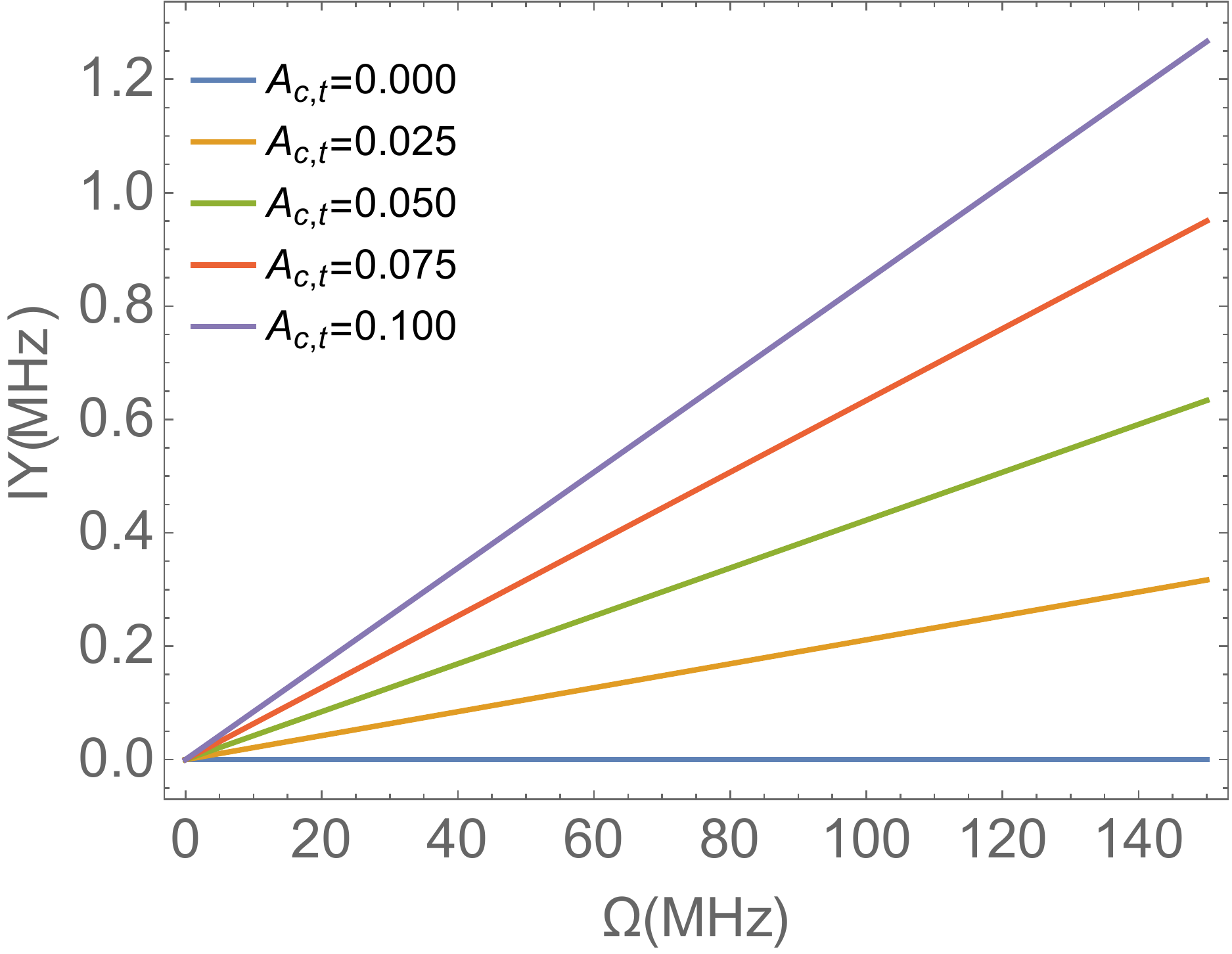}%
} \\
\subfloat[\label{subfig:CrossTalk-IZCrossTalkFuncOfAWSarahsParams}]{%
\includegraphics[scale=0.210]{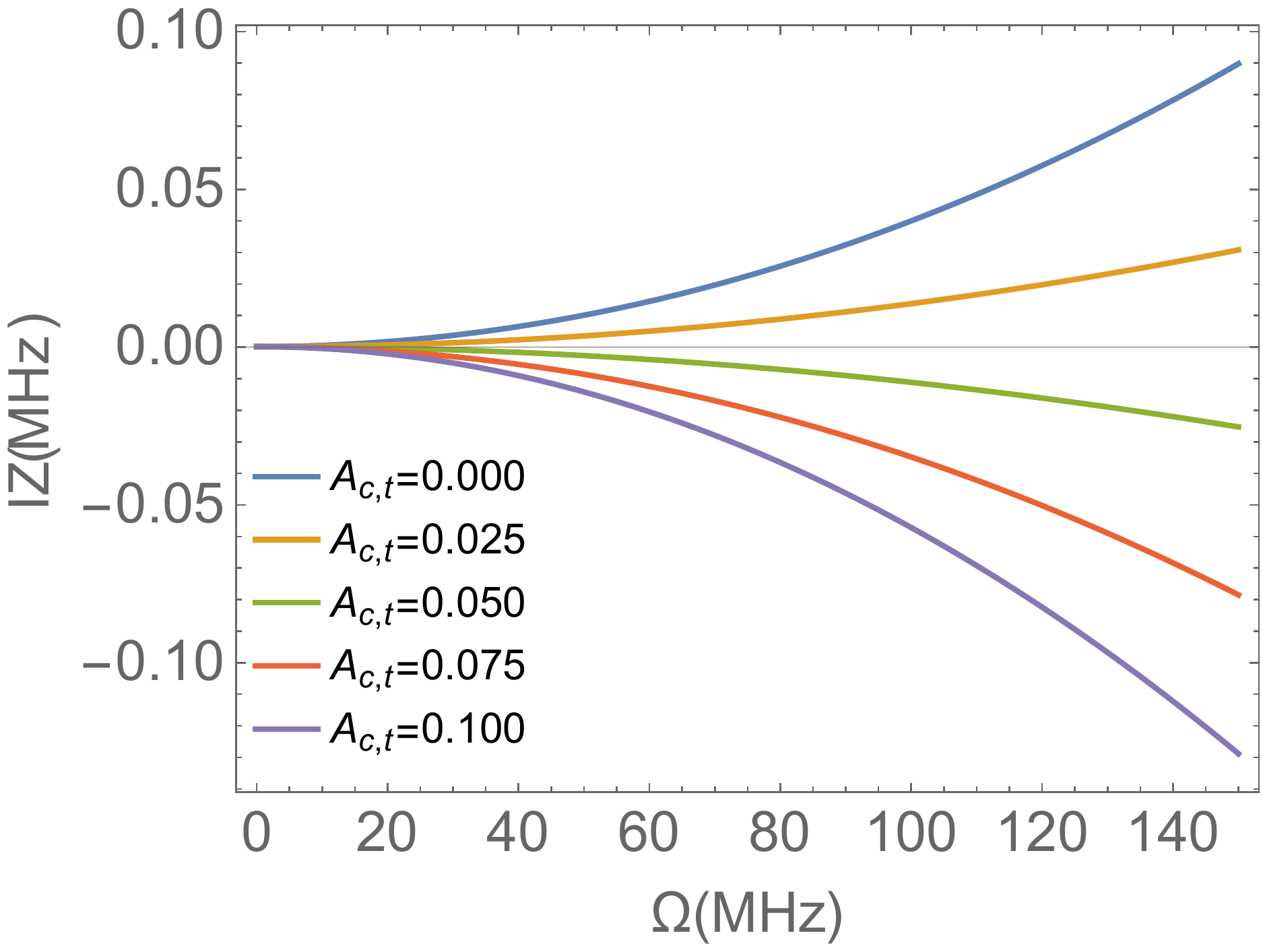}%
}
\subfloat[\label{subfig:CrossTalk-ZICrossTalkFuncOfAWSarahsParams}]{%
\includegraphics[scale=0.210]{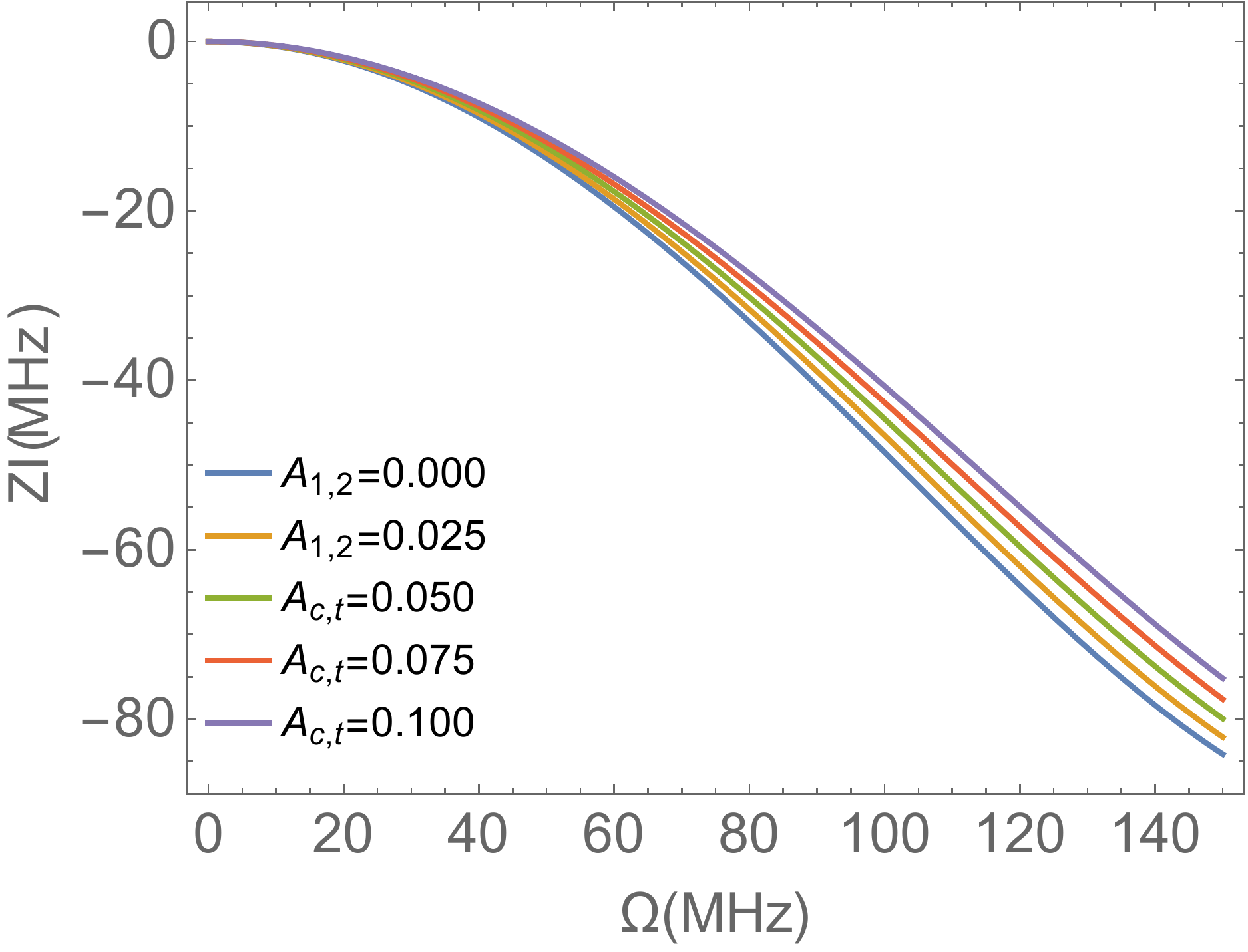}%
}\\
\subfloat[\label{subfig:CrossTalk-ZXCrossTalkFuncOfAWSarahsParams}]{%
\includegraphics[scale=0.210]{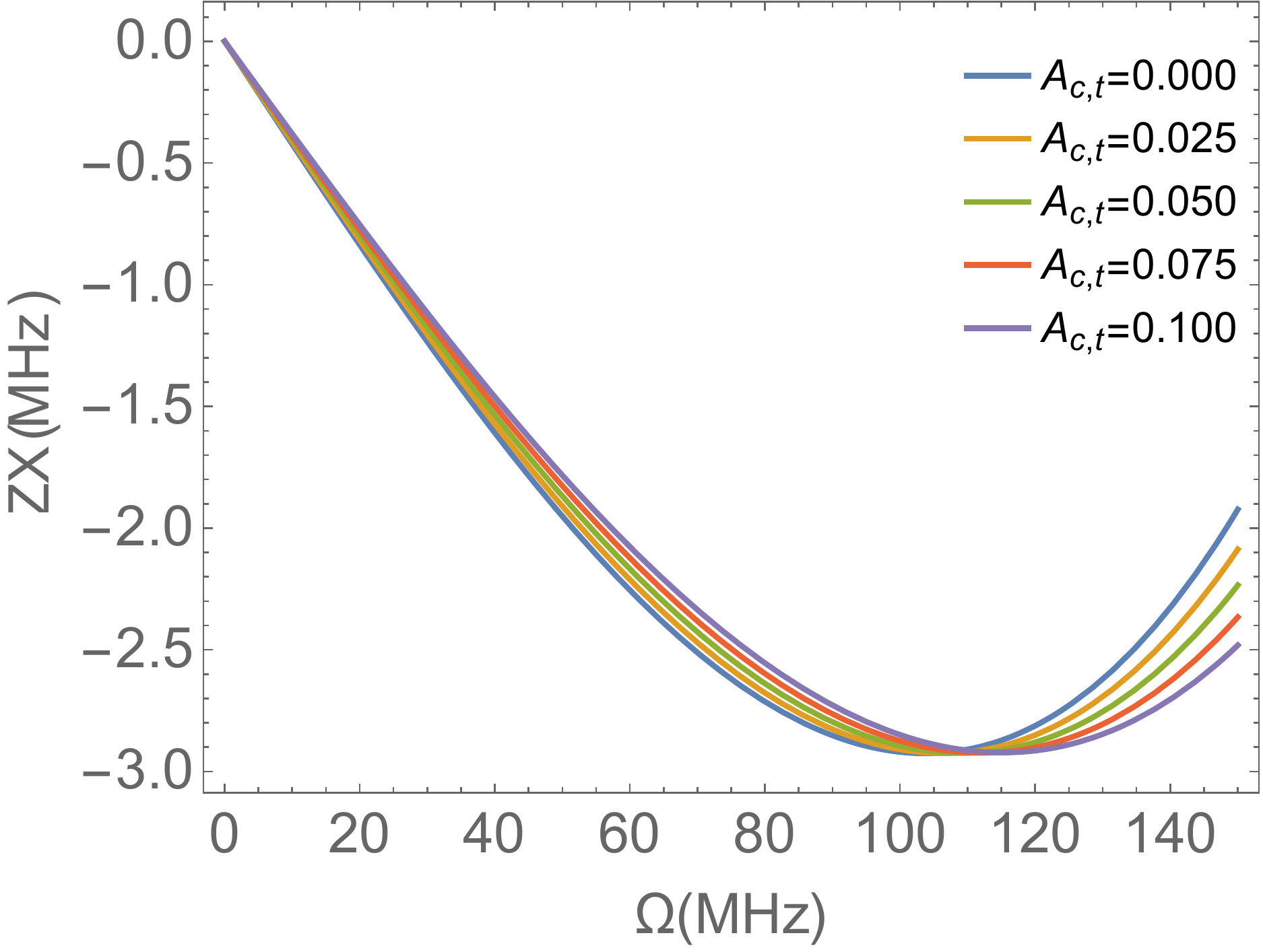}%
} 
\subfloat[\label{subfig:CrossTalk-ZYCrossTalkFuncOfAWSarahsParams}]{%
\includegraphics[scale=0.210]{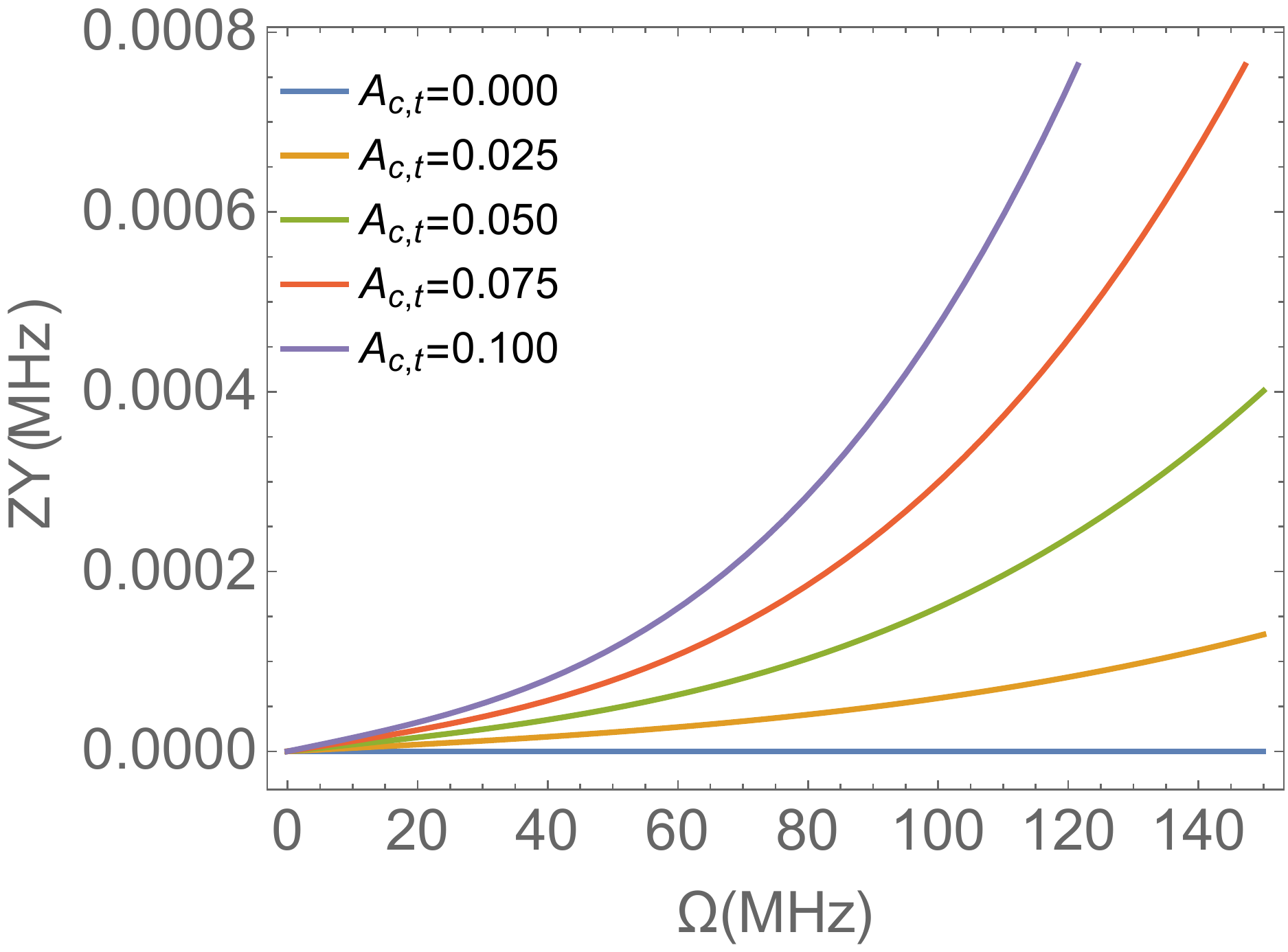}%
} \\
\subfloat[\label{subfig:CrossTalk-ZZCrossTalkFuncOfAWSarahsParams}]{%
\includegraphics[scale=0.210]{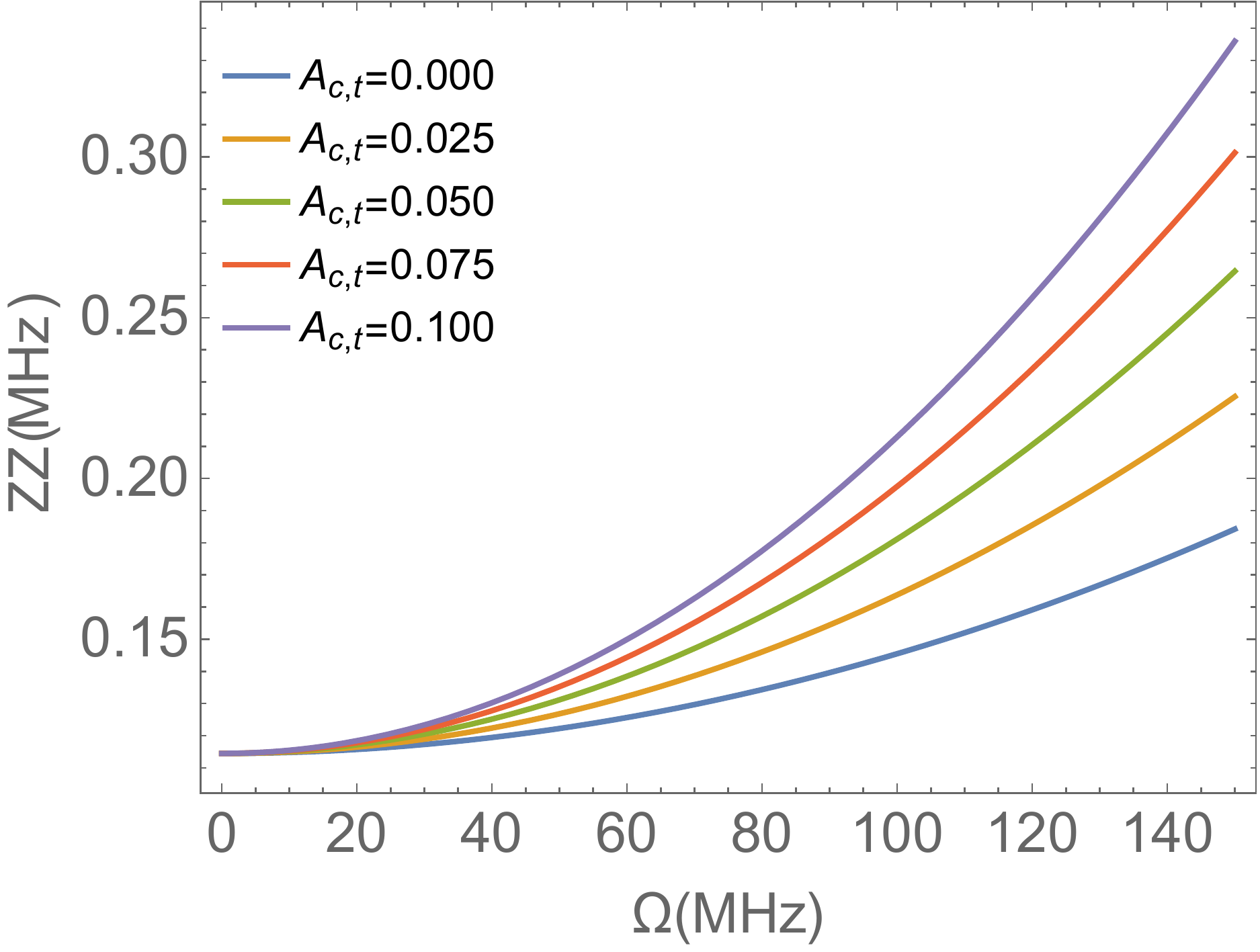}%
}
\caption{CR gate parameters as a function of drive and cross-talk. a) $IX$, b) $IY$, c) $IZ$ d), $ZI$, e) $ZX$, f) $ZY$ and g) $ZZ$. Circuit parameters are the same as those in Table~\ref{tab:LoOrAnal-CRGateParams}. Cross-talk parameters are $A_c=A_t$ that range in $[0,0.1]$ and $\phi_t=0$.}
\label{Fig:CrossTalk-GateParamsFuncOfDr}
\end{figure}

\begin{table}[t!]
  \begin{tabular}{|c|c|c|}
  \hline	
   Op. & Coeff. (energy basis) & Est. (MHz) \\
  \hline\hline  
    $\frac{1}{2}\hat{I}\hat{X}$ & $ \nu_{t,01}A_t\Omega\cos(\phi_t)-\frac{\nu_{t,01}\nu_{c,12}^2}{2 \left(\Delta_{ct}+\alpha_c\right)}J(1-A_c)\Omega$ & 3.603\\
  \hline\hline
  $\frac{1}{2}\hat{I}\hat{Y}$ & $ \nu_{t,01}A_t\Omega\sin(\phi_t)$ & 0.211\\
  \hline\hline
     $\frac{1}{2}\hat{Z}\hat{I}$  & $\Bigg[\frac{\nu _{c,12}^2}{4 \left(\Delta_{ct}+\alpha_c\right)}-\frac{\nu _{c,01}^2}{2 \Delta_{ct}}\Bigg](1-A_c)^2\Omega^2	$ & $-12.969$  \\
  \hline\hline
  $\frac{1}{2}\hat{Z}\hat{X}$  & $ \frac{1}{2}\left(\frac{\nu_{t,01}\nu_{c,12}^2}{\Delta_{ct}+\alpha_c}-\frac{2 \nu_{t,01}\nu_{c,01}^2}{\Delta_{ct}}\right)J (1-A_c)\Omega$ & -2.012	\\
   \hline\hline
  $\frac{1}{2}\hat{Z}\hat{Z}$ & $ \frac{1}{2}\left(\frac{\nu _{c,01}^2 \nu_{t,12}^2}{\Delta_{ct}-\alpha_t}-\frac{\nu_{t,01}^2 \nu_{c,12}^2}{\Delta_{ct}+\alpha_c}\right) J^2$ & 0.114 \\
  \hline  
  \end{tabular}
  \caption{Lowest order CR gate parameters in the presence of classical cross-talk based on Eq.~(\ref{eqn:CrossTalk-Def of Hd}). System parameters are the same as those in Table~\ref{tab:LoOrAnal-CRGateParams}, with the cross-talk parameters chosen as $A_c=A_t=0.05$ and $\phi_t=\pi/36$.}
\label{Tab:CrossTalk-LowestOrGateParams}
\end{table}

In this appendix, we study the dependence of CR gate parameters on classical cross-talk between the control and target qubits. To model cross-talk, we assume that a portion of the CR drive will act directly on the target qubit due to unwanted microwave channels in the circuit. We consider a modified drive Hamiltonian of the form
\begin{align}
\begin{split}
\HO_{d}(t)&=-(1-A_c)\Omega\hat{y}_c\sin(\omega_d t)-A_t\Omega\hat{y}_t\sin(\omega_d t+\phi_t)\\
&\approx\frac{(1-A_c)\Omega}{2}\left(\hat{y}_c^{-}e^{i\omega_d t}+\hat{y}_c^{+}e^{-i\omega_d t}\right)\\
&+\frac{A_t\Omega}{2}\left(\hat{y}_t^{-}e^{i(\omega_d t+\phi_t)}+\hat{y}_t^{+}e^{-(i\omega_d t+\phi_t)}\right),	
\end{split}
\label{eqn:CrossTalk-Def of Hd}
\end{align}
where $A_c$ denotes the suppression in the supposed drive on the control qubit, $A_t$ denotes the relative strength on the target qubit and $\phi_t$ is the phase difference as a result of the distance between the control and the target qubits.

Following the results for SWPT from the previous Appendix, Eqs.~(\ref{eqn:CREffHam-O(lambda) HIeff}--\ref{eqn:CREffHam-O(lambda^4) HIeff}), and substituting the drive Hamiltonian~(\ref{eqn:CrossTalk-Def of Hd}), we solve for the CR gate parameters order by order. The lowest order results for gate parameters are summarized in Table~\ref{Tab:CrossTalk-LowestOrGateParams}. We find that the lowest order expressions for gate parameters in the presence of cross-talk can be inferred from table~\ref{tab:LoOrAnal-CRGateParams} (no cross-talk) in the following manner. Dynamic contributions (dependent on $\Omega$) to gate parameters can be found by replacing $\Omega \rightarrow (1-A_c)\Omega$, which is the ratio by which the drive on the control qubit is suppressed. On top of this, there is also direct contributions coming from the drive on the target qubit to the $IX$ and $IY$ rates as $\nu_{t,01}A_t\Omega\cos(\phi_t)$ and $\nu_{t,01}A_t\Omega\sin(\phi_t)$, respectively. Higher order corrections to gate parameters are studied in Fig.~\ref{Fig:CrossTalk-GateParamsFuncOfDr}.  

\section{Non-local invariants}
\label{App:NonLocInv}

Here, we provide estimates for the non-local gate fidelity \cite{Watts_Optimizing_2015} and entangling power \cite{Zanardi_Entangling_2000, Rezakhani_Characterization_2004} in terms of Makhlin invariants \cite{Makhlin_Nonlocal_2002, Zhang_Geometric_2003} in Secs.~\ref{SubApp:NonLocGateFid} and~\ref{SubApp:EntPow}, respectively.	

\subsection{Non-local gate fidelity in terms of Makhlin invariants}
\label{SubApp:NonLocGateFid}
We define the non-local fidelity as the overlap between the non-local parts of the ideal and implemented CR echo unitary as
\begin{align}
\begin{split}
F_{\text{echo}}^{(nl)} \equiv F(\hat{A}_{\text{ech}},\hat{A}_{\text{ide}})&=\frac{\Tr\left(\hat{A}_{\text{ech}}^{\dag}\hat{A}_{\text{ech}}\right)}{d(d+1)}\\
&+\frac{\left|\Tr\left(\hat{A}_{\text{ech}}^{\dag}\hat{A}_{\text{ide}}\right)\right|^2}{d(d+1)}\;,
\end{split}
\label{Eq:NonLocInv-Def of F_opt}
\end{align}
where $\hat{A}$ denotes the non-local part according to the decomposition $\hat{U}=\hat{K}_L \hat{A} \hat{K}_R	$ introduced in Eqs.~(\ref{eqn:NonLocInv-Canonical Form of U}--\ref{eqn:NonLocInv-Def of A}) and $d=4$ is the dimension of the two-qubit Hilbert space. In the following, we first provide an estimate for Eq.~(\ref{Eq:NonLocInv-Def of F_opt}) in terms of the difference between the Cartan coordinates of $\hat{A}_{\text{ech}}$ and $\hat{A}_{\text{ide}}$. Next, using the one-to-one correspondence between the Makhlin and Cartan coordinates, we rewrite our expression in terms of the Makhlin invariants. The results of this section is valid given that the two unitary transformations under consideration, i.e. $\hat{A}_{\text{ech}}$ and $\hat{A}_{\text{ide}}$, are sufficiently close such that $|\Delta \textbf{c}| \ll 1$. 

The non-local operators $\hat{A}_{\text{ide}}$ and $\hat{A}_{\text{ech}}$ can be represented in the canonical form as 
\begin{subequations}
\begin{align}
&\hat{A}_{\text{ide}}=e^{-i\frac{\pi}{4}\hat{X}_c\hat{X}_t} \;,
\label{Eq:NonLocInv-Def of A_ideal}\\
&\hat{A}_{\text{ech}}=e^{-\frac{i}{2}\left[(\pi/2+\Delta c_x)\hat{X}_c\hat{X}_t+\Delta c_y \hat{Y}_c\hat{Y}_t+\Delta c_z \hat{Z}_c\hat{Z}_t\right]} \;,
\label{Eq:NonLocInv-Def of A_echo}
\end{align}
\end{subequations}
where for $\hat{A}_{\text{ide}}$ we have substituted the Cartan coordinates $(\pi/2,0,0)$ denoting the CNOT class and we have assumed that $\hat{A}_{\text{ech}}$ only slightly deviates from this configuration with coordinates $(\pi/2+\Delta c_x,\Delta c_y,\Delta c_z)$ such that $|\Delta \textbf{c}|\ll 1$. Under this assumption, we first re-express the non-local fidelity~(\ref{Eq:NonLocInv-Def of F_opt}) in terms of $\Delta c_x$, $\Delta c_y$ and $\Delta c_z$. Since $\hat{A}_{\text{ech}}$ is unitary, by construction, the first term in Eq.~(\ref{Eq:NonLocInv-Def of F_opt}) is found as
\begin{align}
\frac{\Tr\left(\hat{A}_{\text{ech}}^{\dag}\hat{A}_{\text{ech}}\right)}{d(d+1)}=\frac{d}{d+1}=\frac{1}{5}\;,
\label{Eq:NonLocInv-1st term in F_opt}
\end{align}
with $d=4$ being the dimension of the two-qubit Hilbert space. Substituting Eqs.~(\ref{Eq:NonLocInv-Def of A_ideal}--\ref{Eq:NonLocInv-Def of A_echo}) into the second term of Eq.~(\ref{Eq:NonLocInv-Def of F_opt}) we obtain
\begin{align}
\begin{split}
\Tr\left(\hat{A}_{\text{ech}}^{\dag}\hat{A}_{\text{ide}}\right)&=4\cos\left(\frac{\Delta c_x}{2}\right)\cos\left(\frac{\Delta c_y}{2}\right)\cos\left(\frac{\Delta c_z}{2}\right) \\
&-4i\sin\left(\frac{\Delta c_x}{2}\right)\sin\left(\frac{\Delta c_y}{2}\right)\sin\left(\frac{\Delta c_z}{2}\right)\\
&=4\left[1-\frac{1}{8}\left(\Delta c_x^2+\Delta c_y^2+\Delta c_z^2\right)\right]\\
&+ O(\Delta \textbf{c}^4)\;.
\end{split}
\label{Eq:NonLocInv-Tr(C_echo,C_ideal) ITO DelC}
\end{align}
Therefore, we find the non-local gate fidelity in terms of the difference in the Cartan coordinates as
\begin{align}
\begin{split}
F(\hat{A}_{\text{ech}},\hat{A}_{\text{ide}})&=\frac{1}{5}+\frac{16}{20}\left[1-\frac{1}{4}\left(\Delta c_x^2+\Delta c_y^2+\Delta c_z^2\right)\right]\\
&=1-\frac{1}{5}\left(\Delta c_x^2+\Delta c_y^2+\Delta c_z^2\right)+ O(\Delta \textbf{c}^4)\;.
\end{split}
\label{Eq:NonLocInv-Fid ITO DelC}
\end{align}

Since the Makhlin invariants are more straightforward to compute for a given two-qubit unitairy, it is beneficial to rewrite Eq.~(\ref{Eq:NonLocInv-Fid ITO DelC}) in terms of the new coordinates. The Cartan and Makhlin coordinates are related as
\begin{subequations}
\begin{align}
g_x&=\frac{1}{4}\left[\cos(2c_x)+\cos(2c_y)+\cos(2c_z)\right. \label{Eq:NonLocInv-gx ITO c}\\
&\left. +\cos(2c_x)\cos(2c_y)\cos(2c_z)\right]\;, \nonumber \\
g_y&=\frac{1}{4}\sin(2c_x)\sin(2c_y)\sin(2c_z)\;,
\label{Eq:NonLocInv-gy ITO c}\\
g_z&=\cos(2c_x)+\cos(2c_y)+\cos(2c_z)\;.
\label{Eq:NonLocInv-gz ITO c}
\end{align}
\end{subequations}
Expanding $c_x$, $c_y$ and $c_z$ around the CNOT class as $c_x=\pi/2+\Delta c_x$, $c_y=\Delta c_y$ and $c_z=\Delta c_z$, we can simplify Eqs.~(\ref{Eq:NonLocInv-gx ITO c}--\ref{Eq:NonLocInv-gz ITO c}) up to the second order in $\Delta \text{c}$ as 
\begin{subequations}
\begin{align}
&g_x=\Delta c_x^2+ O(\Delta \textbf{c}^4)\;,
\label{Eq:NonLocInv-gx ITO c simple}\\
&g_y=-2\Delta c_x\Delta c_y\Delta c_z+ O(\Delta \textbf{c}^5)\;,
\label{Eq:NonLocInv-gy ITO c simple}\\
&g_z=1+2\left(\Delta c_x^2- \Delta c_y^2-\Delta c_z^2\right)+ O(\Delta \textbf{c}^4)\;.
\label{Eq:NonLocInv-gz ITO c simple}
\end{align}
Employing Eqs.~(\ref{Eq:NonLocInv-gx ITO c simple}) and~(\ref{Eq:NonLocInv-gz ITO c simple}) we rewrite $\Delta c_x^2+\Delta c_y^2+\Delta c_z^2$ in terms of the Makhlin invariants as
\end{subequations}
\begin{align}
\Delta c_x^2+\Delta c_y^2+\Delta c_z^2+O(\Delta \textbf{c}^4)=2g_x+\frac{1-g_z}{2}\;.
\label{Eq:NonLocInv-Delc^2 ITO g}
\end{align}
Lastly, using Eqs.~(\ref{Eq:NonLocInv-Delc^2 ITO g}) and~(\ref{Eq:NonLocInv-Fid ITO DelC}) we obtain the non-local fidelity (error) as
\begin{subequations}
\begin{align}
&F_{\text{ech}}^{(nl)}\approx 1-\frac{1}{10}(4g_x+1-g_z) +O(\Delta \textbf{c}^4)\;,
\label{Eq:NonLocInv-F_opt ITO Delc}\\
&E_{\text{ech}}^{(nl)}\equiv 1-F_{\text{ech}}^{(nl)}\approx\frac{1}{10}(4g_x+1-g_z)+O(\Delta \textbf{c}^4) \;.
\label{Eq:NonLocInv-E_opt ITO Delc}
\end{align}
\end{subequations}

\subsection{Entangling power in terms of Makhlin invariants}
\label{SubApp:EntPow}

Entangling power of a unitary operator $\hat{U}$ over a bipartite Hilbert space is defined as the average entanglement that the operator can produce when acting on separable states \cite{Zanardi_Entangling_2000},
\begin{subequations}
\begin{align}
e_p(\hat{U})\equiv \overline{\text{Ent} \left(\hat{U}\ket{\psi_c} \otimes \ket{\psi_t}\right)}^{\ket{\psi_c},\ket{\psi_t}} \;,
\label{Eq:NonLocInv-Def of EntPow}	
\end{align}
where the bar denotes an average over separable states and $\text{Ent}(\ket{\psi})$ is the entanglement measure of choice. It is common to employ the linear entanglement measure, as oppose to the von Neumann entropy, defined as
\begin{align}
&\text{Ent}(\ket{\psi})=1-\Tr_c\left(\hat{\rho}_c^2\right) \;,
\label{Eq:NonLocInv-Def of Ent(Psi)}\\
&\hat{\rho}_c \equiv \Tr_t\left(\ket{\psi}\bra{\psi}\right) \;.
\label{Eq:NonLocInv-Def of rho_c}
\end{align}
\end{subequations}

It can be shown that the entangling power of a two-qubit unitary operator depends only on its non-local properties, hence can be directly expressed in terms of the corresponding Cartan coordinates \cite{Rezakhani_Characterization_2004}
\begin{align}
\begin{split}
e_p(\hat{U})&=\frac{1}{18}\big[3-\cos(2c_x)\cos(2c_y)\\
&-\cos(2c_y)\cos(2c_z)-\cos(2c_z)\cos(2c_x)\big] \;.
\end{split}
\label{Eq:NonLocInv-Def of e_p(U)}
\end{align}
Here, we derive an estimate for $e_p(\hat{U}_{\text{ech}})$. Inserting Ansatz~(\ref{Eq:NonLocInv-Def of A_echo}) for $\hat{U}_{\text{ech}}$ into Eq.~(\ref{Eq:NonLocInv-Def of e_p(U)}) and expanding in terms of $\Delta \textbf{c}$ we find
\begin{align}
\begin{split}
e_p(\hat{U}_{\text{ech}})&=\frac{1}{18}\big[3+\cos(2\Delta c_x)\cos(2 \Delta c_y)\\
&-\cos(2c_y)\cos(2 \Delta c_z)+\cos(2 \Delta c_z)\cos(2\Delta c_x)\big]\\
&=\frac{2}{9}-\frac{2}{9}\Delta c_x^2+O(\Delta \textbf{c}^4)\\
&=\frac{2}{9}-\frac{2}{9}g_x+O(\Delta \textbf{c}^4) \;,
\end{split}
\label{Eq:NonLocInv-e_p(U_ech)}
\end{align}
where in the last step we employed Eq.~(\ref{Eq:NonLocInv-gx ITO c simple}) to replace $\Delta c_x^2$ with the Makhlin $g_x$ invariant. Therefore, the difference between the entangling power of an ideal CNOT and the implemented CR echo unitary operators is obtained as
\begin{align}
e_p(\hat{U}_\text{ide})-e_p(\hat{U}_{\text{ech}})=\frac{2}{9}g_x+O(\Delta \textbf{c}^4) \;.
\label{Eq:NonLocInv-[e_p(U_ide)-e_p(U_ech)]}
\end{align}	

Equations~(\ref{Eq:NonLocInv-F_opt ITO Delc}--\ref{Eq:NonLocInv-E_opt ITO Delc}) and~(\ref{Eq:NonLocInv-[e_p(U_ide)-e_p(U_ech)]}) are the main results of this appendix and have been used in Sec.~\ref{SubSec:NonLocInv} and Fig.~\ref{fig:CREcho-GateError}.

\section{Saturation of cross-resonance gate parameters at strong drive}
\label{App:Saturation}

SWPT provides reliable estimates for the gate parameters up to medium (50 MHz) drive amplitude. On the other hand, it is also crucial to understand the behavior of rates at strong drive power. Some important questions are: 1) Is there an upper bound for the $ZX$ rate? 2) How does this bound depend on the circuit parameters, especially qubit-qubit detuning and qubit anharmonicity? 3) What is the lowest drive amplitude at which we can reach a reasonable fraction of this bound? Note that any perturbation theory in drive amplitude $\Omega$ is unable to predict a saturation behavior, by construction, due to the fact the results are always polynomials of $\Omega$ which will inevitably diverge at strong drive. Reference~\cite{Tripathi_Operation_2019} introduced a semi-analytical method designed specifically to study strong-drive behavior of the $ZX$ rate. In the following, we apply this method to the modified CR Hamiltonian in terms of energy basis [Eqs.~(\ref{eqn:CRHamInEnBasis-H0 2}--\ref{eqn:CRHamInEnBasis-Hd in new basis})].

\begin{figure}[t!]
\centering
\subfloat[\label{subfig:Saturation-ZXSaturationSarahsParams}]{%
\includegraphics[scale=0.325]{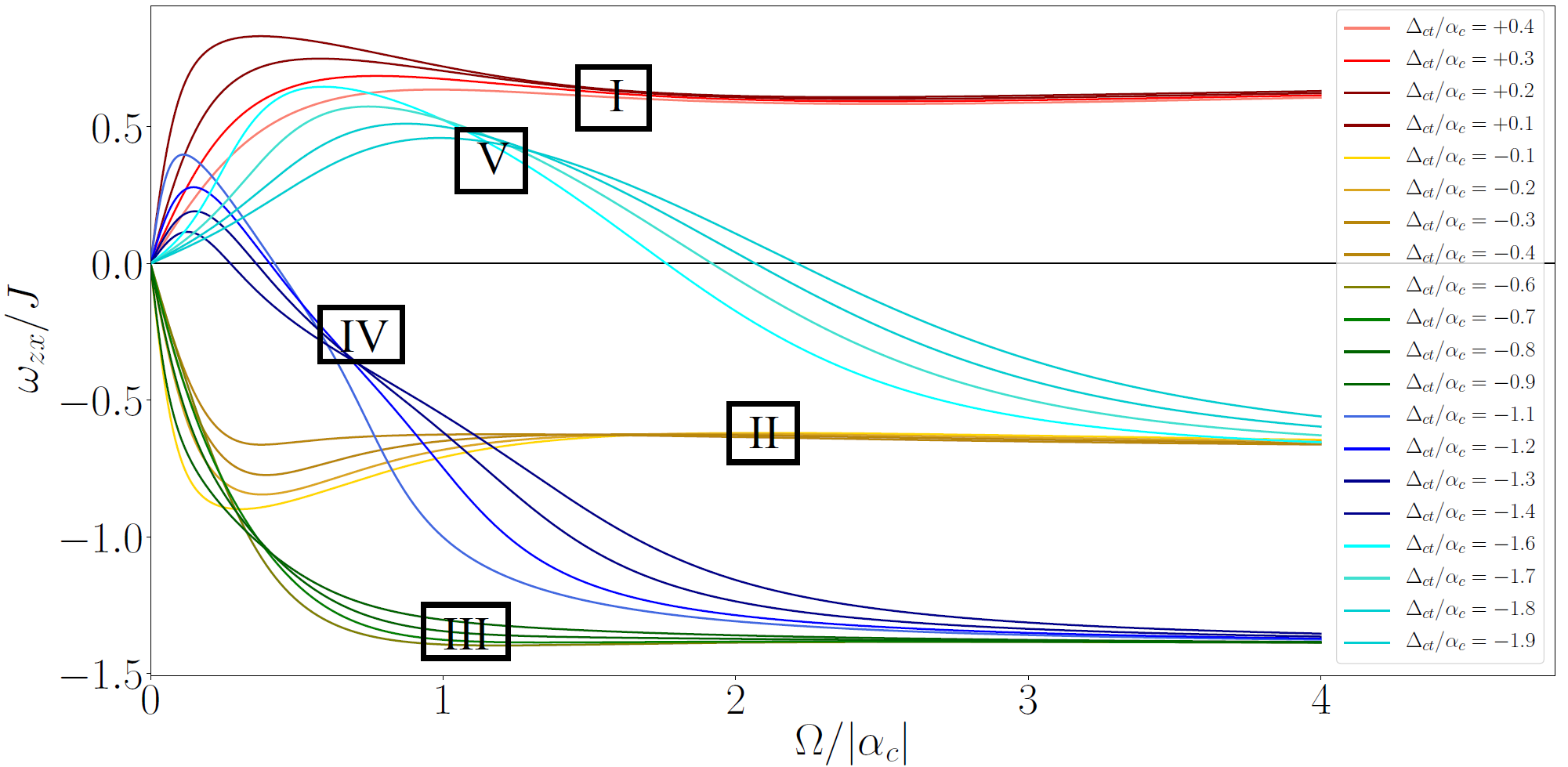}%
} \\
\subfloat[\label{subfig:Saturation-IXSaturationSarahsParams}]{%
\includegraphics[scale=0.325]{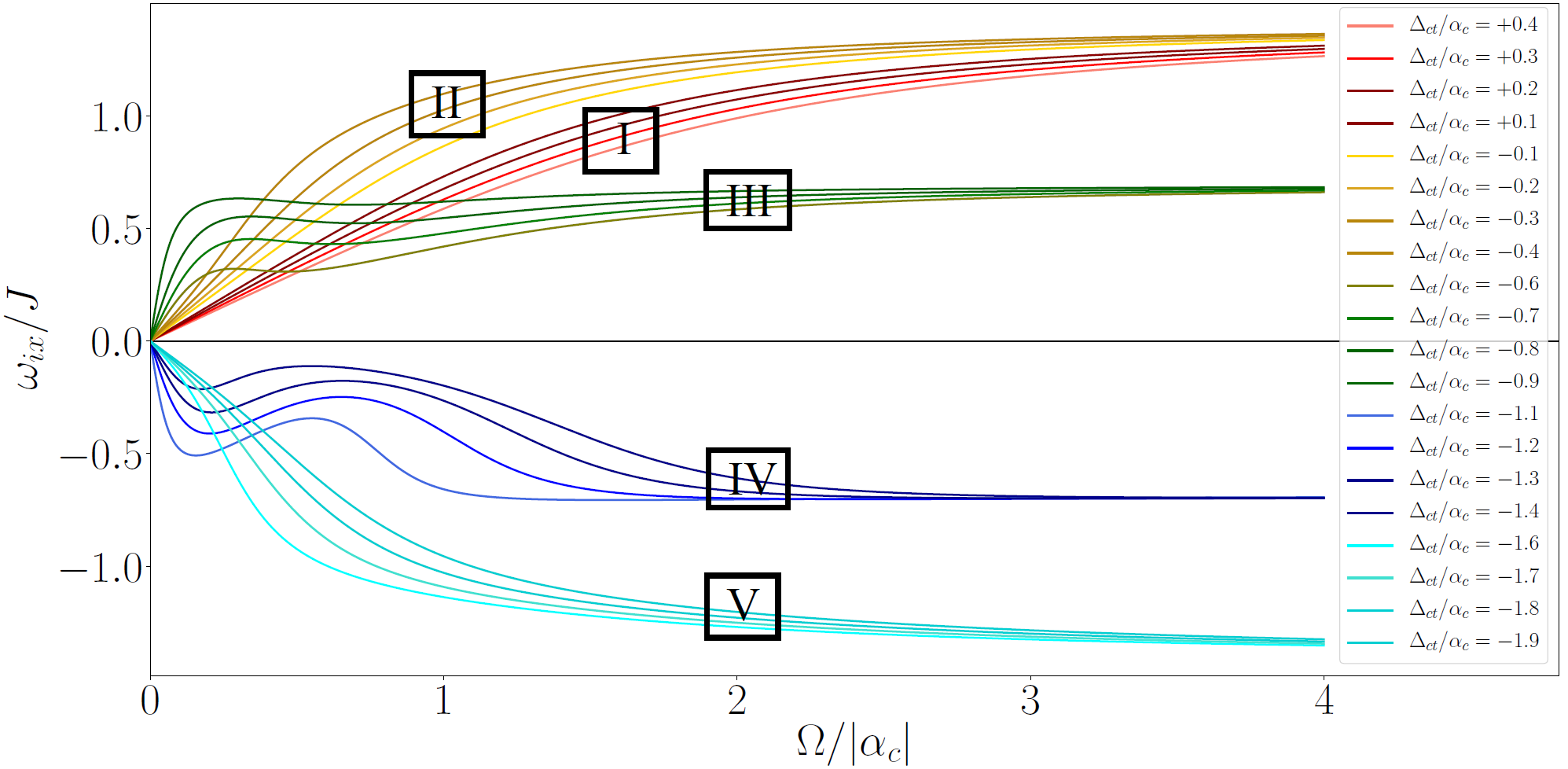}%
}
\caption{a) Normalized $ZX$ rate, b) normalized $IX$ rate as a function of drive amplitude with sample ratios of qubit-qubit detuning over control qubit anharmonicity. Other parameters are the same as Table~\ref{tab:LoOrAnal-CRGateParams}. The results are obtained by applying the semi-analytical method of Tripathi et al.~\cite{Tripathi_Operation_2019} to our energy-basis model of Eqs.~(\ref{eqn:CRHamInEnBasis-H0 2}--\ref{eqn:CRHamInEnBasis-Hd in new basis}). One observes a formation of bands depending on in which of the five regions the detuning is set: I) $-\alpha_t<\Delta_{ct}<0$ (shades of red), II) $0<\Delta_{ct}<-\alpha_c/2$ (shades of yellow), III) $-\alpha_c/2<\Delta_{ct}<-\alpha_c$ (shades of green), IV) $-\alpha_c<\Delta_{ct}<-3\alpha_c/2$ (shades of blue) and V) $-3\alpha_c/2<\Delta_{ct}<-2\alpha_c$ (shades of cyan).} 
\label{fig:Saturation-ZXSaturationSarahsParams}
\end{figure}

The idea of semi-analytical method is to use the \textit{analytical} expression for $ZX$ and $IX$ interactions in terms of interaction constants $a_n \equiv \prescript{}{J}{\bra{\psi_{n1}}}\HO_d \ket{\psi_{n0}}_J$ as
\begin{subequations}
\begin{align}
\omega_{ix}&=a_0+a_1 \;,
\label{eqn:Saturation-w_i1x2 ITO a_n}\\
\omega_{zx}&=a_0-a_1 \;,
\label{eqn:Saturation-w_z1x2 ITO a_n}
\end{align}
\end{subequations}
while calculating these interaction rates \textit{numerically}. In principle, $a_n$ quantifies the interaction that is induced by the drive between the two-qubit states in the frame dressed by the exchange interaction $J$ (See Fig.~\ref{fig:CRHamInEnBasis-JCLadderForCR}). However, at sufficiently strong drive, the exchange interaction $J$ is significantly smaller (at least one order of magnitude) than $\Omega$. Hence, in practice, the drive needs to be accounted for non-perturbatively, while the exchange interaction acts as a small correction between the resulting states dressed by the drive. Employing this \textit{interchangability}, the semi-analytical method proposes the following expression for the interaction rates $a_n$ as  
\begin{align}
a_n = \prescript{}{J}{\bra{\psi_{n1}}}\HO_d \ket{\psi_{n0}}_J\approx \prescript{}{\Omega}{\bra{\psi_{n1}}}\HO_J \ket{\psi_{n0}}_{\Omega} \;,
\label{eqn:Saturation-SemiAnal Expression for a_n}
\end{align}
where the first expression shows the exact definition in terms of the dressed states by $J$ and the second provides the semi-analytical approximation in terms of the dressed states by $\Omega$. Although no formal proof is presented for the validity of this approximation, Tripathi et al.~\cite{Tripathi_Operation_2019} show a good agreement between the semi-analytical and full numerical calculation of the $ZX$ rate. It can be shown that the drive will \textit{only} substantially dress the states of the control qubit and hence one can approximate the two-qubit problem by disentangling the target qubit as   
\begin{align}
\ket{\psi_{nm}}_{\Omega}\approx \ket{\psi_{c,n}}_{\Omega}\ket{\psi_{t,m}} \;.
\label{eqn:Saturation-SemiAnal Expression for a_n}
\end{align}
The dressed control qubit eigenstates are then obtained numerically by solving a 1D Schrodinger equation as
\begin{align}
(\HO_{q_c}+\HO_d)\ket{\psi_{c,n}}_{\Omega}=E_{c,n}(\Omega)\ket{\psi_{c,n}}_{\Omega} \;.
\label{eqn:Saturation-SemiAnal Def of Eig}
\end{align}

Figure~\ref{fig:Saturation-ZXSaturationSarahsParams} presents $ZX$ over a wide range of drive amplitude $\Omega$. Firstly, the results confirm our understanding from SWPT, where depending on the relation qubit-qubit detuning and control qubit anharmonicity completely distinct behavior is observed. In particular, separate bands are formed for each of the five parameter regions (See Sec.~\ref{SubSec:NeOrAnal}). Secondly, in all cases, the saturation limit of the $ZX$ rate is the same order as exchange interaction $J$. Importantly, curves belonging to the same band asymptote to the same limit. Thirdly, in terms of achieving the largest $ZX$ rate at a smaller drive amplitude $\Omega$, we find that parameter region III (shades of green in Fig.~\ref{fig:Saturation-ZXSaturationSarahsParams}) serves as the best operating point. In particular, as was found from SWPT, the ratio $\Delta_{ct}\approx -0.61 \alpha_c$ leads to the largest $ZX$ rate.


\section{Experimental signatures of the energy-basis corrections}
\label{App:ExpSign}

\begin{figure}[t!]
\centering
\includegraphics[scale=0.5]{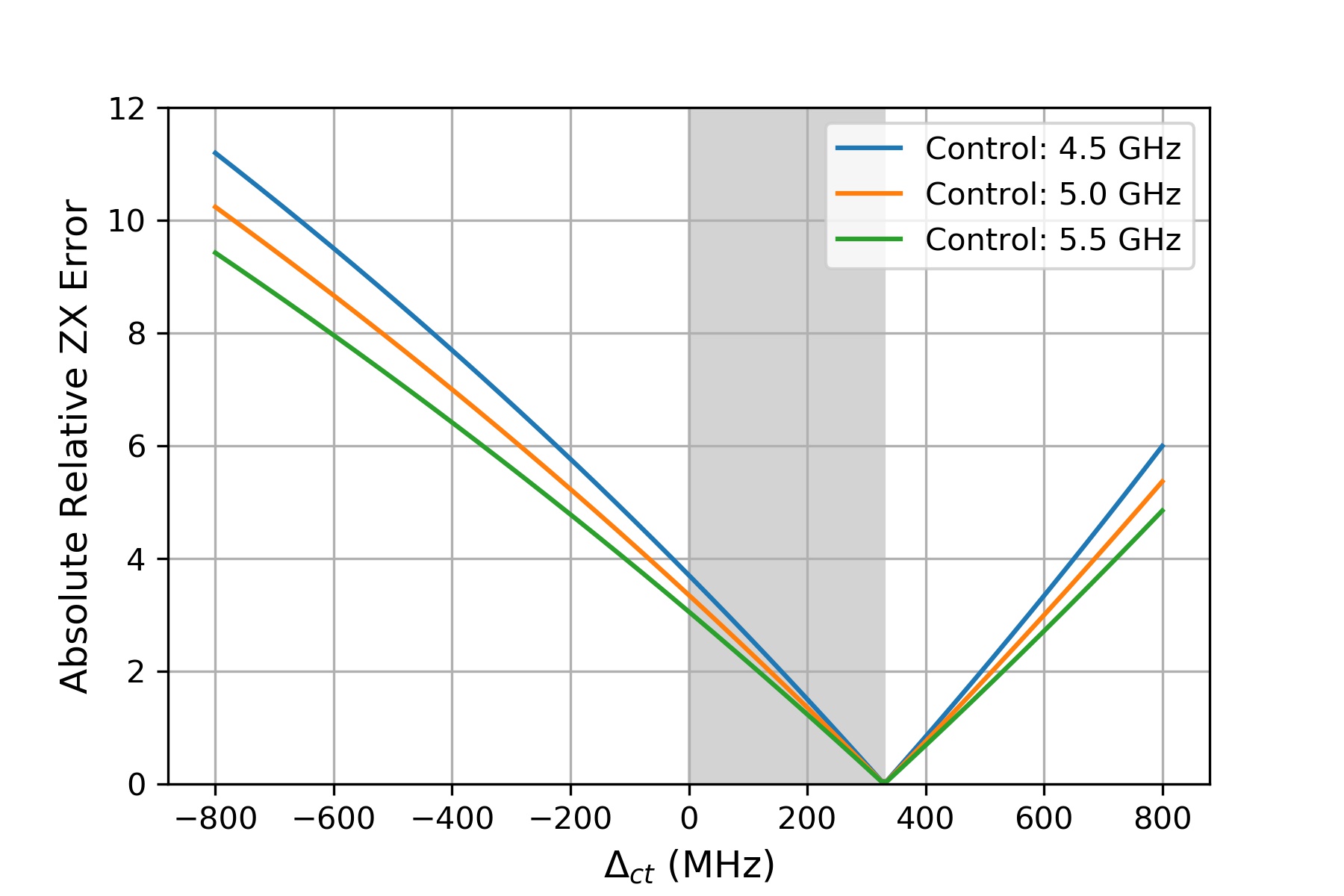}%
\caption{Plot of the absolute value of Eq.~(\ref{eqn:ExpSign-zx_err}) versus control-target detuning for different fixed values of the control qubit frequency (varying the target frequency). The anharmonicity is fixed to -330~MHz. The shaded region shows the typical operating regime for CR, where the difference in the theories does not lead to a large error in the observable measured quantities. However, outside this regime the error can be close to 10\%.} 
\label{fig:exp_relative_zx_error}
\end{figure}

The energy-basis corrections described in this manuscript lead up to 15\% relative correction in measured quantities such as $ZI$ (the Stark shift), $ZZ$, $ZX$ and $IX$ for a given $J$ and $\Omega$ (see Table~\ref{tab:LoOrAnal-CRGateParams}). However, experimentally, $J$ and $\Omega$ are inferred quantities and are not known a priori. Hence, we can only look at consistency between the measured quantities. Typically this set of consistent quantities includes $ZI$, $ZZ$ and $ZX$, while $IX$ is susceptible to classical crosstalk.  Therefore, we want to quantify the error in predicting one of the measured quantities if we adopt the Kerr vs the energy basis theory. To lowest order, $ZX=AJ\Omega$, $ZZ=BJ^2$, $ZI=C\Omega^2$, where $A,B,C$ are the pre-factors in Table~\ref{tab:LoOrAnal-CRGateParams} and we use coefficients $A,B,C$ for the energy basis and $\tilde{A},\tilde{B},\tilde{C}$ for the Kerr theory. Given the measurement for $ZZ$ and $ZI$, the predicted value of $ZX$ in terms of the measured quantities reads
\begin{align}
ZX = \frac{A \sqrt{ZZ \times ZI}}{\sqrt{BC}} \;.
\end{align}
Therefore, the relative \textit{experimental} error between the energy basis and Kerr can be defined as
\begin{align}
\frac{ZX - \tilde{ZX}}{ZX} = 1- \frac{\tilde{A} \sqrt{BC}}{A \sqrt{\tilde{B}\tilde{C}}} \;.
\label{eqn:ExpSign-zx_err}
\end{align}
Substituting in the values from Table~\ref{tab:LoOrAnal-CRGateParams} for the parameters, we plot the absolute relative error in Fig.~\ref{fig:exp_relative_zx_error}. We find that the error is less for higher frequency qubits as they are better approximated by the Kerr theory.

\bibliographystyle{unsrt}
\bibliography{CRGateEffHamiltonian}

\begin{thebibliography}{10}

\bibitem{Shor_Fault_1996}
Peter~W Shor.
\newblock ``fault-tolerant quantum computation''.
\newblock In {\em Proceedings of 37th conference on foundations of computer
  science}, pages 56--65. IEEE, 1996.

\bibitem{Gottesman_Theory_1998}
Daniel Gottesman.
\newblock ``theory of fault-tolerant quantum computation''.
\newblock {\em Physical Review A}, 57(1):127, 1998.

\bibitem{Kitaev_Fault_2003}
A~Yu Kitaev.
\newblock ``fault-tolerant quantum computation by anyons''.
\newblock {\em Annals of Physics}, 303(1):2--30, 2003.

\bibitem{Raussendorf_Fault_2007}
Robert Raussendorf and Jim Harrington.
\newblock ``fault-tolerant quantum computation with high threshold in two
  dimensions''.
\newblock {\em Physical review letters}, 98(19):190504, 2007.

\bibitem{Nielsen_Quantum_2002}
Michael~A Nielsen and Isaac Chuang.
\newblock {\em ``Quantum computation and quantum information''}.
\newblock American Association of Physics Teachers, 2002.

\bibitem{Nakamura_Coherent_1999}
Yu~Nakamura, Yu~A Pashkin, and JS~Tsai.
\newblock Coherent control of macroscopic quantum states in a
  single-cooper-pair box.
\newblock {\em Nature}, 398(6730):786--788, 1999.

\bibitem{Blais_Cavity_2004}
Alexandre Blais, Ren-Shou Huang, Andreas Wallraff, S.~M. Girvin, and R.~J.
  Schoelkopf.
\newblock Cavity quantum electrodynamics for superconducting electrical
  circuits: {An} architecture for quantum computation.
\newblock {\em Phys. Rev. A}, 69(6):062320, June 2004.

\bibitem{Wallraff_Strong_2004}
A.~Wallraff, D.~I. Schuster, A.~Blais, L.~Frunzio, R.-S. Huang, J.~Majer,
  S.~Kumar, S.~M. Girvin, and R.~J. Schoelkopf.
\newblock Strong coupling of a single photon to a superconducting qubit using
  circuit quantum electrodynamics.
\newblock {\em Nature}, 431(7005):162--167, September 2004.

\bibitem{Majer_Coupling_2007}
J~Majer, JM~Chow, JM~Gambetta, Jens Koch, BR~Johnson, JA~Schreier, L~Frunzio,
  DI~Schuster, AA~Houck, Andreas Wallraff, et~al.
\newblock ``coupling superconducting qubits via a cavity bus''.
\newblock {\em Nature}, 449(7161):443--447, 2007.

\bibitem{Koch_Charge_2007}
Jens Koch, Terri~M. Yu, Jay Gambetta, A.~A. Houck, D.~I. Schuster, J.~Majer,
  Alexandre Blais, M.~H. Devoret, S.~M. Girvin, and R.~J. Schoelkopf.
\newblock Charge-insensitive qubit design derived from the {Cooper} pair box.
\newblock {\em Phys. Rev. A}, 76(4):042319, October 2007.

\bibitem{Schreier_Suppressing_2008}
JA~Schreier, Andrew~A Houck, Jens Koch, David~I Schuster, BR~Johnson, JM~Chow,
  Jay~M Gambetta, J~Majer, L~Frunzio, Michel~H Devoret, et~al.
\newblock Suppressing charge noise decoherence in superconducting charge
  qubits.
\newblock {\em Physical Review B}, 77(18):180502, 2008.

\bibitem{McKay_Efficient_2017}
David~C. McKay, Christopher~J. Wood, Sarah Sheldon, Jerry~M. Chow, and Jay~M.
  Gambetta.
\newblock ``efficient $z$ gates for quantum computing''.
\newblock {\em Phys. Rev. A}, 96:022330, Aug 2017.

\bibitem{Dicarlo_Demonstration_2009}
Leonardo DiCarlo, Jerry~M Chow, Jay~M Gambetta, Lev~S Bishop, Blake~R Johnson,
  DI~Schuster, J~Majer, Alexandre Blais, Luigi Frunzio, SM~Girvin, et~al.
\newblock ``demonstration of two-qubit algorithms with a superconducting
  quantum processor''.
\newblock {\em Nature}, 460(7252):240--244, 2009.

\bibitem{Barends_Coherent_2013}
Rami Barends, Julian Kelly, Anthony Megrant, Daniel Sank, Evan Jeffrey,
  Yu~Chen, Yi~Yin, Ben Chiaro, Josh Mutus, Charles Neill, et~al.
\newblock ``coherent josephson qubit suitable for scalable quantum integrated
  circuits''.
\newblock {\em Physical review letters}, 111(8):080502, 2013.

\bibitem{Mckay_Universal_2016}
David~C McKay, Stefan Filipp, Antonio Mezzacapo, Easwar Magesan, Jerry~M Chow,
  and Jay~M Gambetta.
\newblock ``universal gate for fixed-frequency qubits via a tunable bus''.
\newblock {\em Physical Review Applied}, 6(6):064007, 2016.

\bibitem{Caldwell_Parametrically_2018}
S.~A. Caldwell, N.~Didier, C.~A. Ryan, E.~A. Sete, A.~Hudson, P.~Karalekas,
  R.~Manenti, M.~P. da~Silva, R.~Sinclair, E.~Acala, N.~Alidoust, J.~Angeles,
  A.~Bestwick, M.~Block, B.~Bloom, A.~Bradley, C.~Bui, L.~Capelluto,
  R.~Chilcott, J.~Cordova, G.~Crossman, M.~Curtis, S.~Deshpande, T.~El
  Bouayadi, D.~Girshovich, S.~Hong, K.~Kuang, M.~Lenihan, T.~Manning,
  A.~Marchenkov, J.~Marshall, R.~Maydra, Y.~Mohan, W.~O'Brien, C.~Osborn,
  J.~Otterbach, A.~Papageorge, J.-P. Paquette, M.~Pelstring, A.~Polloreno,
  G.~Prawiroatmodjo, V.~Rawat, M.~Reagor, R.~Renzas, N.~Rubin, D.~Russell,
  M.~Rust, D.~Scarabelli, M.~Scheer, M.~Selvanayagam, R.~Smith, A.~Staley,
  M.~Suska, N.~Tezak, D.~C. Thompson, T.-W. To, M.~Vahidpour, N.~Vodrahalli,
  T.~Whyland, K.~Yadav, W.~Zeng, and C.~Rigetti.
\newblock ``parametrically activated entangling gates using transmon qubits''.
\newblock {\em Phys. Rev. Applied}, 10:034050, Sep 2018.

\bibitem{Chow_Universal_2012}
Jerry~M Chow, Jay~M Gambetta, AD~C{\'o}rcoles, Seth~T Merkel, John~A Smolin,
  Chad Rigetti, S~Poletto, George~A Keefe, Mary~B Rothwell, JR~Rozen, et~al.
\newblock ``universal quantum gate set approaching fault-tolerant thresholds
  with superconducting qubits''.
\newblock {\em Physical review letters}, 109(6):060501, 2012.

\bibitem{Corcoles_Demonstration_2015}
Antonio~D C{\'o}rcoles, Easwar Magesan, Srikanth~J Srinivasan, Andrew~W Cross,
  Matthias Steffen, Jay~M Gambetta, and Jerry~M Chow.
\newblock ``demonstration of a quantum error detection code using a square
  lattice of four superconducting qubits''.
\newblock {\em Nature communications}, 6(1):1--10, 2015.

\bibitem{Takita_Demonstration_2016}
Maika Takita, Antonio~D C{\'o}rcoles, Easwar Magesan, Baleegh Abdo, Markus
  Brink, Andrew Cross, Jerry~M Chow, and Jay~M Gambetta.
\newblock ``demonstration of weight-four parity measurements in the surface
  code architecture''.
\newblock {\em Physical review letters}, 117(21):210505, 2016.

\bibitem{Gambetta_Building_2017}
Jay~M Gambetta, Jerry~M Chow, and Matthias Steffen.
\newblock ``building logical qubits in a superconducting quantum computing
  system''.
\newblock {\em npj Quantum Information}, 3(1):1--7, 2017.

\bibitem{Sheldon_Procedure_2016}
Sarah Sheldon, Easwar Magesan, Jerry~M Chow, and Jay~M Gambetta.
\newblock ``procedure for systematically tuning up cross-talk in the
  cross-resonance gate''.
\newblock {\em Physical Review A}, 93(6):060302, 2016.

\bibitem{Paraoanu_Microwave_2006}
GS~Paraoanu.
\newblock ``microwave-induced coupling of superconducting qubits'.
\newblock {\em Physical Review B}, 74(14):140504, 2006.

\bibitem{Rigetti_Fully_2010}
Chad Rigetti and Michel Devoret.
\newblock ``fully microwave-tunable universal gates in superconducting qubits
  with linear couplings and fixed transition frequencies''.
\newblock {\em Physical Review B}, 81(13):134507, 2010.

\bibitem{Magesan_Effective_2018}
Easwar Magesan and Jay~M Gambetta.
\newblock ``effective hamiltonian models of the cross-resonance gate''.
\newblock {\em arXiv preprint arXiv:1804.04073}, 2018.

\bibitem{Tripathi_Operation_2019}
Vinay Tripathi, Mostafa Khezri, and Alexander~N Korotkov.
\newblock ``operation and intrinsic error budget of two-qubit cross-resonance
  gate''.
\newblock {\em arXiv preprint arXiv:1902.09054}, 2019.

\bibitem{Chow_Simple_2011}
Jerry~M Chow, AD~C{\'o}rcoles, Jay~M Gambetta, Chad Rigetti, BR~Johnson, John~A
  Smolin, JR~Rozen, George~A Keefe, Mary~B Rothwell, Mark~B Ketchen, et~al.
\newblock ``simple all-microwave entangling gate for fixed-frequency
  superconducting qubits''.
\newblock {\em Physical review letters}, 107(8):080502, 2011.

\bibitem{Corcoles_Process_2013}
Antonio~D C{\'o}rcoles, Jay~M Gambetta, Jerry~M Chow, John~A Smolin, Matthew
  Ware, Joel Strand, Britton~LT Plourde, and Matthias Steffen.
\newblock ``process verification of two-qubit quantum gates by randomized
  benchmarking''.
\newblock {\em Physical Review A}, 87(3):030301, 2013.

\bibitem{Kirchhoff_Optimized_2018}
Susanna Kirchhoff, Torsten Ke{\ss}ler, Per~J Liebermann, Elie Ass{\'e}mat, Shai
  Machnes, Felix Motzoi, and Frank~K Wilhelm.
\newblock ``optimized cross-resonance gate for coupled transmon systems''.
\newblock {\em Physical Review A}, 97(4):042348, 2018.

\bibitem{Schrieffer_Relation_1966}
JR~Schrieffer and PA~Wolff.
\newblock ``relation between the anderson and kondo hamiltonians''.
\newblock {\em Physical Review}, 149(2):491, 1966.

\bibitem{Bravyi_Schrieffer_2011}
Sergey Bravyi, David~P DiVincenzo, and Daniel Loss.
\newblock ``schrieffer--wolff transformation for quantum many-body systems''.
\newblock {\em Annals of physics}, 326(10):2793--2826, 2011.

\bibitem{Boissonneault_Dispersive_2009}
Maxime Boissonneault, J.~M. Gambetta, and Alexandre Blais.
\newblock Dispersive regime of circuit qed: Photon-dependent qubit dephasing
  and relaxation rates.
\newblock {\em Phys. Rev. A}, 79:013819, Jan 2009.

\bibitem{Malekakhlagh_Lifetime_2020}
Moein Malekakhlagh, Alexandru Petrescu, and Hakan~E. T\"ureci.
\newblock ``lifetime renormalization of weakly anharmonic superconducting
  qubits. \text{I}. role of number nonconserving terms''.
\newblock {\em Phys. Rev. B}, 101:134509, Apr 2020.

\bibitem{Petrescu_Lifetime_2020}
Alexandru Petrescu, Moein Malekakhlagh, and Hakan~E. T\"ureci.
\newblock ``lifetime renormalization of driven weakly anharmonic
  superconducting qubits. \text{II}. the readout problem''.
\newblock {\em Phys. Rev. B}, 101:134510, Apr 2020.

\bibitem{Makhlin_Nonlocal_2002}
Yuriy Makhlin.
\newblock ``nonlocal properties of two-qubit gates and mixed states, and the
  optimization of quantum computations''.
\newblock {\em Quantum Information Processing}, 1(4):243--252, 2002.

\bibitem{Malekakhlagh_Cutoff-Free_2017}
Moein Malekakhlagh, Alexandru Petrescu, and Hakan~E. T\"ureci.
\newblock ``cutoff-free circuit quantum electrodynamics''.
\newblock {\em Phys. Rev. Lett.}, 119:073601, Aug 2017.

\bibitem{Didier_Analytical_2018}
Nicolas Didier, Eyob~A Sete, Marcus~P da~Silva, and Chad Rigetti.
\newblock ``analytical modeling of parametrically modulated transmon qubits''.
\newblock {\em Physical Review A}, 97(2):022330, 2018.

\bibitem{Zhang_Geometric_2003}
Jun Zhang, Jiri Vala, Shankar Sastry, and K~Birgitta Whaley.
\newblock ``geometric theory of nonlocal two-qubit operations''.
\newblock {\em Physical Review A}, 67(4):042313, 2003.

\bibitem{Watts_Optimizing_2015}
Paul Watts, Ji{\v{r}}{\'\i} Vala, Matthias~M M{\"u}ller, Tommaso Calarco,
  K~Birgitta Whaley, Daniel~M Reich, Michael~H Goerz, and Christiane~P Koch.
\newblock ``optimizing for an arbitrary perfect entangler. i. functionals''.
\newblock {\em Physical Review A}, 91(6):062306, 2015.

\bibitem{Zanardi_Entangling_2000}
Paolo Zanardi, Christof Zalka, and Lara Faoro.
\newblock ``entangling power of quantum evolutions''.
\newblock {\em Physical Review A}, 62(3):030301, 2000.

\bibitem{Pedersen_Fidelity_2007}
Line~Hjortsh{\o}j Pedersen, Niels~Martin M{\o}ller, and Klaus M{\o}lmer.
\newblock ``fidelity of quantum operations''.
\newblock {\em Physics Letters A}, 367(1-2):47--51, 2007.

\bibitem{Zhang_Minimum_2004}
Jun Zhang, Jiri Vala, Shankar Sastry, and K~Birgitta Whaley.
\newblock ``minimum construction of two-qubit quantum operations''.
\newblock {\em Physical review letters}, 93(2):020502, 2004.

\bibitem{Rezakhani_Characterization_2004}
AT~Rezakhani.
\newblock ``characterization of two-qubit perfect entanglers''.
\newblock {\em Physical Review A}, 70(5):052313, 2004.

\end{thebibliography}
\end{document}